\documentclass[letter, 10pt]{article}

\usepackage{graphicx,amsmath,amsfonts,amscd,amssymb,amsthm,bm,bbm,latexsym,mathdots,mathtools,fge}
\usepackage{algorithm,algorithmicx,algpseudocode}
\usepackage[T1]{fontenc}
\usepackage{tgpagella}
\usepackage{fullpage}
\usepackage[small,bf]{caption}
\usepackage{microtype}
\usepackage{hyperref}
\usepackage{ulem}
\usepackage[toc, page]{appendix}
\usepackage{scalerel,stackengine} % For wide check
\usepackage[nameinlink]{cleveref}
\usepackage{comment}
\usepackage{wrapfig}
\usepackage{tikz}
\usepackage{pgfplots}

\numberwithin{equation}{section}
\allowdisplaybreaks
\hypersetup{
    colorlinks=true,%
    citecolor=blue,%
    filecolor=blue,%
    linkcolor=blue,%
    urlcolor=blue
}

\newcommand{\vsni}{\vspace{.1in} \noindent}

\newtheorem{theorem}{Theorem}[section]
\newtheorem{lemma}[theorem]{Lemma}
\newtheorem{corollary}[theorem]{Corollary}

\newtheorem{definition}[theorem]{Definition}

\newtheorem{fact}[theorem]{Fact}

\newtheorem{problem}[theorem]{Problem}

\newtheorem{remark}[theorem]{Remark}

\renewenvironment{proof}{\noindent {\bf Proof.} }{\endprf\par}
\def \endprf{\hfill {\vrule height6pt width6pt depth0pt}\medskip}

\renewcommand{\mathbf}{\boldsymbol}
\newcommand{\mb}{\mathbf}
\newcommand{\mr}{\mathrm}
\newcommand{\mc}{\mathcal}

\newcommand{\bb}{\mathbb}

\newcommand{\R}{\bb R}

\newcommand{\N}{\bb N}
\newcommand{\Sp}{\bb S}

\newcommand{\set}[1]{\left\{ #1 \right\}}

\newcommand{\Brac}[1]{\left\lbrace #1 \right\rbrace}
\newcommand{\brac}[1]{\left[ #1 \right]}
\newcommand{\paren}[1]{ \left( #1 \right) }

\newcommand{\floor}[1]{\left\lfloor #1 \right\rfloor}

\newcommand{\wh}{\widehat}
\newcommand{\wt}{\widetilde}
\newcommand{\ol}{\overline}
\newcommand{\ul}{\underline}

\stackMath
\newcommand\widecheck[1]{%
\savestack{\tmpbox}{\stretchto{%
  \scaleto{%
    \scalerel*[\widthof{\ensuremath{#1}}]{\kern-.6pt\bigwedge\kern-.6pt}%
    {\rule[-\textheight/2]{1ex}{\textheight}}%WIDTH-LIMITED BIG WEDGE
  }{\textheight}%
}{0.5ex}}%
\stackon[1pt]{#1}{\scalebox{-1}{\tmpbox}}%
}
\newcommand{\wc}{\widecheck}

\DeclareMathOperator{\prox}{prox}

\DeclareMathOperator{\trace}{tr}
\DeclareMathOperator{\supp}{supp}

\DeclareMathOperator{\diag}{diag}

\DeclareMathOperator{\poly}{poly}
\DeclareMathOperator{\sign}{sign}
\DeclareMathOperator{\grad}{grad}

\DeclareMathOperator*{\argmin}{argmin}

\newcommand{\norm}[2]{\left\| #1 \right\|_{#2}}
\newcommand{\abs}[1]{\left| #1 \right|}
\newcommand{\innerprod}[2]{\left\langle #1,  #2 \right\rangle}
\newcommand{\prob}[1]{\bb P\left[ #1 \right]}
\newcommand{\expect}[1]{\bb E\left[ #1 \right]}
\newcommand{\E}{\bb E}

\newcommand{\soft}[2]{\mathcal S_{#2}\left[#1 \right]}

\newcommand{\convmtx}[1]{\mb C_{#1}}
\newcommand{\checkmtx}[1]{\widecheck{\mb C}_{#1}}

\newcommand{\shift}[2]{s_{#2}[#1]}
\newcommand{\injector}{\mb \iota}
\newcommand{\ip}{\injector}

\newcommand{\eps}{\varepsilon}
\newcommand{\event}{\mc E}
\newcommand{\simiid}{\sim_{\mr{i.i.d.}}}

\newcommand{\1}{\mathbf 1}

\newcommand{\cross}{\times}
\newcommand{\caseof}[1]{\left\{ \begin{array}{ll} #1 \end{array} \right.}

\newcommand{\goodregion}{\mathfrak R(\mc S_{\mb\tau},\gamma(c_\mu))}
\newcommand{\clog}{\theta_{\mr{log}}}
\newcommand{\epsnet}{\mc N_\eps}
\newcommand{\eventlip}{\event_{\mr{Lip}}}
\newcommand{\eventnet}{\event_{\mr{Net}}}
\newcommand{\mut}{\wt{\mu}}
\newcommand{\lambdat}{\wt{\lambda}}

\begin{document}
\title{Geometry and Symmetry in Short-and-Sparse Deconvolution}
\author{Han-Wen Kuo$^{1,2}$, Yuqian Zhang$^{3}$,  Yenson Lau$^{1,2}$, John Wright$^{1,2,4}$ \\ \\ $^1$Department of Electrical Engineering, Columbia University \\
 $^2$Data Science Institute, Columbia University \\
 $^3$Department of Computer Science, Cornell University \\
 $^4$Department of Applied Physics and Applied Mathematics, Columbia University }
\date{January 3, 2019 \hspace{.15in} Revised April 10, 2019  }
\maketitle 
  
\begin{abstract} We study the \emph{Short-and-Sparse (SaS) deconvolution} problem of recovering a short signal $\mb a_0$ and a sparse signal $\mb x_0$ from their convolution. We propose a method based on nonconvex optimization, which under certain conditions recovers the target short and sparse signals, up to a signed shift symmetry which is intrinsic to this model. This symmetry plays a central role in shaping the optimization landscape for deconvolution. We give a {\em regional analysis}, which characterizes this landscape geometrically, on a union of subspaces. Our geometric characterization holds when the length-$p_0$ short signal $\mb a_0$ has shift coherence $\mu$, and $\mb x_0$ follows a random sparsity model with sparsity rate $\theta \in \Bigl[\frac{c_1}{p_0}, \frac{c_2}{p_0\sqrt\mu + \sqrt{p_0}}\Bigr]\cdot\frac{1}{\log^2p_0}$. Based on this geometry, we give a provable method that successfully solves SaS deconvolution with high probability.
\end{abstract}

%-Sections
% !TEX root = ../BD_DQ.tex

\section{Introduction}

Datasets in a wide range of areas, including neuroscience \cite{lewicki1998review}, microscopy \cite{Cheung17-Nature} and astronomy \cite{saha2007diffraction}, can be modeled as superpositions of translations of a basic motif. Data of this nature can be modeled mathematically as a convolution $\mb y = \mb a_0 \ast \mb x_0$, between a {\em short} signal $\mb a_0$ (the motif) and a longer {\em sparse} signal $\mb x_0$, whose nonzero entries indicate where in the sample the motif is present. A very similar structure arises in image deblurring \cite{chan1998total}, where $\mb y$ is a blurry image, $\mb a_0$ the blur kernel, and $\mb x_0$ the (edge map) of the target sharp image. 

\begin{figure}[t!]
	\centering
	\includegraphics[trim = {6cm 6cm 6cm 6cm}, width = 0.45\textwidth]{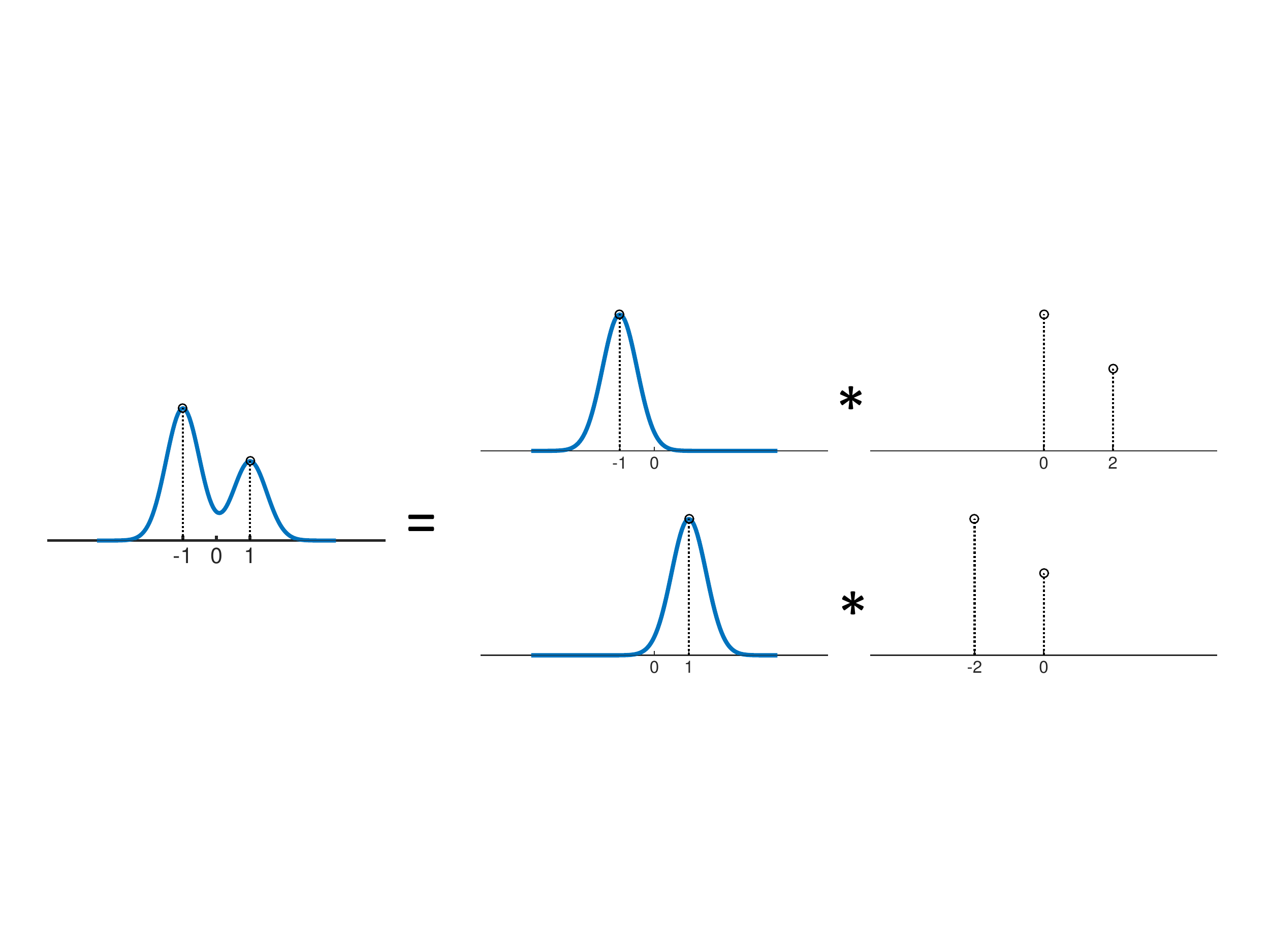}
	\caption{\textbf{Shift symmetry in Short-and-Sparse deconvolution.} An observation $\mb y$ (left) which is a convolution of a short signal $\mb a_0$ and a sparse signal $\mb x_0$ (top right) can be equivalently expressed as a convolution of $\shift{\mb a_0}{\ell}$ and $\shift{\mb x_0}{-\ell}$, where $\shift{\cdot}{\ell}$ denotes a shift $\ell$ samples. The ground truth signals $\mb a_0$ and $\mb x_0$ can only be identified up to a scaled shift. 
	} \label{fig:symmetry}
\end{figure}

Motivated by these and related problems in imaging and scientific data analysis, we study the 
{\em Short-and-Sparse (SaS) Deconvolution} problem of recovering a short signal $\mb a_0 \in \R^{p_0}$ and a sparse signal $\mb x_0 \in \R^n$ ($n \gg p_0$) from their length-$n$ cyclic convolution $\mb y = \mb a_0 \ast \mb x_0 \in \R^n$. This SaS model exhibits a basic {\em scaled shift symmetry}: for any nonzero scalar $\alpha$ and cyclic shift $s_\ell[\cdot]$, 
\begin{equation} \label{eqn:sss}
 \Bigl(\alpha \, \shift{\mb a_0}{\ell}\Bigr) \; \ast \; \Bigl(\tfrac{1}{\alpha} \, \shift{\mb x_0}{-\ell} \Bigr) \;=\; \mb y.
\end{equation}
Because of this symmetry, we only expect to recover $\mb a_0$ and $\mb x_0$ up to a signed shift (see  \Cref{fig:symmetry}). Our problem of interest can be stated more formally as:

\vspace{0.01in}
 
\begin{problem}[Short-and-Sparse Deconvolution] 
Given the cyclic convolution $\mb y = \mb a_0 \ast \mb x_0 \in \R^n$ of $\mb a_0 \in \bb R^{p_0}$ short ($p_0 \ll n$), and $\mb x_0 \in \R^n$  sparse, recover $\mb a_0$ and $\mb x_0$, up to a scaled shift.
\end{problem}

\par\vspace{0.05in} 

Despite a long history and many applications, until recently very little algorithmic theory was available for SaS deconvolution. Much of this difficulty can be attributed to the scale-shift symmetry: natural convex relaxations fail, and nonconvex formulations exhibit a complicated optimization landscape, with many equivalent global minimizers (scaled shifts of the ground truth) and additional local minimizers (scaled shift truncations of the ground truth), and a variety of critical points \cite{zhang2017global,zhang2018structured}. Currently available theory guarantees approximate recovery of a truncation\footnote{I.e., the portion of the shifted signal $s_\ell[\mb a_0]$ that falls in the window $\set{0,\ldots,p_0-1}$.}  of a shift $s_\ell[\mb a_0]$, rather than guaranteeing recovery of $\mb a_0$ as a whole, and requires certain (complicated) conditions on the convolution matrix associated with $\mb a_0$ \cite{zhang2018structured}. 

In this paper, describe an algorithm which, under simpler conditions, {\em exactly} recovers a scaled shift of the pair $(\mb a_0,\mb x_0)$. Our algorithm is based on a formulation first introduced in \cite{zhang2017global}, which casts the deconvolution problem as (nonconvex) optimization over the sphere. We characterize the geometry of this objective function, and show that near a certain union of subspaces, every local minimizer is very close to a signed shift of $\mb a_0$. Based on this geometric analysis, we give provable methods for SaS deconvolution that exactly recover a scaled shift of $(\mb a_0,\mb x_0)$ whenever $\mb a_0$ is {\em shift-incoherent} and $\mb x_0$ is a sufficiently sparse random vector.  Our geometric analysis highlights the role of symmetry in shaping the objective landscape for SaS deconvolution. 

\paragraph{Organization of this paper.} The remainder of this paper is organized as follows. \Cref{sec:formulation_assumption} introduces our optimization approach and modeling assumptions. \Cref{sec:main-results} introduces our main results --- both geometric and algorithmic --- and compares them to the literature. \Cref{sec:geometry}-\ref{sec:alg} describes the main ideas of our analysis. Finally, \Cref{sec:discussion} discusses two main limitations of our analysis and describes directions for future work.

% !TEX root = ../BD_DQ.tex
\section{Formulation and Assumptions}\label{sec:formulation_assumption}

\subsection{Nonconvex SaS over the Sphere} \label{sec:nonconvex-for-short-and-sparse}  

\paragraph{Bilinear Lasso.} Our starting point is the (natural) formulation 
\begin{equation}\label{eqn:bilinear-lasso}
	\min_{\mb a, \mb x}  \;  \underset{\color{purple} \text{\bf Data Fidelity}}{\tfrac{1}{2}\norm{\mb a \ast \mb x - \mb y}2^2} + \underset{\color{purple} \text{\bf Sparsity}}{\lambda \norm{ \mb x }{1}} \quad \text{s.t.} \quad \norm{\mb a}2=1.
\end{equation}
We term this optimization problem the {\em Bilinear Lasso}, for its resemblance to the Lasso estimator in statistics. Indeed, letting 
\begin{equation} \label{eqn:lasso-cost}
\varphi_{\mr{lasso}}( \mb a ) \equiv \min_{\mb x} \set{ \tfrac{1}{2}\norm{\mb a \ast \mb x - \mb y}2^2 + \lambda \norm{ \mb x }{1} }
\end{equation} 
denote the optimal Lasso cost, we see that \eqref{eqn:bilinear-lasso} simply optimizes $\varphi_{\mr{lasso}}$ with respect to $\mb a$:
\begin{equation} \label{eqn:bilinear-lasso-sphere} 
\min_{\mb a} \; \varphi_{\mr{lasso}}( \mb a ) \quad \text{s.t.} \quad \norm{\mb a}2=1.
\end{equation} 
In \eqref{eqn:bilinear-lasso}-\eqref{eqn:bilinear-lasso-sphere}, we constrain $\mb a$ to have unit $\ell^2$ norm. This constraint breaks the scale ambiguity between $\mb a$ and $\mb x$. Moreover, the choice of constraint manifold has surprisingly strong implications for computation: if $\mb a$ is instead constrained to the simplex, the problem admits trivial global minimizers.
In contrast, local minima of the sphere-constrained formulation often correspond to shifts (or shift truncations \cite{zhang2017global}) of the ground truth $\mb a_0$.  
 
\paragraph{Simplifications and approximations.} The problem \eqref{eqn:bilinear-lasso-sphere} is defined in terms of the optimal Lasso cost. This function is challenging to analyze, especially far away from $\mb a_0$. \cite{zhang2017global} analyzes the local minima of a simplification of \eqref{eqn:bilinear-lasso-sphere}, obtained by approximating\footnote{For a generic $\mb a$, we have $\innerprod{ \shift{\mb a}{i} }{ \shift{\mb a}{j} } \approx 0$ and hence $\norm{\mb a \ast \mb x}{2}^2=\mb x^*\mb C_{\mb a}^*\mb C_{\mb a}\mb x\approx\mb x^*\mb I\mb x=\norm{\mb x}{2}^2$.} the data fidelity term as 
\begin{eqnarray}
\tfrac{1}{2} \norm{ \mb a \ast \mb x - \mb y}{2}^2 &=& \tfrac{1}{2} \norm{\mb a \ast \mb x}{2}^2 - \innerprod{ \mb a \ast \mb x }{\mb y } + \tfrac{1}{2} \norm{\mb y}{2}^2, \nonumber \\
&\approx& \tfrac{1}{2} \norm{\mb x}{2}^2 - \innerprod{ \mb a \ast \mb x }{\mb y } + \tfrac{1}{2} \norm{\mb y}{2}^2. \label{eqn:drop-quadratic-appx}
\end{eqnarray} 
This yields a simpler objective function 
\begin{equation}\label{eqn:dq-l1}
\varphi_{\ell^1}(\mb a) = \min_{\mb x}\set{ \tfrac{1}{2}\norm{\mb x}2^2 - \innerprod{\mb a*\mb x}{\mb y} + \tfrac{1}{2}\norm{\mb y}2^2 + \lambda\norm{\mb x}{1} }.
\end{equation}
We make one further simplification to this problem, replacing the nondifferentiable penalty $\norm{\cdot}{1}$ with a smooth approximation $\rho(\mb x)$.\footnote{The objective $\varphi_{\ell^1}$ is not twice differentiable everywhere, and hence cannot be minimized using conventional second order methods. } Our analysis allows for a variety of smooth sparsity surrogates $\rho(\mb x)$; for concreteness, we state our main results for the particular penalty\footnote{This particular surrogate is sometimes being  named as the pseudo-Huber function.}  
\begin{equation}
\rho(\mb x) = \textstyle\sum_{i} \left( \mb x_i^2 + \delta^2 \right)^{1/2}.
\end{equation}
For $\delta > 0$, this is a smooth function of $\mb x$; as $\delta \searrow 0$ it approaches $\norm{\mb x}{1}$. Replacing $\norm{\cdot}{1}$ with $\rho(\cdot)$, we obtain the objective function which will be our main object of study, 
\begin{equation} \label{eqn:lasso-dq}
\varphi_\rho(\mb a) = \min_{\mb x}\set{ \tfrac{1}{2}\norm{\mb x}2^2 - \innerprod{\mb a*\mb x}{\mb y} + \tfrac{1}{2}\norm{\mb y}2^2 + \lambda\rho(\mb x) }.
\end{equation}

\paragraph{Core optimization problem.} As in \cite{zhang2017global}, we optimize $\varphi_\rho(\mb a)$ over the sphere $\Sp^{p-1}$:
\vspace{0.05in}
\begin{equation} \label{eqn:pnc-dq}
\boxed{\boxed{
\quad \min_{\mb a} \; \varphi_\rho(\mb a ) \quad \text{s.t.} \quad \mb a \in \bb S^{p-1}. \quad  }}  
\end{equation} 
\vspace{0.05in}
Here, we set $p = 3p_0 -2$. As we will see, optimizing over this slightly higher dimensional sphere enables us to recover a (full) shift of $\mb a_0$, rather than a {\em truncated} shift. Our approach will leverage the following fact: if we view $\mb a \in \bb S^{p-1}$ as indexed by coordinates $W = \set{-p_0+1, \, \dots, \, 2p_0 - 1}$ , then for any shifts $\ell \in \set{ - p_0 +1, \dots, p_0-1}$, the support of $\ell$-shifted short signal $s_\ell[ \mb a_0 ]$ is entirely contained in interval $W$. We will give a provable method which recovers a scaled version of one of these canonical shifts.  
   
\subsection{Analysis Setting and Assumptions}

For convenience, we assume that $\mb a_0$ has unit $\ell^2$ norm, i.e., $\mb a_0 \in \bb S^{p_0-1}.$\footnote{This is purely a technical convenience. Our theory guarantees recovery of a signed shift $(\pm s_{\ell}[\mb a_0], \pm s_{-\ell}[\mb x_0])$ of the truth. If $\mb a_0$ does not have unit norm, identical reasoning implies that our method recovers a scaled shift $\left( \alpha s_\ell[\mb a_0], \alpha^{-1} s_{-\ell}[\mb x_0] \right)$ with $\alpha = \pm \frac{1}{\norm{\mb a_0}{2}}$.} Our analysis makes two main assumptions, on the short motif $\mb a_0$ and the sparse map $\mb x_0 $, respectively:

\paragraph{Shift incoherence of $\mb a_0$.} The first is that distinct shifts $\mb a_0$ have small inner product. We define the {\em shift coherence} of $\mu(\mb a_0)$ to be the largest inner product between distinct shifts:
\begin{equation}
\mu(\mb a_0) = \max_{\ell \ne 0} \left| \innerprod{ \mb a_0 }{ \shift{\mb a_0}{\ell} } \right|
\end{equation}

\begin{figure}[t!]	
	\centering 
	\input{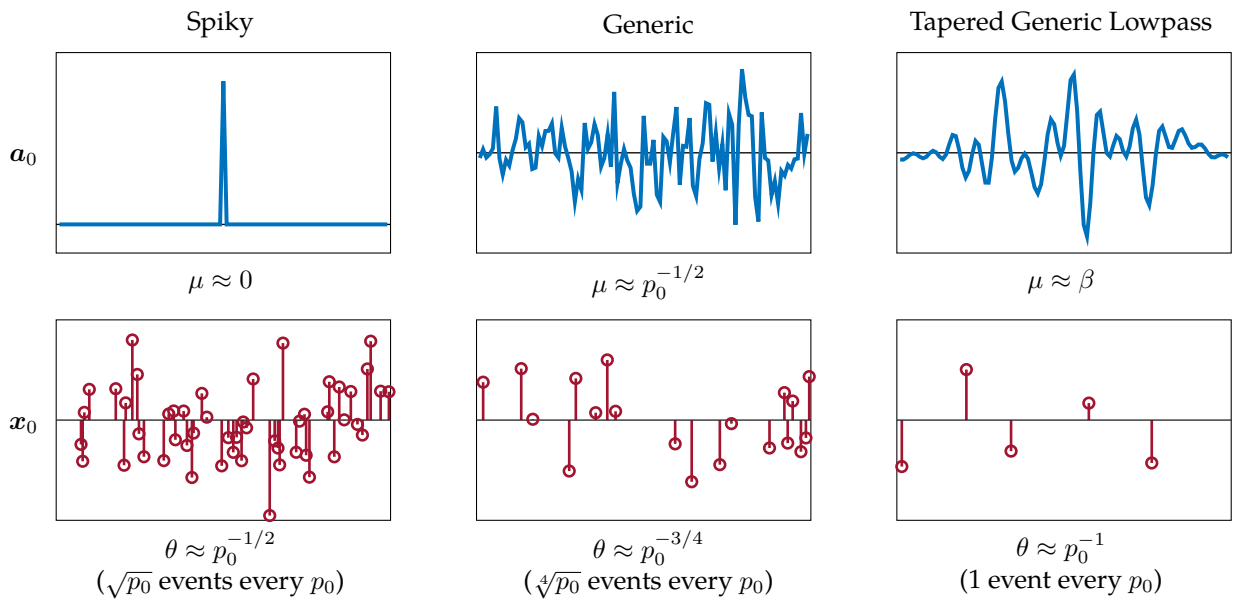}
	\par\vspace{.1in}
	\caption{ \textbf{Sparsity-coherence tradeoff:} Top: three families of motifs $\mb a_0$ with varying coherence $\mu$. Bottom: maximum allowable sparsity $\theta$ and number of copies $\theta p_0$ within each length-$p_0$ window. Here, we suppress constants and logarithmic factors. When the target motif has smaller shift-coherence $\mu$, our result  allows larger $\theta$, and vise versa. This sparsity-coherence tradeoff is made precise in our main result \Cref{thm:main}, which, loosely speaking, asserts that when $\theta\lessapprox 1/(p_0\sqrt\mu + \sqrt{p_0}) $, our method succeeds. } 
	\label{fig:sparsity-coherence}   
\end{figure} 

The quantity $\mu(\mb a_0)$ is bounded between $0$  and $1$. Our theory allows any $\mu$ smaller than some numerical constant.  \Cref{fig:sparsity-coherence} shows three examples of families of $\mb a_0$ that satisfy this assumption: 
\begin{itemize}
	\item \emph{Spiky.}  When $\mb a_0$ is close to the Dirac delta $\mb\delta_0$, the shift coherence $\mu(\mb a_0)\approx 0$.\footnote{The use of ``$\approx$'' here suppresses constant and logarithmic factors.} Here, the observed signal $\mb y$ consists of a superposition of sharp pulses. This is arguably the easiest instance of SaS deconvolution.
	\item \emph{Generic.} If $\mb a_0$ is chosen uniformly at random from the sphere $\Sp^{p_0-1}$, its coherence is bounded as $\mu(\mb a_0) \lessapprox  \sqrt{ 1/p_0 } $ with high probability. 
	\item \emph{Tapered Generic Lowpass.} Here, $\mb a_0$ is generated by taking a random conjugate symmetric superposition of the first $L$ length-$p_0$ Discrete Fourier Transform (DFT) basis signals, windowing (e.g., with a Hamming window) and normalizing to unit $\ell^2$ norm. When $L = p_0 \sqrt{1 -\beta}$, with high probability $\mu(\mb a_0) \lessapprox \beta$. In this model, $\mu$ does not have to diminish as $p_0$ grows -- it can be a fixed constant.\footnote{The upper right panel of \Cref{fig:sparsity-coherence} is generated using random DFT components with frequencies smaller then one-third Nyquist. Such a kernel is incoherent, with high probability. Many commonly occurring low-pass kernels have $\mu(\mb a_0)$ larger -- very close to one. One of the most important limitations of our results is that they do not provide guarantees in this highly coherent situation.}
\end{itemize}   
Intuitively speaking, problems with smaller $\mu$ are easier to solve, a claim which will be made precise in our technical results.

\paragraph{Random sparsity model on $\mb x_0$.} We assume that $\mb x_0$ is a sparse random vector. More precisely, we assume that $\mb x_0$ is Bernoulli-Gaussian, with rate $\theta$:
\begin{equation} 
\mb x_{0i} = \mb \omega_i \mb g_i, 
\end{equation} 
where $\mb \omega_i \sim \mr{Ber}(\theta)$, $\mb g_i \sim \mc N(0,1)$ and all random variables are jointly independent. We write this as 
\begin{equation}
\mb x_0 \simiid \mr{BG}(\theta).
\end{equation}
Here, $\theta$ is the probability that a given entry $\mb x_{0i}$ is nonzero. 
Problems with smaller $\theta$ are easier to solve. In the extreme case, when $\theta \ll 1/p_0$, the observation $\mb y$ contains many isolated copies of the motif $\mb a_0$, and $\mb a_0$ can be determined by direct inspection. Our analysis will focus on the nontrivial scenario, when $\theta \gtrapprox 1/p_0$.

\paragraph{Sparsity-Coherence tradeoffs.} Our technical results will articulate {\em sparsity-coherence} tradeoffs, in which smaller coherence $\mu$ enables larger $\theta$, and vice-versa.  More specifically, in our main theorem, the sparsity-coherence relationship is captured in the form 
\begin{align}
	\theta \;\lessapprox\; 1/(p_0\sqrt\mu +\sqrt{p_0}).
\end{align} 
When the target $\mb a_0$ is highly shift-incoherent ($\mu \approx 0$), our method succeeds when each length-$p_0$ window contains about $\sqrt{p_0}$ copies of $\mb a_0$. When $\mu$ is larger (as in the generic lowpass model), our method succeeds as long as relatively few copies of $\mb a_0$ overlap in the observed signal. In \Cref{fig:sparsity-coherence}, we illustrate these tradeoffs for the three models described above.
 
% !TEX root = ../BD_DQ.tex

\section{Main Results: Geometry and Algorithms} \label{sec:main-results}

In this section, we introduce our main results -- on the geometry of $\varphi_\rho$ (\Cref{sec:main-results-geometry}) and its algorithmic implications (\Cref{sec:main-results-algorithms}). Finally, in \Cref{sec:main-results-comparison}, we compare these results with the literature on deconvolution. 
%In \Cref{sec:main-results-geometry} we start from pictorial example of $\varphi_\rho$, which leads to our main geometrical theorem in \Cref{thm:main}. In \Cref{sec:main-results-algorithms}, with regard to the observation of geometry, a minimizing  algorithm is proposed in \Cref{alg:ssbd} with proven convergence result presented in \Cref{thm:alg}, which exactly recovers both $\mb a_0$ and $\mb x_0$ up to signed shift symmetry. 
%We will close this section by introducing the current status of algorithmic theory in SaS deconvolution in \Cref{sec:main-results-comparison}, in which we also compare the result to other existing literatures. 

% !TEX root = ../BD_DQ.tex

\subsection{Geometry of the Objective $\varphi_\rho$} \label{sec:main-results-geometry}
The goal in SaS deconvolution is to recover $\mb a_0$ (and $\mb x_0$) up to a signed shift --- i.e., we wish to recover some $\pm s_{\ell}[\mb a_0]$. The shifts $\pm\shift{\mb a_0}{\ell}$ play a key role in shaping the landscape of $\varphi_\rho$. In particular, we will argue that over a certain subset of the sphere, {\em every local minimum of $\varphi_{\rho}$ is close to some $\pm\shift{\mb a_0}{\ell}$}.

\begin{wrapfigure}[12]{r}{2.5in}
\par\vspace{-0.15in}
\centerline{
\begin{tikzpicture}
\node at (0,0) {\includegraphics[width=1.5in]{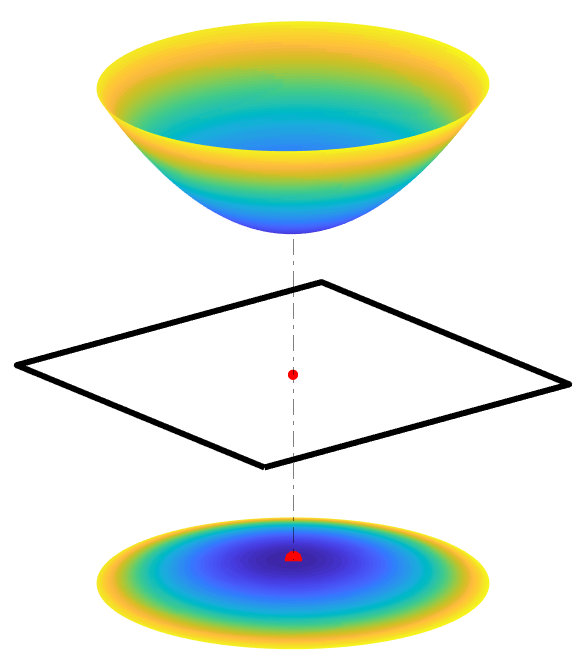}};
\node at (0,-2.35) {$ \mc B_{\ell^2,r}(\shift{\mb a_0}{\ell})\cap \bb S^{p-1}$};
\node at (-0.6,-1.1) {\color{red} $s_{\ell}[\mb a_0]$};
\node at (1.1,.7) {\color{blue} $\varphi_{\rho}(\mb a)$};
\end{tikzpicture}
}
\par\vspace{-0.1cm}
\caption{{\bf Geometry of $\varphi_\rho$ near a shift of $\mb a_0$.} Bottom: a portion of the sphere $\bb S^{p-1}$, colored according to $\varphi_\rho$. Top: $\varphi_\rho$ visualized as height.} $\varphi_\rho$ is strongly convex in this region, and it has a minimizer very close to $s_{\ell}[\mb a_0]$.
\label{fig:one-shift}
\end{wrapfigure}

\paragraph{Geometry near a single shift.} To gain intuition into the properties of $\varphi_\rho$, we first visualize this function in the vicinity of a single shift $s_{\ell}[\mb a_0]$ of the ground truth $\mb a_0$. In \Cref{fig:one-shift}, we plot the function value of $\varphi_\rho$ over
\begin{align}
	  \mc B_{\ell^2,r}(\shift{\mb a_0}{\ell})\cap \bb S^{p-1},\notag
\end{align}
where $\mc  B_{\ell^2,r}(\mb a)$ is a ball of radius $r$ around $\mb a$. We make two observations:
\begin{itemize}
	\item The objective function $\varphi_\rho$ is strongly convex on this neighborhood of $\shift{\mb a_0}{\ell}$.
	\item There is a local minimizer very close to $s_{\ell}[\mb a_0]$.
\end{itemize}

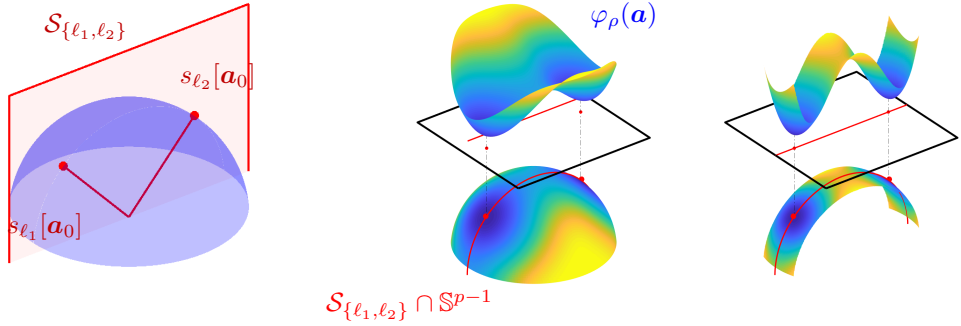
\begin{figure}[H]
\centerline{
\begin{tikzpicture}
\node at (-5,0) {\includegraphics[width=1.6in]{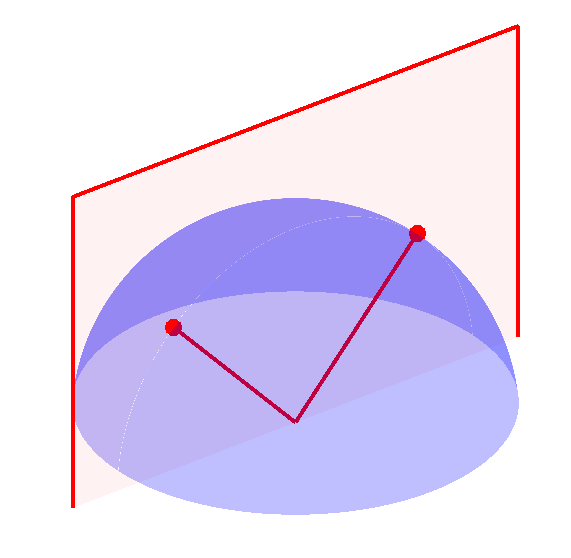}};
\node at (2.5,0) {\includegraphics[width=2.9in]{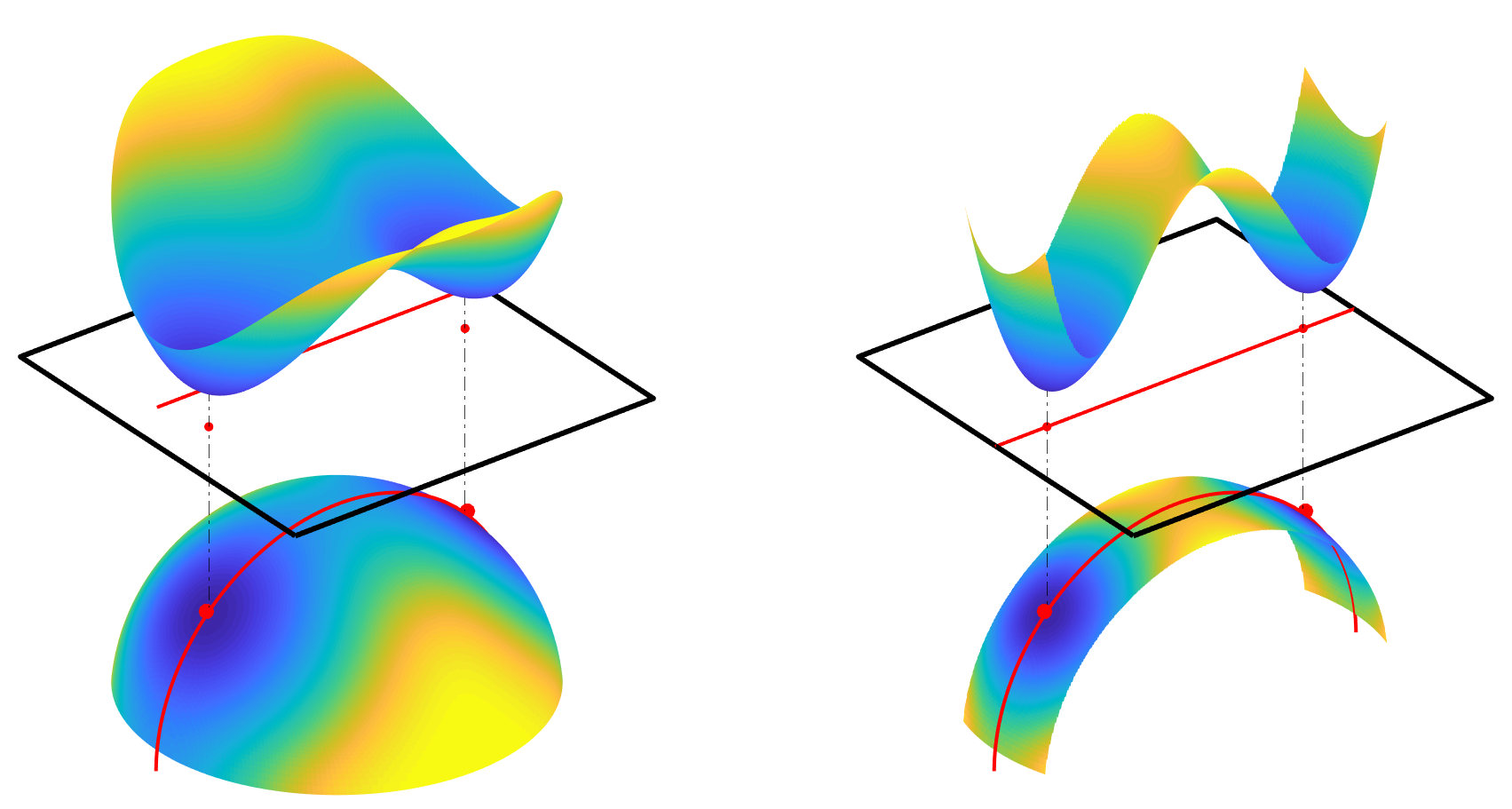}};
\node at (-3.75,.85) {\color[rgb]{.75,0,0} $s_{\ell_2}[\mb a_0]$};
\node at (-6.05,-1.15) {\color[rgb]{.75,0,0} $s_{\ell_1}[\mb a_0]$};
\node at (-5.5,1.5) {\color[rgb]{.75,0,0} $\mc S_{\set{\ell_1,\ell_2}}$};
\node at (-1.2,-2.2) {\color{red} $\mc S_{\set{\ell_1,\ell_2}} \cap \bb S^{p-1}$};
\node at (1.65,1.65) {\color{blue} $\varphi_{\rho}(\mb a)$};
\end{tikzpicture}
}
\caption{{\bf Geometry of $\varphi_\rho$ near the span $\mc S_{\set{\ell_1,\ell_2}}$ of two shifts of $\mb a_0$.} Left: each pair of shifts $s_{\ell_1}[\mb a_0]$, $s_{\ell_2}[\mb a_0]$ defines a linear subspace $\mc S_{\set{\ell_1,\ell_2}}$ of $\R^p$. Center/right: every local minimum of $\varphi_\rho$ near $\mc S_{\set{\ell_1,\ell_2}}$ (red line) is close to either $ s_{\ell_1}[\mb a_0]$ or $ s_{\ell_2}[\mb a_0]$; there is a negative curvature in the middle of $\shift{\mb a_0}{\ell_1}$, $\shift{\mb a_0}{\ell_2}$, and $\varphi_\rho$ is convex in direction away from $\mc S_{\ell_1,\ell_2}$.}
\label{fig:two-shifts}
\end{figure}

\paragraph{Geometry near the span of two shifts.} We next visualize the objective function $\varphi_\rho$ near the linear span of {\em two} different shifts $s_{\ell_1}[\mb a_0]$ and $s_{\ell_2}[\mb a_0]$. More precisely, we plot $\varphi_\rho$ near the intersection (\Cref{fig:two-shifts}, left) of the sphere $\bb S^{p-1}$ and the linear subspace
\begin{equation}
\mc S_{\set{\ell_1,\ell_2}} = \set{ \; \mb\alpha_1 s_{\ell_1}[\mb a_0] + \mb\alpha_2 s_{\ell_{2}}[\mb a_0] \; \middle | \, \mb\alpha_1, \mb\alpha_2 \in \R \, }.\notag
\end{equation}

\vspace{-0.2in}
\vspace{-0.1in} 

\noindent We make three observations:
\begin{itemize}
	\item Again, there is a local minimizer near each  shift $s_{\ell}[\mb a_0]$. 
	\item These are the {\em only} local minimizers in the vicinity of $\mc S_{\set{\ell_1,\ell_2}}$. In particular, the objective function $\varphi$ exhibits {\em negative curvature} along $S_{\set{\ell_1,\ell_2}}$ at any superposition $\mb\alpha_1 s_{\ell_1}[\mb a_0] + \mb\alpha_2 s_{\ell_2}[\mb a_0]$ whose weights $\mb\alpha_1$ and $\mb\alpha_2$ are balanced, i.e., $|\mb\alpha_1| \approx |\mb\alpha_2|$.
	\item Furthermore, the function $\varphi_\rho$ exhibits {\em positive curvature} in directions away from the subspace $\mc S_{\ell_1,\ell_2}$.
\end{itemize}

\begin{figure}[h]
\centerline{
\begin{tikzpicture}
\node at (-6,-.5) {\includegraphics[width=2.0in]{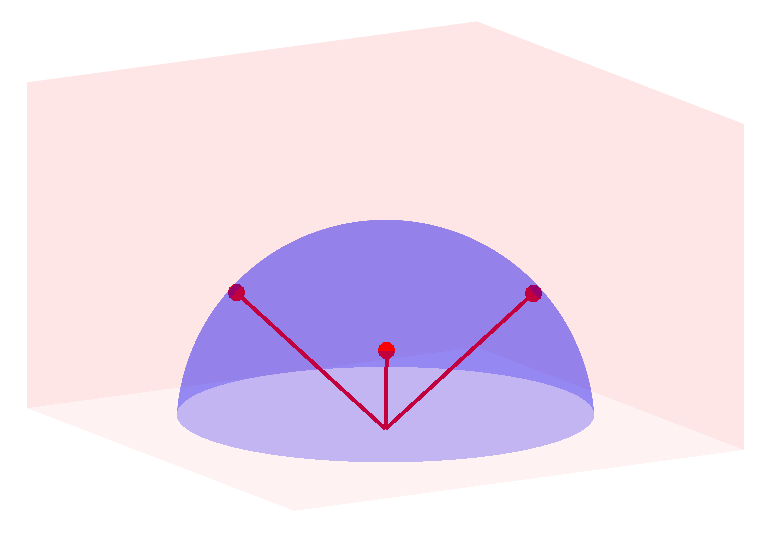}};
\node at (0,0) {\includegraphics[width=1.3in]{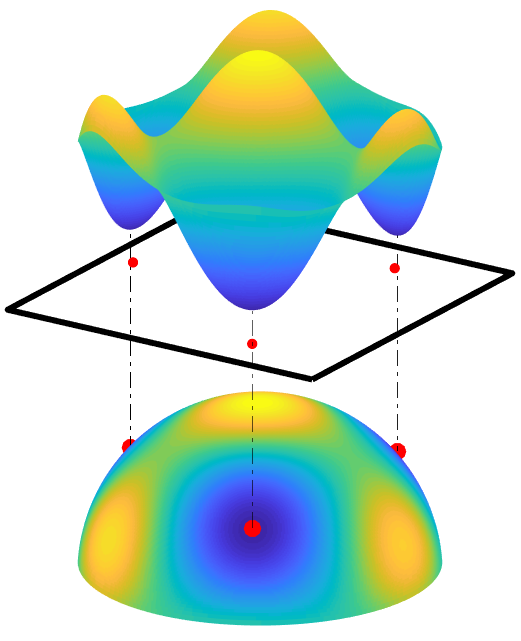}};
\node at (0,-2.2) {\color{black} $\mc S_{\set{\ell_1,\ell_2,\ell_3}} \cap \bb S^{p-1}$};
\node at (-0.9,-1.50) {\color[rgb]{.75,0,0} $s_{\ell_1}[\mb a_0]$};
\node at (1.45,-.6) {\color[rgb]{.75,0,0} $s_{\ell_2}[\mb a_0]$};
\node at (-1.65,-.6) {\color[rgb]{.75,0,0} $s_{\ell_3}[\mb a_0]$};
\node at (1.85,1.65) {\color{blue} $\varphi_{\rho}(\mb a)$};
\node at (-6.2,-1.85) {\color[rgb]{.75,0,0} $s_{\ell_1}[\mb a_0]$};
\node at (-4.4,-.4) {\color[rgb]{.75,0,0} $s_{\ell_2}[\mb a_0]$};
\node at (-7.65,-.6) {\color[rgb]{.75,0,0} $s_{\ell_3}[\mb a_0]$};
\node at (-7,1.25) {\color[rgb]{.75,0,0} $\mc S_{\set{\ell_1,\ell_2,\ell_3}}$};
\end{tikzpicture}
}
\caption{{\bf Geometry of $\varphi_\rho$ over the span $\mc S_{\set{\ell_1,\ell_2,\ell_3}}$ of three shifts of $\mb a_0$.} The subspace $\mc S_{\set{\ell_1,\ell_2,\ell_3}}$ is three-dimensional; its intersection with the sphere $\bb S^{p-1}$ is isomorphic to a two-dimensional sphere. On this set, $\varphi_{\rho}$ has local minimizers near each of the $s_{\ell_i}[\mb a_0]$, and are the only minimizers near $\mc S_{\ell_1,\ell_2,\ell_3}$.} \label{fig:three-shifts}
\end{figure}
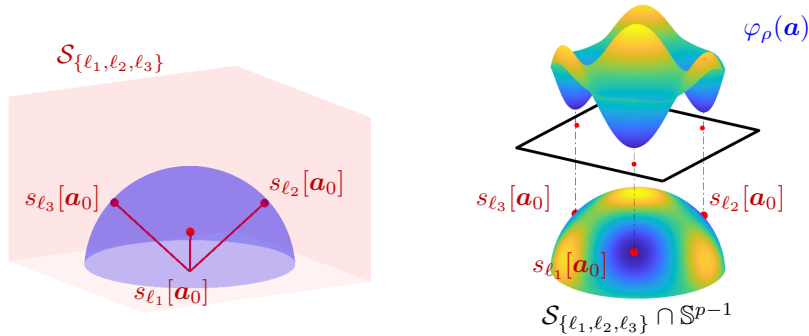

\paragraph{Geometry in the span of multiple shifts.} Finally, we visualize $\varphi_\rho$ over the intersection (\Cref{fig:three-shifts},  left) of the sphere $\bb S^{p-1}$ with the linear span of three shifts $s_{\ell_1}[\mb a_0]$, $s_{\ell_2}[\mb a_0]$, $s_{\ell_3}[\mb a_0]$ of the true kernel $\mb a_0$: 
\begin{equation}
\mc S_{\set{\ell_1,\ell_2,\ell_3}} = \set{ \; \mb\alpha_1 s_{\ell_1} [\mb a_0] + \mb\alpha_2 s_{\ell_2}[\mb a_0]  + \mb\alpha_3 s_{\ell_3}[\mb a_0] \;  | \, \mb\alpha_1, \mb\alpha_2, \mb\alpha_3 \in \R  \, }\notag
\end{equation}

\vspace{-0.3in}

\noindent Again, {\em there is a local minimizer near each signed shift}. At roughly balanced superpositions of shifts, the objective function exhibits negative curvature. As a result, again, the {\em only} local minimizers are close to signed shifts.

\pagebreak

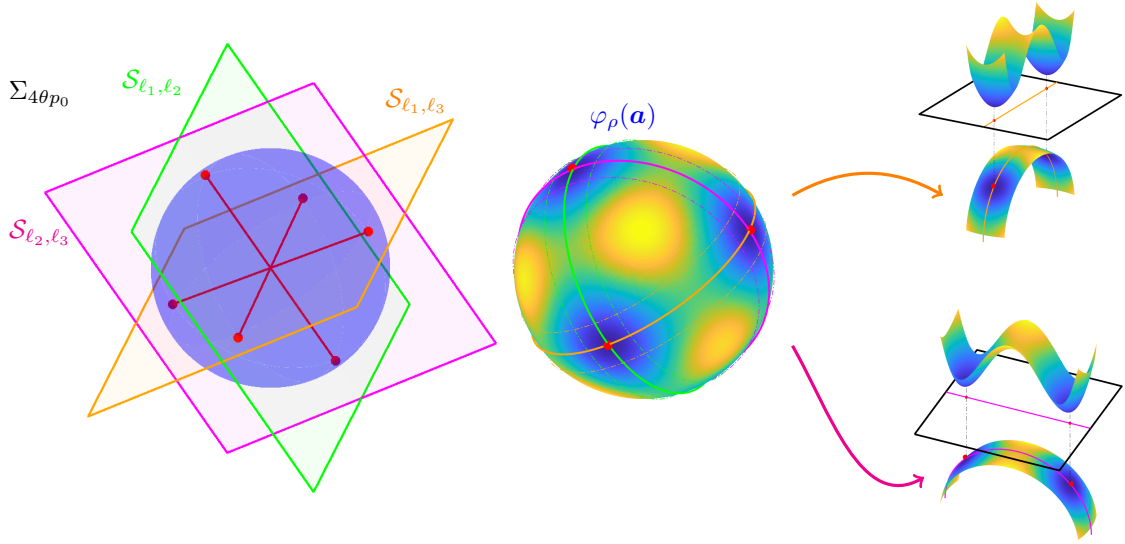
\begin{figure}[t]
\centerline{
\begin{tikzpicture}
\node at (-5,0) {\includegraphics[width=2.5in]{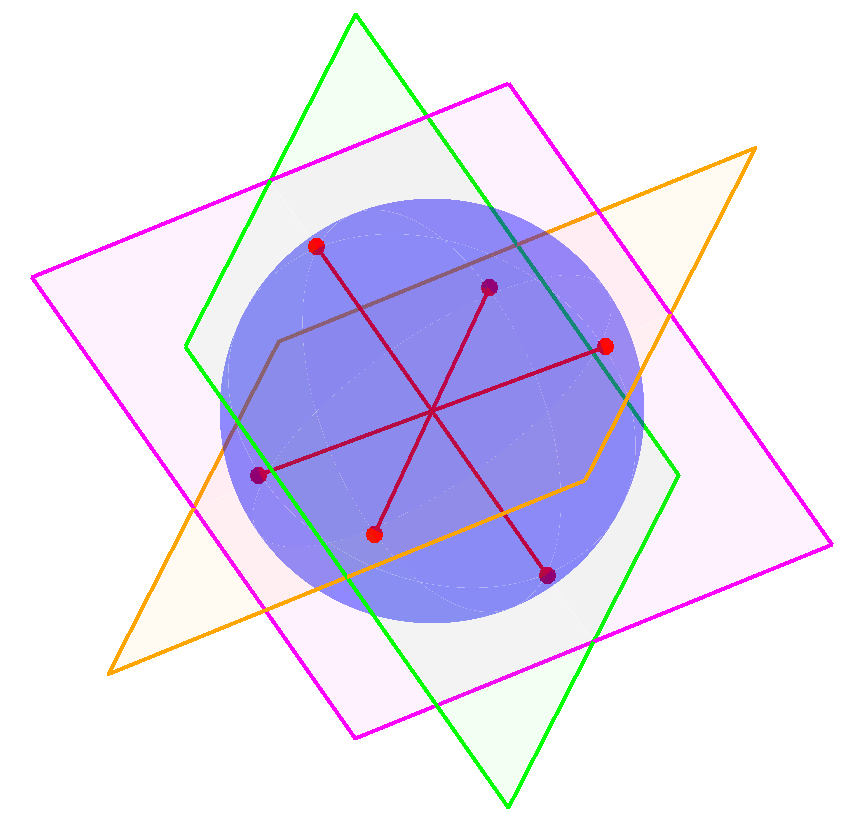}};
\node at (-6.5,2.5) {\color{green} $\mc S_{\ell_1,\ell_2}$};
\node at (-3,2.25) {\color{orange} $\mc S_{\ell_1,\ell_3}$};
\node at (-8,.5) {\color{magenta} $\mc S_{\ell_2,\ell_3}$};
\node at (-8,2.35) {\color{black} $\Sigma_{4 \theta p_0}$};
\node at (0,0) {\includegraphics[width=1.5in]{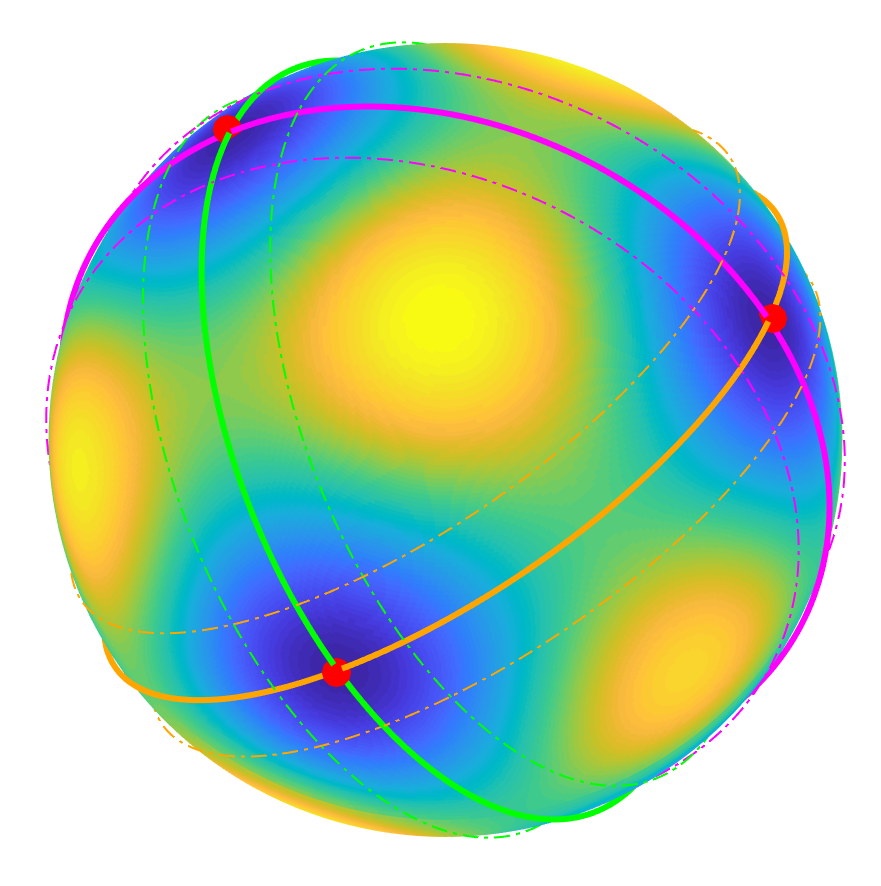}};
\node at (-.25,2.05) {\color{blue} $\varphi_{\rho}(\mb a)$};
\node at (5,2) {\includegraphics[width=1.1in]{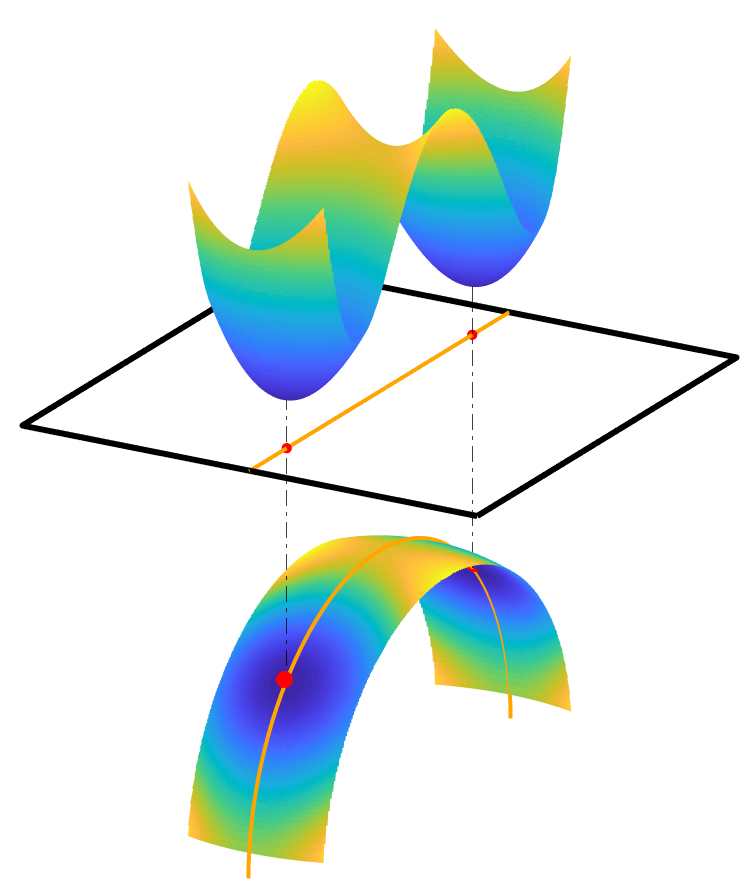}};
\node at (5,-2) {\includegraphics[width=1.1in]{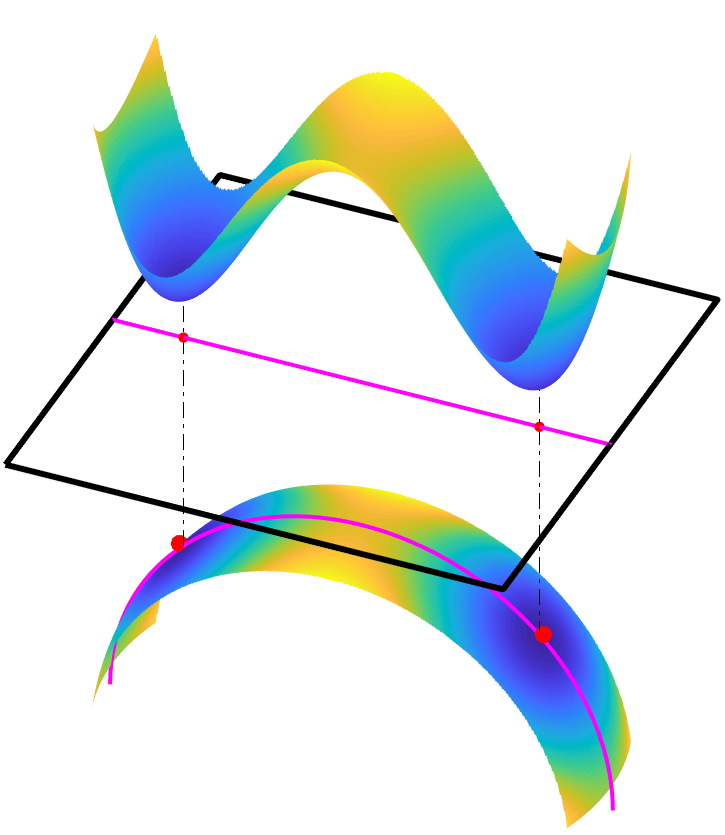}};
\draw [->,very thick, orange] (2,1) to [out = 30,in = 150] (4,1);
\draw [->,very thick, magenta] (2,-1) to [out = -60,in = 1650] (3.75,-2.75);
\end{tikzpicture}
}
	\caption{\textbf{Geometry of $\varphi_\rho$ over the union of subspaces $\Sigma_{4\theta p_0}$.} Left: schematic representation of the union of subspaces $\Sigma_{4 \theta p_0}$. For each set $\mb \tau$ of at most $4 \theta p_0$ shifts, we have a subspace $\mc S_{\mb \tau}$. Right: $\varphi_\rho$ has good geometry near this union of subspaces.}
	\label{fig:geo}
\end{figure}

\paragraph{Geometry of $\varphi_\rho$ over a union of subspaces.} Our main geometric result will show that these properties obtain on {\em every} subspace spanned by a few shifts of $\mb a_0$. Indeed, for each subset
\begin{equation}
\mb \tau \subseteq \set{ -p_0 +1, \dots, p_0 -1 },
\end{equation}
define a linear subspace
\begin{equation}\label{eqn:s_tau}
\mc S_{\mb \tau} = \set{ \sum_{\ell \in \mb \tau} \mb\alpha_\ell s_{\ell}[\mb a_0] \, \middle | \, \mb\alpha_{-p_0+1}, \dots, \mb\alpha_{p_0-1} \in \R }.
\end{equation}
The subspace $\mc S_{\mb \tau}$ is the linear span of the shifts $s_{\ell}[\mb a_0]$ indexed by $\ell$ in the set $\mb \tau$. Our geometric theory will show that with high probability the function $\varphi_{\rho}$ has no spurious local minimizers near any $\mc S_{\mb \tau}$ for which $\mb \tau$ is not too large -- say, $|\mb \tau | \le 4 \theta p_0$. Combining all of these subspaces into a single geometric object, define the union of subspaces
\begin{equation}\label{eqn:union_subspaces_4tp0}
\Sigma_{4 \theta p_0} = \bigcup_{|\mb \tau| \le 4 \theta p_0} \mc S_{\mb \tau}.
\end{equation}
\Cref{fig:geo} (left) gives a schematic representation of this set. We claim:
\begin{itemize}
	\item In the neighborhood of $\Sigma_{4 \theta p_0} $, all local minimizers are near signed shifts.
	\item The value of $\varphi_\rho$ grows in any direction away from $\Sigma_{4 \theta p_0}$.
\end{itemize}

\pagebreak

\paragraph{Main Geometric Result.} Our main result formalizes the above observations, under two key assumptions: first, that the sparsity rate $\theta$ is sufficiently small (relative to the shift coherence $\mu$ of $p_0$), and, second, the signal length $n$ is sufficiently large:

\begin{theorem}[Main Geometric Theorem]
\label{thm:main}  Let $\mb y = \mb a_0 * \mb x_0$  with $\mb a_0\in\bb S^{p_0-1}$ $\mu$-shift coherent  and $\mb x_0\simiid\mr{BG}(\theta)\in\R^n$ with sparsity rate
\begin{align}\label{eqn:sparsity_rate}
	\theta \;\in\; \brac{ \frac{c_1}{p_0}, \frac{c_2}{p_0\sqrt\mu + \sqrt{p_0}}}\cdot\frac{1}{\log^2p_0}.
\end{align}
Choose $\rho(x) =\sqrt{x^2+\delta^2}$ and set $\lambda = 0.1/\sqrt{p_0\theta}$ in $\varphi_\rho$. Then there exists $\delta > 0$ and numerical constant $c$ such that if $n \ge \poly(p_0)$, with high probability, every local minimizer $\bar{\mb a}$ of $\varphi_\rho$ over $\Sigma_{4\theta p_0}$ satisfies $\norm{\bar{\mb a} - \sigma\shift{\mb a_0}{\ell}}2 \leq c\max\set{\mu,p_0^{-1}}$ for some signed shift $\sigma s_\ell[\mb a_0]$ of the true kernel. Above, $c_1, c_2 > 0$ are positive numerical constants.
\end{theorem}
\begin{proof}
	This follows from \Cref{thm:three_regions}. 
\end{proof}

The upper bound on $\theta$ in \eqref{eqn:sparsity_rate} yields the tradeoff between coherence and sparsity described in  \Cref{fig:sparsity-coherence}. Simply put, when $\mb a_0$ is better conditioned (as a kernel), its coherence $\mu$ is smaller and $\mb x_0$ can be denser.

At a technical level, our proof of \Cref{thm:main} shows that (i) $\varphi_\rho(\mb a)$ is strongly convex in the vicinity of each signed shift, and that at every other point $\mb a$  near $\Sigma_{4\theta p_0}$, there is either (ii) a nonzero gradient or (iii) a direction of strict negative curvature; furthermore (iv) the function $\varphi_\rho$ grows away from  $\Sigma_{4\theta p_0}$.  Points (ii)-(iii) imply that near $\Sigma_{4\theta p_0}$ there are no ``flat'' saddles: every saddle point has a direction of strict negative curvature. We will leverage these properties to propose an efficient algorithm for finding a local minimizer near $\Sigma_{4 \theta p_0}$. Moreover, this minimizer is close enough to a shift (here, $\norm{\bar{\mb a} - \shift{\mb a_0}\ell}2\lessapprox \mu$) for us to exactly recover $\shift{\mb a_0}\ell$: we will give a refinement algorithm that produces $(\pm\shift{\mb a_0}{\ell},\pm\shift{\mb x_0}{-\ell})$.

% !TEX root = ../BD_DQ.tex

\subsection{Provable Algorithm for SaS Deconvolution} 
\label{sec:main-results-algorithms}

The objective function $\varphi_\rho$ has good geometric properties on (and near!) the union of subspaces $\Sigma_{4 \theta p_0}$. In this section, we show how to use give an efficient method that exactly recovers $\mb a_0$ and $\mb x_0$, up to shift symmetry. Although our geometric analysis only controls $\varphi_\rho$ near $\Sigma_{4 \theta p_0}$, we will give a descent method which, with appropriate initialization $\mb a^{(0)}$, produces iterates $\mb a^{(1)}, \dots, \mb a^{(k)}, \dots$ that remain close to $\Sigma_{4 \theta p_0}$ for all $k$. In short, it is easy to {\em start} near $\Sigma_{4 \theta p_0}$ and easy to {\em stay} near $\Sigma_{4 \theta p_0}$. After finding a local minimizer $\bar{\mb a}$, we refine it to produce a signed shift of $(\mb a_0,\mb x_0)$ using alternating minimization.

The next two paragraphs give the main ideas behind the main steps of the algorithm. We then describe its components in more detail (\Cref{alg:ssbd}) and state our main algorithmic result (\Cref{thm:alg}), which asserts that under appropriate conditions this method produces a signed shift of $(\mb a_0, \mb x_0)$. 

\paragraph{Minimization: Starting and staying near $\Sigma_{4 \theta p_0}$.}  
Our algorithm starts with a initialization scheme which generates $\mb a^{(0)}$ near the union of subspaces $\Sigma_{4\theta p_0}$, which   consists of linear combinations of just a few shifts of $\mb a_0$. How can we find a point near this union? Notice that {\em the data $\mb y$ also consists of a linear combination of just a few shifts of $\mb a_0$} Indeed:
\begin{eqnarray}
\mb y &=& \mb a_0 \ast \mb x_0 \quad = \quad \sum_{\ell \in \mathrm{supp}(\mb x_0)} {\mb x_0}_\ell s_{\ell}[\mb a_0].
\end{eqnarray}
A length-$p_0$ segment of data $\mb y_{0,\dots, p_0-1} = [\mb y_0, \dots, \mb y_{p_0-1}]^*$ captures portions of roughly $2 \theta p_0 \ll 4 \theta p_0$ shifts $s_{\ell}[\mb a_0]$. 

\begin{figure}[t!]	
	\centering
	\input{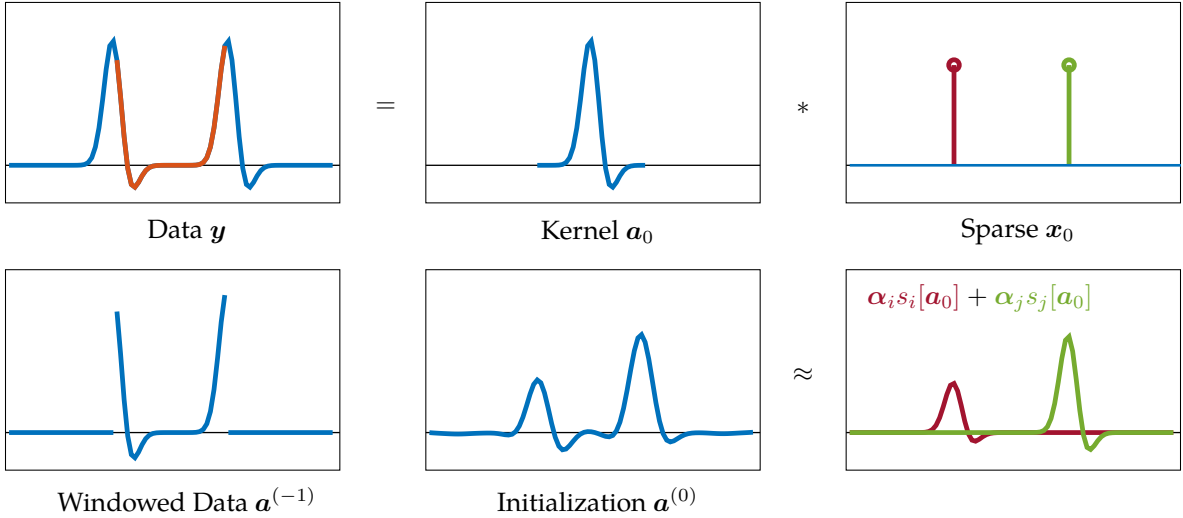}	
	\caption{ \textbf{Data-driven initialization:} using a piece of the observed data $\mb y$ to generate an initial point $\mb a^{(0)}$ that is close to a superposition of shifts $s_{\ell}[\mb a_0]$ of the ground truth.  Top: data $\mb y = \mb a_0 \ast \mb x_0$ is a superposition of shifts of the true kernel $\mb a_0$. Bottom: a length-$p_0$ window contains pieces of just a few shifts. Bottom middle: one step of the generalized power method approximately fills in the missing pieces, yielding a near superposition of shifts of $\mb a_0$ (right).
	}   
	\label{fig:init}   
\end{figure}

Many of these copies of $\mb a_0$ are truncated by the restriction to $\set{0,\dots, p_0-1}$. A relatively simple remedy is as follows: first, we zero-pad $\mb y_{0,\dots, p_0-1}$ to length $p = 3 p_0 - 2$, giving 
\begin{align}   
	\brac{\mb 0^{p_0-1}; \mb y_0; \cdots ;\mb y_{p_0-1}; \mb 0^{p_0-1} }.
\end{align}
Zero padding provides enough space to accommodate any shift $s_{\ell}[\mb a_0]$ with $\ell \in \mb \tau$. We then perform one step of the generalized  power method\footnote{The power method for minimizing a quadratic form $\xi(\mb a) = \tfrac{1}{2} \mb a^* \mb M \mb a$ over the sphere consists of the iteration $\mb a \mapsto -\mb P_{\bb S^{p-1}} \mb M \mb a$. Notice that in this mapping, $-\mb M \mb a = -\nabla \xi(\mb a)$. The generalized power method, for minimizing a function $\varphi$ over the sphere consists of repeatedly projecting $-\nabla \varphi$ onto the sphere, giving the iteration $\mb a \mapsto -\mb P_{\bb S^{p-1}} \nabla \varphi(\mb a)$. \eqref{eqn:gpm} can be interpreted as one step of the generalized power method for the objective function $\varphi_{\rho}$.}, writing 
\begin{equation} \label{eqn:gpm}
\mb a^{(0)} = -\mb P_{\bb S^{p-1}} \nabla \varphi_{\ell^1}\left( \mb P_{\bb S^{p-1}} \brac{\mb 0^{p_0-1};\mb y_0;\cdots;\mb y_{p_0-1};\mb 0^{p_0-1} } \right),
\end{equation}
where $\mb P_{\bb S^{p-1}}$ projects onto the sphere. The reasoning behind this construction may seem obscure. We will explain it at a more technical level in \Cref{sec:alg} after interpreting the gradient $\nabla \varphi_\rho$ in terms of its action on the shifts $s_\ell[\mb a_0]$ in  \Cref{sec:geometry}. For now, we note that this operation has the effect of (approximately) filling in the missing pieces of the truncated shifts $s_{\ell}[\mb a_0]$ -- see \Cref{fig:init} for an example. We will prove that with high probability $\mb a^{(0)}$ is indeed close to $\Sigma_{4 \theta p_0}$. 

\begin{wrapfigure}[14]{r}{1.8in}
\centerline{
\begin{tikzpicture}
\node at (0,0) {\includegraphics[trim={0 0 0 7cm}, width=1.5in]{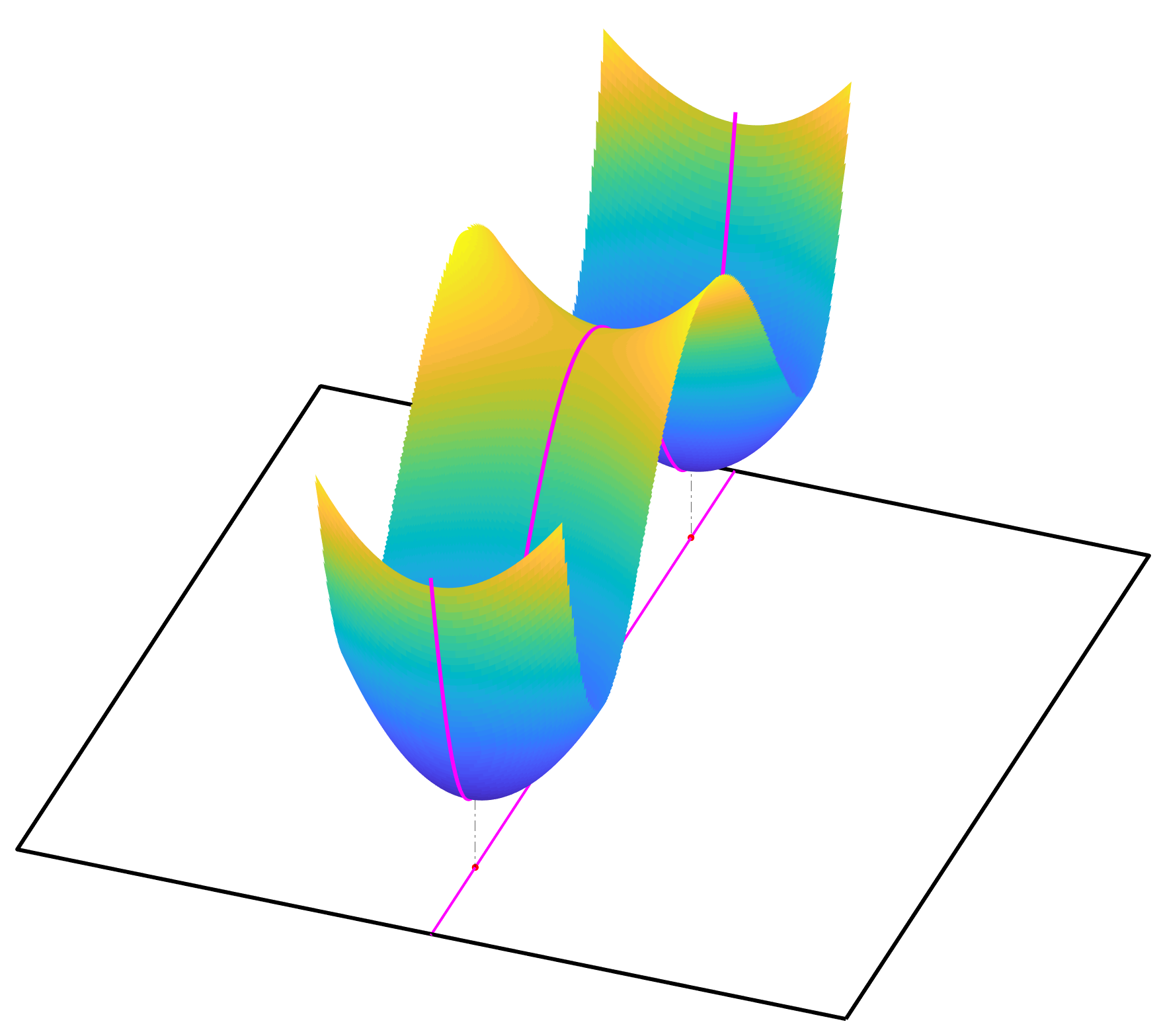}};
\node at (.5,-.5) {\color{magenta} $\mc S_{\mb \tau}$};
\node at (1.25,1) {\color{blue} $\varphi_\rho$}; 
\end{tikzpicture}
}
\caption{{\bf Growth of $\varphi_\rho$ away from $\mc S_{\mb \tau}$.} Because $\varphi_\rho$ grows away from $\mc S_{\mb \tau}$, small-stepping descent methods stay near $\mc S_{\mb \tau}$.} \label{fig:growth}
\end{wrapfigure}

The next key observation is that the function $\varphi_\rho$ grows as we move away from the subspace $\mc S_{\mb \tau}$ -- see \Cref{fig:growth}. Because of this, a small-stepping descent method will not move far away from $\Sigma_{4\theta p_0}$. For concreteness,  we will analyze a variant of the curvilinear search method \cite{goldfarb1980curvilinear,goldfarb2017using},  which moves in a linear combination of the negative gradient direction $-\mb g$ and a negative curvature direction $-\mb v$. At the $k$-th iteration, the algorithm updates $\mb a^{(k+1)}$ as
\begin{align}
	\mb a^{(k+1)} \gets \mb P_{\Sp^{p-1}}\big[\mb a^{(k)} - t\mb g^{(k)} - t^2\mb v^{(k)}\big]
\end{align}
with appropriately chosen step size $t$.  The inclusion of a negative curvature direction allows the method to avoid stagnation near saddle points. Indeed, we will prove that starting from initialization $\mb a^{(0)}$, this method produces a sequence $\mb a^{(1)},\mb a^{(2)},\ldots  $ which efficiently converges to a local minimizer $\bar{\mb a}$ that is near some signed shift $\pm s_{\ell}[\mb a_0]$ of the ground truth. 

\paragraph{Refinement: Rounding a near-solution with homotopy alternating minimization.}   
The second step of our algorithm {\em rounds} the local minimizer $\bar{\mb a} \approx \sigma s_{\ell}[\mb a_0]$ to produce an exact solution $\wh{\mb a} = \sigma s_{\ell}[\mb a_0]$. As a byproduct, it also exactly recovers the corresponding signed shift of the true sparse signal, $\wh{\mb x} = \sigma s_{-\ell}[\mb x_0]$. 

Our rounding algorithm is an alternating minimization scheme, which alternates between minimizing the Lasso cost over $\mb a$ with $\mb x$ fixed, and minimizing the Lasso cost over $\mb x$ with $\mb a$ fixed. We make two modifications to this basic idea, both of which are important for obtaining exact recovery. First, unlike the standard Lasso cost, which penalizes all of the entries of $\mb x$, we maintain a running estimate $I^{(k)}$ of the support of $\mb x_0$, and only penalize those entries that are not in $I^{(k)}$:
\begin{equation}
 \tfrac{1}{2} \norm{ \mb a \ast \mb x - \mb y}{2}^2 + \lambda\sum_{i\not\in I^{(k)}}\abs{\mb x_i}.
\end{equation}
This can be viewed as an extreme form of {\em reweighting} \cite{candes2008enhancing}. Second, our algorithm gradually decreases  penalty variable $\lambda$ to $0$, so that eventually
\begin{align}
	\wh{\mb a}*\wh{\mb x} \approx \mb y.
\end{align}  
This can be viewed as a \emph{homotopy} or \emph{continuation} method \cite{osborne2000new, efron2004least}.  For concreteness, at $k$-th iteration the algorithm reads:
\begin{align}
	\text{Update $\mb x$:}&\qquad \mb x^{(k+1)} \;\gets\; \argmin_{\mb x}\tfrac12\|\mb a^{(k)}*\mb x - \mb y\|_2^2 + \lambda^{(k)}\sum_{i\not\in I^{(k)}}\abs{\mb x_i},\\
	\text{Update $\mb a$:}&\qquad \mb a^{(k+1)} \;\gets\; \mb P_{\Sp^{p-1}}\big[\argmin_{\mb a}\tfrac12\|\mb a*\mb x^{(k+1)}-\mb y\|_2^2\big],\\
	\text{Update $\lambda$ and $I$:}&\qquad   \lambda^{(k+1)} \;\gets\; \tfrac12\lambda^{(k)}, \qquad  I^{(k+1)} \;\gets\;  \supp\big(\mb x^{(k+1)}\big). 
\end{align}    
We prove that the iterates produced by this sequence of operations converge to the ground truth at a linear rate, as long as the initializer $\bar{\mb a}$ is sufficiently nearby.  

\begin{figure}[t!]
	\centering
	\input{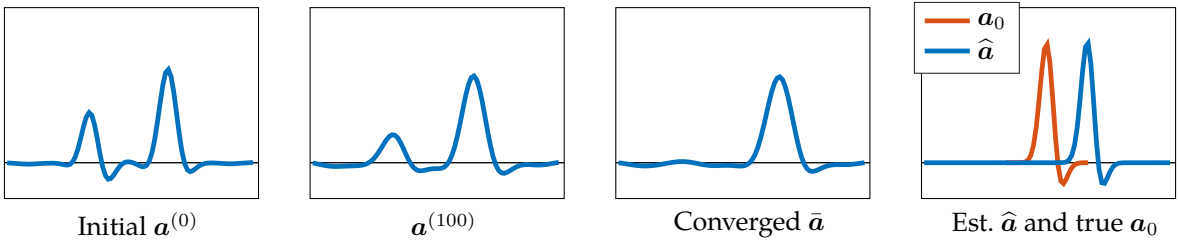}
	\caption{ \textbf{Local minimization and refinement.} Left: data-driven initialization $\mb a^{(0)}$ consisting of a near-superposition of two shifts. Middle: minimizing $\varphi_\rho$ produces a near shift of $\mb a_0$. Right: rounded solution $\wh{\mb a}$ using the Lasso. $\wh{\mb a}$ is very close to a shift of $\mb a_0$. 
	}   
	\label{fig:alg_demo} 
\end{figure} 

\renewcommand{\algorithmicrequire}{\textbf{Input:}}
\renewcommand{\algorithmicensure}{\textbf{Output:}}

\paragraph{Algorithm and Main Algorithmic Result.} Our overall algorithm is summarized as  \Cref{alg:ssbd}. \Cref{fig:alg_demo} illustrates the main steps of this algorithm. Our main algorithmic result states that under closely related hypotheses as above,  \Cref{alg:ssbd} produces a signed shift of the ground truth $(\mb a_0, \mb x_0)$:

\begin{algorithm}[h!]
	\caption{Short and Sparse Deconvolution}
	\label{alg:ssbd}
	\begin{algorithmic}
		\Require Observation $\mb y$, motif length $p_0$, sparsity $\theta$, shift-coherence $\mu$, and curvature threshold $-\eta_v$. 
		\vspace{0.04in}  
		\State \textbf{\ul{Minimization}}:
		\State  Set $\mb a^{(0)}\gets -\mb P_{\bb S^{p-1}} \nabla \varphi_\rho\left( \mb P_{\bb S^{p-1}} \brac{\mb 0^{p_0-1};\mb y_0;\cdots;\mb y_{p_0-1};\mb 0^{p_0-1} } \right)$.
		\State Set $\lambda = 0.1/\sqrt{p_0\theta}$ \footnotemark and $\delta>0$ in $\varphi_\rho$. For $k = 1,2,\ldots,K_1$, let \vspace{-0.11in}
		\begin{align}
			\mb a^{(k+1)} \gets \mb P_{\Sp^{p-1}}[\mb a^{(k)} - t\mb g^{(k)} -t^2\mb v^{(k)}]
		\end{align}\par\vspace{-0.12in} 
		\State where $\mb g^{(k)}$ is the Riemannian gradient; $\mb v^{(k)}$ is the eigenvector of smallest Riemannian Hessian eigenvalue  if less then $-\eta_v$ with $\innerprod{\mb v^{(k)}}{\mb g^{(k)}} \geq 0$, otherwise let $\mb v^{(k)}=\mb 0$; and $t\in(0,0.1/n\theta]$ satisfies \vspace{-0.11in}
		\begin{align}
			\varphi_\rho(\mb a^{(k+1)}) <  \varphi_\rho(\mb a^{(k)}) - \tfrac12 t\|\mb g^{(k)}\|_2^2 -\tfrac14 t^4\eta_v\|\mb v^{(k)}\|_2^2
		\end{align}\par\vspace{-0.12in}
		\State to obtain a near local minimizer $\bar{\mb a}\gets \mb a^{(K_1)}$. 
		\vspace{0.04in} 
		\State \textbf{\ul{Refinement}}: 
		\State Set $\mb a^{(0)}\gets \bar{\mb a}$, $\;\lambda^{(0)} \gets 10(p\theta+\log n)(\mu+1/p)$ and $\;I^{(0)} \gets \soft{\supp(\wc{\mb y}*\bar{\mb a}}{\lambda^{(0)}})$. For $k = 1,2,\ldots,K_2$, let  \vspace{-0.11in}
		\begin{align} 
			&\mb x^{(k+1)} \gets \argmin\nolimits_{\mb x}\tfrac12\|\mb a^{(k)}*\mb x - \mb y\|_2^2 + \lambda^{(k)}\textstyle\sum_{i\not\in I^{(k)}}\abs{\mb x_i},\\[-0.5ex]
			&\mb a^{(k+1)} \gets \mb P_{\Sp^{p-1}}\big[\argmin\nolimits_{\mb a}\tfrac12\|\mb a*\mb x^{(k+1)}-\mb y\|_2^2\big],\\[-0.5ex]
			&\lambda^{(k+1)}\gets\lambda^{(k)}/2,\quad \qquad I^{(k+1)}\gets\supp(\mb x^{(k+1)}),
		\end{align}\par\vspace{-0.13in}
		\State to obtain $(\wh{\mb a},\wh{\mb x})\gets (\mb a^{(K_2)},\mb x^{(K_2)})$.
		\vspace{0.05in}   
		\Ensure Return $(\wh{\mb a}, \wh{\mb x})$. 
	\end{algorithmic} 
\end{algorithm}  

\vspace{-0.05in} 

\begin{samepage}
 
\begin{theorem}[Main Algorithmic Theorem]
\label{thm:alg}  
Suppose $\mb y = \mb a_0*\mb x_0$ where $\mb a_0\in\bb S^{p_0-1}$ is $\mu$-truncated shift coherent such that $\max_{i\neq j}\abs{\innerprod{\ip_{p_0}^*\shift{\mb a_0}{i}}{\ip_{p_0}^*\shift{\mb a_0}{j}}}\leq{\mu}$ and $\mb x_0\simiid\mr{BG}(\theta)\in\R^n$ with $\theta,\mu$ satisfying
\begin{align}
	\theta \in \brac{ \frac{c_1}{p_0}, \frac{c_2}{\paren{p_0\sqrt{\mu} + \sqrt{p_0}}\log^2p_0 }}, \qquad \mu\leq\frac{c_3}{\log^2n}   
\end{align}   
for some constant $c_1,c_2,c_3>0$. If the signal lengths $n,p_0$ satisfy $n>\poly(p_0)$ and $p_0>\mr{polylog}(n)$, then there exist  $\delta,\eta_v > 0$ such that with high probability, \Cref{alg:ssbd} produces $(\wh{\mb a},\wh{\mb x})$ that are equal to the ground truth up to signed shift symmetry: 
\begin{equation}
\norm{\bigl(\wh{\mb a},\,\wh{\mb x}\bigr)  -  \sigma\bigl( s_{\ell}[\mb a_0],\, s_{-\ell}[\mb x_0] \bigr)}2 \leq \eps
\end{equation}
for some $\sigma\in\set{-1,1}$ and $\ell\in\set{-p_0+1,\ldots,p_0-1}$ if $K_1 > \poly(n,p_0)$ and  $K_2 >\mr{polylog}(n,p_0,\eps^{-1})$.     
\end{theorem}  
\begin{proof} 
	See \Cref{thm:minimization} and  \Cref{thm:altmin}.
\end{proof}

\end{samepage}

\footnotetext{In practice, we suggest setting $\lambda = c_\lambda/\sqrt{p_0\theta}$ with $c_\lambda\in\brac{0.5,0.8}$.}

\subsection{Relationship to the Literature} \label{sec:main-results-comparison} 

Blind deconvolution is a classical problem in signal processing \cite{stockham1975blind,cannon1976blind}, and has been studied under a variety of hypotheses. In this section, we first discuss the relationship between our results and the existing literature on the short-and-sparse version of this problem, and then briefly discuss other deconvolution variants in the theoretical literature. 
 
 \paragraph{Applications of SaS Deconvolution.} The short-and-sparse model arises in a number of applications. One class of applications involves finding basic motifs (repeated patterns) in datasets. This {\em motif discovery} problem arises in extracellular spike sorting {\cite{lewicki1998review, Ekanadham2011-NIPS} and calcium imaging \cite{pnevmatikakis2016simultaneous}, where the observed signal exhibits repetitive \emph{short} neuron excitation patterns occurring \emph{sparsely} across time and/or space. Similarly, electron microscopy images \cite{Cheung17-Nature} arising in study of nanomaterials often exhibit repeated motifs. 
 
 Another significant application of SaS deconvolution is {\em image deblurring}. Typically, the blur kernel is small relative to the image size ({\em short}) \cite{ayers1988iterative, you1996anisotropic, carasso2001direct, levin2007deconvolution, levin2011understanding}. In natural image deblurring, the target image is often assumed to have relatively few sharp edges \cite{fergus2006removing, joshi2008psf, levin2011understanding}, and hence have {\em sparse} derivatives. In scientific image deblurring, e.g., in astronomy \cite{lane1992blind,harmeling2009online,briers2013laser} and geophysics \cite{kaaresen1998multichannel}, the target image is often sparse, either in the spatial or wavelet domains, again leading to variants of the SaS model. The literature on blind image deconvolution is large; see, e.g.,  \cite{kundur1996blind, campisi2016blind} for surveys.  
  
Variants of the SaS deconvolution problem arise in many other areas of engineering as well. Examples include \emph{blind equalization} in communications \cite{sato1975method,  shalvi1990new, johnson1998blind}, \emph{dereverberation} in sound engineering \cite{miyoshi1988inverse, naylor2010speech} and image \emph{super-resolution}  \cite{baker2002limits, shtengel2009interferometric, yang2010image}.

 \paragraph{Algorithmic theory for SaS deconvolution.} 
  These applications have motivated a great deal of algorithmic work on variants of the SaS problem \cite{lane1987automatic, bones1995deconvolution,bell1995information,  kundur1996blind,markham1999parametric, campisi2016blind, walk2017blind}. In contrast, relatively little theory is available to explain when and why algorithms succeed.  
Our algorithm minimizes $\varphi_\rho$ as an approximation to the Lasso cost over the sphere. Our formulation and results have strong precedent in the literature. Lasso-like objective functions have been widely used in image deblurring \cite{you1996anisotropic, chan1998total,fergus2006removing, levin2007deconvolution, shan2008high,xu2010two, dong2011image,krishnan2011blind, levin2011understanding, wipf2014revisiting, perrone2014total, zhang2017global}. A number of insights have been obtained into the geometry of sparse deconvolution -- in particular, into the effect of various constraints on $\mb a$ on the presence or absence of spurious local minimizers. In image deblurring, a simplex constraint ($\mb a \ge \mb 0$ and $\| \mb a \|_1 = 1$) arises naturally from the physical structure of the problem \cite{you1996anisotropic, chan1998total}. Perhaps surprisingly, simplex-constrained deconvolution admits trivial global minimizers, at which the recovered kernel $\mb a$ is a spike, rather than the target blur kernel \cite{levin2011understanding,benichoux2013fundamental}. 

\cite{wipf2014revisiting} imposes the $\ell^2$ regularization on $\mb a$ and observes that this alternative constraint gives more reliable algorithm. \cite{zhang2017global} studies the geometry of the simplified objective $\varphi_{\ell^1}$ over the sphere, and proves that in the dilute limit in which $\mb x_0$ has one nonzero entry, all strict local minima of $\varphi_{\ell^1}$ are close to signed shifts truncations of $\mb a_0$. By adopting a different objective function (based on $\ell^4$ maximization) over the sphere, \cite{zhang2018structured} proves that on a certain region of the sphere every local minimum is near a {\em truncated} signed shift of $\mb a_0$, i.e., the restriction of $s_{\ell}[\mb a_0]$ to the window $\set{ 0, \dots, p_0 - 1}$. The analysis of \cite{zhang2018structured} allows the sparse sequence $\mb x_0$ to be denser ($\theta \sim p_0^{-2/3}$ for a generic kernel $\mb a_0$, as opposed to $\theta \lesssim p_0^{-3/4}$ in our result). Both \cite{zhang2017global} and \cite{zhang2018structured} guarantee {\em approximate} recovery of a portion of $s_{\ell}[\mb a_0]$, under  complicated conditions on the kernel $\mb a_0$. Our core optimization problem is very similar to \cite{zhang2017global}. However, we obtains {\em exact} recovery of both $\mb a_0$ and relatively dense $\mb x_0$, under the much simpler assumption of shift incoherence.

\paragraph{Identifiability in SaS deconvolution.} Other aspects of the SaS problem have been studied theoretically. One basic question is under what circumstances the problem is identifiable, up to the scaled shift ambiguity.  \cite{choudhary2015fundamental} shows that the problem ill-posed for worst case $(\mb a_0,\mb x_0)$ -- in particular, for certain support patterns in which $\mb x_0$ does not have any isolated nonzero entries. This demonstrates that {\em some} modeling assumptions on the support of the sparse term are needed. At the same time, this worst case structure is unlikely to occur, either under the Bernoulli model, or in practical deconvolution problems.

\paragraph{Other low dimensional deconvolution models.} 

Motivated by a variety of applications, many low-dimensional deconvolution models have been studied in the theoretical literature. In communication applications, the signals $\mb a_0$ and $\mb x_0$ either live in known low-dimensional subspaces, or are sparse in some known dictionary \cite{ahmed2014blind, li2016identifiability, Chi2016-TIP,Ling2015-IP, Li17-IT,Ling2017-IT, kech2017optimal}. These theoretical works assume that the subspace / dictionary are chosen at random. This low-dimensional deconvolution model does not exhibit the signed shift ambiguity; nonconvex formulations for this model exhibit a different structure from that studied here. In fact, the variant in which both signals belong to known subspaces can be solved by convex relaxation \cite{ahmed2014blind}. The SaS model does not appear to be amenable to convexification, and exhibits a more complicated nonconvex geometry, due to the shift ambiguity. The main motivation for tackling this model lies in the aforementioned applications in imaging and data analysis.

\cite{wang2016blind, Li18-multiBD} study the related {\em multi-instance} sparse blind deconvolution problem (MISBD), where there are $K$ observations $\mb y_i = \mb a_0 * \mb x_i$ consisting of multiple convolutions $i = 1,\ldots,K$ of a kernel $\mb a_0$ and different sparse vectors $\mb x_i$. Both works develop provable algorithms. There are several key differences with our work. First, both the proposed algorithms and their analysis require the kernel to be invertible. Second, despite the apparent similarity between the SaS model and MISBD, these problems are not equivalent. It might seem possible to reduce SaS to MISBD by dividing the single observation $\mb y$ into $K$ pieces; this apparent reduction fails due to boundary effects.

\subsection{Notations}

\paragraph{Vectors and indices.} All vectors/matrices are written in bold font $\mb a$/$\mb A$; indexed values are written as $\mb a_i$, $\mb A_{ij}$. Zeros or ones vectors are defined as $\mb 0$ or $\1$, and $i$-th canonical basis vector defined as $\mb e_i$. The indices for vectors/matrices all start from 0 and is taking modulo-$n$, thus a vector of length $n$ should has its indices labeled as $\Brac{0,1,\ldots,n-1}$. We write $[n] = \Brac{0,\ldots,n-1}$. We often use capitial italic symbols $I,J$ for subsets of $[n]$. We abuse notation slightly and write $[-p] = \set{n-p+1,\ldots,n-1,0}$ and $[\pm p] = \Brac{n-p+1,\ldots,n-1,0,1,\ldots,p-1}$. Index sets can be labels for vectors; $\mb a_I\in\R^{\abs I}$ denotes the restriction of the vector $\mb a$ to coordinates $I$. Also, we use check symbol for reversal operator on index set $\wc{I} = -I$ and vectors $\wc{\mb a}_i = \mb a_{-i} $.
  
\paragraph{Operators.} We let $\mb P_C$ denote the projection operator associated with a compact set $C$.  The zero-filling operator $\mb\iota_I:\R^{\abs{I}} \to \R^n$ injects the input vector to higher dimensional Euclidean space, via $(\mb\iota_I\mb x)_i = \mb x_{I^{-1}(i)}$ for $i\in I$ and $0$ otherwise. Its adjoint operator $\mb\iota^*_I$ can be understood as subset selection operator which picks up entries of coordinates $I$. A common zero-filling operator through out this paper $\ip$ is abbreviation of $\mb\iota_{[p]}$, which is often being addressed as zero-padding operator and its adjoint $\ip^*$ as truncation operator. 

\paragraph{Convolution} The convolution operator are all circular with modulo-$n$: $(\mb a*\mb x)_i = \sum_{j\in[n]}\mb a_{j}\mb x_{i-j}$, also, the convolution operator works on index set: $I*J = \supp{(\1_{I}*\1_J)}$. Similarly, the shift operator $ \shift{\cdot}{\ell}:\R^{p}\to\R^n$ is circular with modulo-n without specification: $(\shift{\mb a}{\ell})_j = (\ip_{[p]}\mb a)_{j-\ell}$. Notice that here $\mb a$ can be shorter $p\leq n$. Let $\mb C_{\mb a}\in\R^{n\cross n}$ denote a circulant matrix (with modulo-$n$) for vector $\mb a$, whose $j$-th column is the cyclic shift of $\mb a$ by $j$: $\mb C_{\mb a}\mb e_j = \shift{\mb a}{j}$. It satisfies for any $b\in\R^n$, 
\begin{align}
	\convmtx{\mb a}\mb b = \mb a*\mb b.
\end{align}
The correlation between $\mb a$ and $\mb b$ can be also written in similar form of convolution operator which reverse one vector before convolution. Define two correlation matrices $\convmtx{\mb a}^*$ and $\checkmtx{\mb a}$ as $\convmtx{\mb a}^*\mb e_j = \shift{\wc{\mb a}}{j}$ and $\checkmtx{\mb a}\mb e_j = \shift{\mb a}{-j}$. The two operators will satisfy
\begin{align}\label{eqn:crosscorr_y_a}
	\convmtx{\mb a}^*\mb b = \wc{\mb a}*\mb b,\quad \checkmtx{\mb a}\mb b = \mb a*\wc{\mb b}.
\end{align}

% !TEX root = ../BD_DQ.tex

\section{Geometry of $\varphi_\rho$ in Shift Space} \label{sec:geometry}

Underlying our main geometric and algorithmic results is a relationship between the geometry of the function $\varphi_\rho$ and the symmetries of the deconvolution problem. In this section, we describe this relationship at a more technical level, by interpreting the gradient and hessian of the function $\varphi_\rho$ in terms of the shifts $s_{\ell}[\mb a_0]$ and stating a key lemma which asserts that a certain neighborhood of the union of subspaces $\Sigma_{4 \theta p_0}$ can be decomposed into regions of negative curvature, strong gradient, and strong convexity near the target solutions $\pm s_{\ell}[\mb a_0]$.

\subsection{Shifts and Correlations}
The set $\Sigma_{4\theta p_0}$ is a union of subspaces. Any point $\mb a$ in one of these subspaces $\mc S_{\mb \tau}$ is a superposition of shifts of $\mb a_0$:
\begin{equation}
\mb a =\sum_{\ell \in \mb \tau} \mb\alpha_\ell s_{\ell}[\mb a_0].
\end{equation}
This representation can be extended to a general point $\mb a \in \bb S^{p-1}$ by writing \begin{equation}
\mb a =\sum_{\ell \in \mb \tau} \mb\alpha_\ell s_{\ell}[\mb a_0]+\sum_{\ell \notin \mb \tau} \mb\alpha_\ell s_{\ell}[\mb a_0].
\label{eqn:alph-general}
\end{equation}
The vector $\mb\alpha$ can be viewed as the coefficients of a decomposition of $\mb a$ into different shifts of $\mb a_0$. This representation is not unique. For $\mb a$ close to $\mc S_{\mb \tau}$, we can choose a particular $\mb \alpha$ for which $\mb \alpha_{\tau^c}$ is small, a notion that we will formalize below.

For convenience, we introduce a closely related vector $\mb\beta\in\R^n$, whose entries are the inner products between $\mb a$ and the shifts of $\mb a_0$: $\mb \beta_{\ell} = \innerprod{ \mb a }{ s_{\ell}[\mb a_0] }$. Since the columns of $\convmtx{\mb a_0}$ are the shifts of $\mb a_0$, we can write
\begin{align} \label{eqn:coef-beta}
	\mb\beta &= \convmtx{\mb a_0}^*\ip\mb a \\
&=\convmtx{\mb a_0}^*\ip\ip^*\convmtx{\mb a_0}\mb\alpha =: \mb M\mb\alpha.
\end{align}
The matrix $\mb M$ is the Gram matrix of the truncated shifts $\injector^* s_{\ell}[\mb a_0]$: $\mb M_{ij} = \innerprod{ \injector^* s_{i}[\mb a_0] }{ \injector^* s_j[\mb a_0] }$. When $\mu$ is small, the off-diagonal elements of $\mb M$ are small. In particular, on $\mc S_{\tau}$ we may take $\mb \alpha_{\tau^c} = \mb 0$, and $\mb \beta \approx \mb \alpha$, in the sense that $\mb \beta_\tau \approx \mb \alpha_\tau$ and the entries of $\mb \beta_{\tau^c}$ are small. For detailed elaboration, see \Cref{sec:vector_shifts}.

\subsection{Shifts and the Calculus of $\varphi_{\ell^1}$}
\label{sec:chi}

Our main geometric claims pertain to the function $\varphi_\rho$, which is based on a smooth sparsity surrogate $\rho( \cdot ) \approx \norm{\cdot}{1}$. In this section, we sketch the main ideas of the proof as if $\rho(\cdot) = \| \cdot \|_1$, by relating the geometry of the function $\varphi_{\ell^1}$ to the vectors $\mb \alpha$, $\mb \beta$ introduced above.
Working with $\varphi_{\ell^1}$ simplifies the exposition; it is also faithful to the structure of our proof, which relates the derivatives of the smooth function $\varphi_\rho$ to similar quantities associated with the nonsmooth function $\varphi_{\ell^1}$.

The function $\varphi_{\ell^1}$ has a relatively simple closed form:
\begin{equation}
\varphi_{\ell^1}(\mb a) = -\tfrac{1}{2} \norm{ \mc S_{\lambda}\left[ \, \wc{\mb y} \ast \mb a \, \right] }{2}^2.
\end{equation} 
Here, $\mc S_\lambda$ is the {\em soft thresholding operator}, which is defined for scalars $t$ as $\mc S_\lambda[t] = \mr{sign}(t) \max\set{|t| - \lambda, 0}$, and is extended to vectors by applying it elementwise. The operator $\mc S_\lambda[\mb x]$ shrinks the elements of $\mb x$ towards zero. Small elements become identically zero, resulting in a sparse vector.

\subsection*{Gradient: Sparsifying the Correlations \texorpdfstring{$\mb \beta$}{beta}}

\paragraph{Gradient over Euclidean space.} Our goal is to understand the local minimizers of the function $\varphi_{\ell^1}$ over the sphere. The function $\varphi_{\ell^1}$ is differentiable. Clearly, any point $\mb a$ at which its gradient (over the sphere) is nonzero cannot be a local minimizer. We first give an expression for the gradient of $\varphi_{\ell^1}$ over Euclidean space $\R^p$, and then extend it to the sphere $\bb S^{p-1}$. Using $\mb y = \mb a_0 \ast \mb x_0$ and calculus gives
\begin{align}\label{eqn:euc_grad}
\nabla\varphi_{\ell^1}(\mb a) &= -\ip^* \mb C_{\mb a_0} \checkmtx{\mb x_0} \mc S_\lambda \left[ \checkmtx{\mb x_0} \mb C_{\mb a_0}^* \injector \mb a \right]  \nonumber \\
&= -\ip^* \mb C_{\mb a_0} \, \checkmtx{\mb x_0} \mc S_\lambda\left[ \checkmtx{\mb x_0} \mb \beta \right] \nonumber \\
&= -\ip^* \mb C_{\mb a_0} \mb \chi[\mb \beta],
\end{align}
where we have simplified the notation by introducing an operator $\mb \chi : \R^n \to \R^n$ as $\mb\chi[\mb\beta]=  \checkmtx{\mb x_0}\soft{\checkmtx{\mb x_0}\mb\beta}{\lambda}$. This representation exhibits the (negative) gradient as a superposition of shifts of $\mb a_0$ with coefficients given by the entries of $\mb \chi[\mb \beta]$:
\begin{align} \label{eqn:neggrad}
-\nabla \varphi_{\ell^1}(\mb a) &=  \sum_\ell  \mb \chi[\mb \beta]_\ell \, s_\ell[\mb a_0].
\end{align}
The operator $\mb \chi$  appears complicated. However, its effect is relatively simple: {\em when $\mb x_0$ is a long random vector, $\mb \chi[\mb \beta]$ acts like a soft thresholding operator on the vector $\mb \beta$}. That is,
\begin{align}\label{eqn:chi_approx_soft_thresh}
	\frac{1}{n\theta}\cdot\mb\chi[\mb\beta]_\ell \approx\caseof{\mb\beta_\ell-\lambda,&\qquad \mb\beta_\ell  >  \lambda \\ \mb\beta_\ell + \lambda, &\qquad \mb\beta_\ell < -\lambda\\ 0, &\qquad \text{otherwise} }.  
\end{align} 
We show this rigorously below, in the proof of our main theorems. Here, we support this claim pictorially, by plotting the $\ell$-th entry $\mb \chi[\mb \beta]_\ell$ as $\mb \beta_{\ell}$ varies -- see \Cref{fig:gradient} (middle left) and compare to \Cref{fig:gradient} (left). Because $\mb \chi[\mb \beta]$ suppresses small entries of $\mb \beta$, the strongest contributions to $-\nabla {\varphi_{\ell^1}}$ in \eqref{eqn:neggrad} will come from shifts $s_\ell[\mb a_0]$ with large $\mb \beta_{\ell}$. {\em In particular, the Euclidean gradient is large whenever there is a single preferred shift $s_{\ell}[\mb a_0]$, i.e., the largest entry of $\mb \beta$ is significantly larger than the second largest entry. }

\begin{figure}[t]
\centering
\input{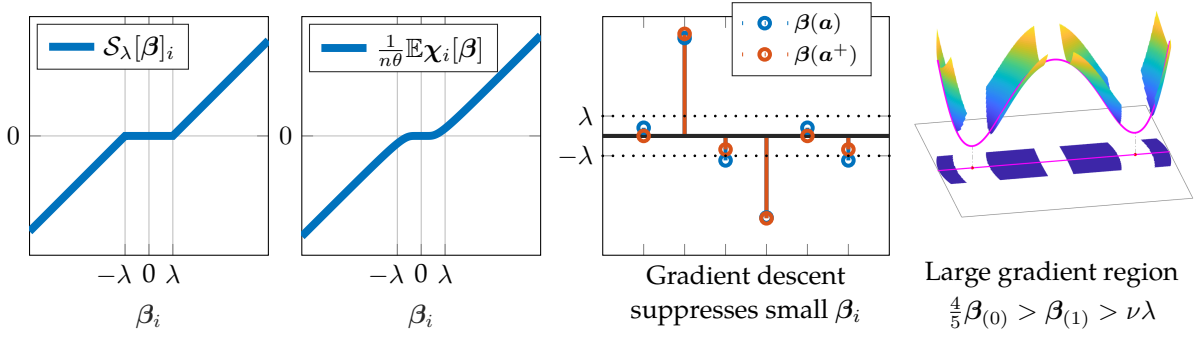}
\caption{{\bf Gradient Sparsifies Correlations.}
 Left: the soft thresholding operator $\mc S_\lambda[\mb \beta]$ shrinks the entries of $\mb \beta$ towards zero, making it sparser.
 Middle left: the negative gradient $- \nabla \varphi_{\ell^1}$ is a superposition of shifts $s_{\ell}[\mb a_0]$, with coefficients $\mb \chi_{\ell}[\mb \beta] \approx \mc S_{\lambda}[\mb \beta]_{\ell}$. Because of this, gradient descent sparsifies $\mb \beta$. Middle right: $\mb \beta(\mb a)$ before, and $\mb \beta(\mb a^+)$ after, one projected gradient step $\mb a^+ = \mb P_{\bb S^{p-1}}[ \mb a - t \cdot \mr{grad}[\varphi_{\ell^1}](\mb a) ]$. Notice that the small entries of $\mb \beta$ are shrunk towards zero. Right: the gradient  $\mr{grad}[\varphi_{\ell^1}](\mb a)$ is large whenever it is easy to sparsify $\mb \beta$; in particular, when the largest entry $\mb\beta_{(0)} \gg \mb\beta_{(1)} \gg 0$.
} \label{fig:gradient} 
\end{figure}

\paragraph{Gradient over Sphere.} The (Euclidean) gradient $\nabla {\varphi_{\ell^1}}$ measures the slope of $\varphi_{\ell^1}$ over $\R^n$. We are interested in the slope of $\varphi_{\ell^1}$ over the sphere $\bb S^{p-1}$, which is measured by the Riemannian gradient
\begin{align}
\mr{grad}[\varphi_{\ell^1}](\mb a ) &= \mb P_{\mb a^\perp} \nabla \varphi_{\ell^1}(\mb a) \nonumber \\
 &= -\mb P_{\mb a^\perp} \sum_{\ell} \mb \chi_{\ell}[\mb \beta] \, s_{\ell}[\mb a_0].
\end{align}
The Riemannian gradient simply projects the Euclidean gradient onto the tangent space $\mb a^\perp$ to $\bb S^{p-1}$ at $\mb a$. The Riemannian gradient is large whenever
\begin{itemize}
\item[(i)] {\bf Negative gradient points to one particular shift}: there is a single preferred shift $s_{\ell}[\mb a_0]$ so that the Euclidean gradient is large {\em and}
\item[(ii)] {\bf $\mb a$ is not too close to any shift}: it is possible to move in the tangent space in the direction of this shift.\footnote{...so the projection of the Euclidean gradient onto the tangent space does not vanish.} Since the tangent space consists of those vectors orthogonal to $\mb a$, this is possible whenever $s_{\ell}[\mb a_0]$ is not too aligned with $\mb a$, i.e., $\mb a$ is not too close to $s_{\ell}[\mb a_0]$.
\end{itemize}
Our technical lemma quantifies this situation in terms of the ordered entries of $\mb \beta$. Write $| \mb\beta_{(0)}| \ge |  \mb\beta_{(1)} | \ge \dots$, with corresponding shifts $s_{(0)}[\mb a_0], s_{(1)}[\mb a_0], \dots$. There is a strong gradient whenever $|\mb\beta_{(0)}|$ is significantly larger than $|\mb\beta_{(1)}|$ and $|\mb\beta_{(1)}|$ is not too small compared to $\lambda$: in particular, when $\tfrac{4}{5} |\mb\beta_{(0)}| > |\mb\beta_{(1)}| > \frac{\lambda}{4\log^2\theta^{-1}} $. In this situation, gradient descent drives $\mb a$ toward $s_{(0)}[\mb a_0]$, reducing $|\mb\beta_{(1)}|, \dots$, and making the vector $\mb \beta$ sparser. We establish the technical claim that the (Euclidean) gradient of $\varphi_{\ell^1}$ sparsifies vectors in shift space in \Cref{sec:grad_soft_thresh}.

\subsection*{Hessian: Negative Curvature Breaks Symmetry} % $\,$ \& $\,$ Positive Curvature away from $\mc S_{\mb\tau}$} 

When there is no  single preferred shift, i.e., when $|  \mb\beta_{(1)} |$ is close to $| \mb\beta_{(0)}|$, the gradient can be small. Similarly, when $\mb a$ is very close to $\pm s_{(0)}[\mb a_0]$, the gradient can be small. In either of these situations, we need to study the curvature of the function $\varphi$ to determine whether there are local minimizers.

\paragraph{Nonsmoothness.} Strictly speaking, the function $\varphi_{\ell^1}$ is not twice differentiable, due to the nonsmoothness of the soft thresholding operator $\mc S_\lambda[t]$ at $t = \pm \lambda$. Indeed, $\varphi_{\ell^1}$ is nonsmooth at any point $\mb a$ for which some entry of $\wc{\mb y} \ast \mb a$ has magnitude $\lambda$. At other points $\mb a$, $\varphi_{\ell^1}$ is twice differentiable, and its Hessian is given by
\begin{align} \label{eqn:euc_hess}
\wt{\nabla}^2\varphi_{\ell^1}(\mb a) = -\ip^*\mb C_{\mb a_0} \checkmtx{\mb x_0}\mb P_{I}\checkmtx{\mb x_0} \mb C_{\mb a_0}^* \ip,
\end{align}
with $I  = \supp\paren{\soft{\checkmtx{\mb y}\ip\mb a}{\lambda}}$. We (formally) extend this expression to {\em every} $\mb a \in \R^n$, terming $\wt{\nabla}^2 \varphi_{\ell^1}$ the {\em pseudo-Hessian} of $\varphi_{\ell^1}$. For appropriately chosen smooth sparsity surrogate $\rho$, we will see that the (true) Hessian of the smooth function $\nabla^2 \varphi_\rho$ is close to $\wt{\nabla}^2 \varphi_{\ell^1}$, and so $\wt{\nabla}^2 \varphi_{\ell^1}$ yields useful information about the curvature of $\varphi_\rho$.

\paragraph{Curvature over Euclidean Space.} As with the gradient, the Hessian is complicated, but becomes simpler when the sample size is large. The following approximation
\begin{align}
\wt{\nabla}^2 \varphi_{\ell^1}(\mb a) &\approx - \sum_{\ell} s_{\ell}[\mb a_0]  s_{\ell}[\mb a_0]^* \left( \frac{\partial}{\partial \mb \beta_{\ell}} \mb \chi_{\ell}[\mb \beta] \right)  \label{eqn:euc-hess-appx}
\end{align}
can be obtained from \eqref{eqn:neggrad} noting that $\frac{\partial}{\partial \mb a} \mb \chi_{\ell}[\mb \beta] = \sum_j s_j[\mb a_0] \frac{\partial}{\partial \mb \beta_j } \mb \chi_\ell[\mb \beta]$, that $\frac{\partial}{\partial \mb \beta_j } \mb \chi_\ell[\mb \beta] \approx 0$ for $j \ne \ell$, and  that
\begin{equation}\label{eqn:hessian_approx_logic}
\frac{1}{n\theta}\cdot\frac{\partial \mb \chi_{\ell}[\mb \beta]}{\partial \mb \beta_{\ell}}  \approx \begin{cases} 0 & |\mb \beta_{\ell}| \ll \lambda \\ 1 & |\mb \beta_{\ell}| \gg \lambda \end{cases}
\end{equation}
Again, we corroborate this approximation pictorially -- see \Cref{fig:hessian}.

From this approximation, we can see that the quadratic form $\mb v^* \wt{\nabla}^2 \varphi_{\ell^1} \mb v$ takes on a large negative value whenever $\mb v$ is a shift $s_{\ell}[\mb a_0]$ corresponding to some $|\mb \beta_{\ell}| \ge \lambda$, or whenever $\mb v$ is a linear combination of such shifts. {\em In particular, if for some $j$, $|\mb \beta_{(0)}|, |\mb \beta_{(1)}|, \dots, | \mb \beta_{(j)}|  \gg \lambda$, then $\varphi_{\ell^1}$ will exhibit negative curvature in any direction $\mb v \in \mr{span}( s_{(0)}[\mb a_0], s_{(1)}[\mb a_0], \dots, s_{(j)}[\mb a_0] )$.}

\begin{figure}[t]
\centering
\input{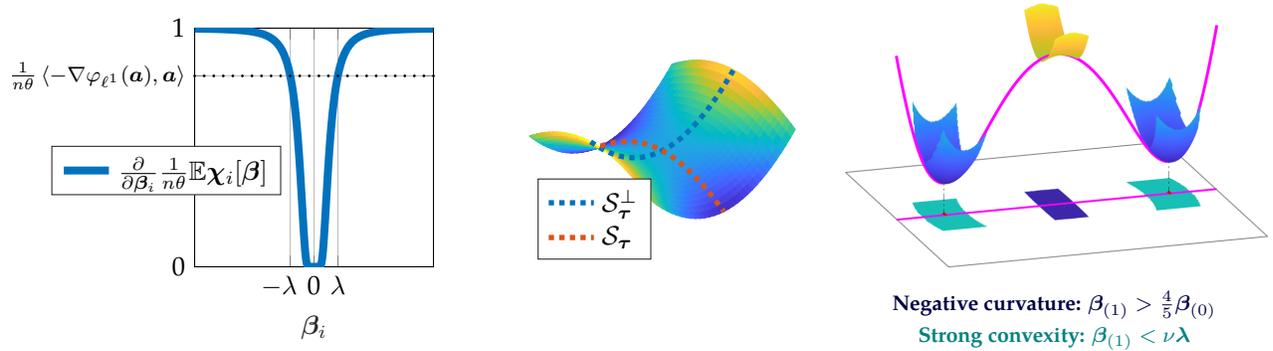}
\caption{{\bf Hessian Breaks Symmetry.} Left: contribution of $-s_i[\mb a_0] s_i[\mb a_0]^*$ to the Euclidean hessian. If $|\mb\beta_i| \gg \lambda$ the Euclidean hessian exhibits a strong negative component in the $s_i[\mb a_0]$ direction. The Riemannian hessian exhibits negative curvature in directions spanned by $s_i[\mb a_0]$ with corresponding $|\mb\beta_i| \gg \lambda$ and positive curvature in directions spanned  by $s_i[\mb a_0]$ with $|\mb\beta_i| \ll \lambda$. Middle: this creates negative curvature along the subspace $\mc S_{\mb \tau}$ and positive curvature orthogonal to this subspace. Right: our analysis shows that there is always a direction of negative curvature when $\mb\beta_{(1)} > \tfrac{4}{5} \mb\beta_{(0)}$; conversely when $\mb\beta_{(1)} \ll \lambda$ there is positive curvature in every feasible direction and the function is strongly convex.}\label{fig:hessian}
\end{figure}

\paragraph{Curvature over the Sphere.} The (Euclidean) Hessian measures the curvature of the function $\varphi_{\ell^1}$ over $\R^n$. The Riemannian Hessian  \begin{equation}
\wt{\mr{Hess}} [\varphi_{\ell^1}](\mb a) = \mb P_{\mb a^\perp} \left( \underset{\text{\color{purple} \bf \footnotesize Curvature of $\varphi_{\ell^1}$}}{\wt{\nabla}^2 \varphi_{\ell^1}(\mb a)} \quad + \quad   \underset{\text{\color{purple} \bf \footnotesize Curvature of the sphere}}{  \innerprod{ -\nabla \varphi_{\ell^1}(\mb a ) }{\mb a } \cdot \mb I } \right) \mb P_{\mb a^\perp}.
\end{equation}
measures the curvature of $\varphi_{\ell^1}$ over the sphere. The projection $\mb P_{\mb a^\perp}$ restricts its action to directions $\mb v \perp \mb a$ that are tangent to the sphere. The additional term $\innerprod{ -\nabla \varphi_{\ell^1}(\mb a) }{\mb a }$ accounts for the curvature of the sphere. This term is always positive. The net effect is that directions of strong negative curvature of $\varphi_{\ell^1}$ over $\R^n$ become directions of moderate negative curvature over the sphere. Directions of nearly zero curvature over $\R^n$ become directions of positive curvature over the sphere. This has three implications for the geometry of $\varphi_{\ell^1}$ over the sphere:
\begin{itemize}
\item[{\bf (i)}] {\bf Negative curvature in symmetry breaking directions}: If $| \mb\beta_{(0)}|, | \mb\beta_{(1)}|, \dots, |  \mb\beta_{(j)}|  \gg \lambda$, $\varphi_{\ell^1}$ will exhibit negative curvature in any tangent direction $\mb v \perp \mb a$ which is in the linear span $$\mr{span}( s_{(0)}[\mb a_0], s_{(1)}[\mb a_0], \dots, s_{(j)}[\mb a_0] )$$ of the corresponding shifts of $\mb a_0$. 
\item[{\bf (ii)}] {\bf Positive curvature in directions away from $\mc S_{\mb \tau}$}: The Euclidean Hessian quadratic form $\mb v^* \wt{\nabla}^2 \varphi_{\ell^1} \mb v$ takes on relatively small values in directions orthogonal to the subspace $\mc S_{\mb \tau}$. The Riemannian Hessian is positive in these directions, creating positive curvature orthogonal to the subspace $\mc S_{\mb \tau}$.
    
\item[{\bf (iii)}] {\bf Strong convexity around minimizers}: Around a minimizer $s_{\ell}[\mb a_0]$, only a single entry $\mb \beta_{\ell}$ is large. Any tangent direction $\mb v \perp \mb a$ is nearly orthogonal to the subspace $\mr{span}( s_{\ell}[\mb a_0] )$, and hence is a direction of positive (Riemannian) curvature. The objective function $\varphi_{\rho}$ is strongly convex around the target solutions $\pm s_{\ell}[\mb a_0]$.  
\end{itemize} 
\Cref{fig:hessian} visualizes these regions of negative and positive curvature, and the technical claim of positivity/negativity of curvature in shift space is presented in detail in \Cref{sec:hessian_logic}.

\subsection{Any Local Minimizer is a Near Shift}

We close this section by stating a key theorem, which makes the above discussion precise. We will show that a certain neighborhood of any subspace $\mc S_{\mb \tau}$ can be covered by regions of \emph{negative curvature}, \emph{large gradient}, and regions of \emph{strong convexity} containing target solutions $\pm s_{\ell}[\mb a_0]$. Furthermore, at the boundary of this neighborhood, the negative gradient points back---\emph{retracts}---toward the subspace $\mc S_{\mb\tau}$, due to the (directional) convexity of $\varphi_\rho$ away from the subspace.
 
\paragraph{Widened subspace region.} To formally state the result, we need a way of measuring how close $\mb a$ is to the subspace $\mc S_{\mb \tau}$. For technical reasons, it turns out to be convenient to do this in terms of the coefficients $\mb \alpha$ in the representation
\begin{equation} \label{eqn:a-rep}
\mb a = \sum_{\ell \in \mb \tau} \mb\alpha_{\ell} s_{\ell}[\mb a_0] + \sum_{\ell' \in \mb \tau^c} \mb\alpha_{\ell'} s_{\ell'}[\mb a_0].
\end{equation}
If $\mb a \in \mc S_{\mb \tau}$, we can take $\mb \alpha$ with $\mb \alpha_{\tau^c} = \mb 0$. We can view the energy $\| \mb \alpha_{\tau^c} \|_2$ as a measure of the distance from $\mb a$ to $\mc S_{\mb \tau}$. A technical wrinkle arises, because the representation \eqref{eqn:a-rep} is not unique. We resolve this issue by choosing the $\mb \alpha$ that minimizes $\| \mb \alpha_{\tau^c} \|_2$, writing:
\begin{align}\label{eqn:dist_s_tau}
	d_\alpha(\mb a,\mc S_{\mb\tau}) = \inf\set{\norm{\mb\alpha_{\mb\tau^c}}2 \, : \, \textstyle\sum_{\ell} \mb\alpha_{\ell} s_{\ell}[\mb a_0] = \mb a }.
\end{align}
The distance $d_\alpha( \mb a, \mc S_{\mb \tau})$ is zero for $\mb a \in \mc S_{\mb \tau}$. Our analysis controls the geometric properties of $\varphi_{\rho}$ over the set of $\mb a$ for which $d_\alpha( \mb a, \mc S_{\mb \tau})$ is not too large. Similar to \eqref{eqn:union_subspaces_4tp0}, we define an object which contains all points that are close to some $\mc S_{\mb \tau}$, in the above sense:
\begin{align}
	\Sigma_{4\theta p_0}^\gamma  := \bigcup_{\abs{\mb\tau}\leq 4\theta p_0}\set{\mb a\,:\, d_\alpha(\mb a,\mc S_{\mb\tau}) \leq \gamma}.  
\end{align} 
The aforementioned geometric properties hold over this set:

\begin{theorem}[Three subregions]\label{thm:three_regions} Suppose that $\mb y = \mb a_0*\mb x_0$ where $\mb a_0\in\bb S^{p_0-1}$ is $\mu$-shift coherent and $\mb x_0\simiid\mr{BG}(\theta)\in\R^n$ satisfying
\vspace{-0.1in}
\begin{align}
	\theta  \in \brac{\frac{c'}{p_0}, \frac{c}{p_0\sqrt\mu + \sqrt{p_0}}}\cdot\frac{1}{\log^2 p_0}   
\end{align}  
for some constants $c',c > 0$. Set $\lambda = 0.1 / \sqrt{p_0\theta}$ in $\varphi_\rho$ where $\rho(x) = \sqrt{x^2 + \delta^2}$. There exist numerical constants $C,c'',c''',c_1$-$c_4>0$ such that if $\delta \leq \tfrac{c''\lambda\theta^8}{p^2\log^2n} $ and  $ n > Cp_0^5\theta^{-2}\log p_0$, then with probability at least $1-c'''/n$, for every $\mb a \in \Sigma_{4\theta p_0}^\gamma$, we have:   
	\begin{itemize}
		\item (Negative curvature): If $\abs{\mb\beta_{(1)}} \geq  \nu_1\abs{\mb\beta_{(0)}}$, then
	\begin{align}
		\lambda_{\mr{min}}\paren{\mr{Hess}[\varphi_{\rho}](\mb a)} \,\leq\, -c_1n\theta\lambda;
	\end{align}
		\item (Large gradient): If $\nu_1\abs{\mb\beta_{(0)}}\geq\abs{\mb\beta_{(1)}}\geq \nu_2(\theta)\lambda $, then
	\begin{align}
		\norm{\mr{grad}[\varphi_\rho](\mb a)}2 \,\geq\,  c_2n\theta\tfrac{\lambda^2}{\log^2\theta^{-1}};
	\end{align}
		\item (Convex near shifts): If $\nu_2(\theta)\lambda\geq \abs{\mb\beta_{(1)}} $, then
	\begin{align}
		\mr{Hess}[\varphi_{\rho}](\mb a) \,\succ\,  c_3n\theta\mb P_{\mb a^\perp};
	\end{align}
		\item (Retraction to subspace): If $\tfrac{\gamma}{2} \leq d_\alpha(\mb a,\mc S_{\mb\tau})\leq \gamma $, then for every $\mb\alpha$ satisfying $\mb a = \ip^*\convmtx{\mb a_0}\mb\alpha$, there exists $\mb\zeta$ satisfying $\grad[\varphi_\rho](\mb a) = \ip^*\convmtx{\mb a_0}\mb\zeta$, such that  
	\begin{align}\label{eqn:thm_three_region_retract}
		\innerprod{\mb\zeta_{\mb\tau^c}}{\mb \alpha_{\mb\tau^c}} \,\geq\, c_4\norm{\mb\zeta_{\mb\tau^c}}2\norm{\mb\alpha_{\mb\tau^c}}2; 
	\end{align}
		\item (Local minimizers):  If $\mb a$ is a local minimizer,
		\begin{align}\label{eqn:main_thm_local_min}
			\min_{\substack{\ell\in[\pm p] \\\sigma\in\set{\pm 1}}} \norm{\mb a - \sigma\,\shift{\mb a_0}{\ell}}2  \,\leq\,  \tfrac12\max\set{\mu, p_0^{-1}},
		\end{align}
	\end{itemize}\vspace{-0.1in}
	where  $\nu_1 = \tfrac45$,  $\nu_2(\theta) = \frac{1}{4\log^2\theta^{-1}}$ and $\gamma = \frac{c\cdot\poly(\sqrt{1/\theta},\,\sqrt{1/\mu})}{\log^2\theta^{-1}}\cdot \frac{1}{\sqrt{p_0}}$.  
\end{theorem}   
\begin{proof}
	See \Cref{sec:proof_three_region}.
\end{proof}
The retraction property elaborated in \eqref{eqn:thm_three_region_retract} implies that the negative gradient at $\mb a$ points in a direction that decreases $d_\alpha(\mb a,\mc S_{\mb\tau})$. This is a consequence of positive curvature away from $\mc S_{\mb\tau}$. It essentially implies that the gradient is monotone in $\mb\alpha_{\mb\tau^c}$ space: choose any $\ul{\mb a}\in\mc S_{\mb\tau}\cap\Sp^{p-1}$, write $\ul{\mb\alpha}$ to be its coefficient, and let $\ul{\mb\zeta}$ be the coefficient of $\grad[\varphi_\rho](\ul{\mb a})$. Then $\ul{\mb\alpha}{}_{\mb\tau^c} = \mb 0$, $\ul{\mb\zeta}{}_{\mb\tau^c}\approx \mb 0 $ and  
\begin{align} 
	\langle \mb\zeta_{\mb\tau^c}-\ul{\mb\zeta}{}_{\mb\tau^c},\,\mb\alpha_{\mb\tau^c}-\ul{\mb\alpha}_{\mb\tau^c}\rangle \approx \innerprod{\mb\zeta_{\mb\tau^c}-\mb 0}{\,\mb\alpha_{\mb\tau^c}-\mb 0}   = \innerprod{\mb \zeta_{\mb\tau^c}}{\mb\alpha_{\mb\tau^c}}  >    0.\notag 
\end{align} 

Our main geometric claim in  \Cref{thm:main} is a direct consequence of \Cref{thm:three_regions}. Moreover, it suggests that as long as we can minimize $\varphi_\rho$ within the region $\Sigma^\gamma_{4 \theta p_0}$, we will solve the SaS deconvolution problem. 

% !TEX root = ../BD_DQ.tex

\section{Provable Algorithm} \label{sec:alg}
In light of \Cref{thm:three_regions}, in this section we introduce a two-part algorithm \Cref{alg:ssbd}, which first applies the curvilinear descent method to find a local minimum of $\varphi_\rho$ within $\Sigma_{4\theta p_0}^\gamma$, followed by refinement algorithm that uses alternating minimization to exactly recover the ground truth. This algorithm exactly solves SaS deconvolution problem.

\subsection{Minimization}
\label{sec:alg_minimization}
There are three major issues in finding a local minimizer within $\Sigma_{4\theta p_0}^\gamma$. We want \dots
\begin{itemize}
	\item[] {\bf (i) Initialization.} the initializer $\mb a^{(0)}$ to reside within $\Sigma_{4\theta p_0}^\gamma$,
	\item[] {\bf (ii) Negative curvature.}  the method to avoid stagnating near the saddle points of $\varphi_\rho$,
	\item[] {\bf (iii) No exit.} the descent method to remain inside $\Sigma_{4\theta p_0}^\gamma$.
\end{itemize}
In the following paragraphs, we describe how our proposed algorithm achieves the above desiderata.

\paragraph{Initialization within $\Sigma_{4\theta p_0}^\gamma$.}  
Our data-driven initialization scheme produces $\mb a^{(0)}$, where 
\begin{align}
\mb a^{(0)} &= -\mb P_{\bb S^{p-1}} \nabla \varphi_\rho\left( \mb P_{\bb S^{p-1}} \brac{\mb 0^{p_0-1};\mb y_0;\cdots;\mb y_{p_0-1};\mb 0^{p_0-1} } \right) \notag  \\
	&= -\mb P_{\Sp^{p-1}}\nabla{\varphi_\rho}\mb P_{\Sp^{p-1}}\brac{\mb P_{[p_0]}(\mb a_0*\mb x_0)}, \notag \\
	&\approx -\mb P_{\Sp^{p-1}}\nabla{\varphi_\rho}\brac{\mb P_{[p_0]}(\mb a_0*\wt{\mb x}_0)}, \notag 
\end{align}
is the normalized gradient vector from a chunk of data $\mb a^{(-1)}:=\mb P_{[p_0]}(\mb a_0*\wt{\mb x}_0)$ with $\wt{\mb x}_0$ a normalized Bernoulli-Gaussian random vector of length $2p_0-1$. Since $\nabla\varphi_\rho\approx\nabla\varphi_{\ell^1}$, expand the gradient $\nabla{\varphi_{\ell^1}}$ and rewrite the gradient $\nabla_{\ell^1}(\mb a^{(-1)})$ in shift space, we get  
\begin{align}
	-\nabla\varphi_{\rho^1}(\mb a^{(-1)}) &\approx  \ip^* \mb C_{\mb a_0} \checkmtx{\mb x_0} \mc S_\lambda \left[ \checkmtx{\mb x_0} \mb C_{\mb a_0}^*\mb P_{[p_0]}(\mb a_0*\wt{\mb x}_0)  \right]  \nonumber \\
	&= \ip^* \mb C_{\mb a_0} \mb \chi\brac{\,\convmtx{\mb a_0}^*\mb P_{[p_0]}\convmtx{\mb a_0}\wt{\mb x}_0\,} \notag \\
	&\approx \ip^*\convmtx{\mb a_0}\mb\chi\brac{\wt{\mb x}_0} \notag \\
	&\approx n\theta\cdot \ip^*\convmtx{\mb a_0}\soft{\wt{\mb x}_0}{\lambda}, \notag 
\end{align}  
where the approximation in the third equation is accurate if the truncated shifts are incoherent
\begin{align}
	\max_{i\neq j}\abs{\innerprod{\injector_{p_0}^*\shift{\mb a_0}i}{\injector_{p_0}^*\shift{\mb a_0}{j}}} \leq \mu \ll 1.
\end{align} 
With this simple approximation, it comes clear that the coefficients (in shift space) of initializer $\mb a^{(0)}$, 
\begin{align}
	\mb a^{(0)} \,\approx\, \mb P_{\Sp^{p-1}}\ip^*\convmtx{\mb a_0}\soft{\wt{\mb x}_0}{\lambda},  
\end{align}
approximate $\soft{\wt{\mb x}_0}{\lambda}$, which resides near the subspace $\mc S_{\mb\tau}$, in which $\mb\tau$ contains the nonzero entries of $\wt{\mb x}_0$ on $\set{-p_0+1,\ldots,p_0-1}$. With high probability, the number of non-zero entries is $\abs{\mb\tau} \lessapprox 4 \theta p_0$, we therefore conclude that our initializer $\mb a^{(0)}$ satisfies  
\begin{align} 
	\mb a^{(0)} \in \Sigma_{4\theta p_0}^\gamma. 
\end{align} 
Furthermore, since $\wt{\mb x}_0$ is normalized, the largest magnitude for entries of $\abs{\wt{\mb x}_0}$ is likely to be around  $1/\sqrt{2p_0\theta}$. To ensure that $\soft{\wt{\mb x}_0}{\lambda}$ does not annihilate all nonzero entries of $\wt{\mb x}_0$ (otherwise our initializer $\mb a^{(0)}$ will become $\mb 0$), the ideal $\lambda$ should be slightly less then the largest magnitude of $\abs{\wt{\mb x}_0}$. We suggest setting $\lambda$ in $\varphi_\rho$ as
\begin{align}
	\lambda = \frac{c}{\sqrt{p_0\theta}}.
\end{align}
for some $c\in(0,1)$.

\paragraph{Minimize $\varphi_\rho$ within $\Sigma_{4\theta p_0}^\gamma$.} Many methods have been proposed to optimize functions whose saddle points exhibit strict negative curvature, including the noisy gradient method \cite{ge2015escaping}, trust region methods \cite{absil2009optimization,sun2017complete} and curvilinear search \cite{wen2013feasible}. Any of the above methods can be adapted to minimize $\varphi_\rho$. In this paper, we use \emph{curvilinear method with restricted stepsize} to demonstrate how to analyze an optimization problem using the geometric properties of $\varphi_\rho$ over $\Sigma_{4 \theta p_0}^\gamma$ -- in particular, negative curvature in symmetry-breaking directions and positive curvature away from $\mc S_{\mb \tau}$.

Curvilinear search uses an update strategy that combines the gradient $\mb g$ and a direction of negative curvature $\mb v$, which here we choose as an eigenvector of the hessian $\mb H$ with smallest eigenvalue, scaled such that $\mb v^* \mb g \ge 0$. In particular, we set
\begin{align}
	\mb a^+ \gets \mb P_{\Sp^{p-1}}\brac{\mb a - t \mb g - t^2\mb v}
\end{align}
For small $t$,
\begin{align}
	\varphi(\mb a^+) \approx \varphi(\mb a) +  \innerprod{\mb g}{\mb\xi} + \tfrac12\mb\xi^*\mb H \mb\xi.
\end{align}
Since $\mb\xi$ converges to $\mb 0$ only if $\mb a$ converges to the local minimizer (otherwise either gradient $\mb g$ is nonzero or there is a negative curvature direction $\mb v$), this iteration produces a local minimizer for $\varphi_\rho$, whose saddle points near any $\mc S_{\mb \tau}$ has negative curvature, we just need to ensure all iterates stays near some such subspace. We prove this by showing:
\begin{itemize}
	\item When $d_\alpha(\mb a,\mc S_{\mb\tau}) \leq \gamma$, curvilinear steps move a small distance away from the subspace:
\begin{align}
 	\abs{d_\alpha\paren{\mb a^+,\mc S_{\mb\tau}} - d_\alpha\paren{\mb a,\mc S_{\mb\tau}} }\leq\tfrac\gamma2. 
\end{align}
\item When  $d_\alpha(\mb a,\mc S_{\mb\tau})\in\brac{\tfrac\gamma2,\gamma}$, curvilinear steps retract toward subspace:
\begin{align}
		d_\alpha\paren{\mb a^+,\mc S_{\mb\tau}} \leq d_\alpha\paren{\mb a,\mc S_{\mb\tau}}.
\end{align}  
\end{itemize}
Together, we can prove that the iterates  $\mb a^{(k)}$ converge to a minimizer, and   
\begin{align}
	\forall\,k = 1,2,\ldots,\quad  \mb a^{(k)}\in \Sigma_{4\theta p_0}^\gamma.
\end{align}
We conclude this section with the following theorem:

\begin{theorem}[Convergence of retractive curvilinear search]\label{thm:minimization} Suppose signals $\mb a_0, \mb x_0$ satisfy the conditions of \Cref{thm:three_regions}, $\theta > 10^3c/p_0 $ ($c>1$), and $\mb a_0$ is $\mu$-truncated shift coherent $\max_{i\neq j}\abs{\innerprod{\ip_{p_0}^*\shift{\mb a_0}{i}}{\ip_{p_0}^*\shift{\mb a_0}j}}\leq \mu$. Write $\mb g = \grad[\varphi_\rho](\mb a)$ and $\mb H = \mr{Hess}[\varphi_\rho](\mb a)$. When the smallest eigenvalue of $\mb H$ is strictly smaller than $-\eta_v$ let $\mb v$ be the unit eigenvector of smallest eigenvalue, scaled so $\mb v^* \mb g \ge 0$; otherwise let $\mb v = \mb 0$. Define a sequence $\set{\mb a^{(k)}}_{k\in\N}$ where $\mb a^{(0)}$ equals \eqref{eqn:gpm} and for $k = 1,2,\ldots,K_1$:      
\begin{align}\label{eqn:curvi_steps}
	\mb a^{(k+1)} \gets \mb P_{\Sp^{p-1}}\brac{\mb a^{(k)}  - t\mb g^{(k)} - t^2\mb v^{(k)}}
\end{align}
with largest $t \in\left(0,\frac{0.1}{n\theta}\right]$ satisfying Armijo steplength:  
\begin{align}
 	\varphi_\rho(\mb a^{(k+1)}) < \varphi_{\rho}(\mb a^{(k)}) -  \tfrac12\paren{t\|\mb g^{(k)}\|_2^2 + \tfrac12 t^4 \eta_v\|\mb v^{(k)}\|_2^2 },\label{eqn:conv_curvi_armijo} 
\end{align}
then with probability at least $1- 1/c$, there exists some signed shift $\bar{\mb a} = \pm\shift{\mb a_0}{i}$ where $i\in[\pm p_0]$ such that $\norm{\mb a^{(k)} - \bar{\mb a}}2 \leq \mu + 1/p$ for all $k \ge K_1 = \poly(n,p)$. Here, $\eta_v = c'n\theta\lambda$ for some $c' < c_1$ in \Cref{thm:three_regions}. 
\end{theorem}
\begin{proof}
	See \Cref{sec:proof_of_minimization}.
\end{proof}

\subsection{Local Refinement}
In this section, we describe and analyze an algorithm which refines an estimate $\bar{\mb a} \approx \mb a_0$ of the kernel to exactly recover $(\mb a_0,\mb x_0)$. Set 
\begin{align}
	\mb a^{(0)}\gets\bar{\mb a},\qquad \lambda^{(0)} \gets C(p\theta+\log n)(\mu + 1/p),   \qquad  I^{(0)} \gets \supp(\soft{\convmtx{\bar{\mb a}}^*\mb y}{\lambda}).
\end{align}
We alternatively minimize the Lasso objective with respect to $\mb a$ and $\mb x$:
\begin{align}
	 \mb x^{(k+1)} &\;\gets\; \argmin_{\mb x}\tfrac12\|\mb a^{(k)}*\mb x - \mb y\|_2^2 + \lambda^{(k)}\sum_{i\not\in I^{(k)}}\abs{\mb x_i}, \label{eqn:refine_altmin_1}\\
	 \mb a^{(k+1)} &\;\gets\; \mb P_{\Sp^{p-1}}\big[\argmin_{\mb a}\tfrac12\|\mb a*\mb x^{(k+1)}-\mb y\|_2^2\big], \label{eqn:refine_altmin_2} \\
	 \lambda^{(k+1)} &\;\gets\; \tfrac12\lambda^{(k)}, \qquad  I^{(			k+1)} \;\gets\;  \supp\big(\mb x^{(k+1)}\big).\label{eqn:refine_altmin_3}
\end{align}   
One departure from standard alternating minimization procedures is our use of a continuation method, which (i) decreases $\lambda$ and (ii) maintains a running estimate $I^{(k)}$ of the support set. Our analysis will show that $\mb a^{(k)}$ converges to one of the signed shifts of $\mb a_0$ at a linear rate, in the sense that
\begin{align}
	\min_{\sigma\in\pm 1,\,\ell\in[\pm p_0]}\big\|\mb a^{(k)}-\sigma\cdot \shift{\mb a_0}{\ell}\big\|_2 \leq C'2^{-k}.
\end{align}
 
 \paragraph{Modified coherence and support density assumptions} It should be clear that exact recovery is unlikely if $\mb x_0$ contains many consecutive nonzero entries: in fact in this situation, even {\em non-blind} deconvolution fails. Therefore to obtain exact recovery it is necessary to put an upper bound on signal dimension $n$. Here, we introduce the notation $\kappa_I$ as an upper bound for number of nonzero entries of $\mb x_0$ in a length-$p$ window:
\begin{equation}
\kappa_I := 6\max\set{ \theta p, \log n},
\end{equation}
where the indexing and addition should be interpreted modulo $n$. We will denote the support sets of true sparse vector $\mb x_0$ and recovered $\mb x^{(k)}$ in the intermediate $k$-th steps as 
\begin{align}
 	 I = \supp(\mb x_0) ,\qquad \qquad I^{(k)} = \supp(\mb x^{(k)}),  
\end{align}  
then in the Bernoulli-Gaussian model, with high probability,
\begin{equation}\label{eqn:refine_mu_kappaI}
\max_{\ell} \big| I \cap \paren{[p]+\ell} \big| \leq\kappa_I. 
 \end{equation}
The $\log n$ term reflects the fact that as $n$ becomes enormous (exponential in $p$) eventually it becomes likely that some length-$p$ window of $\mb x_0$ is densely occupied. In our main theorem statement, we preclude this possibility by putting an upper bound on signal length $n$ with respect to window length $p$ and shift coherence $\mu$. We will assume
\begin{align}
	(\mu + 1/p)\cdot \kappa_I^2 < c
\end{align}
for some numerical constant $c\in(0,1)$.

\paragraph{Alternating minimization produces $\mb a$ that contracts toward $\mb a_0$.} Recall that \eqref{eqn:theta_mu_bound}} in \Cref{thm:three_regions} provides that 
\begin{align}
	\norm{\bar{\mb a}-\mb a_0}2 \leq \paren{\mu + 1/p},
\end{align}
which is sufficiently close to $\mb a_0$ as long as \eqref{eqn:refine_mu_kappaI} holds true. Here, we will elaborate this by showing a single iteration of alternating minimization algorithm \eqref{eqn:refine_altmin_1}-\eqref{eqn:refine_altmin_3}  is a contraction mapping for $\mb a$ toward $\mb a_0$.

To this end, at $k$-th iteration, write  $T = I^{(k)}$, $\,J = I^{(k+1)}$ and $\mb\sigma^{(k)} = \sign\paren{\mb x^{(k)}}$, then first observe that the solution to the reweighted Lasso problem \eqref{eqn:refine_altmin_1} can be written as 
\begin{align}\label{eqn:refine_nextx}
	\mb x^{(k+1)} = \injector_{J}\paren{\injector_{J}^*\convmtx{\mb a^{(k)}}^*\convmtx{\mb a^{(k)}}\injector_{J}}^{-1}\injector_{J}^*\paren{\convmtx{\mb a^{(k)}}^*\convmtx{\mb a_0}\mb x_0 - \lambda^{(k)}\mb P_{J\setminus T}\mb\sigma^{(k+1)}},
\end{align}
and the solution to least squares problem \eqref{eqn:refine_altmin_2} will be
\begin{align}\label{eqn:refine_nexta}
	\mb a^{(k+1)} \;=\;\paren{\ip^*\convmtx{\mb x^{(k+1)}}^*\convmtx{\mb x^{(k+1)}}\ip}^{-1}\paren{\ip^*\convmtx{\mb x^{(k+1)}}^*\convmtx{\mb x_0}\ip\mb a_0 }.
\end{align}
Here, we are going to illustrate the relationship between $\mb a^{(k+1)}-\mb a_0$ and $\mb a^{(k)}-\mb a_0$ using simple approximations. First, let us assume that  $\mb a^{(k)}\approx\mb a_0$, $\,\convmtx{\mb a_0}^*\convmtx{\mb a_0} \approx \mb I$, and $I\approx J\approx T$. Then  \eqref{eqn:refine_nextx} gives
\begin{align}
	\mb x^{(k+1)} &\;\approx\;\mb x_0,\\
 	(\mb x^{(k+1)}-\mb x_0) &\;\approx\; \mb P_I\paren{\convmtx{\mb a_0}^*\convmtx{\mb a_0}\mb x_0-\convmtx{\mb a_0}^*\convmtx{\mb a^{(k)}}\mb x_0} \notag \\
 	&\;\approx\; \mb P_I\brac{\,\convmtx{\mb a_0}^*\convmtx{\mb x_0}\ip(\mb a_0-\mb a^{(k)})\,},
\end{align}
which implies, while assuming $\convmtx{\mb x_0}^*\convmtx{\mb x_0} \approx n\theta\mb I$, that from \eqref{eqn:refine_nexta}:
\begin{align} 
	(\mb a^{(k+1)}-\mb a_0)  &\;\approx\; (n\theta)^{-1}\,\injector^*\convmtx{\mb x^{(k+1)}}^*\convmtx{\mb x_0}\injector\mb a_0 - \injector^*\convmtx{\mb x^{(k+1)}}^*\convmtx{\mb x^{(k+1)}}\injector\mb a_0  \notag \\
	&\;\approx\; (n\theta)^{-1}\,\injector^*\convmtx{\mb x_0}^*\convmtx{\mb a_0}(\mb x_0-\mb x^{(k+1)}) \notag \\
	&\;\approx\; (n\theta)^{-1} \,\ip^*\convmtx{\mb x_0}^*\convmtx{\mb a_0}\mb P_I\convmtx{\mb a_0}^*\convmtx{\mb x_0}\ip\,(\mb a^{(k)}-\mb a_0).
\end{align}
Now since $\convmtx{\mb x_0}^*\mb P_I\convmtx{\mb x_0} \approx n\theta\, \mb e_0 \mb e_0^*$, this suggests that $(n\theta)^{-1} \,\ip^*\convmtx{\mb x_0}^*\convmtx{\mb a_0}\mb P_I\convmtx{\mb a_0}^*\convmtx{\mb x_0}\ip$ approximates a contraction mapping with fixed point $\mb a_0$, as follows:
\begin{align}
	(n\theta)^{-1} \,\ip^*\convmtx{\mb x_0}^*\convmtx{\mb a_0}\mb P_I\convmtx{\mb a_0}^*\convmtx{\mb x_0}\ip &\;\approx\;\ip^*\convmtx{\mb a_0}\mb e_0\mb e_0^*\convmtx{\mb a_0}^*\ip \notag \\
	&\;\approx\; \mb a_0\mb a_0^*.    
\end{align} 
Hence, if we can ensure all above approximation is sufficiently and increasingly accurate as the iterate proceeds, the alternating minimization essentially is a power method which finds the leading eigenvector of matrix $\mb a_0\mb a_0^*$---and the solution to this algorithm is apparently $\mb a_0$. Indeed, we prove that the iterates produced by this sequence of operations converge to the ground truth at a linear rate, as long as it is initialized sufficiently nearby:

\begin{theorem}[Linear rate convergence of alternating minimization]\label{thm:altmin} Suppose $\mb y = \mb a_0*\mb x_0$ where $\mb a_0$ is $\mu$-shift coherent and $\mb x_0\sim\mr{BG}(\theta)$, then there exists some constants $C,c,c_\mu$ such that if $(\mu+1/p)\,\kappa_I^2 < c_\mu $  and $n > C \theta^{-2}p^2\log n$, then with probability at least $1-c/n$, 
for any starting point $\mb a^{(0)}$ and $\lambda^{(0)}$, $I^{(0)}$ such that
\begin{align}
	 \big\|\mb a^{(0)} - \mb a_0 \big\|_2 \le \mu + 1/p,\qquad  \lambda^{(0)} = 5\kappa_I(\mu+1/p) ,\qquad I^{(0)} = \supp\paren{\convmtx{\mb a^{(0)}}^*\mb y},
\end{align}  
and for $k=1,2,\ldots,$:
\begin{align}
	 \mb x^{(k+1)} &\;\gets\; \textstyle\argmin_{\mb x}\tfrac12\|\mb a^{(k)}*\mb x - \mb y\|_2^2 + \lambda^{(k)}\textstyle\sum_{i\not\in I^{(k)}}\abs{\mb x_i},\\
	 \mb a^{(k+1)} &\;\gets\; \mb P_{\Sp^{p-1}}\big[\textstyle\argmin_{\mb a}\tfrac12\|\mb a*\mb x^{(k+1)}-\mb y\|_2^2\big],  \\
	 \lambda^{(k+1)} &\;\gets\; \tfrac12\lambda^{(k)}, \qquad  I^{(			k+1)} \;\gets\;  \supp\big(\mb x^{(k+1)}\big)
\end{align}   
then
\begin{align}\label{eqn:refine_main_bound}
	\big\|\mb a^{(k+1)}-\mb a_0\big\|_2\;\leq\; (\mu + 1/p)2^{-k}
\end{align} 
for every $k = 0,1,2,\dots$.
\end{theorem}
\begin{proof}
	See \Cref{sec:proof_altmin}.	
\end{proof} 
 
\begin{remark} The estimates $\mb x^{(k)}$ also converges to the ground truth $\mb x_0$ at a linear rate.
\end{remark}

% !TEX root = ../BD_DQ.tex
\section{Experiments}\label{sec:experiment}

We demonstrate that the tradeoffs between the motif length $p_0$ and sparsity rate $\theta$ produce a transition region for successful SaS deconvolution under generic choices of $\bm a_0$ and $\bm x_0$. For fixed values of $\theta \in [10^{-3}, 10^{-2}]$ and $p_0 \in [10^{3},10^{4}]$, we draw 50 instances of synthetic data by choosing $\bm a_0 \sim \mr{Unif}(\bb S^{p_0-1})$ and $\mb x_0 \in \bb R^n$ with $\mb x_{0} \simiid \mr{BG}(\theta)$ where $n = 5\times 10^5$. Note that choosing $\mb a_0$ this way implies $\mu(\mb a_0) \approx \frac1{\sqrt{p_0}}$.

For each instance, we recover $\mb a_0$ and $\mb x_0$ from $\mb y = \mb a_0 \ast \mb x_0$ by minimizing problem \eqref{eqn:dq-l1}. For ease of computation, we modify \Cref{alg:ssbd} by replacing curvilinear search with \emph{accelerated Riemannian gradient descent} method (\Cref{alg:agd}), which is an adaptation of accelerated gradient descent \cite{beck2009fast} to the sphere. In particular, we apply momentum and increment by the Riemannian gradient via the exponential and logarithmic operators
\begin{align}
  \text{Exp}_{\bm a}(\bm u)\ &:= \  \cos(\norm{\bm u}2 )\cdot \bm a + \sin(\norm{\bm u}2)\cdot \tfrac{\bm u}{\norm{\bm u}2}, \\
  \text{Log}_{\bm a}(\bm b)\ &:= \ \arccos(\innerprod{\bm a}{\bm b})\cdot \tfrac{\mb P_{\bm a^\perp}(\bm b - \bm a)}{\norm{\mb P_{\bm a^\perp}(\bm b - \bm a)}2},
\end{align}
derived from \cite{absil2009optimization}. Here $\text{Exp}_{\bm a}: \bm a^\perp \rightarrow \bb S^{p-1}$ takes a tangent vector of $\bm a$ and produces a new point on the sphere, whereas $\text{Log}_{\bm a}: \bb S^{p-1} \rightarrow \bm a^\perp$ takes a point $\bm b \in \bb S^{p-1}$ and returns the tangent vector which points from $\bm a$ to $\bm b$.

\begin{figure}[t!]
  \centering
\begin{tikzpicture}

\begin{axis}[%
width=3in,
height=3in,
at={(1.007in,0.713in)},
scale only axis,
axis on top,
xmin=0.5, 
xmax=11.5,
xticklabel style={font=\Large, yshift=-0.5ex},
xtick={1,3,5,7,9,11}, 
xticklabels={{$\text{10}^{\text{-3}}$},{$\text{10}^{\text{-2.8}}$},{$\text{10}^{\text{-2.6}}$},{$\text{10}^{\text{-2.4}}$},{$\text{10}^{\text{-2.2}}$},{$\text{10}^{\text{-2}}$}},
xlabel style={font=\Large, yshift=-1ex},
xlabel={$\theta\text{ (log scale)}$},
y dir=reverse,
ymin=0.5,
ymax=11.5,  
yticklabel style={font=\Large, xshift=-0.5ex},
ytick={1,3,5,7,9,11},
yticklabels={{$\text{10}^\text{4}$},{$\text{10}^{\text{3.8}}$},{$\text{10}^{\text{3.6}}$},{$\text{10}^{\text{3.4}}$},{$\text{10}^{\text{3.2}}$},{$\text{10}^{\text{3}}$}},
ylabel style={font=\Large, yshift=2ex}, 
ylabel={$ p_0 \text{ (log scale)}$},
axis background/.style={fill=white},
legend style={legend cell align=left, align=left, draw=white!15!black}
]
\addplot [forget plot] graphics [xmin=0.5, xmax=11.5, ymin=0.5, ymax=11.5] {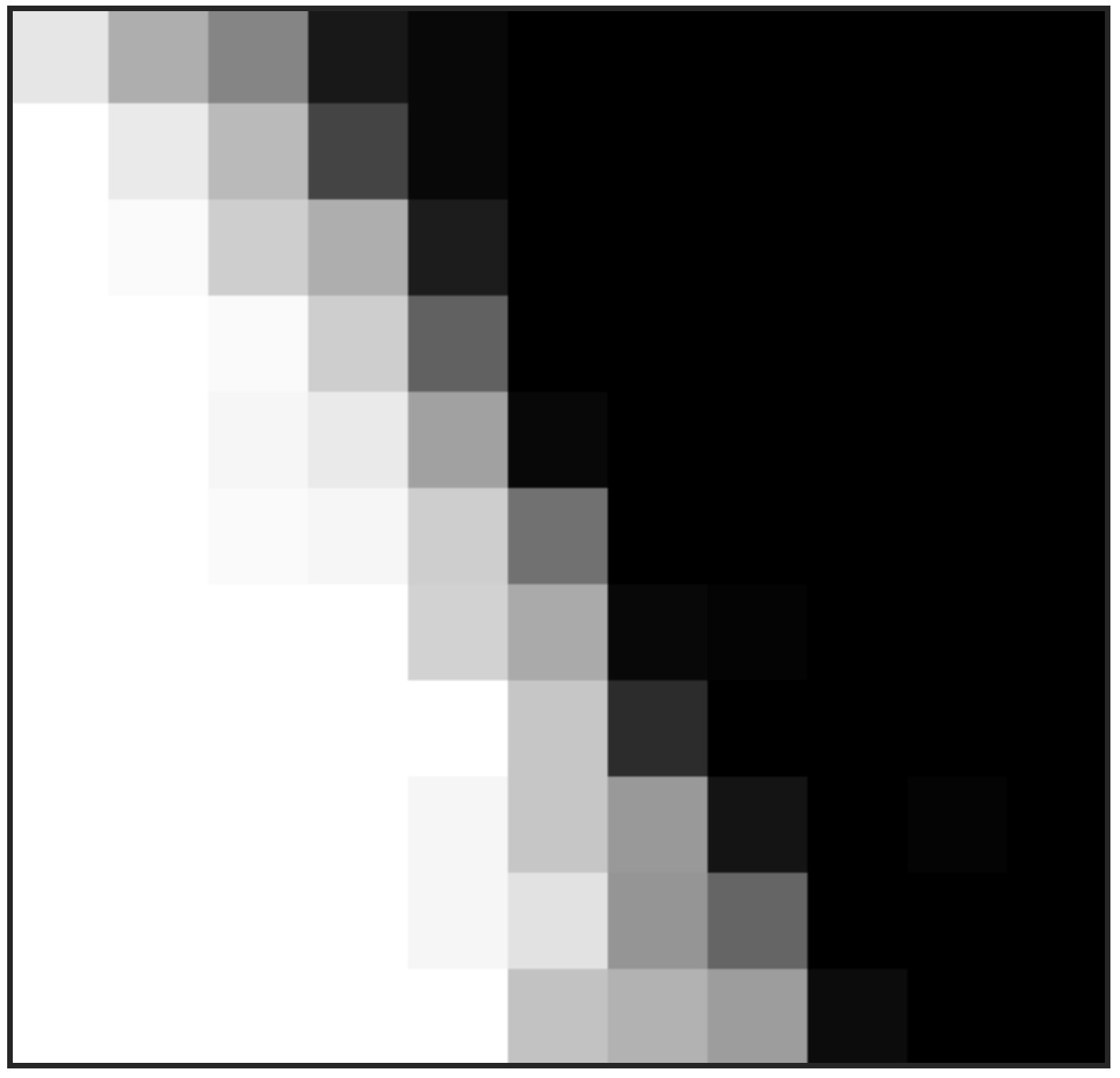}; 
\end{axis}  
\end{tikzpicture}%
  \caption{
    {\bf Success probability of SaS deconvolution under generic $\bm a_0$, $\bm x_0$ with varying kernel length $p_0$, and sparsity rate $\theta$}. When sparsity rate decreases sufficiently with respect to kernel length, successful recovery becomes very likely (brighter), and vice versa (darker). A transition line is shown with slope $\frac{\log p_0}{\log\theta}\approx -2$, implying \Cref{alg:agd} works with high probability when $\theta \lessapprox \frac{1}{\sqrt{p_0}}$ in generic case. }
  \label{fig:experiments}
\end{figure}

For each recovery instance, we say the local minimizer $\mb a_{\mr{min}}$ generated from \Cref{alg:agd} is sufficiently close to a solution of SaS deconvolution problem, if
\begin{equation}
  \text{success}(\mb a_{\mr{min}},; \mb a_0) \;:=\; \set{\,\textstyle\max_\ell\abs{\innerprod{\shift{\mb a_0}{\ell}}{\mb a_{
  \mr{min}}}} > 0.95\,}.
\end{equation}
The result is shown in \Cref{fig:experiments}. Our source code can be accessed via the following address:
\begin{quotation}
	\centering
	\url{https://github.com/sbdsphere/sbd_experiments.git}
\end{quotation}

\begin{algorithm}[h]
  \caption{SaS deconvolution with Accelerated Riemannian gradient descent}
  \label{alg:agd}
  \begin{algorithmic}
    \Require Observation $\bm y$, sparsity penalty $\lambda = 0.5/\sqrt{p_0\theta}$, momentum parameter $\eta \in [0,1)$.

    \State Initialize $\mb a^{(0)}\gets -\mb P_{\Sp^{p-1}}\nabla\varphi_\rho\paren{\mb P_{\Sp^{p-1}}\brac{\mb 0^{p_0-1};[\mb y_0,\cdots,\mb y_{p_0-1}];\mb 0^{p_0-1} }  } $,

    \For{$k=1,2,\ldots,K$}
      \State Get momentum: $\bm w \gets \text{Exp}_{\mb a^{(k)}}\big(\eta\cdot\text{Log}_{\bm a^{(k-1)}}(\mb a^{(k)})\big)$.
      \State Get negative gradient direction:  $\mb g \gets -\grad[\varphi_\rho](\mb w) $.
      \State Armijo step $\mb a^{(k+1)} \gets \text{Exp}_{\mb w}(t\mb g)$,\ \ choosing $t\in(0,1)$ s.t. $\varphi_\rho(\mb a^{(k+1)})-\varphi_\rho(\mb w) < -t\norm{\mb g}2^2$.
    \EndFor
   \Ensure Return $\mb a^{(K)}$.
  \end{algorithmic}
\end{algorithm}

% !TEX root = ../BD_DQ.tex
\section{Discussion}\label{sec:discussion}

In this section, we close by discussing several of the most important limitations of our results, and highlighting corresponding directions for future work.

\paragraph{Minimizing $\varphi_\rho$ does not accurately recover coherent kernels.} The main drawback of our proposed method is that it does not succeed when the target motif $\mb a_0$ has shift coherence very close to $1$. For instance, a common scenario in image blind deconvolution involves deblurring an image with a smooth, low-pass point spread function (e.g., Gaussian blur).  Both our analysis and numerical experiments show that in this situation minimizing $\varphi_\rho$ does not find the generating signal pairs $(\mb a_0,\mb x_0)$ consistently---the minimizer of $\varphi_\rho$ is often spurious and is not close to any particular shift of $\mb a_0$. We do not suggest minimizing $\varphi_\rho$ in this situation. On the other hand, minimizing the bilinear lasso objective $\varphi_{\mr{lasso}}$ over the sphere often succeeds  even if the true signal pair $(\mb a_0,\mb x_0)$ is coherent and dense.

\paragraph{Relation of $\varphi_\rho$ to Bilinear Lasso.} In light of the above observations, we view the analysis of the bilinear lasso as the most important direction for future theoretical work on SaS deconvolution. The drop quadratic formulation studied here has commonalities with the bilinear lasso: both exhibit local minima at signed shifts, and both exhibit negative curvature in symmetry breaking directions. A major difference (and hence, major challenge) is that gradient methods for bilinear lasso do not retract to a union of subspaces -- they retract to a more complicated, nonlinear set. 

\paragraph{Suboptimality in the analysis.} Finally, there are several directions in which our analysis could be improved. Our lower bounds on the length $n$ of the random vector $\mb x_0$ required for success are clearly suboptimal. We also suspect our sparsity-coherence tradeoff between $\mu,\theta$ (roughly, $\theta \lessapprox 1/(\sqrt{\mu}p_0)$) is suboptimal, even for the $\varphi_\rho$ objective. Articulating optimal sparsity-coherence tradeoffs for is another interesting direction for future work.

\section*{Acknowledgement}
The authors gratefully acknowledge support from NSF 1343282, NSF CCF 1527809, and NSF IIS 1546411.

%-Bibliography
\InputIfFileExists{BD_DQ.bbl}
%{\small
%\bibliographystyle{alpha}
%\bibliography{deconv}
%}

%-Appendix
\newpage
\appendix
  
\newpage
% !TEX root = ../../BD_DQ.tex
\section{Basic bounds for Bernoulli-Gaussian vectors}\label{sec:basic_bg_bounds}
In this section, we prove several lemmas pertaining to the sparse random vector $\mb x_0\simiid\mr{BG}(\theta)$.

\begin{lemma}[Support of $\mb x_0$]\label{lem:x0_supp} Let $\mb x_0 \simiid \mr{BG}(\theta)$ and $I_0 = \mr{supp}(\mb x_0)\subseteq[n]$.  Suppose $n > 10\theta^{-1} $, then for any $\eps\in\paren{0,\frac1{10}}$, with probability at least $1-\eps$ we have 
\begin{align}
\abs{\abs{I_0}-n\theta} \leq  2\sqrt{n\theta}\log\eps^{-1}.
\end{align}
And suppose $n\geq C\theta^{-2}\log p $ and $\theta$, then with probability at least $1-2/n$,  we have 
\begin{align}\label{eqn:support_x0_2}  
	\forall\,t\in[2p]\setminus\set{0},\quad  \tfrac{1}{2} n\theta^2 \leq \abs{I_0\cap(I_0 + t)} \leq 2n\theta^2
\end{align}
where $C$ is a numerical constant.
\end{lemma}
\begin{proof} Let $\mb x_0 = \mb \omega \cdot \mb g \simiid \mr{BG}(\theta)$, notice that the support of the Bernoulli-Gaussian vector $\mb x_0$ is almost surely equal to the support of the Bernoulli vector $\mb \omega$. Applying Bernstein inequality  \Cref{lem:mc_bernstein_scalar} with $(\sigma^2,
R) = (1,1)$, then if $n\theta > 10$ we have
\begin{align}
	\prob{\abs{\sum_{k\in[n]}\mb\omega_k - n\theta} > 2\sqrt{n\theta}\log \eps^{-1}} &\leq 2\exp\left({\frac{-4n\theta\log^2 \eps^{-1}}{2n\theta + 4\sqrt{n\theta}\log \eps^{-1}}}\right) \leq \eps.
	\notag 
\end{align}
For \eqref{eqn:support_x0_2}, let $J_t := I_0\cap\paren{I_0+t}$. The cardinality of $J_t$ is an inner product between shifts of $\mb\omega$:
\begin{align}
	\abs{J_t} = \sum_{k\in[n]}\mb\omega_k\mb\omega_{k-t},
\end{align}
and define two subset $J_{t1}\uplus J_{t2} = J_t $, as follows:
\begin{align}\label{eqn:supp_x0_Jt12}
	\caseof{J_{t1} = J_t \cap \mc K_1,\quad  \mc K_1 :=[n]\cap\set{0,\ldots,t-1,2t,\ldots,3t-1,\ldots} \\ J_{t2} = J_t\cap\mc K_2,\quad \mc K_2 := [n]\cap\set{t,\ldots,2t-1,3t,\ldots,4t-1,\ldots}  }.
\end{align} 
Here, the size of sets $\mc K_1,\mc K_2$ has two-side bounds $0.4n\leq \paren{n-2p}/2 \leq \abs{\mc K_2} \leq \abs{\mc K_1} \leq \paren{n+2p}/2 \leq 0.6n$, thus the size of sets $J_{t1},J_{t2}$ can be derived using Bernstein inequality \Cref{lem:mc_bernstein_scalar} with $n> C\theta^{-2}\log p$ as
\begin{align}\label{eqn:cross_x0_J1_2}
	\prob{\max_{t\in [2p]\setminus\set{0}}\abs{J_{t_1}} \geq n\theta^2 } &= \prob{\max_{t\in[2p]\setminus \set 0}\sum_{k\in\mc K_1}\mb\omega_k\mb\omega_{k-t} \geq n\theta^2 } \leq 2p \cdot\prob{\sum_{k\in\mc K_1}\mb\omega_k\mb\omega_{k+1} \geq  n\theta^2} \notag \\
	&\leq 2p\cdot \prob{\sum_{k\in\mc K_1}\mb\omega_k\mb\omega_{k+1} - \E\sum_{k\in\mc K_1}\mb\omega_k\mb\omega_{k+1} \geq n\theta^2 - 0.6n\theta^2  }\notag \\ 
	&\leq 4p\cdot\exp\paren{\frac{-\paren{0.4n\theta^2}^2}{2\cdot 0.6n\theta^2 + 2\cdot0.4n\theta^2}} = \exp\paren{\log (4p)-0.08n\theta^2}\leq 1/n,
\end{align}
where the last two inequalities hold with $C > 10^5$. The lower bound can also derived as follows 
\begin{align} \label{eqn:cross_x0_J2_2}
	\prob{\min_{t\in [2p]\setminus\set{0}}\abs{J_{t_1}} \leq n\theta^2/4 } &= \prob{\min_{t\in[2p]\setminus \set 0}\sum_{k\in\mc K_1}\mb\omega_k\mb\omega_{k-t} \leq n\theta^2/4 } \leq 2p \cdot\prob{\sum_{k\in\mc K_1}\mb\omega_k\mb\omega_{k+1} \leq  n\theta^2/4} \notag \\
	&\leq 2p\cdot \prob{\sum_{k\in\mc K_1}\mb\omega_k\mb\omega_{k+1} - \E\sum_{k\in\mc K_1}\mb\omega_k\mb\omega_{k+1} \leq  n\theta^2/4 - 0.4n\theta^2  }\notag \\ 
	&\leq 4p\cdot\exp\paren{\frac{-\paren{0.15n\theta^2}^2}{2\cdot 0.6n\theta^2 + 2\cdot0.15n\theta^2}} = \exp\paren{\log(4p)-0.0015n\theta^2}\leq 1/n.
\end{align}
The bound for $\abs{J_2}$ can derived similarly to \eqref{eqn:cross_x0_J1_2}-\eqref{eqn:cross_x0_J2_2}.
\end{proof}

\begin{lemma}[Norms of $\mb x_0$]\label{lem:x0_bound} Let $\mb x_0\sim_{\mr{i.i.d.}}\mr{BG}(\theta)\in\R^n$. If $n\geq 10\theta^{-1}$, then for any $\eps \in \paren{0,\tfrac{1}{10}}$, with probability at least $1-\eps$, 
\begin{align}
	\abs{\norm{\mb x_0}1 - \sqrt{2/\pi}n\theta}\leq 2\sqrt{n\theta}\log \eps^{-1},\quad  \abs{\norm{\mb x_0}2^2 -n\theta} \leq 3\sqrt{n\theta}\log \eps^{-1}
\end{align}	
\end{lemma}

\begin{proof}
	To bound $\norm{\mb x_0}1$, using Bernstein inequality with $(\sigma^2,R) = (\theta,1)$ and with $n\theta\geq 10$ we have
	\begin{align}
		\prob{\abs{\norm{\mb x_0}1 -\sqrt{\frac{2}{\pi}} n\theta }\geq 2\sqrt{n\theta}\log \eps^{-1} }\leq 2\exp\left(\frac{-4n\theta\log^2 \eps^{-1}}{2n\theta + 4\sqrt{n\theta}\log \eps^{-1}}\right) \leq \eps \notag 
	\end{align}
Similarly for $\norm{\mb x_0}2^2$, from Gaussian moments \Cref{lem:gaussian_moment}  , we know the 2-norm $\sum_{i\in[n]}\E\abs{x_{0i}}^4 = 3n\theta$ and $q$-norm $\sum_{i\in[n]}\bb E\abs{x_{0i}}^{2p} \leq (n\theta)(2q-1)!! \leq \frac12 (3n\theta) 2^{q-2}q! $ for $q\geq 3$.  Let $(\sigma^2,R) = (3\theta,2)$ in Bernstein inequality form \Cref{lem:mc_bernstein_scalar},  $n\theta \geq 10 $ we have
\begin{align}
	\prob{\abs{\norm{\mb x_0}2^2 - n\theta } \geq 3\sqrt{n\theta}\log \eps^{-1} } \leq 2\exp \left( \frac{-9n\theta\log^2 \eps^{-1}}{2(3n\theta) + 12\sqrt{n\theta}\log \eps^{-1}}\right) \leq \eps, \notag
\end{align}
completing the proof.
\end{proof}  

\begin{lemma}[Norms of $\mb x_0$ subvectors]\label{lem:x0_subvec_bound} Let $\mb x_0\sim_{\mr{i.i.d.}}\mr{BG}(\theta)\in\R^n$ and $n>10$, then with probability at least $1-3/n$,  we have
\begin{align}
	\max_{\substack{U=[2p]+j \\ j\in[n] } }\norm{\mb P_U\mb x_0}2^2 \leq  2p\theta + 6\paren{\sqrt{p\theta}+\log n}
\end{align}
and if $\mb a_0$ is $\mu$-shift coherent and there exists a constance $c_\mu $ such that both $\theta^2p <c_\mu $ and $\mu p^2\theta < c_\mu$, then
\begin{align}
	\max_{\substack{U=[p]+j \\ j\in[n] } }\norm{\mb P_U\brac{\mb a_0*\mb x_0}}2^2 \leq  p\theta + \log n.
\end{align}

\end{lemma} 
\begin{proof} Use Bernstein inequality with $(\sigma^2,R) = (3\theta,2)$ and $t = \max\set{\sqrt{p\theta},\log n}$, with union bound we obtain:
\begin{align}
	\prob{\max_{\substack{U = [2p]+j\\ j\in[n]}} \norm{\mb P_U\mb x_0}2^2 \,\geq\, 2p\theta + 6\paren{\sqrt{p\theta}+\log n} } &\;\leq\; 2n\exp\paren{- \frac{36\paren{\sqrt{p\theta}+\log n}^2 }{6p\theta + 12\paren{\sqrt{p\theta} + \log n}} }\notag \\
	&\;\leq\; 2\exp\paren{\log n - \frac{36t^2}{6t^2 + 12t}}\leq \frac{2}{n}.
\end{align}
For the second inequality, first we know calculate the expectation
\begin{align}
	\E\norm{\mb P_U\brac{\mb a_0*\mb x_0} }2^2 &\;=\; \E\brac{\mb x_0^*\convmtx{\mb a_0}^*\mb P_U\convmtx{\mb a_0}\mb x_0} \notag\\
	&\;=\; \theta\cdot\trace\paren{\convmtx{\mb a_0}^*\mb P_U\convmtx{\mb a_0}} \norm{\mb a_0}2^2 \;+\; \theta\cdot \sum_{i=1}^{p-1}\norm{\ip^*\shift{\mb a_0}{i}}2^2 \notag\\
	&\;=\; p\theta. 
\end{align}
Then apply Henson Wright inequality \Cref{lem:Hanson-wright} with $\norm{ \convmtx{\mb a_0}^*\mb P_U\convmtx{\mb a_0}}F^2 =  \norm{\ip^*\convmtx{\mb a_0}^*\convmtx{\mb a_0}\ip}F^2  \leq   p\paren{1+\mu p}$ and also  $\norm{ \convmtx{\mb a_0}^*\mb P_U\convmtx{\mb a_0}}2 = \norm{\convmtx{\mb a_0}\ip}2^2 = 1+\mu p $, we can derive
\begin{align} 
	\prob{\max_{\substack{  U = [p]+j \\ j\in[n]}}\norm{\mb P_U\brac{\mb a_0*\mb x_0}}2^2 \,\geq\, p\theta + \log n} &\;\leq\; n\exp\paren{-\min\set{\frac{\log^2 n}{64\theta^2p\paren{1+\mu p}},\frac{\log n}{8\sqrt 2\theta\paren{1+\mu p}}}} \notag \\
	&\;\leq\; \exp\paren{\log n-\min\set{\frac{\log^2n}{128c_\mu},\frac{\log n}{32c_\mu }} } \leq\frac{1}{n} 
\end{align} 
when $c_\mu < \frac{1}{300}$.
\end{proof}

\begin{lemma}[Inner product between shifted $\mb x_0$]\label{lem:x0_innerprod} Let $\mb x_0\sim_{\mr{i.i.d.}}\mr{BG}(\theta)\in\R^n$. There exists a numerical constant $C$ such that if $n > C\theta^{-2}\log p$ and $p\theta\log^2 \theta^{-1}>1$, with probability at least $1-4/n$, the following two statements hold simultaneously:
	\begin{align}\label{eqn:innerprod_x0}
		\max_{i\neq j\in[2p]}\innerprod{\shift{\mb x_0}{i}}{\shift{\mb x_0}{j}} \leq 6\sqrt{n\theta^2\log n};
	\end{align} 
and for $\mb x_i = |\mb x_{0,i}| \in \R^n_+$ the vector of magnitudes of $\mb x_0$, 
	\begin{align}\label{eqn:innerprod_x0_abs}
		\max_{i\neq j\in[2p]}\innerprod{\shift{\mb x}{i}}{\shift{\mb x}{j}} \leq 4n\theta^2.
	\end{align}
\end{lemma}
\begin{proof} We will start from proving \eqref{eqn:innerprod_x0_abs}. Write $\mb x = \abs{\mb g}\circ \mb\omega$ where $\mb g$ / $\mb\omega$ are Gaussian/Bernoulli random vectors respectively. Let $I_0$ denote the support of $\mb \omega$ and $t = \abs{j-i}$ with $0<t<p$.
Then \eqref{eqn:innerprod_x0_abs} can be written as summation of Gaussian r.v.s. on intersection of support set between shifts:
\begin{align}\label{eqn:inprod_x0_sum1}
	\innerprod{\shift{\mb x}{i}}{\shift{\mb x}{j}} = \sum_{k \in I_0\cap\paren{I_0+t}}  \abs{\mb g_k}\abs{\mb g_{k-t}}   
\end{align}
Define $J_t := I_0\cap\paren{I_0+t} = J_{t1}\uplus J_{t2}$ same as \eqref{eqn:supp_x0_Jt12}.  Notice that both $\sum_{k\in J_{t1}}\abs{\mb g_k}\abs{\mb g_{k-t}}$ and $\sum_{k\in J_{t2}}\abs{\mb g_k}\abs{\mb g_{k-t}}$ are sum of independent r.v.s.. We are left to consider the upper bound of $\sum_{j\in {J_{ti}}}\abs{\mb g_j}\abs{\mb g_j'}$ where $\mb g$, $\mb g'$ are independent Gaussian vectors. 
We condition on the following event 
\begin{align}
	\event_J := \set{\forall t\in[2p]\setminus \set 0,\; n\theta^2/4 \leq \abs{J_{t1}},\abs{J_{t2}} \leq n\theta^2},
\end{align}
which holds w.p. at least $1-2/n$ from \Cref{lem:x0_supp}.  Since $\sum_{j\in J_{t1}}\abs{\mb g_j}\abs{\mb g_j'} \leq \norm{\mb g_{J_{t1}}}2\norm{\mb g_{J_{t1}}'}2 $, we use Gaussian concentration   \Cref{lem:gaussian_concentration} and union bound to obtain 
\begin{align}
	\prob{\max_{t\in[2p]\setminus\set0}\sum_{j\in J_{t1}}\abs{\mb g_j\mb g_j'} > 2\abs{J_{t1}}}  &\leq 2p\cdot \prob{\norm{\mb g_{J_{t1}}}2\norm{\mb g_{J_{t1}}'}2  - \E\norm{\mb g_{J_{t1}}}2\norm{\mb g_{J_{t1}}'}2 > \abs{J_{t1}} } \notag \\
	&\leq 4p\cdot\prob{\norm{\mb g_{J_{t1}}}2 -\E \norm{\mb g_{J_{t1}}}2 > \sqrt{\abs{J_{t1}}}/3 }\notag \\
	& \leq 4p\exp\paren{-(\abs{J_{t1}}/9)/2} \leq 4p\exp\paren{-n\theta^2/72} \leq 1/n
\end{align}
where the last inequality is derived simply via assuming $n = C\theta^{-2}\log p$ for some $C > 10^4$, such that
\begin{align}
	C > 400*(4C)^{1/5} &\implies  C\log p > 400 \log((4C)^{1/5}p) \implies C\log p > 72 \log(4C p^5) > 72\log(4Cp^2\log^3p)  \notag \\
	&\implies n\theta^2 > 72\log(p\cdot 4C\theta^{-2}\log p)  = 72 \log(4np). \notag
\end{align} 
Likewise for sum on set $J_{t2}$,  we collect all above result and conclude for every $i\neq j\in [2p]$,
\begin{align}\label{eqn:innerprod_x0_res1}
	\innerprod{\shift{\mb x}{i}}{\shift{\mb x}{j}}  = \sum_{k\in J_{t1}}\abs{\mb g_k}\abs{\mb g'_{k-t}} + \sum_{k\in J_{t2}}\abs{\mb g_k}\abs{\mb g'_{k-t}} \leq 2\paren{\abs{J_{t_1}} + \abs{J_{t_2}} } \leq 4n\theta^2.
\end{align} 
For \eqref{eqn:innerprod_x0} similarly condition on event $\event_J$, using Bernstein inequality \Cref{lem:mc_bernstein_scalar} with $(\sigma^2,R) = (1,1)$:
\begin{align}
	\prob{\max_{t\in[2p]\setminus\set 0}\abs{\sum_{j\in J_{t1}}\mb g_j\mb g_j'} > 3\sqrt{n\theta^2\log n} } &\leq p\cdot\exp\paren{\frac{-9n\theta^2\log n}{2\abs{J_{t1}} + 6\sqrt{n\theta^2\log n}}} \leq p\cdot\exp\paren{\frac{-9n\theta^2\log n}{3n\theta^2}} \leq \frac1n
\end{align}
thus for every $i\neq j\in[2p]$, 
\begin{align}\label{eqn:innerprod_x0_res2}
	\abs{\innerprod{\shift{\mb x_0}{i}}{\shift{\mb s_0}{j}}} \leq \abs{\sum_{k\in J_{t1}}\mb g_k\mb g'_{k-t} }+ \abs{\sum_{k\in J_{t2}}\mb g_k\mb g'_{k-t}} \leq 6\sqrt{n\theta^2\log n}.
\end{align}
Finally, both \eqref{eqn:innerprod_x0_res1},\eqref{eqn:innerprod_x0_res2} holds simultaneously  with probability at least
\begin{align}
	1-2/n-1/n-1/n = 1-4/n
\end{align}

\end{proof}

\begin{lemma}[Convolution of $\mb x_0$]\label{lem:y_bound} Given $\mb y = \mb x_0*\mb a_0$ where $\mb x_0 \simiid \mr{BG}(\theta)\in\R^n$ and $\mb a_0\in\R^{p_0}$ is $\mu$-shift coherent. Suppose $n \geq C\theta^{-2}\log p$ for some numerical constant $C>0$, with probability at least $1-7/n$,  we have the following two statement simultaneously hold:
\begin{align}
\norm{\convmtx{\mb y}\ip}2^2 &\leq 3(1+\mu p)n\theta
\end{align}
and for all $J\subseteq[n]$,
\begin{align}
	\norm{\mb P_J\convmtx{\mb y}\ip}2^2 \leq 14\abs J(1+\mu p)\paren{ p\theta+\log n}
\end{align} 
\end{lemma}

\begin{proof} Given any $\mb a\in\Sp^{p-1}$, write $\mb\beta = \convmtx{\mb a_0}^*\ip\mb a $ where $\abs{\mb\beta} \leq 2p$ . Apply $\norm{\mb x_0}2^2\leq 2n\theta$ from \Cref{lem:x0_bound} by choosing $\eps = 1/n$, also $\abs{\innerprod{\shift{\mb x_0}{i}}{\shift{\mb x_0}{j}}}\leq 6\sqrt{n\theta^2\log n}$ from \Cref{lem:x0_innerprod} we get:
\begin{align}
	\norm{\convmtx{\mb y}\ip\mb a}2^2 &= \norm{\convmtx{\mb x_0}\mb \beta}2^2\leq \norm{\mb\beta}2^2\norm{\mb x_0}2^2 + \sum_{ i\neq j\in[\pm p]}\abs{\beta_i\beta_j\innerprod{\shift{\mb x_0}{i}}{\shift{\mb x_0}{j}} } \notag \\
	&\leq \norm{\mb\beta}2^2\norm{\mb x_0}2^2 + \norm{\mb\beta}1^2\max_{i\neq j\in[\pm p]}\abs{\innerprod{\shift{\mb x_0}{i}}{\shift{\mb x_0}{j}}} \notag \\
	&\leq \norm{\mb\beta}2^2\cdot 2n\theta +p\norm{\mb\beta}2^2 \cdot 6\sqrt{n\theta^2\log n} \leq  3\norm{\mb\beta}2^2 n\theta \notag  
\end{align}
where $n = C \theta^{-2}\log p$ with $C\geq 10^4$, and the statement holds with probability at least $1-5/n$.

For the bound of $\norm{\mb P_J\convmtx{\mb y}\ip\mb a}2^2$. Simply apply \Cref{lem:x0_subvec_bound} and utilize norm bound of $\norm{\mb\beta}2^2$, with probability at least $1-2/n$ we have:
\begin{align} 
	\norm{\mb P_J\convmtx{\mb y}\ip\mb a}2^2 &= \sum_{i\in J} \abs{\innerprod{\shift{\mb x_0}{i}}{\mb \beta}}^2 \leq \abs{J}\max_{\substack{U = [2p]+j \\ j\in [n]}} \norm{\mb P_U\mb x_0}2^2\norm{\mb\beta}2^2 \leq \abs{J}\cdot14\paren{p\theta + \log n}\cdot\norm{\mb\beta}2^2 \notag
\end{align}
Finally apply \Cref{fact:M_entries} and Gershgorin disc theorem obtain
\begin{align}
	\norm{\mb\beta}2^2 = \norm{\convmtx{\mb a_0}^*\ip\mb a}2^2 \leq \norm{\convmtx{\mb a_0}^*\ip}2^2  = \sigma_{\mr{max}}\paren{\mb M}  \leq 1+\mu p.
\end{align} 

\begin{remark}\label{rmk:convolution_x0} When $\mb a_0$ is a basis vector $\mb e_0$, the result of \Cref{lem:y_bound} gives upper bound of $\norm{\convmtx{\mb x_0}}2 < 3n\theta$, whose lower bound can be derived similarly with $\norm{\convmtx{\mb x_0}\ip}2\geq \frac23n\theta$ 
	
\end{remark}

\end{proof}

% !TEX root = ../../BD_DQ.tex

%% ../sections/appendix/app_signalprop.tex
\section{Vectors in shift space}\label{sec:vector_shifts}
  
In this section, we will establish a number of properties of the coefficient vectors $\mb\alpha$ and correlation vector $\mb\beta$. Generally speaking, when $\mb a$ is close to the subspace $\mc S_{\mb\tau}$, then both vectors $\mb \alpha$,$\mb\beta$ have most of their energy concentrated on the entries $\mb\tau$. In this section, we derive upper bounds on $\mb\alpha_{\mb\tau^c}$ and $\mb\beta_{\mb\tau^c}$ under various assumptions. 

In particular, we will introduce a relationship between the sparsity rate $\theta$, coherence $\mu$ and size $|\mb \tau|$, which we term the sparsity-coherence condition. In \Cref{lem:d_alpha_norm} we prove that measuring the distance from $\mb a$ to subspace $\mc S_{\mb\tau}$ in terms of $\|\mb\alpha_{\mb\tau^c}\|_2$ gives a seminorm. We then use this distance to characterize a region $\goodregion$ around the subspace $\mc S_{\mb \tau}$. Later, in \Cref{fact:M_entries}  we illustrate the relationship between $\mb\alpha$ and $\mb\beta$, where $\mb\beta =  \convmtx{\mb a_0}^*\ip\ip^*\convmtx{\mb a_0}\mb\alpha$. Finally in  \Cref{lem:alpha_beta_tau_lb} and \Cref{cor:tail_beta_x0_tauc}, controls the magnitude of $\mb \alpha_{\mb \tau^c}$ and $\mb \beta_{\mb \tau^c}$ near $\mc S_{\mb \tau}$.

\begin{definition}[Sparsity-coherence condition]\label{asm:theta_mu} 
Let $\mb a_0 \in \bb S^{p_0-1}$ with shift coherence $\mu$. We say that $(\mb a_0,\theta,|\mb \tau|)$ satisfies the sparsity-coherence condition $\mr{SCC}(c_\mu)$ with constant $c_\mu$, if 
\begin{align}\label{eqn:theta_mu_bound}
	\theta  \in \brac{\frac{1}{p},\frac{c_\mu }{4\max\set{\abs{\mb\tau},\sqrt p}}}\cdot\frac{1}{\log^2\theta^{-1}} ,\quad  \mu\cdot\max\set{\abs{\mb\tau}^2, p^2\theta^2}\cdot \log^2\theta^{-1} \leq \frac{c_\mu}4,
\end{align}
where $p = 3p_0-2$. 
\end{definition}

\begin{lemma}[$d_\alpha$ is a seminorm]\label{lem:d_alpha_norm} For every solution subspace $\mc S_{\mb\tau}$, the function $d_{\alpha}(\cdot,\mc S_{\mb\tau}): \R^p\to \R_+$  defined as
\begin{align}\label{eqn:dist_s_tau_apen}
	d_\alpha(\mb a,\mc S_{\mb\tau}) = \inf\set{\norm{\mb\alpha_{\mb\tau^c}}2 \; \middle | \; \mb a = \ip^*\convmtx{\mb a_0}\mb\alpha}.
\end{align} 
is a seminorm, and for all $\mb a\in\mc S_{\mb \tau}$, $d_\alpha(\mb a,\mc S_{\mb\tau})= 0$. 
\end{lemma}
\begin{proof} It is immediate from definition that $d(\cdot,\mc S_{\mb\tau})$ is nonnegative and $\mc S_{\mb \tau} \subseteq \set{\mb a:d_\alpha(\mb a,\mc S_{\mb\tau}) = 0}$. Subadditivity can be shown from simple norm inequalities and our definition of $d_{\alpha}$, for all $\mb a_1$, $\mb a_2$ we have
\begin{align}
	d_\alpha(\mb a_1 + \mb a_2,\mc S_{\mb\tau}) &= \inf\set{\norm{\mb\alpha_{\mb\tau^c}}2 \; \middle | \; \mb a_1+\mb a_2 = \ip^*\convmtx{\mb a_0}\mb\alpha}  \notag \\
	& = \inf\set{\norm{\mb\alpha_{1\mb\tau^c} + \mb\alpha_{2\mb\tau^c}}2 \; \middle | \; \mb a_1 = \ip^*\convmtx{\mb a_0}\mb\alpha_1,\quad \mb a_2 = \ip^*\convmtx{\mb a_0}\mb\alpha_2} \notag \\
	&\leq \inf\set{\norm{\mb\alpha_{1\mb\tau^c}}2 + \norm{\mb\alpha_{2\mb\tau^c}}2 \; \middle | \; \mb a_1 = \ip^*\convmtx{\mb a_0}\mb\alpha_1,\quad \mb a_2 = \ip^*\convmtx{\mb a_0}\mb\alpha_2} \notag\\
	& =  \inf\set{\norm{\mb\alpha_{1\mb\tau^c}}2 \; \middle | \; \mb a_1 = \ip^*\convmtx{\mb a_0}\mb\alpha_1} + \inf\set{\norm{\mb\alpha_{2\mb\tau^c}}2 \; \middle | \; \mb a_2 = \ip^*\convmtx{\mb a_0}\mb\alpha_2} \notag\\
	& = d_\alpha(\mb a_1,\mc S_{\mb\tau}) + d_\alpha(\mb a_2,\mc S_{\mb\tau}).\notag
\end{align}
Similarly the absolute homogeneity, for any $c\in\R$:  
\begin{align}
	d_\alpha(c\cdot\mb a,\mc S_{\mb\tau}) &= \inf\set{\norm{\mb\alpha'_{\mb\tau^c}}2\; \middle | \ c\cdot \mb a = \ip^*\convmtx{\mb a_0}\mb\alpha'} = \inf\set{\norm{c\cdot \mb\alpha_{\mb\tau^c}}2\; \middle | \ \mb a = \ip^*\convmtx{\mb a_0}\mb\alpha}  \notag \\   
	 &= \abs c\cdot  \inf\set{\norm{\mb\alpha_{\mb\tau^c}}2 \; \middle | \ \mb a = \ip^*\convmtx{\mb a_0}\mb\alpha}  = \abs c\cdot d_\alpha(\mb a,\mc S_{\mb\tau}),\notag
\end{align}
which completes the proof that $d_\alpha$ is a seminorm.
\end{proof}  

\begin{definition}[Widened subspace]\label{def:gamma} For subspace $\mc S_{\mb \tau}$ let 
\begin{align}\label{eqn:goodregion_s_tau_apen}
	\goodregion := \set{\mb a\in\Sp^{p-1}\,\big|\, d_\alpha(\mb a,\mc S_{\mb\tau}) \leq \gamma}
\end{align}   
denote its widening by $\gamma$, in the seminorm $d_\alpha$. 
\end{definition}  
Our analysis works with a specific choice of width $\gamma(c_\mu)$, which depends on the problem parameters $\mb a_0, \theta, |\tau|$ and a constant $c_\mu$, via
\begin{align}\label{eqn:gamma}
	\gamma(c_\mu) = \frac{c_\mu}{4\log^2\theta^{-1}}\min\set{\frac{1}{\sqrt{\abs{\mb\tau}}},\frac{1}{\sqrt{\mu p}},\frac{1}{\mu p\sqrt\theta \abs{\mb\tau}}}.  
\end{align} 
  
\begin{lemma}[Properties of $\convmtx{\mb a_0}^*\ip\ip^*\convmtx{\mb a_0}$]\label{fact:M_entries} Let  $\mb M = \convmtx{\mb a_0}^*\ip\ip^*\convmtx{\mb a_0}$, with $\mb a_0\in\Sp^{p_0-1}$ $\mu$-shift coherent. The diagonal entries of $\mb M$ satisfy
\begin{align}
	\begin{cases}  \mb M_{ii} = 1 &\quad i\in [-p_0+1,p_0-1] = [\pm p_0], \\ 0\leq \mb M_{ii}\leq 1 &\quad i \in [-2p_0+2,-p_0]\cup [p_0,2p_0-2], \\
	\mb M_{ii} = 0 &\quad \text{otherwise},  
	\end{cases}
\end{align} 
and the off-diagonal entries satisfy
\begin{align}
	\begin{cases} \abs{\mb M_{ij}} \leq \mu  &\quad 0<\abs{i-j}<p_0,\;\; \set{i\in[-p_0+1,p_0-1]} \cup \set{j\in[-p_0+1,p_0-1]}, \\
	\abs{\mb M_{ij}}<1 &\quad \set{i,j\in[-2p_0+2,-p_0]}\cup \set{i,j\in[p_0,2p_0-2]},\\
	0&\quad \text{otherwise}.  
	\end{cases}
\end{align}
Furthermore, let $\mb\tau\subset[\pm p_0]$, and $\mb \tau^c = [\pm 2p_0-1] \setminus \mb \tau$. The singular values of submatrix $\injector_{\mb \tau}^* \mb M \injector_{\mb \tau}$ can be bounded as: 
\begin{align}
	\begin{cases} 1-\mu\abs{\mb\tau}\leq \sigma_{\mr{min}}\paren{\injector_{\mb\tau}^*\mb M\injector_{\mb\tau}} \leq \sigma_{\mr{max}}\paren{\injector_{\mb\tau}^*\mb M\injector_{\mb\tau}}\leq 1+\mu\abs{\mb\tau}, \\
	\sigma_{\mr{max}}\paren{\injector_{\mb\tau^c}^*\mb M\injector_{\mb\tau}} \leq \mu\sqrt{p\abs{\mb\tau}}, \\
	\sigma_{\mr{max}}\paren{\injector_{\mb\tau^c}^*\mb M\injector_{\mb\tau^c}}   \leq  1+\mu p. 
	\end{cases}
\end{align}  
\end{lemma}
\begin{proof}
	Recall the definition of $\ip$,  which selects the entries $\set{-p_0+1,\ldots,2p_0-2}$. The entrywise properties of $\mb M$ can be derived by carefully counting the entries of the shifted support. The submatrix $\mb M$ on support $\set{-2p_0+2,\ldots,2p_0-2}$ has an upper bound to be characterized as follows:
	\begin{align}
		\abs{\injector^*_{[\pm 2p_0-1]}\mb M\injector_{[\pm 2p_0-1]}} \leq  \begin{bmatrix}
			\mb J & \mu\cdot\1 & \begin{bmatrix} 0\\ \vdots \\ 0 
			\end{bmatrix}    & \mb 0     & \mb 0 \\
			\mu\cdot\1 & \mb I + \mu \cdot \1_o & \begin{bmatrix} \mu\\ \vdots \\ \mu 
			\end{bmatrix}  & \mu\cdot\1     & \mb 0 \\
			\begin{bmatrix} 0 \cdots   0 
			\end{bmatrix}  & \begin{bmatrix} \mu \cdots   \mu 
			\end{bmatrix}  & 1 & \begin{bmatrix} \mu \cdots   \mu 
			\end{bmatrix}  & \begin{bmatrix} 0 \cdots   0 
			\end{bmatrix}  \\   
			\mb 0 & \mu\cdot\1  & \begin{bmatrix} \mu\\ \vdots \\ \mu 
			\end{bmatrix}    & \mb I + \mu \cdot  \1_o & \mu\cdot\1 \\
			\mb 0 & \mb 0  &\begin{bmatrix} 0\\ \vdots \\ 0 
			\end{bmatrix}    & \mu\cdot\1     & \mb J 
		\end{bmatrix}.  
	\end{align} 
	Here, the center row/column vector is indexed at $0$, the matrices $\mb J,\mb I,\1$ and $\1_o$ are square and of size $(p_0-1)^2$. Among which, $\mb I$ is the identity matrix, $\1$ is the ones matrix whereas $\1_o$ has all off diagonal entries equal 1. Also $\abs{\mb J}$  has property $ \abs{\mb J_{ij}} < 1$ for all $i,j$.
	
	As for the singular values, notice that the first and second inequalities consider submatrix not containing  $\mb J$ since $\mb\tau \subseteq[\pm p_0]$; thus the first inequality can be derived with Gershgorin disc theorem directly, and the second inequality with the upper bound with its Frobenius norm:
	\begin{align}
		\sigma_{\mr{max}}\paren{\injector_{\mb\tau^c}^*\mb M\injector_{\mb\tau}} \leq\mu\sqrt{(2p_0-1)\abs{\mb\tau}} < \mu \sqrt{p\abs{\mb\tau}}.
	\end{align} 
	Finally by recalling $p = 3p_0-2 > 2p_0-1$.  The last inequality is direct from bound of $\ip^*\convmtx{\mb a_0}$:
	\begin{align}
		\sigma_{\mr{max}}\paren{\injector_{\mb\tau^c}^*\mb M\injector_{\mb\tau^c}} \leq  \norm{\convmtx{\mb a_0}^*\ip\ip^*\convmtx{\mb a_0}}2  =   \norm{\ip^*\convmtx{\mb a_0}\convmtx{\mb a_0}^*\ip}2 = \norm{\ip^*\convmtx{\mb a_0}^*\convmtx{\mb a_0}\ip}2   \leq  1+\mu p   
	\end{align} 
where the third equality is derived via commutativity of convolution.
\end{proof}  
   
\begin{lemma}[Shift space vectors in widened subspace]\label{lem:alpha_beta_tau_lb} Let $(\mb a_0,\theta,\abs{\mb\tau})$ satisfy the sparsity-coherence condition $\mr{SCC}(c_\mu)$. Then for every $\mb a\in\goodregion$, every $\mb\alpha$ satisfying $\mb a = \ip^*\convmtx{\mb a_0}\mb\alpha$ and $\norm{\mb\alpha_{\mb\tau^c}}2\leq \gamma(c_\mu)$ has
\begin{align} 
	\abs{ \norm{\mb\alpha_{\mb\tau}}2-1}\leq c_\mu;
\end{align}
moreover, $\mb\beta =  \convmtx{\mb a_0}^*\ip\mb a$ satisfies
\begin{align}
	1-3c_\mu \leq \norm{\mb\beta_{\mb\tau}}2^2 \leq 1+\frac{c_\mu}{\abs{\mb\tau}\log^2\theta^{-1}},\quad \norm{\mb\beta_{\mb\tau^c}}\infty \leq \frac{c_\mu}{\sqrt{\abs{\mb\tau}} \log^2\theta^{-1} } ,\quad \norm{\mb\beta_{\mb\tau^c}}2\leq \frac{c_\mu}{\abs{\mb\tau}\theta\log\theta^{-1}} \min\set{\sqrt\theta, \gamma(c_\mu) }.   
\end{align}

\end{lemma}

\begin{proof}
	Write $-1/\log\theta = \clog$ and $\gamma = \gamma(c_\mu)$ for convenience.  First, by using bounds on $\gamma$ in \eqref{eqn:gamma} and $\mu\abs{\mb\tau}<1$ we obtain: 
\begin{align}\label{eqn:alpha_beta_tau_lb_gamma2_ub}
	\begin{dcases} 
		\gamma\cdot\sqrt{1+\mu p} \,\leq\, \gamma\, (1+\sqrt{\mu p})\,\leq\, c_\mu\clog^2/2 \\
		\gamma\cdot \sqrt{1+\mu^2p} \,\leq\, \gamma \paren{1+\sqrt{\mu^2 p} }\,\leq\,  \frac{c_\mu\clog^2}4 \paren{\frac{1}{\sqrt{\abs{\mb\tau}}}+ \sqrt\mu} \,\leq\, \frac{c_\mu\clog^2}{2\sqrt{\abs{\mb\tau}}} \\
		\gamma \cdot \mu\sqrt{p\abs{\mb\tau}} \,\leq\, \gamma \cdot \sqrt{\mu p} \cdot \sqrt{\mu \abs{\mb\tau}} \,\leq\, c_\mu\clog^2/4
	\end{dcases}
\end{align}   
     
\vsni Let $\mb a = \ip^*\convmtx{\mb a_0}\mb\alpha$ with $\norm{\mb\alpha_{\mb\tau^c}}2 < \gamma$. Utilize properties of $\ip^*\convmtx{\mb a_0}$ from \Cref{fact:M_entries} and $\mu\abs{\mb\tau} < c_\mu/4$ and \eqref{eqn:alpha_beta_tau_lb_gamma2_ub}, we have: 
\begin{align}
	\norm{\mb\alpha_{\mb\tau}}2 &\,\geq\, \norm{\ip^*\convmtx{\mb a_0}\injector_{\mb\tau}}2^{-1}\paren{\norm{\mb a}2 - \norm{\ip^*\convmtx{\mb a_0}\mb\alpha_{\mb\tau^c}}2 } \,\geq\, \norm{\ip^*\convmtx{\mb a_0}\injector_{\mb\tau}}2^{-1}\paren{1 - \norm{\ip^*\convmtx{\mb a_0}}2\norm{\mb\alpha_{\mb\tau^c}}2 } \notag\\
	 &\,\geq\,  \frac{1}{\sqrt{1+\mu\abs{\mb\tau}}}\paren{1-\gamma\cdot \sqrt{1+\mu p} } \,\geq\, \frac{1-c_\mu/2 }{\sqrt{1+c_\mu/4}}\,\geq\, 1-c_\mu,
\end{align}
and similarly, the upper bound can be derived as:
 \begin{align}
	\norm{\mb\alpha_{\mb\tau}}2 &\,\leq\, \sigma^{-1}_{\mr{min}}\paren{\ip^*\convmtx{\mb a_0}\injector_{\mb\tau}}\paren{\norm{\mb a}2 + \norm{\ip^*\convmtx{\mb a_0}\mb\alpha_{\mb\tau^c}}2 } \,\leq\,  \sigma^{-1}_{\mr{min}}\paren{\ip^*\convmtx{\mb a_0}\injector_{\mb\tau}} \paren{1 + \norm{\ip^*\convmtx{\mb a_0}}2\norm{\mb\alpha_{\mb\tau^c}}2 } \notag\\
	 &\,\leq\,  \frac{1}{\sqrt{1-\mu\abs{\mb\tau}}}\paren{1+\gamma\cdot \sqrt{1+ \mu p} } \,\leq\, \frac{1+c_\mu/2}{\sqrt{1-c_\mu/4}} \,\leq\, 1+c_\mu.
\end{align} 
The bound of $\norm{\mb\beta_{\mb\tau}}2^2$ can be simply obtained using $\mu\abs{\mb\tau}< c_\mu/4$ and $\gamma$ bound from \eqref{eqn:alpha_beta_tau_lb_gamma2_ub} as: 
\begin{align}
\norm{\mb\beta_{\mb\tau}}2^2 &\,\leq\,\sigma^2_{\mr{max}}\paren{\injector^*_{\mb\tau}\convmtx{\mb a_0}\ip} \,\leq\, 1+\mu\abs{\mb\tau} \,\leq\,  1+\frac{c_\mu\clog^2}{\abs{\mb\tau}} \\
\norm{\mb\beta_{\mb\tau}}2^2&\,\geq\, \paren{\sigma_{\mr{min}}\paren{\injector^*_{\mb\tau}\mb M\injector_{\mb\tau}}\norm{\mb\alpha_{\mb\tau}}2 - \sigma_{\mr{max}}\paren{\injector_{\mb\tau}^*\mb M\injector_{\mb\tau^c}}\norm{\mb\alpha_{\mb\tau^c}}2}^2  \notag \\
&\,\geq\,  \paren{\paren{1-\mu\abs{\mb\tau}}\paren{1-c_\mu} -\mu\sqrt{p\abs{\mb\tau}}\cdot\gamma }^2 \,\geq\, 1-3c_\mu.
 \end{align}
As for the upper bound of  and $\norm{\mb\beta_{\mb\tau^c}}\infty$, follow from \eqref{eqn:alpha_beta_tau_lb_gamma2_ub}, we have: 
\begin{align}
	\norm{\mb\beta_{\mb\tau^c}}\infty &\,\leq\, \norm{\injector^*_{\mb\tau^c}\mb M\mb\alpha_{\mb\tau}}\infty + \norm{\injector^*_{\mb\tau^c}\mb M\mb\alpha_{\mb\tau^c}}\infty \,\leq\, \mu\sqrt{\abs{\mb\tau}}  \norm{\mb\alpha_{\mb\tau}}2 +  \sqrt{1+\mu^2p}\norm{\mb\alpha_{\mb\tau^c}}2 \notag \\
	&\,\leq\, \frac{c_\mu\clog^2(1+c_\mu)}{4\abs{\mb\tau}} + \gamma\cdot \sqrt{1+\mu^2p}  \,\leq\, \frac{c_\mu\clog^2}{\sqrt{\abs{\mb\tau}}};   
\end{align}  
the bound for $\norm{\mb\beta_{\mb\tau^c}}2$ requires two inequalities,  we know
\begin{align}
	\norm{\mb\beta_{\mb\tau^c}}2 &\,\leq\, \norm{\injector^*_{\mb\tau^c}\mb M\mb\alpha_{\mb\tau}}2 + \norm{\injector^*_{\mb\tau^c}\mb M\mb\alpha_{\mb\tau^c}}2 \,\leq\, \mu\sqrt{p\abs{\mb\tau}}\norm{\mb\alpha_{\mb\tau}}2 + \paren{1+\mu p}\norm{\mb\alpha_{\mb\tau^c}}2, \label{eqn:alpha_beta_prop_beta_2_1}
\end{align}  
for the first inequality, use $\big(\mu\abs{\mb\tau}^2\big)^{3/4}\paren{\mu p^2\theta^2}^{1/4} = \mu\sqrt{p\theta}\abs{\mb\tau}^{3/2} < c_\mu\clog^2/4$ , definition of $\gamma$ and $\theta\abs{\mb\tau}\leq c_\mu\clog^2/4$ we have:  
\begin{align}       
	\eqref{eqn:alpha_beta_prop_beta_2_1} &\,\leq\, \frac{\mu\sqrt{p\theta}\abs{\mb\tau}^{3/2}}{\sqrt\theta\abs{\mb\tau}}\paren{1+c_\mu} + \frac{\sqrt{\theta\abs{\mb\tau}}\cdot \sqrt{\abs{\mb\tau}}\gamma}{\sqrt\theta\abs{\mb\tau}} + \frac{\mu p\sqrt{\theta}\abs{\mb\tau}\gamma}{\sqrt\theta\abs{\mb\tau}} \notag \\
	&\,\leq\, \frac{2c_\mu\clog^2 + c_\mu\clog^3 + c_\mu\clog^2}{4\sqrt\theta\abs{\mb\tau}} \,\leq\, \frac{c_\mu\clog^2 }{\sqrt\theta\abs{\mb\tau}},
\end{align}  
and similarly for the second inequality, use both conditions of $\mu$, we have:
\begin{align}
	\eqref{eqn:alpha_beta_prop_beta_2_1} & \,\leq\, \frac{\gamma}{\theta\abs{\mb\tau}}\cdot\frac{\mu\sqrt p\theta\abs{\mb\tau}^{3/2}}{\gamma}(1+c_\mu) + \gamma + \mu p\gamma\notag \\
	&\,\leq\, \frac{\gamma}{\theta\abs{\mb\tau}}\cdot \frac{4\mu \sqrt p\theta\abs{\mb\tau}^{3/2}}{c_\mu\clog^2}\cdot\max\set{\sqrt{\abs{\mb\tau}},\,\sqrt{\mu p},\,\mu p\sqrt\theta \abs{\mb\tau}} + \frac{\gamma}{\theta\abs{\mb\tau}}\cdot \theta\abs{\mb\tau} + \frac{\gamma}{\theta\abs{\mb\tau}}\cdot \mu p\theta\abs{\mb\tau}
	\notag \\
	&\,\leq\, \frac{\gamma}{\theta\abs{\mb\tau}}\cdot\paren{\frac{4}{c_\mu\clog^2}\cdot \max\set{\mu\abs{\mb\tau}^2\cdot \sqrt p\theta,\,\mu(p\theta)\abs{\mb\tau}\cdot\sqrt{\mu\abs{\mb\tau}},\,\mu\sqrt{p\theta}\abs{\mb\tau}^{3/2}\cdot\mu p\theta\abs{\mb\tau}} +  \frac{c_\mu\clog^2}{4} + \frac{c_\mu\clog^2}{4} }\notag\\
	&\,\leq\, \frac{\gamma}{\theta\abs{\mb\tau}}\paren{\frac{c_\mu\clog}4 + \frac{c_\mu\clog^2}4 + \frac{c_\mu\clog^2}4} \,\leq\, \frac{c_\mu\clog\gamma}{\theta\abs{\mb\tau}},
\end{align}
which completes the proof.
\end{proof}

\begin{corollary}[ $\abs{\innerprod{\mb\beta_{\mb\tau^c}}{\mb x_{0,\mb\tau^c}}}$ is small]\label{cor:tail_beta_x0_tauc}  Given  $\mb x_0\simiid  \mr{BG}(\theta)$ in $\R^n$ and $\abs{\mb\tau}, c_\mu$ such that $(\mb a_0,\theta,\abs{\mb\tau})$ satisfies the sparsity-coherence condition $\mr{SCC}(c_\mu)$. Write $\lambda = c_\lambda/\sqrt{\abs{\mb\tau}}$ with some $c_\lambda \geq 1/5$, then if $c_\mu  \leq \tfrac{c_\lambda}{25}$,  
\begin{align}
	\prob{\abs{\sum_{i\in\mb\tau^c}\mb\beta_i\mb x_{0i}} > \frac\lambda{10}} \leq 2\theta,\qquad  \prob{\abs{\sum_{i}\mb\beta_i\mb x_{0i}} > \frac{\lambda}{10}} \leq \theta\abs{\mb\tau}  + 2\theta.
\end{align}
\end{corollary}

\begin{proof}
	We bound tail probability of the first result with Gaussian moments \Cref{lem:gaussian_moment} and Bernstein inequality \Cref{lem:mc_bernstein_scalar}. Via H\"older's inequality, $\sum_{i\in\mb\tau^c}\E(\beta_i x_i)^q = \E x_0^q \norm{\mb\beta_{\mb\tau^c}}q^q \leq \theta(q-1)!!\norm{\mb\beta_{\mb\tau^c}}2^2\norm{\mb\beta_{\mb\tau^c}}\infty^{q-2}$, thus 
 \begin{align} \label{eqn:tail_bx_tauc1}
 	\prob{\abs{\sum_{i\in\mb\tau^c}\mb\beta_i\mb x_{0i}} > \lambda/10} \leq  2\exp\paren{\frac{-(\lambda/10)^2}{2\theta\norm{\mb\beta_{\mb\tau^c}}2^2 + 2(\lambda/10)\norm{\mb\beta_{\tau^c}}\infty }} 
 \end{align}
Write $\clog = -\frac{1}{\log\theta}$,  \Cref{lem:alpha_beta_tau_lb} imples when $c_\mu \leq \frac{c_\lambda}{25}$, we have $\theta \norm{\mb\beta_{\mb\tau^c}}2^2 \leq \frac{c_\mu^2\clog^2}{\abs{\mb\tau}^2} \leq \frac{\clog\lambda^2}{625} $ and $\norm{\mb\beta_{\mb\tau^c}}\infty \leq \frac{c_\mu\clog}{\sqrt{\abs{\mb\tau}}}\leq  \frac{\clog\lambda}{25}$, therefore,  
\begin{align} 
	\eqref{eqn:tail_bx_tauc1} \leq 2\exp\paren{\frac{-\lambda^2/100}{2\clog\lambda^2/625 + 2(\clog\lambda/25)\cdot(\lambda/10)}} \leq 2\exp\paren{ \log\theta} \leq 2\theta\label{eqn:chi_ct_sj_tail} 
\end{align}
The second tail bound is straight forward from the first tail bound as follows:
\begin{align}
	\prob{\abs{\sum_{i}\mb\beta_i\mb x_{0i}} > \frac{\lambda}{10}} &\leq \prob{\abs{\mb\beta_{\mb\tau}^*\mb x_{\mb\tau}} + \abs{\mb\beta_{\mb\tau^c}^*\mb x_{\mb\tau^c}} > \lambda/10} \notag \\
	& \leq \prob{\mb x_{\mb\tau} \neq \mb 0} + \prob{\mb x_{\mb\tau} = \mb 0}\cdot \prob{\abs{\mb\beta^*_{\mb\tau^c}\mb x_{\mb\tau^c}} > \lambda/10} \notag \\
	&\leq \theta\abs{\mb\tau} + 2\theta.	
\end{align}
\end{proof}

\begin{corollary}[ $\abs{\innerprod{\mb\beta_{\mb\tau\setminus (0)}}{\mb x_{0,\mb\tau\setminus(0)}}}$ is small near shifts]\label{cor:tail_beta_x0_tau}  Suppose that $\mb x_0\simiid  \mr{BG}(\theta)$ in $\R^n$, and $\abs{\mb\tau}, c_\mu$ such that $(\mb a_0,\theta,\abs{\mb\tau})$ satisfies the sparsity-coherence condition $\mr{SCC}(c_\mu)$, then if $c_\mu\leq \frac{1}{10}$, for any $\mb a$ such that $\abs{\mb\beta_{(1)}}\leq \frac{\lambda}{4\log\theta^{-1}}$, we have
\begin{align}
	\prob{\abs{\sum_{i\in\mb\tau\setminus(0)}\mb\beta_i\mb x_{0i}} > \frac{2\lambda}5} \leq 2\theta
\end{align}
\end{corollary}
\begin{proof}
For the last tail bound, write $\mb x = \mb\omega\circ\mb g$. Wlog define $\mb\beta_0$ be the largest correlation $\mb\beta_{(0)}$, define random variables $s' = \innerprod{\mb\beta_{\mb\tau\setminus\set{0}}}{\mb x_{\mb\tau\setminus\set{0}}}$. Firstly most of the entries of $\mb x_{\mb\tau} $ would be zero since via Bernstein inequality with $\theta\abs{\mb\tau} < 0.1$:
\begin{align}
	\prob{\sum_{i\in\mb\tau}\mb\omega_i > \log\theta^{-1}}\leq \prob{\sum_{i\in\mb\tau}\mb\omega_i > \theta\abs{\mb\tau} +  0.9\log \theta^{-1}} \leq \exp\paren{\frac{-0.9^2\log^2\theta^{-1}}{2\paren{\theta\abs{\mb\tau}+0.9\log\theta^{-1}/3}}} \leq \theta   
\end{align}  
thus with probability at least $1-\theta$, we can write $s'$ as a Gaussian r.v. with variation bounded as $\E s'^2 \leq \E\brac{\sum_{i=1}^{\log\theta^{-1}}\mb\beta_i\mb g_i}^2 = \log\theta^{-1}\mb\beta_{(1)}^2 $, then via Gaussian tail bound \Cref{lem:gaussian_tail_bound}: 
\begin{align}
	\prob{\abs{s'} > 0.4\lambda} &\leq  \prob{\abs{g} > \frac{0.4\lambda}{\sqrt{\log\theta^{-1}}\abs{\mb\beta_{(1)}}} } + \prob{\sum_{i\in\mb\tau}\mb\omega_i > \log\theta^{-1}} \leq \frac{2}{\sqrt{2\pi}}\exp\paren{-1.2\log\theta^{-1}} + \theta  \leq 2\theta,
\end{align}
\end{proof}

% !TEX root = ../../BD_DQ.tex
\section{Euclidean gradient as soft-thresholding in shift space}\label{sec:grad_soft_thresh}

In this section, we will study the Euclidean gradient \eqref{eqn:euc_grad}, by deriving bounds showing that the $\mb\chi$ operator approximates a soft-thresholding function in shift space (\Cref{lem:chibeta_ublb} and \Cref{cor:chibeta_ct}). Furthermore, we will show the operator $\mb\chi[\mb\beta_i]$ is monotone in $\abs{\mb\beta_i}$ from \Cref{lem:mono_chi}. A figure of visualized $\mb\chi$ operator is shown in \Cref{fig:expect_chi}.
 
\paragraph{Expectation of $\mb \chi$ operator.} To understand the $\mb\chi$ operator, we shall first consider a simple case---when $\mb x_0$ is highly sparse.
By definition of $\mb\beta$ from \eqref{eqn:coef-beta} we can see that $\mb\beta$ has a short support of size at most $2p-1$, when $\mb x_0$ has support entries separated by at least $2p$, the entries of vector $\mb\chi[\mb\beta]_i$ become sum of independent random variables as:
\begin{align}
	\mb\chi[\mb\beta]_i = \innerprod{\shift{\mb x_0}{-i}}{\soft{\mb x_0*\wc{\mb\beta}}{\lambda}} \underbrace{=}_{\text{$\mb x_0$ sep.}} \innerprod{\shift{\mb x_0}{-i}}{\soft{\mb\beta_i\shift{\mb x_0}{-i}}{\lambda}} = \sum_{j\in\supp(\mb x_0)}\mb g_j\cdot \soft{\mb g_j\cdot \mb\beta_i}{\lambda} \notag
\end{align}
where $\paren{\mb g_j}_{j\in[n]}$ are standard Gaussian r.v.s. 

\vsni The following lemma describes the behavior of the summands in the above expression:

\begin{lemma}[Gaussian smoothed soft-thresholding]\label{lem:expect_soft_gaussian} Let $g\sim \mc N(0,1)$. Then for every $b,s\in\R $ and $\lambda > 0 $,
\begin{align}\label{eqn:expect_soft_gaussian}
\E_g \Bigl[ g\soft{b\cdot g+s}{\lambda} \Bigr] = b\paren{1- \mr{erf}_b(\lambda,s)},
\end{align}
where
\begin{align}\label{eqn:erfb}
	\mr{erf}_b (\lambda,s) = \frac12\mr{erf}\paren{\frac{\lambda+s}{\sqrt 2\abs{b}}} + \frac12\mr{erf}\paren{\frac{\lambda-s}{\sqrt 2 \abs b}}.
\end{align}
Furthermore, for $s =0$, $b\in[-1,1]$ and $\eps\in(0,1/4)$, letting $\sigma = \sign(b)$ we have 
\begin{align}\label{eqn:expect_soft_gaussian_ublb}
	 \sigma \soft{b}{\nu_2'\lambda} \leq\sigma\E_g \Bigl[ g\soft{ b\cdot g}{\lambda} \Bigr] \leq \sigma\soft{b}{\nu_1'(\eps) \lambda} + \eps
\end{align}
where $\nu_1'(\eps) = 1/(2\sqrt{-\log\eps})$ and $\nu_2' = \sqrt{2/\pi}$. 
\end{lemma}

\begin{proof}
Wlog assume $b>0$. 
	Write $f$ as the pdf of standard Gaussian distribution. With integral by parts: 
	\begin{align}
		\int_{-\infty}^t t'f(t')dt' = -f(t), \quad  \int_{-\infty}^t t^{'2} f(t')dt' = \frac{1}{2}\mr{erf}\paren{\frac{t}{\sqrt 2}} - tf(t) \notag 
	\end{align} 
Integrating, we obtain
\begin{align}
	\E \Bigl[ g\soft{b\cdot g+s}{\lambda} \Bigr] = \int_{t\geq \frac{\lambda-s }{b}}\paren{bt^2 -(\lambda-s)t} \, f(t) dt   +  \int_{t\leq -\frac{\lambda+s }{b}}\paren{bt^2 +(\lambda+s) t}  \, f(t) dt, \notag
\end{align}
by writing $L = \lambda-s$, the integral of first summand 
\begin{align}
	 \int_{t\geq \frac{L}{b}}\paren{bt^2 -L t}f(t)dt &= b\brac{\frac12 - \frac12\mr{erf}\paren{\frac{L}{\sqrt 2b}} +  \frac{L}{b}f\paren{\frac{L}{b}}} - L f\paren{\frac{L}b}  = \frac{b}{2} - \frac{b}{2} \mr{erf}\paren{\frac{L}{\sqrt 2b}},\notag
\end{align}
and similarly for the second summand, which gives
\begin{align}
	\E \Bigl[ g\soft{b\cdot g+\mb s}{\lambda} \Bigr] = \frac{b}{2} - \frac{b}{2} \mr{erf}\paren{\frac{\lambda- s}{\sqrt 2b}} + \frac{b}{2} - \frac{b}{2} \mr{erf}\paren{\frac{\lambda+ s}{\sqrt 2b}} = b\paren{1-\mr{erf}_b(\lambda, s)} \notag
\end{align}
For $b<0$, alternatively we have
\begin{align}
	\bb E\Bigl[g S_{\lambda}[-|b| \cdot g + s]\Bigr] = - \bb E[g S_{\lambda}[ |b| \cdot g -  s ] = - |b| ( 1 - \mr{erf}_b( \lambda, - s ) ) = b ( 1 - \mr{erf}_b(\lambda,  s)), \notag
\end{align}
To show \eqref{eqn:expect_soft_gaussian_ublb}, via definition of error function, for $x>0$, we know:
\begin{align}\label{eqn:erf_ublb}
\min\set{1-\eps, \frac{1-\eps}{\sqrt{\log(1/\eps)}} x }\leq \mr{erf}(x) = \frac{2}{\sqrt\pi}\int_0^x e^{-t^2} dt \leq \frac{2x}{\sqrt\pi} 
\end{align}
where the lower bound is derived by first knowing erf is increasing thus for all $x>\sqrt{\log(1/\eps)}$,  
\begin{align}
	\mr{erf}(x) \geq 1-e^{-x^2}\geq 1-e^{\log \eps} = 1-\eps \notag
\end{align}
and from concavity of $\mr{erf}$ we have for $0<x<\sqrt{\log(1/\eps)} = T$, 
\begin{align}
	\mr{erf}(x) \geq \frac{\mr{erf}(T) -\mr{erf}(0)}{T-0}x +\mr{erf}(0) \geq \frac{1-\eps}{\sqrt{\log(1/\eps)}}x.\notag
\end{align}
Lastly plug \eqref{eqn:erf_ublb} into \eqref{eqn:expect_soft_gaussian} and apply condition $\abs{b}\leq 1$ and $\eps < 1/4$ we have
\begin{align}
	\abs b - \sqrt{\frac{2}{\pi}}\lambda \leq \abs{b} - \abs{b}\mr{erf}\paren{\frac{\lambda}{\sqrt 2\abs{b}}} \leq \max\set{\abs{b}\eps, \abs b - \frac{\lambda(1-\eps)}{\sqrt{2\log(1/\eps)}}}\leq \max\set{\eps,\abs b - \frac{\lambda}{2\sqrt{\log(1/\eps)}}},\notag 
\end{align}
which completes the proof.
\end{proof}

\vsni This lemma establishes when $\mb x_0$ is separated, then $\mb\chi$ is soft thresholding operator on $\mb\beta$ with threshold about $\lambda/2$. This phenomenon extends beyond the separated case, as long as when $\mb x_0$ is sufficiently sparse (when \Cref{asm:theta_mu} holds). Recall that $\mb\chi:\R^n\to\R^n$ is defined as
\begin{align}\label{eqn:chi_def}
	\mb\chi[\mb\beta] = \checkmtx{\mb x_0}\soft{\checkmtx{\mb x_0}\mb\beta}{\lambda}.
\end{align}
The following lemma bounds its expectation:

\begin{figure}[t!]
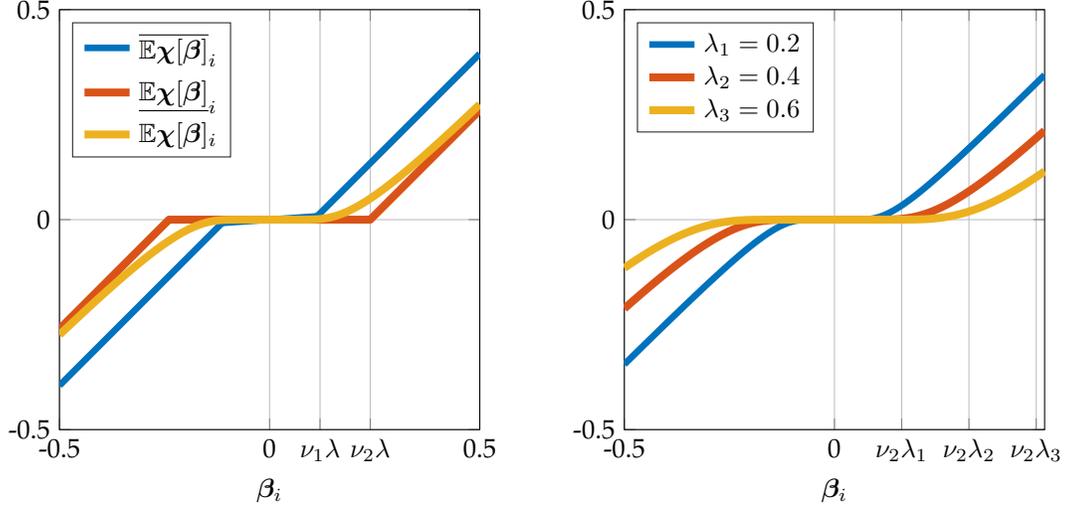

\centering
\begin{minipage}{.45\textwidth}
	\centering
	\input{sections/expect_chi.tex}		
\end{minipage}
\begin{minipage}{.45\textwidth}
	\centering
	\input{sections/expect_chi2.tex}		
\end{minipage}
\caption{{\bf A numerical example of $\E\mb\chi[\mb\beta]_i$.} We provide figures for the expectation of $\mb\chi$ when entries of $\mb x_0$ are $2p$-separated. Left: the yellow line is the function $\mb\beta_i \to \mb\beta_i\paren{1-\mr{erf}_{\mb\beta_i}(\lambda,0)}$ derived from \eqref{eqn:expect_soft_gaussian}, and the blue/red lines are its upper/lower bound \eqref{eqn:expect_soft_gaussian_ublb} utilized in the analysis respectively. Right: functions of  $\mb\beta_i \to \mb\beta_i\paren{1-\mr{erf}_{\mb\beta_i}(\lambda,0)}$ with different $\lambda$, the section of function of $\mb\beta_i > \nu_2\lambda$ are close to linear.
} \label{fig:expect_chi} 
\end{figure}
\begin{lemma}[Expectation of $\mb\chi(\mb\beta)$]\label{lem:chibeta_ublb} Let $\mb x_0\simiid\mr{BG}(\theta)$ and $\lambda>0$, then for every $\mb a\in\Sp^{p-1}$ and every $i\in[n]$, define the operator $\mb\chi$ as in \eqref{eqn:chi_def}, then
\begin{align} 
 n^{-1}\E\mb\chi[\mb\beta]_i = \theta \mb\beta_i\paren{1-\E_{\mb s_i} \mr{erf}_{\mb\beta_i}(\lambda,\mb s_i)} \label{eqn:exp_chibeta}
\end{align}
where $\mb s_i = \sum_{\ell\neq i}\mb\beta_\ell \mb x_{0\ell}$. Suppose $(\mb a_0,\theta,\abs{\mb\tau})$ satisfies the sparsity-coherence condition $\mr{SCC}(c_\mu)$ and $\lambda = c_\lambda/\sqrt{\abs{\mb\tau}}$ for some $c_\lambda > 1/5$ and $\sigma_i = \sign(\mb\beta_i)$, then there exists some numerical constant $\ol c$ such that if $c_\mu \leq  \ol c  $ then for every  $\mb a\in\goodregion$ and every $i\in[n]$, \eqref{eqn:exp_chibeta} has upper bound 
\begin{align} \label{eqn:exp_chi_ub}
	\sigma_i n^{-1}\E\mb\chi[\mb\beta]_i \leq  \sigma_i n^{-1}\ol{\E\mb\chi[\mb\beta]}_i := \begin{cases}
		4\theta^2\abs{\mb\tau}\abs{\mb\beta_i}  &\quad \abs{\mb\beta_i} < \nu_1\lambda\\
		\theta\paren{\abs{\mb\beta_i} - \nu_1\lambda/2}  &\quad \abs{\mb\beta_i} \geq \nu_1\lambda
	\end{cases}, 
\end{align}
and lower bound
\begin{align}
	\sigma_i n^{-1}\E\mb\chi[\mb\beta]_i \geq   \sigma_i n^{-1} \ul{\E\mb\chi[\mb\beta]}_i	=:\theta\soft{\abs{\mb\beta_i}}{\nu_2\lambda},   
\end{align}
where $\nu_1 = 1/ \left( 2\sqrt{\log\theta^{-1}} \right)$,  $\nu_2 = \sqrt{2/\pi}$. 
\end{lemma}

This lemma shows the expectation of $\mb\chi[\mb\beta]_i$ acts like a shrinkage operation on $|\mb\beta_i|$: for large $|\mb\beta_i|$, it acts like a soft thresholding operation, and for small $|\mb\beta_i|$, it reduces $|\mb\beta_i|$ by  multiplying  a very small number $4\theta\abs{\mb\tau} \ll 1$. We rigorously prove this segmentation of $\mb\chi$ operator as follows:

\vsni 

\begin{proof} First, since $\shift{\mb x_0}{i}\equiv_d \shift{\mb x_0}{j}$,
\begin{align}
	\mb\chi[\mb\beta]_i &=  \mb e^*_i\checkmtx{\mb x_0}\soft{\checkmtx{\mb x_0}\mb\beta}{\lambda} = \innerprod{\shift{\mb x_0}{-i}}{\soft{\mb x_0*\wc{\mb\beta}}{\lambda}}\equiv_d \innerprod{\shift{\mb x_0}{-j}}{\soft{\shift{\mb x_0}{i-j}*\wc{\mb\beta}}{\lambda}} = \mb\chi[\shift{\mb\beta}{j-i}]_j \notag
\end{align}
Thus wlog let us consider $i=0$ and write $\mb x $ as $\mb x_0$. The random variable $\mb\chi[\mb\beta]_0$ can be written sum of random variables as:
\begin{align}
	\mb\chi\brac{\mb\beta}_0 &= \innerprod{\mb x}{\soft{\mb\beta_0\mb x_0+ \sum_{\ell\neq  0}\mb\beta_\ell\shift{\mb x}{-\ell}}{\lambda}} = \sum_{j\in[n]} \mb x_j \soft{\mb\beta_0 \mb x_j + \sum_{\ell\neq  0}\mb\beta_\ell \mb x_{j+\ell}}{\lambda},\notag
\end{align}
and a random variable $Z_j(\mb\beta)$ is defined as 
\begin{align}\label{eqn:chibeta_Zj}
	Z_j(\mb\beta) = \mb x_j\mc S_\lambda\brac{\mb\beta_0 \mb x_j + \sum_{\ell\in[\pm p]\setminus 0}\mb\beta_\ell \mb x_{j+\ell} },
\end{align}
gives $\mb\chi[\mb\beta]_0 = \sum_{j\in[n]}Z_j(\mb\beta)$ as sum of r.v.s. of same distribution and thus $n^{-1}\E\mb\chi[\mb\beta]_0 = \E Z_0(\mb\beta)$. Define a random variable $\mb s_0 = \sum_{\ell\neq 0}\mb\beta_\ell \mb x_{\ell}$, which is independent of $\mb x_0$. From \Cref{lem:expect_soft_gaussian}, we can conclude
\begin{align}\label{eqn:exp_chi_is_erf}
	n^{-1}\E\mb\chi[\mb\beta]_0 = \E_{\mb x_0,\mb s_0} \mb x_0\mc S_\lambda\brac{\mb\beta_0 \mb x_0 + \mb s_0 } = \theta \mb\beta_0\paren{1-\E_{\mb s_0} \mr{erf}_{\mb\beta_0}(\lambda,\mb s_0)}  
\end{align}
so that \eqref{eqn:exp_chibeta} holds for $i=0$, and hence for all $i$.

\vsni  1. (\ul{Upper bound of $\E Z$}) Wlog assume $\mb\beta_0 \geq  0$ and write $Z = Z_0$.  We derive the upper bound on $\E Z$ in two pieces.

\vsni (1). First, since $\E \mb x_0\soft{0\cdot \mb x_0+\mb s_0}{\lambda} = 0$, we have  
\begin{align}
	\E Z(\mb\beta) &\leq \mb\beta_0 \sup_{\beta\in[0,\mb\beta_0]} \frac{d}{d\beta} \E_{\mb x_0,\mb s_0}  \mb x_0\soft{\beta \mb x_0 + \mb s_0}{\lambda} = \theta\mb\beta_0 \sup_{\beta\in[0,\mb\beta_0]} \frac{d}{d\beta} \int_{\abs{\beta g + \mb s_0}>\lambda}  g\paren{\beta g+\mb s_0 - \sign(\beta g+\mb s_0)\cdot\lambda}d\mu(g)d\mu(\mb s_0) \notag \\
	&= \theta\mb\beta_0\sup_{\beta\in[0,\mb\beta_0]}\E_{g,\mb s_0}\brac{ g^2 \1_{\set{\abs{\beta g + \mb s_0}>\lambda}} } \leq \theta\mb\beta_0 \sup_{\beta\in[0,\mb\beta_0]} \E_{g,\mb s_0} \brac{g^2\paren{\1_{\set{\abs{\beta g} > \frac{9\lambda}{10}}} + \1_{\set{\abs{\mb s_0} > \frac{\lambda}{10}}}}}  \notag \\
	&\leq \theta\mb\beta_0\paren{\paren{\E g^6 }^{1/3}\prob{\abs{\mb\beta_0 g} > (9\lambda/10)}^{2/3} + \prob{\abs{\mb s_0}>\lambda/10 } }  \label{eqn:chi_ct_exp_ub1_1}
\end{align}   
We bound the tail probability of $\mb s_0$ using  \Cref{cor:tail_beta_x0_tauc} where
 \begin{align} 
 	\prob{\abs{\mb s_0} > \lambda/10} &\leq \prob{\abs{\textstyle\sum_i\mb\beta_{i}\mb x_{i}} > \lambda/10} \leq  \theta\abs{\mb\tau} + 2\theta \leq 3\theta\abs{\mb\tau}.  
\end{align}
On the other hand, the first term in \eqref{eqn:chi_ct_exp_ub1_1} can be derived by pdf of Gaussian r.v. \Cref{lem:gaussian_tail_bound} as:  
\begin{align}
	\paren{\E g^6 }^{1/3}\prob{\abs{\mb\beta_0 g} > (9\lambda/10)}^{2/3}  \leq \sqrt[3]{15}\paren{\frac{10\mb\beta_0}{9\lambda\sqrt{2\pi}}}^{2/3}\exp\paren{-\frac{\lambda^2}{4\mb\beta_0^2}} \leq \frac32\paren{\frac{\mb\beta_0}{\lambda}}^{2/3}\exp\paren{-\frac{\lambda^2}{4\mb\beta_0^2}}. \label{eqn:chi_ct_gausstail} 
\end{align}
Combine \eqref{eqn:chi_ct_sj_tail}, \eqref{eqn:chi_ct_gausstail}, when $\mb\beta_0 < \nu_1\lambda$, we know $e^{-\frac{\lambda^2}{4\mb\beta_0^2}} \leq e^{\log\theta} \leq \theta\abs{\mb\tau}$. The first type of upper bound  $\E Z$ is derived as
\begin{align} 
	\forall\,\mb\beta_0\in\brac{0,\nu_1\lambda}
	,\quad \E Z(\mb\beta)  \leq \theta\mb\beta_0\paren{\frac{3}{2}\nu_1^{2/3}\exp\paren{-\frac{\lambda^2}{4\mb\beta_0^2}} + 3\theta\abs{\mb\tau}} \leq 4\theta^2\abs{\mb\tau}\mb\beta_0.\label{eqn:chi_ct_expbd_1}
\end{align}

\vsni (2). The second type of upper bound can be derived directly from \Cref{lem:expect_soft_gaussian}: 
\begin{align}\label{eqn:chi_ct_exp_ub2_1}
	\E Z(\mb\beta) &\leq  \E_{\mb x_0} \E_{\mb s_0} \mb x_0\soft{\mb\beta_0\mb x_0 + \mb s_0  }{\lambda}\leq \E_{\mb x_0} \mb x_0 \soft{\mb\beta_0\mb x_0}{\lambda} + \E_{\mb x_0}\abs{\mb x_0}\E_{\mb s_0}\abs{\mb s_0}\notag \\
	&\leq \theta\cdot\paren{\soft{\mb\beta_0}{\nu_1'\lambda} + \eps  + \sqrt{2/\pi}\cdot \E \abs{\mb s_0} },
\end{align} 
where $\E\abs{s}$ can be bounded with $\norm{\mb\beta}2$ and $\theta\abs{\mb\tau} < c_\mu\clog$ from \Cref{lem:alpha_beta_tau_lb}. When $c_\mu < \frac{1}{10}$, observe that
\begin{align}
		\E\abs{s} &\leq \sqrt{\sum_{\ell}\E \mb x_{\ell}^2\mb \beta_\ell^2}\leq\sqrt\theta\paren{\norm{\mb\beta_{\mb\tau}}2 + \norm{\mb\beta_{\mb\tau^c}}2 } \leq\sqrt{\theta}  \paren{1+c_\mu} + \frac{c_\mu\clog}{\abs{\mb\tau}} \leq \frac{2c_\mu\clog}{\sqrt{\abs{\mb\tau}}}.\label{eqn:chi_ct_Es_ub} 
\end{align}
Now choose $\eps = \theta \leq \frac{c_\mu\clog}{\abs{\mb\tau}} $, so that $\nu_1' = \nu_1 = \frac{\sqrt{\clog}}{2}$ in  \eqref{eqn:chi_ct_exp_ub2_1}. Since $c_\mu < \frac{c_\lambda}{25}$ we gain
\begin{align}
 	\E Z (\mb\beta) & \leq  \theta \paren{\soft{\mb\beta_0}{\nu_1\lambda} + \frac{c_\mu\clog}{\abs{\mb\tau}} + \sqrt{\frac{2}{\pi}}\cdot \frac{2c_\mu\clog}{\sqrt{\abs{\mb\tau}}} }  \leq \theta\paren{\soft{\mb\beta_0}{\nu_1\lambda} +   \frac{3c_\mu \clog}{\sqrt{\abs{\mb\tau}}} } \notag \\
 	&\leq \theta\paren{\soft{\mb\beta_0}{\nu_1\lambda} + \frac{\sqrt{\clog}}{5}\lambda }  \leq \theta\paren{\soft{\mb\beta_0}{\nu_1\lambda} + \frac{1}{2}\nu_1\lambda } \label{eqn:chi_ct_exp_ub1_2}
\end{align}
  
\vsni (3). Combine both \eqref{eqn:chi_ct_expbd_1} and \eqref{eqn:chi_ct_exp_ub1_2}, we can thus conclude that
\begin{align}
	\E Z(\mb\beta) := \ol{\E Z(\mb\beta)}\leq \begin{cases}
		4\theta^2\abs{\mb\tau}\mb\beta_0 &\quad \mb\beta_0\leq \nu_1\lambda\\
		\theta\paren{\mb\beta_0 - \frac{\nu_1}2\lambda}  &\quad \mb\beta_0 > \nu_1\lambda
	\end{cases}.
\end{align} 

\vsni 2. (\ul{Lower bound of $\E Z$})  On the other hand, for the lower bound for $\E Z$, use the fact that $\mr{erf}_{\mb\beta}(\lambda,s)$ is concave in $\mb s_0$, we have 
\begin{align}
	\E Z(\mb\beta) &= \E_{\mb s_0}\E_{\mb x_0}\mb x_0\soft{\mb\beta_0 \mb x_0 + \mb s_0}{\lambda} = \theta\cdot \E_{\mb s_0}\brac{\mb\beta_0 - \frac{\mb\beta_0}2\cdot\mr{erf}\paren{\frac{\lambda - \mb s_0}{\sqrt 2\abs{\mb\beta_0}}} -\frac{\mb\beta_0}2\cdot\mr{erf}\paren{\frac{\lambda+\mb s_0}{\sqrt 2\abs{\mb\beta_0}}} } \notag \\
	&\geq \theta\paren{\mb\beta_0 - \mb\beta_0\cdot \mr{erf}\paren{\frac{\lambda}{\sqrt 2\abs{\mb\beta_0}}}  }  \geq \theta\cdot  \soft{\mb\beta_0}{\nu_2'\lambda} =: \ul{\E Z(\mb\beta)}.
\end{align}
The proof of $\mb\beta_0 < 0$ is in the same vein. For cases of $i\neq 0$, since $\mb\chi[\mb\beta]_i\equiv_d\mb\chi[\shift{\mb\beta}{-i}]_0$, replace $\mb\beta_{0}$ with $\mb\beta_{i}$ we obtain the desired result. 
\end{proof}

\paragraph{Monotonicity of $\mb\chi$.} Another convenient fact of $\E\mb\chi[\mb\beta]_i$ is that it is monotone increasing w.r.t.\ $\abs{\mb\beta_i}$. The monotonicity is clear in \Cref{fig:expect_chi}; it is demonstrated rigorously with the following lemma:
\begin{lemma}[Monotonicity of $\E\mb\chi(\mb\beta)$]\label{lem:mono_chi} Suppose $\mb x_0\simiid  \mr{BG}(\theta)$ in $\R^n$, and $\abs{\mb\tau}, c_\mu$ such that $(\mb a_0,\theta,\abs{\mb\tau})$ satisfies the sparsity-coherence condition $\mr{SCC}(c_\mu)$. Define $\lambda = c_\lambda/\sqrt{\abs{\mb\tau}}$ in $\varphi_{\ell^1}$ where $c_\lambda\in\brac{0,\frac14}$, then there exists some numerical constant $\ol c>0$, such that if $c_\mu < \ol c$,  the expectation $\abs{\E[\mb\chi[\mb\beta]]_i}$ is monotone increasing in $\abs{\mb\beta_i}$. In other words, if $\abs{\mb\beta_i} > \abs{\mb\beta_j}$ then
\begin{align} 
 	\sigma_i \E\mb\chi[\mb\beta]_i \geq \sigma_j \E\mb\chi[\mb\beta]_j 
\end{align}
where $\sigma_i = \sign(\mb\beta_i)$.
\end{lemma}
The proof first operate simple calculus and then followed by studying cases of $\abs{\mb\beta_i}-\abs{\mb\beta_j}$ when either it is smaller are larger then $\lambda$.  

\vsni 

\begin{proof} 1. (\ul{Monotonicity by gradient negativity}) Wlog assume $\mb\beta_i > \mb\beta_j > 0$,  and from \Cref{lem:chibeta_ublb} we can write $(n\theta)^{-1}\E \mb\chi[\mb\beta]_i = \mb\beta_i\paren{1-\E_{\mb s_i}\mr{erf}_{\mb\beta_i}(\lambda,\mb s_i)}$. Consider $t\in[0,1]$ and define $\ell(t) = t\mb\beta_i - t\mb\beta_j$. Write the random variable $\mb s_{ij} = \sum_{\ell\neq i,j}\mb\beta_\ell \mb x_\ell $. Define $h$ as a function of $t$ such that 
\begin{align}  
	h(t) &= \E_{x,\mb s_{ij}}\brac{\paren{(1-t)\mb\beta_i + t\mb\beta_j}\paren{1-\mr{erf}_{(1-t)\mb\beta_i + t\mb\beta_j}(\lambda, ((1-t)\mb\beta_j+t\mb\beta_i)x  + \mb s_{ij})}} \notag \\
	&=  \E_{x,\mb s_{ij}}\brac{\paren{\mb\beta_i-\ell(t)}\paren{1-\mr{erf}_{\mb\beta_i - \ell(t)} (\lambda, x\cdot (\mb\beta_j+\ell(t)) +\mb s_{ij})}}.
\end{align}
Notice that $\E\mb\chi[\mb\beta]_i = h(0)$ and $\E\mb\chi[\mb\beta]_j = h(1)$ respectively, thus it suffices to prove $h'(t) < 0$ for all $t\in[0,1]$. Write $f$ as pdf of standard Gaussian r.v. where 
\begin{align}
	\mr{erf}_\beta(\lambda,\mb s_{ij}) = \int_0^{\frac{\lambda+\mb s_{ij}}{\beta}}f(z)\,dz + \int_0^{\frac{\lambda-\mb s_{ij}}{\beta}}f(z)\,dz, \notag
\end{align}
and use chain rule:
\begin{align} 
	h'(t) &= \E_{x,\mb s_{ij}}\left[(\mb\beta_j - \mb\beta_i)\paren{1-\mr{erf}_{\mb\beta_i - \ell(t)}(\lambda, x\cdot(\mb\beta_j + \ell(t))+\mb s_{ij})} \right.\notag\\
	&\phantom{=\E_{x,s}[]{}} -\paren{\mb\beta_i - \ell(t)}\cdot \frac{d}{dt} \paren{\frac{\lambda + x\cdot(\mb\beta_j + \ell(t))+ \mb s_{ij}}{ \mb\beta_i - \ell(t) }}\cdot f\paren{\frac{\lambda + x\cdot (\mb\beta_j + \ell(t))+ \mb s_{ij}}{\mb\beta_i - \ell(t)}} \notag \\
	&\phantom{=\E_{x,\mb s_{ij}}[]{}}\left.-\paren{\mb\beta_i - \ell(t)}\cdot\frac{d}{dt}\paren{\frac{\lambda - x\cdot (\mb\beta_j + \ell(t)) - \mb s_{ij}}{\mb\beta_i-\ell(t)}}\cdot f\paren{\frac{\lambda - x\cdot(\mb\beta_j +\ell(t))  - \mb s_{ij}}{\mb\beta_i - \ell(t)}}  \right]  \notag \\
	&=  (\mb\beta_j - \mb\beta_i)\E_{x,\mb s_{ij}}\left[1-\mr{erf}_{\mb\beta_i -  \ell(t)}(\lambda, x\cdot(\mb\beta_j +  \ell(t))+\mb s_{ij})\right.\notag \\ 
	&\phantom{=(\mb\beta_j-\mb\beta_i) \E_{x,\mb s_{ij}}[]{}} +  \paren{ \frac{ \lambda + x(\mb\beta_j +  \ell(t))+\mb s_{ij}}{ \mb\beta_i - \ell(t) } + x} \cdot f\underbrace{\paren{\frac{ \lambda + x (\mb\beta_j + \ell(t))  + \mb s_{ij}}{\mb\beta_i - \ell(t)}}}_{z_{\lambda_+}}  \notag \\
	&\phantom{=(\mb\beta_j-\mb\beta_i)\E_{x,\mb s_{ij}}[]{}} +\paren{ \frac{\lambda - x(\mb\beta_j+\ell(t))-\mb s_{ij}}{ \mb\beta_i  - \ell(t)} -x}\cdot f\underbrace{\paren{\frac{ \lambda - x(\mb\beta_j+   \ell(t))  - \mb s_{ij}}{\mb\beta_i - \ell(t)}}}_{z_{\lambda_-}}\bigg]  \notag \\
	& = (\mb\beta_j - \mb\beta_i)\E_{x,\mb s_{ij}}\brac{1-\int_0^{z_{\lambda_+}} f(z)\,dz -\int_0^{z_{\lambda_-}} f(z)\,dz + (z_{\lambda_+} +x)f(z_{\lambda_+}) + (z_{\lambda_-} -x)f(z_{\lambda_-}) }. \label{eqn:monotone_ht}
\end{align} 
Consider the term only related to $z_{\lambda_+}$, condition on cases that it is either positive or negative, observe that 
\begin{align}
	\begin{cases}
		{\color{black} \mu_{+-} := \E_{x,\mb s_{ij}|z_{\lambda_+ \leq 0}}\brac{ \int_0^{z_{\lambda_+}} f(z) \,dz - z_{\lambda_+} f(z_{\lambda_+})  } =  \E_{x,s|z_{\lambda_+ \leq 0}}\brac{ -\int_0^{-z_{\lambda_+}} f(z) \,dz - z_{\lambda_+} f(z_{\lambda_+})} \leq 0 } \\
		\mu_{++} := \E_{x,\mb s_{ij}|z_{\lambda_+ > 0}}\brac{ \int^{z_{\lambda_+}}_0  f(z) \,dz - z_{\lambda_+} f(z_{\lambda_+}) } \leq \min\set{\frac{1}{2},\frac{1}{\sqrt{2\pi}}\E_{x,\mb s_{ij}|z_{\lambda_+}>0}{\, z_{\lambda_+}} }  
	\end{cases},\notag
\end{align}   
where the negativity of the first equation can be observed by writing  $v = -z_{\lambda_+}$ and take derivative:
\begin{align}
\begin{cases}	
	-\int_0^v f(z)dz + v\cdot f(v) = 0 & v = 0
\\	\frac{d}{dv}\Brac{-\int_0^vf(z)dz + v\cdot f(v)} = -f(v) + f(v) + v\cdot f'(v) < 0 & v>0
	\end{cases}; \notag
\end{align}
and similarly for $z_{\lambda_-}$:
\begin{align}
\begin{cases}  
	{\color{black} \mu_{--} := \E_{x,\mb s_{ij}|z_{\lambda_- \leq 0}}\brac{\int_0^{z_{\lambda_-}} f(z) \,dz - z_{\lambda_-} f(z_{\lambda_-}) } \leq 0 } \\
	\mu_{-+} := \E_{x,\mb s_{ij}|z_{\lambda_- > 0}}\brac{\int^{z_{\lambda_-}}_0  f(z) \,dz - z_{\lambda_-} f(z_{\lambda_-}) } \leq \min\set{\frac{1}{2},\frac{1}{\sqrt{2\pi}}\E_{x,\mb s_{ij}|z_{\lambda_-}>0} z_{\lambda_-} }
\end{cases},\notag
\end{align}
then combine every term to \eqref{eqn:monotone_ht} using tower property and from assumption $\mb\beta_j-\mb\beta_i < 0$ we obtain 
\begin{align}  
 	\eqref{eqn:monotone_ht} &\leq  \paren{\mb\beta_j-\mb\beta_i}\paren{1 - \prob{z_{\lambda_+}>0}\cdot \mu_{++} - \prob{z_{\lambda_-}>0}\cdot \mu_{-+} + \E_{x,\mb s_{ij}} \brac{x(f(z_{\lambda_+}) - f(z_{\lambda_-})) } } \notag \\   
 	&\leq \paren{\mb\beta_j - \mb\beta_i}\paren{1 -\min\set{\frac{\prob{z_{\lambda_+} > 0}}{2}, \frac{\E\abs{z_{\lambda_+}}}{\sqrt{2\pi}}}  -   \min\set{\frac{\prob{z_{\lambda_-} > 0}}{2}, \frac{\E\abs{z_{\lambda_-}}}{\sqrt{2\pi}}}  - \frac{\theta}{\sqrt{2\pi}}\cdot\E\abs{g} }, \label{eqn:mono_grad}
\end{align}
where $g$ is standard Gaussian r.v..
 
\vsni 2. (\ul{Cases of varying $\mb\beta_i,\mb\beta_j$}) { Let $c_\lambda < \frac14$. Suppose $\mb\beta_i - \ell(t) \leq \frac{1}{4\sqrt{\abs{\mb\tau}}}$}. Recall that  $\norm{\mb\beta_{\mb\tau}}2^2\geq 1-3c_\mu$. We are going to show there is at least one of the entry $\mb\beta_*\in \set{\mb\beta_r}_{r\in\mb\tau\neq i,j}\uplus\set{\mb\beta_j+\ell(t)}$ is greater than $\frac{0.85}{\sqrt{|\mb\tau|}}$. First, if both $i,j\not\in \mb\tau$,  the lower bound is immediate since $\mb\beta_*^2 = \norm{\mb\beta_{\mb\tau}}\infty^2 > \frac{1-3c_\mu}{\abs{\mb\tau}}$. On the other hand if at least one of $i,j$ is in $\mb\tau$ and all other $\mb\beta_{\mb\tau}$ entries are small where $\norm{\mb\beta_{\mb\tau\setminus\set{i,j}}}\infty^2<\frac{1-3c_\mu}{\abs{\mb\tau}}$, then we know via norm inequalities,
\begin{align}
	\paren{\mb\beta_i +\mb\beta_j}^2 > \mb\beta_i^2 + \mb\beta_j^2 >\norm{\mb\beta_{\mb\tau}}2^2 - \paren{\abs{\mb\tau}-1}\norm{\mb\beta_{\mb\tau\setminus\set{i,j}}}\infty^2 \geq  \frac{1-3c_\mu}{\abs{\mb\tau}}, 
\end{align}  
which implies if $c_\mu < \frac{1}{100}$,{
\begin{align}
	\mb\beta_* &= \mb\beta_j+\ell(t)  = \paren{\mb\beta_i + \mb\beta_j} - \paren{\mb\beta_i-\ell(t)} \geq \frac{\sqrt{1-3c_\mu}}{\sqrt{\abs{\mb\tau}}}-\frac{1}{4\sqrt{\abs{\mb\tau}}} \geq \frac{0.72}{\sqrt{\abs{\mb\tau}}}.
\end{align}}
In this case, adopt result from \Cref{cor:tail_beta_x0_tauc} such that $\prob{\abs{\sum\mb\beta_\ell x_\ell} > \lambda/10} \leq 3\theta\abs{\mb\tau}\leq .01$, we have 
\begin{align}    
	\prob{z_{\lambda_-} > 0} = \prob{z_{\lambda_+} > 0 }  &= 1-\prob{x(\mb\beta_j+\ell(t)) + \mb s_{ij} < -\lambda } \notag \\
	&\leq {\color{black} 1- \prob{\mb x_*\mb\beta_* < -11\lambda/10} \cdot  \prob{x(\mb\beta_j+\ell(t)) + \mb s_{ij} - \mb x_*\mb\beta_* < \lambda/10} } \notag \\  
	&\leq 1-\theta\cdot\prob{\mb g_*\cdot \frac{0.72}{\sqrt{\abs{\mb\tau}}} <\frac{-11c_\lambda}{10\sqrt{\abs{\mb\tau}}}}\cdot\paren{1-\prob{\sum\mb\beta_\ell \mb x_\ell > \frac{\lambda}{10}}} \notag  \\
	&\leq  { 1- \theta\cdot\prob{0.72\cdot \mb g_*\leq - 1.1 \cdot 0.25} \cdot\paren{1-3c_\mu}}\notag \\
	&\leq 1-0.35\theta.   \label{eqn:mono_ub1}
\end{align} 
{ On the other hand, when $\mb\beta_i-\ell(t) \geq \frac{1}{4\sqrt{\abs{\mb\tau}}}$, both $z_{\lambda_+}$, $z_{\lambda_-}$ are upper bounded via $\abs{\mb\tau}\theta \leq{\frac{1}{800}}$ such as:  
\begin{align}
	\E_{x,\mb s_{ij}}\abs{z_{\lambda_-}} = \E_{x,\mb s_{ij}}\abs{z_{\lambda_+}} &\leq \E_{x,\mb s_{ij}}\frac{\lambda + \abs{x(\mb\beta_j+\ell(t)) -\mb s_{ij}}}{\mb\beta_i-\ell(t)} \leq  1+ 4\sqrt{\abs{\mb\tau}}\cdot \paren{\E_{x,\mb s_{ij}}\abs{x(\mb\beta_j + \ell(t)) - \mb s_{ij}}^2}^{1/2}   \notag \notag \\
	&\leq  1+4\sqrt{\abs{\mb\tau}\theta}\norm{\mb\beta}2 \leq 1+4\sqrt{\abs{\mb\tau}\theta}\paren{1+c_\mu+\frac{c_\mu}{\sqrt\theta\abs{\mb\tau}}} \leq 1.2.  \label{eqn:mono_ub2}
\end{align}}
Combine \eqref{eqn:mono_grad}, \eqref{eqn:mono_ub1} we have
\begin{align}
	h'(t)  \leq (\mb\beta_j-\mb\beta_i)\paren{1-2\cdot \frac{(1-0.35\theta)}2 - \frac{\theta}{\sqrt{2\pi}}\cdot\sqrt{\frac{2}{\pi}} } \leq 0.03\theta(\mb\beta_j-\mb\beta_i) < 0,
\end{align} 
and combine \eqref{eqn:mono_grad}, \eqref{eqn:mono_ub2} and $\theta<c_\mu$ we have
\begin{align}
 	h'(t) \leq (\mb\beta_j-\mb\beta_i)\paren{1-2\cdot\frac{1.2}{\sqrt{2\pi}} -\frac{\theta}{\sqrt{2\pi}}\cdot\sqrt{\frac{2}{\pi}} } \leq 0.03(\mb\beta_j-\mb\beta_i) < 0,
\end{align}
which proves the monotonicity.
\end{proof}
  
\paragraph{Finite sample deviation of $\mb\chi$.} When the signal length  of $\mb y$ is sufficiently large, operator $\mb\chi$ will be enough close to its expected value.  
\begin{corollary}[Finite sample deviation of $\mb\chi(\mb\beta)$]\label{cor:chibeta_ct} Suppose $\mb x_0\simiid  \mr{BG}(\theta)$ in $\R^n$, and $k, c_\mu$ such that $(\mb a_0,\theta,k)$ satisfies the sparsity-coherence condition $\mr{SCC}(c_\mu)$. Define $\lambda = c_\lambda/\sqrt{k}$ in $\varphi_{\ell^1}$ for some $c_\lambda > 1/5$, then there exists some numerical constants $C,c,\ol c > 0$, such that  if $n\geq C p^5\theta^{-2}\log p$ and $c_\mu \leq \ol c$, then with probability at least $1-3/n$, for every $\mb a\in\cup_{\abs{\mb\tau} \leq  k}\goodregion$ and every $i\in[n]$, we have:
\begin{align}
	 \abs{ n^{-1}\mb\chi[\mb\beta]_i -  n^{-1}\E\mb\chi[\mb\beta]_i } \leq   c\theta/p^{3/2},    
\end{align}
\end{corollary} 
\begin{proof}
	See \Cref{sec:proof_chibeta_ct}
\end{proof}

% !TEX root = ../../BD_DQ.tex
\section{Euclidean Hessian  as logic function in shift space}\label{sec:hessian_logic}

We can express the (pseudo) curvature \eqref{eqn:euc_hess} in direction $\mb v\in\Sp^{p-1}$ in terms of the correlation $\mb\gamma = \convmtx{\mb a_0}^*\ip\mb v$ between $\mb v$ and $\mb a_0$, giving
\begin{align}
	\mb v^* \wt{\nabla}^2\varphi_{\ell^1}(\mb a)\mb v = -\mb\gamma^*\checkmtx{\mb x_0}\mb P_I\checkmtx{\mb x_0}\mb \gamma, \notag
\end{align}
where 
\begin{align}\label{eqn:suppI}
	I(\mb a) = \supp\paren{\soft{\checkmtx{\mb x_0}\convmtx{\mb a_0}^*\ip\mb a}{\lambda}} = \set{i\in[n]\big| \abs{\mb x_0 * \wc{\mb\beta}}_i > \lambda }.
\end{align}
The $i$-th diagonal entry of $\checkmtx{\mb x_0}\mb P_{I(\mb a) }\checkmtx{\mb x_0}$ is   
\begin{align}
	-\mb e_i^*\checkmtx{\mb x_0}\mb P_{I(\mb a) }\checkmtx{\mb x_0} \mb e_i = -\norm{\mb P_{I(\mb a)}\checkmtx{\mb x_0}\mb e_i}2^2  = - \norm{\mb P_{I(\mb a)}\shift{\mb x_0}{-i}}2^2, 
\end{align}
which is the core component for us to study the  curvature of objective $\varphi_{\ell^1}$. We illustrate the expectation of diagonal term of Hessian in \Cref{lem:expect_support} and \Cref{cor:ct_support}, whose figure of visualized $\norm{\mb P_{I(\mb a)}\shift{\mb x_0}{-i}}2$ is shown in \Cref{fig:expect_chi}. Lastly, we also prove the off-diagonal terms $\mb e_i^*\checkmtx{\mb x_0}\mb P_{I(\mb a) }\checkmtx{\mb x_0} \mb e_j$ of Hessian is likely  inconsequential in calculation of curvature in \Cref{lem:hess_cross_convex}.

\paragraph{Expectation of Hessian diagonals.} We expect the Hessian to have stronger negative component in the $\shift{\mb a_0}{i}$ direction as $\norm{\mb P_{I(\mb a)}\shift{\mb x_0}{-i}}2^2$ becomes larger. This term can by tremendously simplified when $\mb x_0$ is very sparse: suppose all entries of its  support $I_0$ are  separated by at least $2p-1$ samples, then by implementing the definition of support from  \eqref{eqn:suppI}, we can derive  
\begin{align}  
	-\norm{\mb P_{I(\mb a)}\shift{\mb x_0}{-i}}2^2 =  -\sum_{j\in I_0}\mb x_{0j}^2\1_{\set{\abs{\sum_\ell\mb  \beta_\ell \mb x_{0(\ell+j-i)}} > \lambda } }  \underbrace{=}_{\text{sep.}} -\sum_{j\in I_0}\mb g_j^2\1_{\set{\abs{\mb \beta_i \mb g_j}>\lambda }},  \label{eqn:curve_coef_sumgauss}
\end{align}  
where $\1$ is the indicator function and $\mb g_j$ are independent standard Gaussian r.v.s.. In expectation, the summands in \eqref{eqn:curve_coef_sumgauss} acts like a smoothed logic function on entry $\mb \beta_i$:
  
\begin{lemma}[Gaussian smoothed indicator]\label{lem:guass_smoothed_supp}  Let $g\sim \mc N(0,1)$, then for any $b,s\in\R $ and $\lambda > 0 $.
\begin{align}\label{eqn:gauss_smooth_ind}
\E_g \left[ g^2\1_{\set{\abs{b\cdot g + s} > \lambda  }}\right] = 1- \mr{erf}_b \paren{\lambda,s} + f_b(\lambda,s),
\end{align}
where
\begin{align}
	f_b(\lambda,s) = \frac{1}{\sqrt{2\pi}}\brac{ \paren{\frac{\lambda+s}{\abs b}}e^{-\frac{\paren{\lambda+s}^2}{2b^2}} +  \paren{\frac{\lambda - s}{\abs b}}e^{-\frac{\paren{\lambda - s}^2}{2b^2}} }.
\end{align} 
\end{lemma}
\begin{proof}
	The proof can be derived via same calculation of integrals  in   \Cref{lem:expect_soft_gaussian}.
\end{proof}  
\vsni Although the definition \eqref{eqn:gauss_smooth_ind} seems incomprehensible at first glance, we can actually interpret it as a smoothed indicator function which compares $\abs{b}$ to the threshold $\sqrt{2/\pi}\lambda$.  Once we  assign $s=0$, then we can see that  $\E g^2\1_{\set{ \abs{b\cdot g}>\lambda}}$ is be an increasing function of $\abs{b}$. Moreover by assigning different values for $\abs b$ we obtain:
\begin{align}
	\E g^2\1_{\set{\abs{b\cdot g} > \lambda}} \approx \begin{cases}
		1,&\quad \abs{b} \approx 1\\
		1/2,&\quad \abs{b} \approx \sqrt{2/\pi}\lambda\\
		0,&\quad \abs{b} \approx 0
	\end{cases}.\label{eqn:curv_coeff_gauss_as_logic}
\end{align}
Relate \eqref{eqn:curv_coeff_gauss_as_logic} to   \eqref{eqn:curve_coef_sumgauss}, when $\abs{\mb\beta_i}$ is close to 1 then we expect $-\tfrac{1}{n\theta}\norm{\mb P_I\shift{\mb x_0}{-i}}2^2$ to be close to $-1$, and it increases to $0$ as $\abs{\mb\beta_i}$ decreases, suggests that the Euclidean Hessian at point $\mb a$ has stronger negative component at $\shift{\mb a_0}{i}$ direction if $\abs{\innerprod{\mb a}{\shift{\mb a_0}{i}}}$ is larger. See \Cref{fig:expect_dchi} for a numerical example. This phenomenon can be extend beyond the idealistic separating case as follows:

\begin{figure}[t]
\centering
\input{sections/expect_dchi.tex}		

\caption{\textbf{A numerical example for $\E \norm{\mb P_{I(\mb a)}\shift{\mb x_0}{i}}2^2$.}  We provide a figure to illustrate the expectation of $-\frac{1}{n\theta}\norm{\mb P_{I(\mb a)}\shift{\mb x_0}{i}}2^2$ when entries of $\mb x_0$ are $2p$-separated, as  a function plot of $\mb\beta_i\to 1- \mr{erf}_{\mb\beta_i} \paren{\lambda,0} + f_{\mb\beta_i}(\lambda,0)$ from  \eqref{eqn:gauss_smooth_ind} with different $\lambda$. When $\abs{\mb\beta_i}\approx \nu_2\lambda$ where $\nu_2 = \sqrt{2/\pi}$, then the its function value is close to $0.5$. If $\abs{\mb\beta_i}$ is much larger then $\lambda$ its value grow to $1$, implies there is a negative curvature at $\shift{\mb a_0}{i}$ direction. Similarly if $\abs{\mb\beta_i}$ is much smaller then $\lambda$ the function value is $0$ thus the curvature is positive in $\shift{\mb a_0}{i}$ direction.  } \label{fig:expect_dchi} 
\end{figure}
 
\begin{lemma}[Expected Hessian diagonals] \label{lem:expect_support} Let $\mb x_0\simiid\mr{BG}(\theta)$ and $\lambda > 0$, define the set $I(\mb a)$ in \eqref{eqn:suppI}, write $\mb s_i = \sum_{\ell\neq i}\mb\beta_\ell \mb x_{0\ell}$, then for every $\mb a\in\Sp^{p-1}$ and $i\in[n]$:
\begin{align} 
	n^{-1}\E \norm{\mb P_{I(\mb a)}\shift{\mb x_0}{-i}}2^2 = \theta\brac{1-\E_{\mb s_i}\mr{erf}_{\mb\beta_i}\paren{\lambda,\mb s_i} + \E_{\mb s_i}f_{\mb\beta_i}\paren{\lambda,\mb s_i}} 
\end{align}
\end{lemma}

\begin{proof} Write $\mb x_0$ as $\mb x$. Observe that $
	\mb y*\wc{\mb a} = \mb x_0 * \wc{\mb\beta} = \sum_{\ell}\mb\beta_\ell\shift{\mb x_0}{-\ell} $. Thus for any $j\in [n]$ and $i\in[\pm p]$:
\begin{align}
	\paren{\mb y*\wc{\mb a}}_{j-i} = \Big(\mb\beta_i\shift{\mb x}{-i} + \sum_{\ell\neq i}\mb\beta_\ell\shift{\mb x}{-\ell}\Big)_{j-i} =  \mb\beta_i\mb x_{j} + \sum_{\ell\neq i}\mb\beta_\ell\mb x_{j+\ell-i}   =: \mb\beta_i\mb x_{j} + \mb s_j,   
\end{align}   
where $\mb x_j$ is independent of $\mb s_j$, and both $\mb x_j, \mb s_j$ are symmetric and identically distributed for all $j\in[n]$. Rewrite the random variable using \eqref{eqn:suppI} as 
\begin{align}
	\norm{\mb P_{I(\mb a)}\shift{\mb x_0}{-i}}2^2 = \norm{\mb P_{I(\mb a) }\textstyle\sum_{j\in [n]}\paren{\mb x_{0j}\mb e_{j-i}} }2^2 = \sum_{j\in [n]} \mb x_{0j}^2\1_{\set{\abs{\mb y*\check{\mb a}}_{j-i} > \lambda}} = \sum_{j\in[n]}\mb x_{0j}^2\1_{\set{\abs{\mb\beta_i\mb x_{0j} + \mb s_j} > \lambda}}\notag
\end{align}
Write  $\mb x = \mb g \circ \mb\omega$ as composition of Gaussian/Bernoulli r.v.s., the expectation has a simple form:
\begin{align}
	\E \norm{\mb P_{I(\mb a) }\shift{\mb x_0}{-i}}2^2 = n\theta \cdot  \E \mb g_0^2 \1_{\set{\abs{ \mb\beta_i \mb g_0 + \mb s_0 }>\lambda}} = n\theta\cdot \E\paren{1-\mr{erf}_{\mb\beta_i}(\lambda,\mb s_i) + f_{\mb\beta_i}(\lambda,\mb s_i)}\notag
\end{align}
where $\mb s_i = \sum_{\ell\neq i}\mb x_{0i}\mb\beta_i$ with $\mb x_{0i} \simiid \mr{BG}(\theta)$, yielding the claimed expression.
 \end{proof} 

\paragraph{Finite sample deviation of Hessian diagonals.} When the signal length  of $\mb y$ is sufficiently large, then $i$-th diagonal term for Hessian $\norm{\mb P_{I(\mb a)}\shift{\mb x_0}{-i}}2^2$ will be close enough to its expected value.

\begin{corollary}[Large sample deviation of curvature]\label{cor:ct_support} Suppose $\mb x_0\simiid  \mr{BG}(\theta)$ in $\R^n$, and $k, c_\mu$ such that $(\mb a_0,\theta,k)$ satisfies the sparsity-coherence condition $\mr{SCC}(c_\mu)$. Define $\lambda = c_\lambda/\sqrt{k}$ in $\varphi_{\ell^1}$ for some $c_\lambda > 1/5$, then there exists some numerical constant $C,c,\ol c>0$, such that if $n \geq C p^4\theta^{-1}\log p$ and $c_\mu \leq \ol c$, then with probability at least $1-3/n$, for every $\mb a\in\cup_{\abs{\mb\tau} \leq k}\goodregion$ and every $i\in[n]$, we have:
\begin{align}\label{eqn:ct_support}
\abs{n^{-1}\norm{\mb P_{I(\mb a) }\shift{\mb x_0}{-i}}2^2 - n^{-1}\E\norm{\mb P_{I(\mb a) }\shift{\mb x_0}{-i}}2^2} \leq c\theta/p  
\end{align}
\end{corollary}    
 \begin{proof}
 	 See \Cref{sec:proof_ct_supp}.
 \end{proof}

\paragraph{Hessian off-diagonal terms near solution.} The off-diagonal entries of Hessian in general are much smaller then the diagonal entries; however, it affects the region near sign shifts of $\mb a_0$ the most where we need to show strong convexity in the region. We provide an upper bound for off-diagonal entries in the vicinity of signed shifts. In these regions, only one entry of the correlations $\abs{\mb\beta_{(0)}}$ is large and the rest is small. 
\begin{lemma}[Hessian off-diagonal term near solution]\label{lem:hess_cross_convex} Suppose $\mb x_0\simiid \mr{BG}(\theta)$ in $\R^n$, and $k, c_\mu$ such that $(\mb a_0,\theta,k)$ satisfies the sparsity-coherence condition $\mr{SCC}(c_\mu)$. Let $\lambda = c_\lambda/\sqrt k$ with $c_\lambda > 1/5$, then there exists some numerical constant $C,\ol c>0$ such that if  $n \geq C\theta^{-4}\log p$ and $c_\mu\leq\ol c$, then with probability at least $1-4/n$, for every $\mb a\in\cup_{\abs{\mb\tau}\leq k}\goodregion$,  where $\abs{\mb\beta_{(1)}}\leq \frac{1}{4\log\theta^{-1}}\lambda$ and every $i\neq j\in[\pm p]\setminus\set{(0)}$, we have 
	\begin{align}\label{eqn:cross-term-inprod}
		\abs{\shift{\mb x_0}{i}^*}\mb P_{I(\mb a)}\abs{\shift{\mb x_0}{j}} < 8n\theta^3
	\end{align}
\end{lemma}
\begin{proof} Write $\clog = -1/\log\theta$ and $\mb x_0$ as $\mb x = \mb\omega\circ\mb g$. Wlog let $\mb\beta_0$ be the largest correlation $\mb\beta_{(0)}$. Define random variables $s'= \innerprod{\mb\beta_{\mb\tau\setminus\set{0,i,j}}}{\mb x_{\mb\tau\setminus\set{0,i,j}}}$. Firstly via \Cref{cor:tail_beta_x0_tau} we have $\prob{\abs{s'} > 0.4\lambda} \leq 2\theta$; also define $s = \innerprod{\mb\beta_{\mb\tau^c\setminus\set{0,i,j}}}{\mb x_{\mb\tau^c\setminus\set{0,i,j}}}$, and base on \Cref{cor:tail_beta_x0_tauc} we have $\prob{\abs{s} > \lambda/10} \leq 2\theta$. Expand the $(-i,-j)$-th cross term with $\theta < 0.1$ we have: 
\begin{align}\label{eqn:hess_cross_conv_exp}
	\E\abs{\shift{\mb x}{-i}^*}\mb P_{I(\mb a)}\abs{\shift{\mb x}{-j}} &= \E\textstyle\sum_{k\in[n]}\abs{\mb x_{k+i}\mb x_{k+j}}\1_{\set{\abs{\mb\beta_0\mb x_k + \mb\beta_{i}\mb x_{k+i}+\mb\beta_{j}\mb x_{k+j} + s + s'}>\lambda }}  \notag \\
	&= n\theta^2\cdot \E\abs{\mb g_{i}\mb g_{j}}\1_{\set{\abs{\mb\beta_0\mb x_0 + \mb\beta_{i}\mb g_{i}+\mb\beta_{j}\mb g_{j} + s + s'}>\lambda }}  \notag \\
	&\leq n\theta^2 \cdot\E\brac{\abs{\mb g_i\mb g_j}\paren{2\1_{\set{\abs{\mb\beta_i\mb g_i }>\lambda/4 }} + \prob{\mb x_0\neq 0} + \prob{\abs{s}>0.1\lambda} + \prob{\abs{s'}>0.4\lambda}} } \notag \\ 
	&\leq n\theta^2\cdot\paren{\exp\paren{-\log^2\theta^{-1}} + \theta + 2\theta + 2\theta }  \notag \\
	&\leq 6n\theta^3. 
\end{align} 
Write \eqref{eqn:cross-term-inprod} as two summation of independent random variables with $t = j-i$ by separating sum into two sets $J_{t1},J_{t2}$ defined in \eqref{eqn:supp_x0_Jt12} with both $\abs{J_{t1}},\abs{J_{t2}} < n\theta^2$ with probability at least $1-2/n$ from \Cref{lem:x0_supp}
\begin{align}
	\E\abs{\shift{\mb x}{-i}^*}\mb P_{I(\mb a)}\abs{\shift{\mb x}{-j}} =  \sum_{(k-i)\in I(\mb a)}\abs{\mb x_{k}}\abs{\mb x_{k+t}} = \sum_{(k-i)\in I(\mb a)\cap J_{t1}}\abs{\mb g_k}\abs{\mb g_{k+t}} + \sum_{(k-i)\in I(\mb a) \cap J_{t2}}\abs{\mb g_k}\abs{\mb g_{k+t}},\notag
\end{align} 
whose first summands can be upper bounded w.h.p. via Bernstein inequality \Cref{lem:mc_bernstein_scalar} with $(\sigma^2,R) = (1,1)$ and writes $\mc C := \cup_{\abs{\mb\tau}\leq k}\goodregion\cap\set{\mb a\,\big|\,\abs{\mb\beta_{(1)}}\leq\frac{1}{4\log\theta^{-1}}\lambda}$, then we have
\begin{align}\label{eqn:hess_cross_conv_tail}
	&\quad\; \prob{\max_{\substack{i\neq j\in[\pm p]\setminus\set0 \\ \mb a\in \mc C}}\paren{\sum_{(k-i)\in I(\mb a)\cap J_{t1}}\abs{\mb g_k}\abs{\mb g_{k+t}} - \E\sum_{(k-i)\in I(\mb a)\cap J_{t1}}\abs{\mb g_k}\abs{\mb g_{k+t}}} \geq  n\theta^3 } \notag \\
	&\quad\; \prob{\max_{\substack{i\neq j\in[\pm p]\setminus\set0 }}\paren{\sum_{(k-i)\in\cap J_{t1}}\abs{\mb g_k}\abs{\mb g_{k+t}} - \E\sum_{(k-i)\cap J_{t1}}\abs{\mb g_k}\abs{\mb g_{k+t}}} \geq n\theta^3 } \notag \\
	&\leq 4p^2\cdot   \exp\paren{\frac{-n^2\theta^6}{2\abs{J_{t1}} + 2n\theta^3 }} \leq \exp\paren{8\log p - \frac{-n^2\theta^6}{3n\theta^2} }\leq  \exp\paren{-\frac{n\theta^4}{10}}\leq \frac{1}{n}   
\end{align} 
when $n = C\theta^{-4}\log p$ with $C > 10^4$ and $\theta\log^2\theta^{-1} \geq  1/p $. Thus  for all $i\neq j\in[\pm p]\setminus\set 0$ and $\mb a$ satisfies our condition of lemma, from \eqref{eqn:hess_cross_conv_exp} and \eqref{eqn:hess_cross_conv_tail} we can conclude : 
\begin{align}
	\abs{\shift{\mb x}{-i}^*}\mb P_{I(\mb a)}\abs{\shift{\mb x}{-j}} &\leq \sum_{I(\mb a)\cap J_{t1}}\E\abs{\mb g_k}\abs{\mb g_{k+t}} + \sum_{I(\mb a)\cap J_{t2}}\E\abs{\mb g_k}\abs{\mb g_{k+t}} + 2n\theta^3  \leq 8n\theta^3 \notag
\end{align}   
 which holds with probability at least $1-2/n - 2\cdot 1/n = 1-4/n$ base on \Cref{lem:x0_supp} and  \eqref{eqn:hess_cross_conv_tail}.
\end{proof}

% !TEX root = ../../BD_DQ.tex

\section{Geometric relation between $\rho$ and $\ell^1$-norm}\label{sec:smooth_approx}

In this section, we discuss how to ensure that the smooth sparsity surrogate $\rho$ approximates $\|\cdot\|_1$ accurately enough that guarantees $\varphi_{\rho}$ inherits the good properties of $\varphi_{\ell^1}$. We prove several lemmas which allow us to transfer properties of $\varphi_{\ell^1}$ to $\varphi_\rho$. Our result does not pertain to the suggested pseudo-Huber surrogate $\rho(x)_i = \sqrt{x_i^2+\delta^2}$ in the main script, and is general for a class of function class defined in \Cref{def:smooth_ell1} that is smooth and well approximates $\ell^1$  when the proper smoothing parameter $\delta$ is chosen from the result of \Cref{lem:approx_rho_ell1}. In particular we ask the regularizer $\rho_\delta(x)$ to be uniformly bounded to $\abs{x}$ by $\delta/2$:
\begin{align}
	\forall\, x\in\R,\qquad \abs{\rho_\delta(x)-\abs x} \leq \delta/2
\end{align}
then if $\delta\to 0$ we have for every $\mb a$ near subspace,
\begin{align}
	\norm{\prox_{\lambda\ell^1}[\wc{\mb a}*\mb y]-\prox_{\lambda\rho_\delta}[\wc{\mb a}*\mb y]}2 \to 0,\\ 
	\norm{\nabla{\varphi_{\ell^1}(\mb a)} - \nabla\varphi_{\rho_\delta}(\mb a) }2 \to 0, \\
	\|\wt{\nabla}^2\varphi_{\ell^1}(\mb a)-\nabla^2\varphi_{\rho_\delta}(\mb a)\|_2 \to 0. 
\end{align}
An example choices of eligible smooth sparse surrogate is demonstrated in \Cref{tbl:rho_class}.

\paragraph{Calculus of $\varphi_\rho$.} The marginal minimizer over $\mb x$ in \eqref{eqn:lasso-dq} can be expressed in terms of the proximal operator \cite{BC11} of $\rho$  at point $\wc{\mb a}*\mb y$: 
\begin{align}
	 \prox_{\lambda\rho}\brac{\wc{\mb  a} *\mb y} = \argmin_{\mb x\in\R^n} \set{ \lambda\rho(\mb x) + \tfrac{1}{2}\norm{\mb x}2^2 - \innerprod{\mb a*\mb x}{\mb y} }.\notag
\end{align}\vspace{-.1in}
Plugging in, we obtain
\begin{align}\label{eqn:varphi_moreau}
	\varphi_\rho(\mb a) =  \lambda\rho\bigl(\prox_{\lambda\rho}[\wc{\mb a}*\mb y]\bigr) + \tfrac{1}{2}\norm{\wc{\mb a}*\mb y - \prox_{\lambda\rho}\brac{\wc{\mb a}*\mb y}}2^2 - \tfrac{1}{2}\norm{\wc{\mb a}*\mb y}2^2  +\tfrac{1}{2}\norm{\mb y}2^2
\end{align}
The objective function $\varphi_\rho(\mb a)$ is a differentiable function of $\mb a$. This can be seen, e.g., by noting that 
\begin{align}
\varphi_\rho( \mb a ) = \epsilon( \lambda \rho ) ( \wc{\mb a} * \mb y ) - \tfrac{1}{2}\norm{ \wc{\mb a} * \mb y }{2}^2 + \tfrac12\norm{\mb y}2^2, 
\end{align}
where $\epsilon(g)(\mb z ) = g\left( \prox_{g}( \mb z ) \right) + \tfrac{1}{2} \norm{ \mb z - \prox_g( \mb z ) }{2}^2$ is the {\em Moreau envelope} of a function $g$. The Moreau envelope is differentiable:
\begin{fact}[Derivative of Moreau envelope, \cite{BC11}, Prop.12.29]\label{fact:deriviative_moreau} Let $f$ be a proper lower semicontinuous convex function and $\lambda > 0$ then the Moreau envelope $\epsilon(\lambda f)(\mb z) = \lambda f(\prox_{\lambda f}[\mb z]) + \frac{1}{2}\norm{\mb z - \prox_{\lambda f}[\mb z]}2^2 $
is Fr\'{e}chet differentiable with $\nabla \epsilon(\lambda f)(\mb z) = \mb z - \prox_{\lambda\rho}[\mb z] $.
\end{fact}   Furthermore, $\varphi_\rho$ is twice differentiable whenever $\prox_{\lambda\rho}$ is differentiable. In this case, the (Euclidean) gradient and hessian of $\varphi_\rho$ are given by
\begin{align}
	\nabla \varphi_\rho(\mb a) &=  - \ip^*\checkmtx{\mb y}\prox_{\lambda\rho}\brac{\checkmtx{\mb y}\ip\mb a} \label{eqn:smooth_rho_grad},\\ 
	\nabla^2 \varphi_\rho(\mb a)  &= -\ip^*\checkmtx{\mb y}\nabla\prox_{\lambda\rho}\brac{\checkmtx{\mb y}\ip\mb a} \checkmtx{\mb y}\ip.\label{eqn:smooth_rho_hess}
\end{align}
The Riemannian gradient and hessian over $\Sp^{p-1}$ are
\begin{align}
	\mr{grad}[\varphi_\rho](\mb a) & =  -\mb P_{\mb a^\perp} \ip^*\checkmtx{\mb y} \prox_{\lambda\rho}\brac{\checkmtx{\mb y}\ip\mb a}, \label{eqn:grad_rho}\\
	\mr{Hess}[\varphi_\rho](\mb a) & = -\mb P_{\mb a^\perp}\paren{\ip^*\checkmtx{\mb y}\nabla\prox_{\lambda\rho}\brac{\checkmtx{\mb y}\ip\mb a} \checkmtx{\mb y}\ip - \innerprod{\nabla \varphi_\rho(\mb a)}{\mb a}\mb I } \mb P_{\mb a^\perp}. \label{eqn:hess_rho}   
\end{align}  

\paragraph{Sparse regularizer $\rho$ as smoothed $\ell^1$ function.} Our analysis accommodates any sufficiently accurate smooth approximation $\rho$ to the $\ell^1$ function. The requisite sense of approximation is captured in the following definition: 

\begin{definition}[$\delta$-smoothed $\ell^1$ function]\label{def:smooth_ell1}  We call an additively separable function $\rho(\mb x) = \sum_{i=1}^n\rho_i(\mb x_i) : \R^n \to \R$, a $\delta$-smoothed $\ell^1$ function with $\delta > 0$ if for each $i\in[n]$, $\rho_i $ is even, convex, twice differentiable  and $\nabla^2\rho_i(x)$ being monotone decreasing w.r.t. $\abs x$, where, there exists some constant $c$, such that  for all $ x\in\R$:  
\begin{align}\label{eqn:smooth_rho_derivative}
	\abs{\rho_i(x) - \abs x + c} \leq \delta/2
\end{align} 
\end{definition} 
\vsni 

The proximal operator of the $\ell^1$ norm is the entrywise soft thresholding function $\mc S_\lambda$; the proximal operator associated to a smoothed $\ell^1$ function turns out to be a differentiable approximation to $\mc S_\lambda$. In particular, we will show that it approximates $\mc S_\lambda$ in the following sense: 
\begin{definition}[$\sqrt{\delta}$-smoothed soft threshold]\label{def:soft_thresh} An odd function $\mc S_\lambda^\delta[\cdot]:\R \to \R$ is a $\sqrt{\delta}$-smoothed soft thresholding function with parameter $\delta > 0$ if it is a strictly monotone odd function and is differentiable everywhere, whose function value  satisfies
\begin{align}\label{eqn:smooth_soft} 
	0 \leq \sign(z)\paren{\mc  S_\lambda^\delta[z] - 
		\soft{z}{\lambda}} \leq \sqrt{\lambda\delta},\qquad \forall z\in\R 
\end{align}
and its derivative satisfies for any given $B \in(0,\lambda)$:
\begin{align}\label{eqn:smooth_soft_grad}
	\abs{\nabla\mc S_\lambda^\delta[z] - \nabla\mc S_\lambda[z]} \leq \sqrt{\lambda\delta}/B,\qquad \abs{\abs{z} - \lambda} \geq B.
\end{align}
\end{definition}
\vsni 

\begin{table}[t!]
\centering
{\def\arraystretch{2}
\begin{tabular}{|c|c|c|c|c|}
\hline 
Surrogate class & $\rho_i(x)$ &  $\nabla\rho_i(x)$ & $\nabla^2\rho_i(x)$  \\
\hline    
Log hyperbolic cosine & 
	{$\!\begin{aligned}
	\frac{\delta}2\log\paren{e^{2x/\delta} + e^{-2x/\delta}}  
	\end{aligned}$} & {$\!\begin{aligned}
	\frac{e^{4x/\delta} -1}{e^{4x/\delta} + 1}
	\end{aligned}$} & {$\!\begin{aligned} \frac{4e^{4x/\delta}}{\delta(e^{4x/\delta}+1)^2} \end{aligned}$} \\
\hline
Pseudo Huber & {$\!\begin{aligned} \sqrt{x^2 + \delta^2}  \end{aligned}$} & {$\!\begin{aligned} \frac{x}{\sqrt{x^2+\delta^2}} \end{aligned}$} & {$\!\begin{aligned} \frac{\delta^2}{(x^2 + \delta^2)^{3/2}} \end{aligned}$}\\
\hline
Gaussian convolution  & {$\!\begin{aligned} \int\abs{x-t}f_\delta(t) dt \end{aligned}$}  & {$\!\begin{aligned} \mr{erf}(x/\sqrt 2\delta)    \end{aligned}$}  & {$\!\begin{aligned}2f_\delta(x) \end{aligned}$} \\
\hline
\end{tabular}}
\caption{\textbf{Classes of smooth sparse surrogate $\rho$ and how to set its parameter.}\label{tbl:rho_class} Three common classes are listed with parameter $\delta$ to tune the smoothness. All the listed functions are greater then $\abs{x}$ pointwise and has largest distance to $\abs x$ at origin  where $\rho(0) - \abs x \leq \delta$, satisfies the condition   \eqref{eqn:smooth_rho_derivative}. Also its second order derivatives $\nabla^2\rho_i(x)$ are monotone decreasing w.r.t. $\abs{x}$, hence are certified to be eligible $\delta$-smoothed $\ell^1$ surrogates.}  
\end{table}

If $\rho$ is a $\delta$-smooth $\ell^1$ function, then for all $i\in[n]$,  we have that $\mr{prox}_{\lambda \rho}[\mb z]_i$ is a $\sqrt{\delta}$-smoothed soft threshold function of $\mb z_i$. This can be proven with the following lemma:
\begin{lemma}[Proximal operator for smoothed $\ell^1$]\label{lem:smooth_rho} Suppose $\rho$ is a $\delta$-smoothed $\ell^1$ function, then $\mb  z_i \mapsto \mr{prox}_{\lambda\rho}[\mb z]_i$ is a $\sqrt{\delta}$-smoothed soft threshold function.  
\end{lemma}
\begin{proof}We know that 
	\begin{align}\label{eqn:prox_rho}
		\mb x_z := \mr{prox}_{\lambda\rho}[\mb z] = \argmin_{\mb x\in\R^n} \lambda\rho(\mb x) + \tfrac{1}{2}\norm{\mb x-\mb z}2^2.
	\end{align}
	This optimization problem is strongly convex, and so the minimizer $\mb x_z$ is unique. Using the stationarity condition and since $\rho$ is separable, for all $i\in[n]$, we have $\lambda\nabla\rho_i(\mb x_{zi}) + \mb x_{zi} - \mb z_i = 0$, implies
\begin{align}\label{eqn:prox_stationary}
	\mb x_{zi} = (\mr{Id}+ \lambda\nabla\rho_i)^{-1}(\mb z_i).
\end{align} 
Since $\rho_i$ is convex and even , $\nabla\rho_i$ is monotone increasing and odd. By inverse function theorem, we know that strict monotonicity and differentiability of $\mr{Id} + \lambda\nabla\rho_i$ implies its inverse is differentiable and is a strictly monotone increasing odd function. Furthermore, it implies   $\nabla\mb x_{zi}$ has the form 
\begin{align}
	\nabla\mb x_{zi} = \nabla_i(\mr{Id}+ \lambda\nabla\rho_i)^{-1}(\mb z_i) = \frac{1}{\lambda\nabla^2\rho_i(\mb x_{zi}) + 1}  < 1.
\end{align}
Notice that since $\nabla^2\rho_i(x)$ is monotone decreasing when $x \geq 0$, hence $\nabla\mb x_{zi}$ is monotone increasing in $\mb z_i \geq 0$. 

Now we are left to show that \eqref{eqn:smooth_soft} and \eqref{eqn:smooth_soft_grad} hold, and since $\mr{\prox}_{\lambda\rho}[\cdot]_i$ is an odd function it suffices to consider the case when the input vector $\mb z_i$ is nonnegative. Firstly,  via convexity and entrywise bounded difference  $\abs{\rho_i(x) -\abs{x}} \leq \delta/2$ we are going to show 
\begin{align}\label{eqn:smooth_rho_mvt}
	\abs{\nabla\rho_i(x)} \leq 1\quad \forall\,x\in\R,\qquad  \nabla\rho_i(x) \geq 1-\sqrt{\delta/\lambda}\quad \forall\, x\geq \sqrt{\lambda\delta}.
\end{align} 
Consider a positive $x$ with $\nabla\rho_i(x) > 1+\eps$ for some $\eps>0$, by convexity if $\wt x > x$ then $\nabla\rho_i(\wt x) > 1+\eps$, hence 
\begin{align}
	\rho_i(x + \delta/\eps) \geq \rho_i(x) + \nabla\rho_i(x)\cdot(\delta/\eps) >  x -\delta/2 + (1+\eps)\cdot(\delta/\eps) = (x + \delta/\eps)  + \delta/2,\notag
\end{align}
contradicts the boundedness condition. Secondly, use mean value theorem we know for all $x\geq \sqrt{\lambda\delta}$:
\begin{align}
	\nabla \rho_i(x) \geq  \frac{\rho_i(\sqrt{\lambda\delta}) - \rho_i(0)}{\sqrt{\lambda\delta} - 0} \geq \frac{(\sqrt{\lambda\delta} -\delta/2) - \paren{0+\delta/2}}{\sqrt{\lambda\delta} - 0} \geq 1-\sqrt{\frac{\delta}{\lambda}}.\notag
\end{align}

To prove  \eqref{eqn:smooth_soft}, when $0\leq\mb z_i \leq \lambda$, then $\mc S_{\lambda}[\mb z_i] = 0$ and $\mb x_{zi}\leq \sqrt{\lambda\delta}$ since if $\mb x_{zi} > \sqrt{\lambda\delta}$, by \eqref{eqn:smooth_rho_mvt}:
\begin{align}
	\lambda\nabla\rho_i(\mb x_{zi}) + \mb x_{zi} >  \lambda(1-\sqrt{\delta/\lambda}) + \sqrt{\lambda\delta} = \lambda  \geq \mb z_i\notag
\end{align} 
then  $\mb x_{zi}$ violate the stationary condition in \eqref{eqn:prox_stationary}, resulting $0\leq\mb x_{zi} - \soft{\mb z_i}{\lambda} \leq \sqrt{\lambda\delta} $ whenever $0\leq \mb z_i\leq \lambda$. Likewise in the case of $\mb z_i \geq\lambda$ where $\soft{\mb z_i}{\lambda} = \mb z_i - \lambda $,  \eqref{eqn:smooth_rho_mvt} provides:
\begin{align}
	\begin{cases} \forall\,\mb x_{zi} > \mb z_i - \lambda+ \sqrt{\lambda \delta}, &\qquad \lambda\nabla\rho_i(\mb x_{zi}) + \mb x_{zi} > \lambda(1-\sqrt{\delta/\lambda}) + \mb z_i -\lambda + \sqrt{\lambda\delta} = \mb z_i  \\
	\forall\,\mb x_{zi} < \mb z_i - \lambda, &\qquad \lambda\nabla\rho_i(\mb x_{zi}) + \mb x_{zi} < \lambda + \mb z_i - \lambda = \mb z_i
	\end{cases}\notag
\end{align} 
again violates \eqref{eqn:prox_stationary} and therefore \eqref{eqn:smooth_soft} holds for all $\mb z_i\in\R$.

Lastly \eqref{eqn:smooth_soft_grad} is a direct result of \eqref{eqn:smooth_soft}. For all $\mb z_i \leq \lambda -B$, recall that  $\nabla \mb x_{zi}$ is monotone increasing in $\mb z_i$: 
\begin{align}
	\nabla\mb x_{zi} \leq \min_{y\in\brac{\lambda - B,\lambda}}\nabla\mb x_{yi} \leq  \frac{\mb x_{\lambda i} - \mb x_{(\lambda-B)i}}{\lambda - (\lambda - B)} \leq \frac{(\sqrt{\lambda\delta}+\soft{\lambda}{\lambda}) - \soft{\lambda - B}{\lambda}}{B}= \frac{\sqrt{\lambda\delta}}{B};\notag 
\end{align}
and similarly for all $\mb z_ i > \lambda +B$:
\begin{align}
	\nabla \mb x_{zi} \geq \max_{y\in[\lambda,\lambda + B]} \nabla \mb x_{yi} \geq \frac{\mb x_{(\lambda+B) i} - \mb x_{\lambda i} }{(\lambda + B)-\lambda} \geq \frac{\soft{\lambda + B}{\lambda} - (\soft{\lambda}{\lambda} + \sqrt{\lambda \delta})}{B} = 1-\frac{\sqrt{\lambda\delta}}{B},\notag
\end{align}
implies  \eqref{eqn:smooth_soft_grad} holds.
\end{proof}  
  
\paragraph{Approximate geometry of $\varphi_\rho$ using $\varphi_{\ell^1}$}

Based on \eqref{eqn:grad_rho}-\eqref{eqn:hess_rho} and denote $\checkmtx{\mb y}\ip\mb a = \wc{\mb a}*\mb y$, the only differences of Riemannian gradient and Hessian between $\varphi_\rho$  and $\varphi_{\ell^1}$ comes from the difference of $\prox_{\lambda\rho}\brac{\wc{\mb a}*\mb y}$ and  $\prox_{\lambda\norm{\cdot}1}\brac{\wc{\mb a}*\mb y}$. Thus for the purpose of obtaining good geometric approximation of $\varphi_\rho$ with that of objective $\varphi_{\ell^1}$, we may apply both \Cref{def:soft_thresh} and \Cref{lem:smooth_rho}, together suggest if $\rho$ is a $\delta$-smoothed $\ell^1$ function,  then the $i$-th entry of $\prox_{\lambda\rho}[\wc{\mb a}*\mb y]$ will be $\sqrt{\lambda\delta}$-close to the authentic soft thresholding function $\mc{S}_\lambda\brac{\wc{\mb a}*\mb y}_i$, and its gradient $\nabla\prox_{\lambda\rho}[\wc{\mb a}*\mb y]$ is $\sqrt{\lambda\delta}/B$-close to $\nabla \soft{\wc{\mb a}*\mb y}{\lambda}$ as long as $\paren{\wc{\mb a}*\mb y}_i$ is not close to $\pm \lambda$ by distance $B$.

Firsly, we will show by utilizing the random structure of $\mb y$, such that with high probability,  only a fraction of entries of $\wc{\mb a}*\mb y$ will be close to $\pm\lambda$.
 
\begin{lemma}[Gradients discontinuity entries]\label{lem:num_entries_onlambda}  For each $\mb a \in \bb S^{p-1}$, let
\begin{align}
	J_B(\mb a) :=  \set{i \;\middle | \; \paren{\checkmtx{\mb y}\ip\mb a}_i \in[-\lambda - B,-\lambda + B]\cup [\lambda - B,\lambda + B]}.
\end{align}
Suppose the subspace dimension is at most $k$ and signal $\mb y$ satisfies \Cref{asm:theta_mu}. Let $\lambda = c_\lambda/\sqrt{k}$ and $B\leq c'\lambda\theta^2/p\log n$ for some $c_\lambda ,c' \in (0,1)$, then there is a numerical constant $C>0$ such that if $n \geq Cp^5\theta^{-2}\log p$, then with probability at least $1-3/n$, for every $\mb a \in \bb \cup_{\abs{\mb\tau}\leq k}\goodregion$, we have
\begin{align}
	\abs{J_B(\mb a)} \leq \frac{24c'n\theta^2}{p\log n}
\end{align}
\end{lemma}
\begin{proof}
	See \Cref{sec:proof_entries_onlambda}.
\end{proof} 

The geometric approximation between $\varphi_{\ell^1}$ and $\varphi_\rho$ necessarily consists of three parts: the gradient, the Hessian, and the coefficients. Here we conclude the approximation result with the following lemma:

\begin{lemma}[$\varphi_{\ell^1}$ approximates $\varphi_{\rho}$] \label{lem:approx_rho_ell1}  Suppose $\mb x_0\simiid \mr{BG}(\theta)$ in $\R^n$, and $k, c_\mu$ such that $(\mb a_0,\theta,k)$ satisfies the sparsity-coherence condition $\mr{SCC}(c_\mu)$. Let $\rho\in\R^n\to\R$ be a $\delta$-smoothed $\ell^1$ function with 
\begin{align}\label{eqn:approx_grad_hess_eps_ub}
	\lambda = \frac{c_\lambda}{\sqrt k},\qquad \delta \leq \frac{c'^4\theta^8}{p^2\log^2n} \lambda
\end{align}
with some $c', \,c_\lambda\in (0,1)$, then there is a numerical constant $C,\ol c>0$ such that if  $n>Cp^5\theta^{-2}\log p$ and $c_\mu \leq \ol c$, then with probability at least $1-10/n$, the following statements hold simultaneously for every $\mb a\in\cup_{\abs{\mb\tau}\leq k}\goodregion$:

\vsni (1). The coefficients has norm difference 
\begin{align}\label{eqn:smooth_approx_alpha}
	\norm{\injector_{[\pm p]}^*\checkmtx{\mb x_0}\prox_{\lambda\ell^1}[\wc{\mb a}*\mb y]  - \injector_{[\pm p]}^*\checkmtx{\mb x_0}\prox_{\lambda\rho}[\wc{\mb a}*\mb y]  }2 \leq c' n\theta^4.   
\end{align}
(2). The gradient has norm difference 
\begin{align}\label{eqn:smooth_approx_grad}
	\norm{\nabla\varphi_{\ell^1}(\mb a)- \nabla\varphi_{\rho}(\mb a)}2 & \leq c' n\theta^4.
\end{align}
(3). The (pesudo) Riemmannian curvature difference is bounded in all directions $\mb v \in \Sp^{p-1}$ via
\begin{align}\label{eqn:smooth_approx_hess}
	\forall\,\mb v\in\Sp^{p-1},\quad \abs{\mb v^*\paren{\wt{\mr{Hess}}[\varphi_{\ell^1}](\mb a) -\mr{Hess}[\varphi_{\rho}](\mb a)}\mb v}  \leq 200c' n\theta^2.
\end{align}
\end{lemma}

\begin{proof} 1. (\ul{Coefficients}) From \Cref{lem:smooth_rho}, the proximal $\delta$-smoothed $\ell^1$ function satisfies 
\begin{align} 
\abs{\soft{\wc{\mb a}*\mb y}{\lambda} - \mc S_\lambda^\delta\brac{\wc{\mb a}*\mb y}}_j <\sqrt{\lambda\delta}\qquad  \forall j\in[n]. \notag 
\end{align}
Since the support of  coefficient vectors are contained in $[\pm p]$, using simple norm inequality:  
\begin{align}\label{eqn:coef_approx_ub1}
		& \norm{\injector_{[\pm p]}^*\checkmtx{\mb x_0}\mc S_\lambda\brac{\wc{\mb a}*\mb y}  - \injector_{[\pm p]}^*\checkmtx{\mb x_0}\mc S_{\lambda}^{\delta}\brac{\wc{\mb a}*\mb y}}2 \leq  \sqrt{\lambda \delta n}\cdot \norm{\injector_{[\pm p]}^*\checkmtx{\mb x_0} }2. 
\end{align}  
Apply \Cref{lem:y_bound} by replacing $\mb a_0$ with standard basis $\mb e_0$ and extend support of $\injector$ to $\injector_{[\pm p]}$,
\begin{comment}
\end{comment}
notice that in this case we have $\mu = 0$. Condition on the event 
\begin{align}
		\norm{\injector_{[\pm p]}^*\checkmtx{\mb x_0}}2 \leq \norm{\injector_{[\pm p]}^*\checkmtx{\mb x_0}\convmtx{\mb e_0}^*}2 \leq \sqrt{3(1+2\mu p)n\theta} \leq \sqrt{3n\theta}, \notag
\end{align}
and we gain
\begin{align}
	\eqref{eqn:coef_approx_ub1} &\leq \sqrt{\lambda \delta n} \cdot \sqrt{3n\theta }\leq  n\sqrt{3\lambda\theta\delta} \leq  c' n\theta^4.\notag
\end{align}

\vsni 2. (\ul{Gradient}) From definition of Riemannian gradient \eqref{eqn:grad_rho} and apply similar norm bound of \eqref{eqn:coef_approx_ub1}, and condition on the following events of \Cref{lem:y_bound}  holds, obtain
\begin{align}\label{eqn:approx_phi_ell1_grad_ub1}
	\norm{\nabla\varphi_{\ell^1}(\mb a)- \nabla\varphi_{\rho}(\mb a)}2 &\leq \sqrt{\lambda\delta n }\cdot  \norm{\ip^*\checkmtx{\mb y}}2 \leq n\sqrt{3\lambda\theta(1+\mu p)\delta } \leq c'n\theta^4.
\end{align}

\vsni 3. (\ul{Hessian}) For every realization of $J_B(\mb a)$ from $\mb a\in\cup_{\abs{\mb\tau}\leq k}\goodregion$, base on  \Cref{lem:num_entries_onlambda}, condition on the event such that
\begin{align}\label{eqn:l1_approx_BJ}
	B \leq  \frac{c'\lambda\theta^2}{p\log n},\qquad \abs{J} \leq \frac{24c' n\theta^2}{p\log n};
\end{align}
and rewrite $J_B(\mb a)$ as $J$. Also condition on the event using \Cref{lem:y_bound} and $(1+\mu p) \theta\log\theta^{-1} < 1$ 
\begin{align}\label{eqn:l1_approx_Cy_bd}
	\norm{\ip^*\checkmtx{\mb y}}2 \leq \sqrt{3n},\qquad \norm{\ip^*\checkmtx{\mb y}\mb P_J}2 \leq \sqrt{8\abs Jp\log n}, 
\end{align}    
then the difference of Hessian \eqref{eqn:hess_rho}, in direction $\mb v \in \Sp^{p-1}$ can be bounded as
\begin{align}
	&\phantom{{}=} \abs{\mb v^*\paren{\wt{\mr{Hess}}[\varphi_{\ell^1}](\mb a) -\mr{Hess}[\varphi_{\rho}](\mb a)}\mb v} \notag \\
	&\leq  \abs{\mb v^*\ip^*\checkmtx{\mb y}\paren{\mb P_{I(\mb a)} - \diag\brac{\nabla\mc S_{\lambda}^{\delta}\brac{\checkmtx{\mb y}\ip\mb a}}} \checkmtx{\mb y}\ip\mb v} + \norm{\nabla\varphi_{\ell^1}(\mb a)- \nabla\varphi_{\rho}(\mb a)}2 \label{eqn:eqn_approx_hess_ub1}
\end{align}
where $I(\mb a)$ is defined in \eqref{eqn:suppI}. Let $\mb D = \mb P_{I(\mb a)} - \diag\brac{\nabla\mc S_{\lambda}^{\delta}\brac{\checkmtx{\mb y}\ip\mb a}}$ and notice that $\mb D$ is a diagonal matrix, which suggests \eqref{eqn:eqn_approx_hess_ub1} can be decomposed using 
\begin{align}
	(\mb P_{J} + \mb P_{J^c} )\mb D(\mb P_{J} + \mb P_{J^c} ) = \mb P_{J} \mb D \mb P_{J} + \mb P_{J^c} \mb D \mb P_{J^c},\notag  
\end{align}
where, from with property of $\sqrt\delta$-smoothed $\ell^1$ function  \Cref{lem:smooth_rho}:
\begin{align}
	\max_{j}\abs{\mb P_{J} \mb D \mb P_{J}}_{jj} \leq 1,\qquad  \max_{j}\abs{\mb P_{J^c} \mb D \mb P_{J^c}}_{jj} \leq \sqrt{\lambda\delta}/ B.\notag
\end{align}
Finally, once again apply $\delta$ bound from \eqref{eqn:approx_grad_hess_eps_ub} and bounds for  $B,\abs{J},\mb y$ from \eqref{eqn:l1_approx_BJ}-\eqref{eqn:l1_approx_Cy_bd}, we gain 
\begin{align}
\eqref{eqn:eqn_approx_hess_ub1}	&\leq \norm{\ip^*\checkmtx{\mb y}\mb P_{J}}2^2 +\frac{\sqrt{\lambda\delta}}{B}\norm{\ip^*\checkmtx{\mb y}}2^2 + \norm{\nabla\varphi_{\ell^1}(\mb a)- \nabla\varphi_{\rho}(\mb a)}2 \notag\\
&\leq  8\abs{J}p\log n + \frac{3n\sqrt{\lambda\delta}}{B} + c'n\theta^2 \notag \\
&\leq 8\cdot\frac{24c'n\theta^2}{p \log n}\cdot p\log n  + \frac{3n\paren{c'^4\lambda^2\theta^8/p^2\log^2n }^{1/2} }{c'\lambda \theta^2/p\log p} + c' n\theta^2 \notag \\
& \leq 200c' n\theta^2,\notag
\end{align}
where all above result holds with probability at least $1-10/n$ from \Cref{lem:num_entries_onlambda} and \Cref{lem:y_bound}.
\end{proof}

% !TEX root = ../../BD_DQ.tex 
\section{Analysis of geometry} \label{sec:analysis_geometry}
In this section we prove major geometrical result in \Cref{thm:three_regions}. This lemma consists of three parts of geometry of $\varphi_\rho$; including the negative curvature region \Cref{cor:neg_curve}, large gradient region \Cref{cor:strong_grad}, strong convexity region near shift \Cref{cor:strong_convex}, and retraction to subspace \Cref{cor:retraction_alpha_tauc}, which are respectively base on geometric properties of $\varphi_{\ell^1}$ in \Cref{lem:neg_curve}, \Cref{lem:strong_grad},  \Cref{lem:strong_convex} and \Cref{lem:retraction_alpha_tauc}. We will handle each individual region in the following subsections. To shed light on the technical detail of the proof, we will begin with two figures for illustration of a toy example,  which demonstrate the geometry near a two dimension solution subspace $\mc S_{\set{i,j}}$, as follows: 

\begin{figure}[h!]  
	\centering
	\includegraphics[width = 0.58\textwidth]{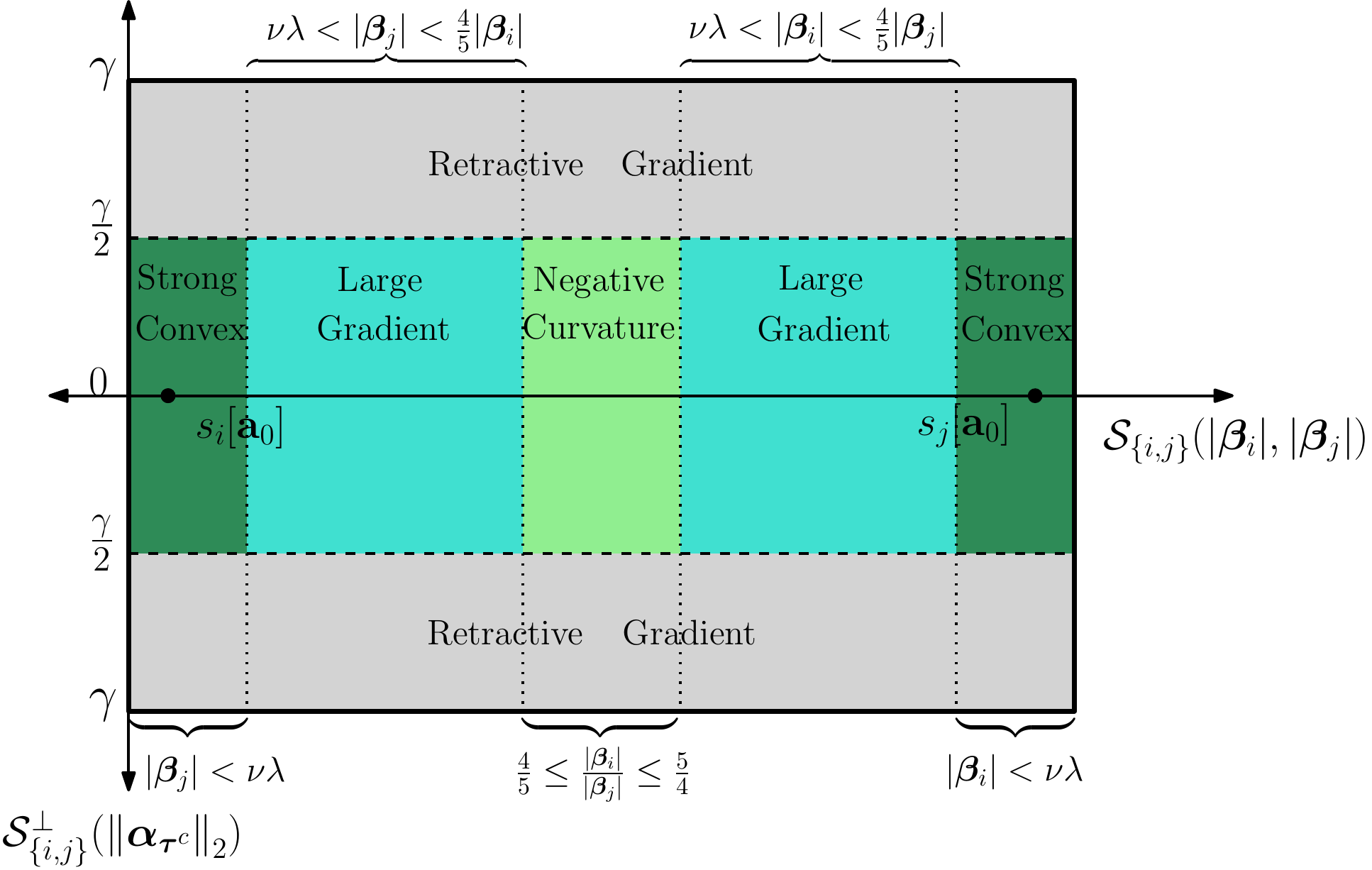}
	\caption{\textbf{The top view of geometry over subspace $\mc S_{\set{i,j}}$}.  We display the geometric properties in the neighborhood of subspace $\mc S_{\set{i,j}}$ (horizontal axis) which contains the solutions $\shift{\mb a_0}{i}$ and $\shift{\mb a_0}{j}$. When $\mb a$ lies near middle of two shifts (light green region) such that $\abs{\mb \beta_i}\approx\abs{\mb\beta_j}$, then there exists a negative curvature direction in subspace $\mc S_{\set{i,j}}$. When $\mb a$ leans closer to one of the shifts $\shift{\mb a_0}{i}$ (blue green region), its negative gradient direction  points at that nearest shift. When $\mb a$ is in the neighborhood of the shift $\shift{\mb a_0}{i}$ (dark green region) such that $\abs{\mb \beta_i}\ll \lambda$, it will be strongly convex at $\mb a$, and the unique minimizer within the convex region will be close to $\shift{\mb a_0}{i}$. Finally, the negative  gradient will be pointing back toward the subspace $\mc S_{\set{i,j}}$ if near boundary (grey region).   }
	\label{fig:top_view}
\end{figure}  
\vspace{-0.2in} 
\begin{figure}[h!]  
	\centering 
	\includegraphics[width = 0.58\textwidth]{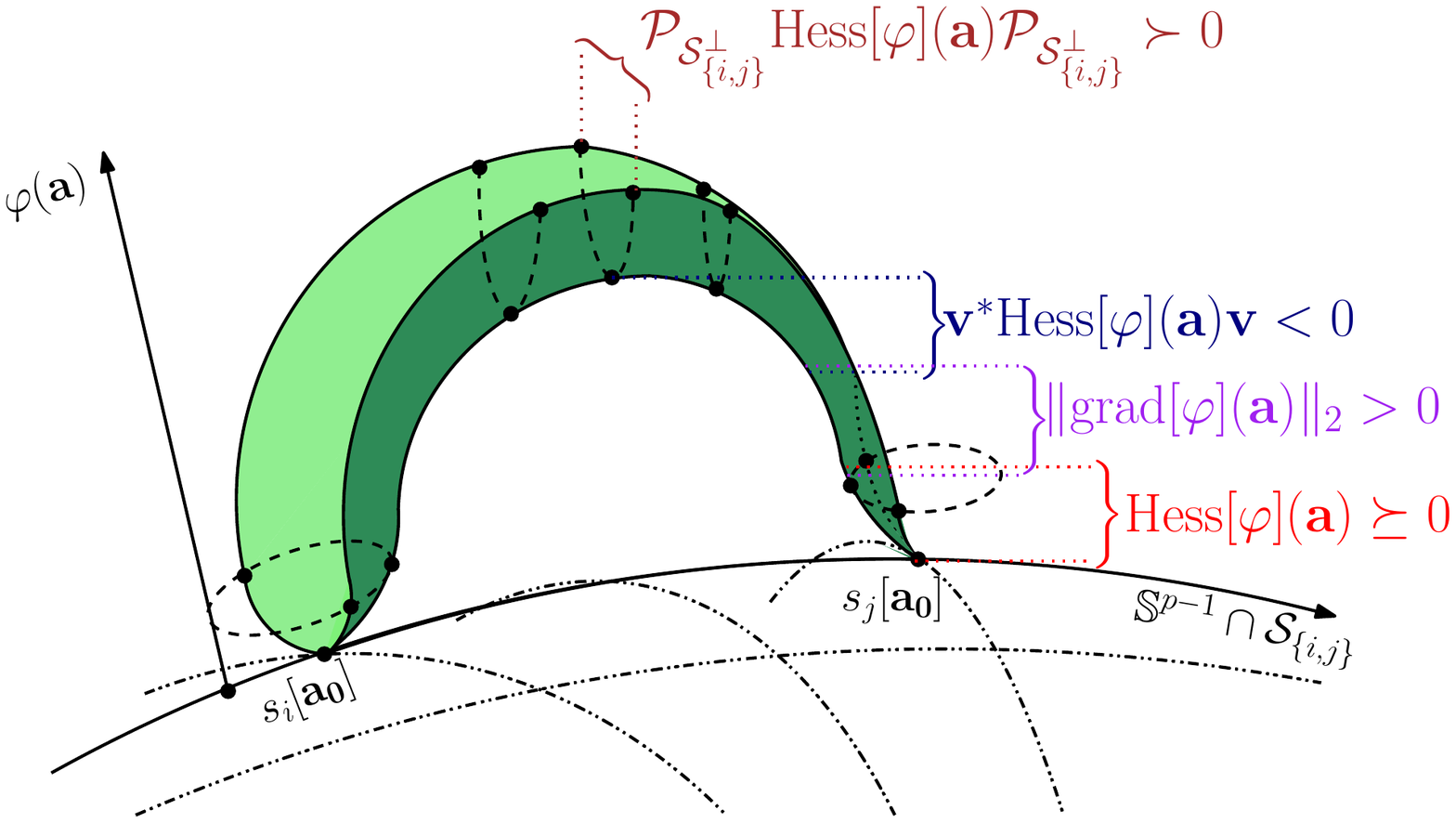}
	\caption{\textbf{The side view of geometry of subspace $\mc S_{\set{i,j}}$ on sphere}. We illustrate  the geometry of $\mc S_{\set{i,j}}$ over the sphere, in which the properties of the three regions are denoted. In negative curvature region, there exists a direction $\mb v$ such that $\mb v^*\mr{Hess}[\varphi](\mb a)\mb v$ is negative. In large gradient region, the norm of Riemannian gradient $\norm{\mr{grad}[\varphi](\mb a)}2$ will be strictly greater then $0$ and pointing at the nearest shift. Finally there is a convex region near all shifts such that $\mr{Hess}[\varphi](\mb a)$ is positive semidefinite. }
	\label{fig:side_view}
\end{figure}     

\pagebreak
    
\subsection{Negative curvature }

For any $\mb a\in\Sp^{p-1}$ near the subspace $\mc S_{\mb \tau}$ such that the entries of leading correlation vector $\mb\beta_{(0)},\mb\beta_{(1)}$ have balanced magnitude, the Hessian of $\varphi_{\rho}(\mb a)$ exhibits negative curvature in the span of $\shift{\mb a_0}{(0)},\shift{\mb a_0}{(1)}$. We will first demonstrate the pseudo negative curvature of $\varphi_{\ell^1}$ in  \Cref{lem:neg_curve}, then show $\varphi_\rho$ approximates $\varphi_{\ell^1}$ in terms of Hessian in \Cref{cor:neg_curve} when $\rho$ is properly defined as in \Cref{sec:smooth_approx}.

\begin{lemma}[Negative curvature for $\varphi_{\ell^1}$]\label{lem:neg_curve} Suppose that $\mb x_0\simiid  \mr{BG}(\theta)$ in $\R^n$, and $k, c_\mu$ such that $(\mb a_0,\theta,k)$ satisfies the sparsity-coherence condition $\mr{SCC}(c_\mu)$. Set $\lambda = c_\lambda/\sqrt{k}$ in $\varphi_{\ell^1}$ with  $c_\lambda\in\brac{\frac{1}{5},\frac14}$. There exist numerical constants $C,c,c',\ol c>0$ such that if  $n>C p^5\theta^{-2}\log p$, and $c_\mu \leq \ol c$, then with probability at least $1-c'/n$ the following holds at every $\mb a\in\cup_{\abs{\mb\tau}\leq k}\goodregion$  satisfying $  \abs{\mb\beta_{(1)}} \geq    \frac45\abs{\mb\beta_{(0)}}$: for $\mb v \in\mc S_{\set{(0),(1)}}\cap\Sp^{p-1} \cap \mb a^\perp$, 
\begin{align}   
	\mb v^*\wt{\mr{Hess}}[\varphi_{\ell^1}](\mb a)\mb v \leq -cn\theta\lambda.
\end{align}
\end{lemma}

\begin{proof} First of all the regional condition $\abs{\frac{\mb\beta_{(0)}}{\mb\beta_{(1)}}}\leq\frac 54$ provides a two side bound for the two leading $\beta$'s
\begin{align}
	0.79 \geq \frac{\abs{\mb\beta_{(0)}}}{\sqrt{\mb\beta_{(0)}^2+\mb\beta_{(1)}^2}}\norm{\mb\beta_{\mb\tau}}2 \geq \abs{\mb\beta_{(0)}} \geq \abs{\mb\beta_{(1)}} \geq \frac{4}{5}\abs{\mb\beta_{(0)}} \geq \frac45\cdot \frac{\norm{\mb\beta_{\mb\tau}}2}{\sqrt{\abs{\mb\tau}}} \geq \frac{0.79}{\sqrt{\abs{\mb\tau}}} \label{eqn:nc_beta_ublb}  
\end{align}  
 Set $J = \set{(0),(1)}$, choose $\mb v = \ip^*\convmtx{\mb a_0}\injector_J\mb\gamma$ with $\norm{\mb v}2 = 1$ then $\abs{\norm{\mb\gamma}2^2 -1} \leq \mu$. There exists such $\mb v$ satisfies condition above with $\mb a\perp\mb v$ by choosing $\mb\gamma$ as
\begin{align}
	\mb a^*\mb v=\mb a^*\ip^*\convmtx{\mb a_0}\injector_J\mb\gamma   = \mb\gamma_{(0)}\mb\beta_{(0)} +  \mb\gamma_{(1)}\mb\beta_{(1)} = 0,\notag
\end{align} 
hence $\abs{\frac{\mb\gamma_{(1)}}{\mb\gamma_{(0)}}} = \abs{\frac{\mb\beta_{(0)}}{\mb\beta_{(1)}}} \leq \frac54 $. This implies $\mb\gamma_{(0)}^2 \geq   \frac{16}{25}\mb\gamma_{(1)}^2 \geq \frac{16}{25}(1-\mu-\mb\gamma_{(0)}^2)$ where  $\mu\leq \frac{c_\mu}4\leq \frac{1}{100}$, it gives the lower bound of $\mb\gamma_{(0)}$ as 
\begin{align}
	\mb\gamma_{(0)}^2 \geq \frac{(1-\mu)\cdot 16}{25+16} \geq 0.385 \label{eqn:nc_gamma0_lb}
\end{align}
  
\vsni 1. (\ul{Expand the Hessian}) The (pseudo) curvature along direction $\mb v$ is written as  
\begin{align}
	\mb v^*\wt{\mr{Hess}}[\varphi_{\ell^1}](\mb a)\mb v &= \mb v^*\wt{\nabla}^2\varphi_{\ell^1}(\mb a)\mb v - \innerprod{\nabla\varphi_{\ell^1}(\mb a)}{\mb a}  =  -\mb \gamma^*\injector_J^*\mb M \checkmtx{\mb x}\mb P_{I(\mb a)}\checkmtx{\mb x} \mb M\injector_J\mb\gamma + \mb\beta^*\mb\chi[\mb\beta] \label{eqn:nc_hess}
\end{align}
expand the first term of \eqref{eqn:nc_hess} we obtain
\begin{align}
& -\mb \gamma^*\injector_J^*\mb M \checkmtx{\mb x}\mb P_{I(\mb a)}\checkmtx{\mb x} \mb M\injector_J\mb\gamma \notag\\ 
&\qquad = -\mb\gamma^*\injector_J^*\mb M\paren{\mb P_{(0)} + \mb P_{(1)} + \mb P_{J^c}} \checkmtx{\mb x}\mb P_{I(\mb a)}\checkmtx{\mb x}\paren{\mb P_{(0)} + \mb P_{(1)} + \mb P_{J^c}} \mb M\injector_J\mb\gamma \notag \\ 
	&\qquad \leq -\sum_{i\in J}\norm{\mb P_{I(\mb a)}\checkmtx{\mb x}\mb e_{i}}2^2\paren{\mb e_{i}^*\mb M \injector_J\mb\gamma  }^2 + 2\sum_{\substack{(i,j)\in\set{J,J^c} \\ (i,j) = ((0),(1))}}\abs{\mb e^*_{i}\checkmtx{\mb x}\mb P_{I(\mb a)}\checkmtx{\mb x}\mb e_j}\abs{\paren{\mb e_{i}^*\mb  M\injector_J\mb\gamma }\paren{\mb e_{j}^*\mb M\injector_J\mb\gamma}}  \notag \\
	&\qquad \leq -\sum_{i\in J}\norm{\mb P_{I(\mb a)}\checkmtx{\mb x}\mb e_{i}}2^2\paren{\abs{\mb\gamma_{i}} - \mu}^2 \notag \\
	&\qquad\qquad + 2\max_{i\neq j\in[\pm p]}\abs{\mb e^*_{i}\checkmtx{\mb x}\mb P_{I(\mb a)}\checkmtx{\mb x}\mb e_j}\paren{\norm{\injector_J^*\mb M\injector_J\mb\gamma}1 \norm{\injector_{J^c}^*\mb M\injector_J\mb\gamma}1 + \paren{ \abs{\mb\gamma_{(0)}}+\mu }\paren{ \abs{\mb\gamma_{(1)}}+\mu} } 
\end{align}	
Consider the following events 
\begin{align}
	\begin{cases}
		\event_{\mr{cross}}:= \set{\forall\,\mb a\in\Sp^{p-1},\,\max_{i\neq j\in[\pm p]}\abs{\mb e^*_{i}\checkmtx{\mb x}\mb P_{I(\mb a) }\checkmtx{\mb x}\mb e_j} <4 n\theta^2} \\  
		\event_{\mr{ncurv}}:= \set{\forall\,\mb a\in\goodregion,\,\min_{i\in J} \norm{\mb P_{I(\mb a) }\shift{\mb x}{-i}}2^2 \geq  n\theta\paren{1-\E_{\mb s_i}(\lambda,\mb s_i) + \E_{\mb s_i}(\lambda,\mb s_i)} - \frac{c_\mu n\theta}{p} }  
	\end{cases},
\end{align}
and from \Cref{fact:M_entries}  we know 
\begin{align}
	\norm{\injector_J^*\mb M\injector_J\mb\gamma}1 \leq \norm{\mb\gamma}1 + 2\mu \leq 1.5,\quad \norm{\injector_{J^c}^*\mb M\injector_J\mb\gamma}1 \leq \mu p \norm{\mb\gamma}1\leq 1.5 \mu p,\notag  
\end{align}
on the event $\event_{\mr{cross}}\cap\event_{\mr{ncurv}}$, we have 
\begin{align}
	&-\mb \gamma^*\injector_J^*\mb M \checkmtx{\mb x}\mb P_{I(\mb a) }\checkmtx{\mb x} \mb M\injector_J\mb\gamma \notag \\
	&\qquad \leq \underbrace{-n\theta\cdot \sum_{i \in J}(\abs{\mb \gamma_i} - \mu)^2\paren{1-\E_{\mb s_i}\mr{erf}_{\mb \beta_i}(\lambda,\mb s_i) + \E_{\mb s_i} f_{\mb \beta_i}(\lambda,\mb s_i)}}_{g_1(\mb\beta)} + \paren{18\mu p + 8}n\theta^2+\frac{2c_\mu n\theta}{\sqrt{\abs{\mb\tau}}}\label{eqn:nc_hess_lb} 
\end{align}  
Meanwhile, for the latter term of \eqref{eqn:nc_hess},  consider the following event $\event_{\ol{\chi}}$ where we write $\sigma_i = \sign(\mb \beta_i)$ as: 
\begin{align} 
	\event_{\ol{\chi}} := \set{ \sigma_i \mb\chi[\mb\beta]_i \leq \begin{cases} n\theta\cdot \abs{\mb \beta_i}\paren{1-\E_{\mb s_i} \mr{erf}_{\mb \beta_i}(\lambda,\mb s_i)} + \frac{c_\mu n\theta}{p}, &\quad \forall\,i\in\mb\tau \\ n\theta\cdot \abs{\mb \beta_i} 4\theta\abs{\mb\tau} + \frac{c_\mu n\theta}{p}, &\quad \forall\,i\in\mb\tau^c \end{cases} }, 
\end{align}
\vsni and use both $\norm{\mb\beta}1\leq \frac{c_\mu p}{\sqrt{\abs{\mb\tau}}}$, $\norm{\mb\beta_{\mb\tau^c}}2^2 \leq \frac{c_\mu}{\theta\abs{\mb\tau}^2}$. On this event we have 

\vspace{-0.1in} 

\begin{align}  
	\mb\beta^*\mb\chi[\mb\beta] &\leq  n\theta\cdot \sum_{i\in\mb\tau}\mb \beta_i^2\paren{1-\E_{\mb s_i}\mr{erf}_{\mb \beta_i}(\lambda,\mb s_i)} + 4n\theta^2\abs{\mb\tau}\norm{\mb\beta_{\mb\tau^c}}2^2 + \frac{c_\mu n\theta}{p}\norm{\mb\beta}1 \notag\\  &\leq \underbrace{n\theta\cdot \sum_{i\in\mb\tau}\mb \beta_i^2\paren{1-\E_{\mb s_i}\mr{erf}_{\mb \beta_i}(\lambda,\mb s_i)}}_{g_2(\mb\beta)} + \frac{5 c_\mu n\theta}{\sqrt{\abs{\mb\tau}}}. \label{eqn:nc_betachibeta_ub}  
\end{align}  

\vsni 2. (\ul{Lower bound $\E f_{\mb \beta_i}$}) Combine the first term from each of the \eqref{eqn:nc_hess_lb} and \eqref{eqn:nc_betachibeta_ub}. Use $\mu\leq c_\mu\leq \frac{1}{300}$ and \eqref{eqn:nc_gamma0_lb} to obtain $\paren{\abs{\mb\gamma_{(0)}} -\mu}^2 > 0.38$, we have 
\begin{align}
	&\frac{1}{n\theta}\paren{g_1(\mb\beta) + g_2(\mb\beta)}  \leq - \sum_{i\in J}\brac{\paren{\abs{\mb \gamma_i} - \mu}^2 - \mb \beta_i^2} \paren{1-\E_{\mb s_i}\mr{erf}_{\mb \beta_i}(\lambda,\mb s_i)} \notag \\
	&\phantom{{}\frac{1}{n\theta}\paren{g_1(\mb\beta) + g_2(\mb\beta)} \leq }  + \sum_{i\in\mb\tau \setminus J}\mb \beta_i^2\paren{1-\E_{\mb s_i}\mr{erf}_{\mb \beta_i}(\lambda,\mb s_i)}  - 0.38\sum_{i\in J} \E_{\mb s_i}f_{\mb \beta_i}(\lambda,\mb s_i), \label{eqn:nc_g1g2}
\end{align} 
now use Taylor expansion \footnote{ Apply $\exp\brac{-x^2/2} > 1 - x^2/2$} for $f_{\mb \beta_i}$ and apply upper bound $\E \mb s_i^2 \leq \theta\norm{\mb\beta}2^2\leq \theta\paren{1+\frac{c_\mu}{\sqrt{\abs{\mb\tau}}} + \frac{c_\mu}{\theta\abs{\mb\tau}^2}} \leq \frac{3c_\mu}{\abs{\mb\tau}}$,  
\begin{align}\label{eqn:nc_fbeta}  
	\E_{\mb s_i}f_{\mb \beta_i}(\lambda,\mb s_i) &\geq \E_{\mb s_i}\frac{1}{\sqrt{2\pi}}\cdot\paren{\frac{2\lambda}{\abs{\mb \beta_i}} - \frac{\lambda^3}{\abs{\mb \beta_i}^3}\paren{1+\frac{3\mb s_i^2}{\lambda^2}}} \geq  \frac{1}{\sqrt{2\pi}}\cdot\underbrace{\paren{\frac{2\lambda}{\abs{\mb \beta_i}} - \frac{1}{\abs{\mb \beta_i}^3}\paren{\lambda^3 + \frac{9c_\mu\lambda}{\abs{\mb\tau}}} }}_{f(\beta)}, \notag
\end{align}
where  $f(\beta)$ is concave at stationary point since 
\begin{align}
	\begin{cases}
		f'(\beta_*) = 0 \implies 2\lambda\beta_*^2 = 3\lambda\paren{\lambda^2 + \frac{9c_\mu}{\abs{\mb\tau}}}  \\
		f''(\beta_*) = \frac{1}{\abs{\beta_*}^3}\paren{4\lambda - \frac{12\lambda}{\beta_*^2}\paren{\lambda^2  +  \frac{9c_\mu}{\abs{\mb\tau}}}} = \frac{1}{\abs{\beta_*}^3}\paren{4\lambda - \frac{12}{3/2}\lambda} < 0    
	\end{cases}, \notag  
\end{align}
then combine with regional condition \eqref{eqn:nc_beta_ublb}, and also apply assumption $ c_\lambda \leq \frac13$ and $c_\mu\leq\frac{1}{300}$, we gain 
\begin{align}
 0.38\sum_{i\in J} \E_{\mb s_i}f_{\mb \beta_i}(\lambda,\mb s_i) &\geq 0.3\min_{\beta = \frac{0.79}{\sqrt{\abs{\mb\tau}}}, 0.79}f(\beta) \notag \\
 &\geq 0.3\min\set{\frac{2c_\lambda}{0.79} - \frac{c_\lambda^3 + 9c_\mu c_\lambda}{0.79^3}, \,\lambda\paren{\frac{2}{0.79} - \frac{c_\lambda^2 + 9c_\mu}{0.79^3} } }  \notag \\
 &\geq 0.3\min\set{2c_\lambda,2\lambda} \geq 0.6\lambda.
\end{align} 
    
\vsni 3. (\ul{Upper bound $\E \mb\chi[\mb\beta]_i$}) When $\mb\beta_{(0)}^2 = \paren{\abs{\mb\gamma_{(0)}} - \mu}^2 -\eta $ for some $\eta > 0$. With monotonicity \Cref{lem:mono_chi}, which implies:
\begin{align}
	\paren{1-\E_{\mb s_{(0)}}\mr{erf}_{\mb\beta_{(0)}}(\lambda,\mb s_{(0)}) }  \geq \paren{1-\E_{\mb s_{(1)}}\mr{erf}_{\mb\beta_{(1)}}(\lambda,\mb s_{(1)}) }\geq  \paren{1-\E_{\mb s_i}\mr{erf}_{\mb \beta_i}(\lambda,\mb s_i) }, \label{eqn:nc_order}
\end{align}
then combine \eqref{eqn:nc_fbeta}-\eqref{eqn:nc_order} and use $\mu \leq \frac{c_\mu}{4\sqrt{\abs{\mb\tau}}}$ from \Cref{lem:alpha_beta_tau_lb}
\begin{align}
	\eqref{eqn:nc_g1g2} &\leq -\underbrace{\paren{\paren{\abs{\mb\gamma_{(0)}}^2 - \mu}^2 -\mb\beta_{(0)}^2 - \eta }}_{ = 0}\paren{1-\E_{\mb s_{(0)}}\mr{erf}_{\mb\beta_{(0)}}(\lambda,\mb s_{(0)}) } \notag \\
	&\phantom{\leq{}} +\paren{\sum_{i\in\mb\tau\setminus(0)}\mb \beta_i^2 - \paren{\abs{\mb\gamma_{(1)}}-\mu}^2 -\eta} \underbrace{\paren{1-\E_{\mb s_{(1)}}\mr{erf}_{\mb\beta_{(1)}}(\lambda,\mb s_{(1)})}}_{<1} -0.38\sum_{i\in J} \E_{\mb s_i}f_{\mb \beta_i}(\lambda,\mb s_i) \notag \\
	&\leq \paren{\norm{\mb\beta_{\mb\tau}}2^2 - \norm{\mb\gamma}2^2 + 2\mu\norm{\mb\gamma}1}- 0.6\lambda \notag \\
	&\leq \frac{2c_\mu}{\sqrt{\abs{\mb\tau}}} - 0.6\lambda.  \label{eqn:nc_g1g2_ub1} 
\end{align}
On the other hand, when $\mb\beta_{(0)}^2 \geq   \paren{\abs{\mb\gamma_{(0)}} - \mu}^2 > 0.38$, combining \eqref{eqn:nc_fbeta}-\eqref{eqn:nc_order} gives:
\begin{align}
	\eqref{eqn:nc_g1g2} &\leq  \paren{\norm{\mb\beta_{\mb\tau}}2^2  - \norm{\mb\gamma}2^2  + 2\mu\norm{\mb\gamma}1}  +\paren{\paren{\abs{\mb\gamma_{(0)}} - \mu}^2 - \mb\beta_{(0)}^2}\E_{\mb s_{(0)}}\mr{erf}_{\mb\beta_{(0)}}(\lambda,\mb s_{(0)})  \notag \\
	&\phantom{\leq{}} +\paren{\paren{\abs{\mb\gamma_{(1)}} - \mu}^2 - \sum_{i\in\mb\tau\setminus(0)}\beta_{i}^2}\E_{\mb s_{(1)}}\mr{erf}_{\mb\beta_{(1)}}(\lambda,\mb s_{(1)})   - 0.38\sum_{i\in J}\E_{\mb s_i}f_{\mb \beta_i}(\lambda,\mb s_i) \notag \\
	&\leq \paren{\frac{c_\mu}{\sqrt{\abs{\mb\tau}}} + 4\mu} + \paren{\mb\gamma_{(1)}^2 - \norm{\mb \beta_{\mb\tau}}2^2 + \mb\beta_{(0)}^2 }\E_{\mb s_{(1)}}\mr{erf}_{\mb\beta_{(1)}}(\lambda,\mb s_{(1)})  - 0.6\lambda, \label{eqn:nc_large_beta0_1}
\end{align} 
where \Cref{lem:chibeta_ublb} provides the upper bound for $\E_{\mb s_{(1)}}\mr{erf}_{\mb\beta_{(1)}}(\lambda,\mb s_{(1)})$ as
\begin{align}
	\E_{\mb s_{(1)}}\mr{erf}_{\mb\beta_{(1)}}(\lambda,\mb s_{(1)}) &= 1-\frac{1}{n\theta\mb\beta_{(1)}}\E\mb\chi[\mb\beta]_{(1)} \leq 1-\frac{\sigma_{(1)}}{n\theta\abs{\mb\beta_{(1)}}}\ul{\E\mb\chi[\mb\beta]}_{(1)} = 1-\frac{1}{\abs{\mb\beta_{(1)}}}\paren{\abs{\mb\beta_{(1)}}- \sqrt{\frac{2}{\pi}}\lambda} \notag \\
	&\leq \sqrt{\frac{2}{\pi}}\cdot\frac{\lambda}{\abs{\mb\beta_{(1)}}}, \label{eqn:nc_erf_ub}  
\end{align}
then calculate the constant for the second term in \eqref{eqn:nc_large_beta0_1} by writing $\kappa = \abs{\frac{\mb\gamma_{(1)}}{\mb\gamma_{(0)}}} = \abs{\frac{\mb\beta_{(0)}}{\mb\beta_{(1)}}} \leq \frac{5}{4} $, which provides $\mb\gamma_{(1)}^2 \leq \frac{(1+\mu)\kappa^2}{\kappa^2+1}$ and $\mb\beta_{(0)}^2 \leq \frac{\norm{\mb\beta_{\mb\tau}}2^2\kappa^2}{\kappa^2+1}$ where $\mu < \frac{c_\mu}{4}$, and by applying $\abs{\mb\beta_{(1)}} > \frac{4}{5}\abs{\mb\beta_{(0)}}\geq 0.3$, we have
\begin{align}
	\frac{ (\mb\gamma_{(1)}^2-1) + c_\mu +  \mb\beta_{(0)}^2 }{\abs{\mb\beta_{(1)}}} &\leq  - \frac{\kappa}{(\kappa^2+1)\abs{\mb\beta_{(0)}}}   +  \kappa \abs{\mb\beta_{(0)}} + \frac{\mu+c_\mu}{0.3} \leq \frac{\kappa^2-1}{\sqrt{\kappa^2+1}} + \kappa\paren{\norm{\mb\beta_{\mb\tau}}2^2-1} + 4.2c_\mu  \leq 0.36  + 6c_\mu \label{eqn:nc_const_ub},
\end{align}
and finally combine \eqref{eqn:nc_erf_ub}-\eqref{eqn:nc_const_ub}, follow from \eqref{eqn:nc_large_beta0_1} and use $c_\lambda \leq \frac13$:  
\begin{align}
	\eqref{eqn:nc_g1g2}  &\leq \frac{2c_\mu}{\sqrt{\abs{\mb\tau}}} + \sqrt{\frac{2}{\pi}}\paren{\mb\gamma_{(1)}^2- 1 + c_\mu + \mb\beta_{(0)}^2}\frac{\lambda}{\abs{\mb\beta_{(1)}}} - 0.6\lambda \notag \\
	&\leq  \frac{2c_\mu}{\sqrt{\abs{\mb\tau}}} +\sqrt{\frac{2}{\pi}}\paren{0.36\lambda + \frac{6c_\mu c_\lambda}{0.3} } - 0.6\lambda \notag \\
	&\leq \frac{4c_\mu}{\sqrt{\abs{\mb\tau}}} - 0.3\lambda \label{eqn:nc_g1g2_ub2}  
\end{align}  

\vsni 3. (\ul{Collect all results}) Combine the components of pseudo Hessian \eqref{eqn:nc_hess_lb},  \eqref{eqn:nc_betachibeta_ub} with bounds for $g_1+g_2$ from \eqref{eqn:nc_g1g2_ub1} and \eqref{eqn:nc_g1g2_ub2}, and use Lemma \ref{lem:alpha_beta_tau_lb} which provides both $\mu p\theta\abs{\mb\tau} < \frac{c_\mu}4$ and $\theta\abs{\mb\tau} <\frac{c_\mu}4 $ where $c_\mu<\frac{1}{300}$ and $c_\lambda \geq \frac{1}{5}$, we can obtain: 
\begin{align}
	\mb v^*\wt{\mr{Hess}}_{\varphi_{\ell^1}}[\mb a]\mb v &\leq g_1(\mb\beta) + g_2(\mb\beta) + \frac{7c_\mu n\theta}{\sqrt{\abs{\mb\tau}}} + \paren{18\mu p + 8}n\theta^2 \notag \\
	&\leq n\theta\cdot\paren{\frac{4c_\mu}{\sqrt{\abs{\mb\tau}}}  -0.3\lambda} + n\theta \cdot \frac{7c_\mu}{\sqrt{\abs{\mb\tau}}} +  n\theta\cdot \frac{6.5c_\mu}{\abs{\mb\tau}}  \notag \\  
	&\leq \frac{n\theta}{\sqrt{\abs{\mb\tau}}}\paren{0.059 - 0.06  } \leq -0.001n\theta\lambda
\end{align}
Finally, the curvature is negative along $\mb v$ direction with probability at least 
\begin{align}
	1-\underbrace{\prob{\event_{\mr{cross}}^c}}_{\text{\Cref{lem:x0_innerprod}}} - \underbrace{\prob{\event_{\mr{ncurv}}^c}}_{\text{\Cref{cor:ct_support}}} - \underbrace{\prob{\event_{\ol{\chi}}^c}}_{\text{\Cref{cor:chibeta_ct}}}.  
\end{align}  

\end{proof}

\vsni Similarly for objective $\varphi_\rho$, we have that

\begin{corollary}[Negative curvature for $\varphi_\rho$]\label{cor:neg_curve} Suppose that $\mb x_0\simiid  \mr{BG}(\theta)$ in $\R^n$, and $k, c_\mu$ such that $(\mb a_0,\theta,k)$ satisfies the sparsity-coherence condition $\mr{SCC}(c_\mu)$. Define  $\lambda = c_\lambda/\sqrt{k}$ in $\varphi_{\rho}$ where $c_\lambda\in\brac{\frac{1}{5},\frac14}$, then there exists some numerical constants $C,c,c',c'',\ol c>0$ such that if $\rho$ is $\delta$-smoothed $\ell^1$ function where $\delta \leq c''\lambda\theta^8/p^2\log^2n $, $n>C p^5\theta^{-2}\log p$ and $c_\mu\leq \ol c$, then with probability at least $1-c'/n$, for every $\mb a\in\cup_{\abs{\mb\tau} \leq  k}\goodregion$  satisfying $\abs{\mb\beta_{(1)}} \geq  \frac45\abs{\mb\beta_{(0)}}$: for $\mb v \in\mc S_{\set{(0),(1)}}\cap\Sp^{p-1}\cap \mb a^\perp$,
\begin{align} 
	\mb v^*\wt{\mr{Hess}}[\varphi_{\rho}](\mb a)\mb v \leq -cn\theta\lambda
\end{align}
\end{corollary}
\begin{proof}
	Choose $\mb v\in\Sp^{p-1}$ according to \Cref{lem:neg_curve} and \eqref{eqn:smooth_approx_hess} from \Cref{lem:approx_rho_ell1} with constant multiplier $\delta$ satisfies $c''^{1/4}< 10^{-3}c$, we gain
\begin{align}
	\mb v^*\mr{Hess}[\varphi_\rho](\mb a)\mb v & \leq -cn\theta\lambda + 200c'n\theta^2 \leq -cn\theta\lambda/2    \label{eqn:main_nc}
\end{align}
\end{proof}

\vspace{-0.2in}

% =================================== %
\subsection{Large gradient}
For any $\mb a\in\Sp^{p-1}$ near subspace and the second largest correlation $\mb\beta_{(1)}$ much smaller then the first correlation $\mb\beta_{(0)}$ while not being near 0, the negative gradient of $\varphi_{\rho}(\mb a)$ will point at the largest shift. We show this in \Cref{lem:strong_grad}, and  the $ \varphi_\rho$ version in \Cref{cor:strong_grad} when $\rho$ is properly defined as in \Cref{sec:smooth_approx}.

\begin{lemma}[Large gradient for $\varphi_{\ell^1}$]\label{lem:strong_grad} Suppose that $\mb x_0\simiid  \mr{BG}(\theta)$ in $\R^n$, and $k, c_\mu$ such that $(\mb a_0,\theta,k)$ satisfies the sparsity-coherence condition $\mr{SCC}(c_\mu)$. Define $\lambda = c_\lambda/\sqrt{k}$ in $\varphi_{\ell^1}$ with some $c_\lambda\in\brac{\frac{1}{5},\frac{1}{4}}$, then there exists some numerical constants $C,c',c,\ol c>0$, such that if $n>C p^5\theta^{-2}\log p$ and $c_\mu\leq \ol c$, then with probability  at least $1-c'/n$, for every $\mb a\in\cup_{\abs{\mb\tau} \leq  k}\goodregion$ satisfying $\tfrac45\abs{\mb\beta_{(0)}}>\abs{\mb\beta_{(1)}} > \frac{1}{4\log\theta^{-1}}\lambda $, 
\begin{align}
	\innerprod{\mb\sigma_{(0)} \ip^*\shift{\mb a_0}{(0)}}{-\mr{grad}[\varphi_{\ell^1}](\mb a)} \geq cn\theta\paren{\log^{-2}\theta^{-1}}\lambda^2 
\end{align}
where $\mb\sigma_i = \sign(\mb\beta_i)$.
\end{lemma}
\begin{proof} 1. (\ul{Properties for $\mb\alpha,\mb\beta$}) Define $\clog = \frac{1}{\log\theta^{-1}}$, we first derive upper bound on the dominant entry $\abs{\mb\beta_{(0)}}$ as follows. Write the geodesic distance between $\mb a$ and $\ip^*\shift{\mb a_0}{i}$ as a function of $\mb\beta_i$ as $d_{\Sp}(\mb a, \pm \ip^*\shift{\mb a_0}{i}) = \cos^{-1}(\mb\beta_i)$, then by triangle inequality we have:  
\begin{align}
	& d_{\Sp}(\mb a, \pm \ip^*\shift{\mb a_0}{(0)}) \geq d_{\Sp}(\pm \ip^*\shift{\mb a_0}{(0)}, \ip^*\shift{\mb a_0}{(1)})  -  d_{\Sp}(\mb a, \ip^*\shift{\mb a_0}{(1)}) \notag \\
	\implies &  \cos^{-1}\pm\mb\beta_{(0)} \geq \cos^{-1}\mu -\cos^{-1}\abs{\mb\beta_{(1)}} \notag\\
	\implies & \pm \mb\beta_{(0)} \leq \cos\paren{\cos^{-1}\mu - \cos^{-1}\abs{\mb\beta_{(1)}}} = \mu\abs{\mb\beta_{(1)}} + \sqrt{\paren{1-\mu^2}\paren{1-\mb\beta^2_{(1)}}} \leq 1-\tfrac12\paren{\abs{\mb\beta_{(1)}} - \mu}^2.\notag   
\end{align}
Use the regional condition $\abs{\mb\beta_{(1)}}\geq \frac{\clog}{4}\lambda$ and since $\mu\abs{\mb\tau}^{3/2} < \frac{c_{\lambda}}{100}\clog$ from \Cref{asm:theta_mu}, implies 
\begin{align}  
	\abs{\mb\beta_{(0)}} &\leq  1-\tfrac{\mb\beta_{(1)}^2}{2}\paren{1-\tfrac{4\mu\sqrt{\abs{\mb\tau}}}{\clog c_\lambda}} \leq 1-0.49\mb\beta_{(1)}^2 =: \beta_{\mr{ub}}.\label{eqn:strong_grad_beta0_ub}
\end{align} 
Meanwhile a lower bound for $\mb\beta_{(0)}$ can be easily determined by the other side of regional condition: 
\begin{align}\label{eqn:strong_grad_beta0_lb}
	\abs{\mb\beta_{(0)}} \geq \tfrac{5}{4}\abs{\mb\beta_{(1)}} =: \beta_{\mr{lb}}.
\end{align}
Also since $\mb\beta = \mb M \mb\alpha$, based on properties of $\mb M$ from \Cref{fact:M_entries}. When $\norm{\mb\alpha_{\mb\tau}}2 \leq 1+c_\mu$ and   $\norm{\mb\alpha_{\mb\tau^c}}2 \leq  \gamma \leq \frac{c_\mu\clog^2}{4\mu\sqrt p\abs{\mb\tau}}$, we gain: 
\begin{align} 
	& \mb\beta_{(0)} = \mb\alpha_{(0)} + \mb e_{(0)}^*\mb M\mb\alpha_{\setminus(0)}    \notag  \\
\implies & \abs{\mb\alpha_{(0)} - \mb\beta_{(0)}} \leq    \mu\sqrt{\abs{\mb\tau}}\norm{\mb\alpha_{\mb\tau}}2 + \mu\sqrt{p}\norm{\mb\alpha_{\mb\tau^c}}2 \leq  \tfrac{c_\mu\clog^2(1+c_\mu)}{4\abs{\mb\tau}} + \mu\sqrt p\gamma  \leq  \tfrac{c_\mu\clog^2}{\abs{\mb\tau}}.    \label{eqn:strong_grad_alpha0_ub}
\end{align}
and therefore 
$\abs{\mb\alpha_{(0)}} \leq \abs{\mb\beta_{(0)}} + \frac{c_\mu\clog^2}{\abs{\mb\tau}} \leq 1-.49\paren{\frac{\clog}{4}\lambda}^2 +   \frac{c_\mu\clog^2}{\abs{\mb\tau}}<1$.

\vsni 2. (\ul{Upper bound of $\mb\beta^*\mb\chi[\mb\beta]$}) Define a piecewise smooth convex upper bound $h$ for $\mb\beta_i\mb\chi[\mb\beta]_i$ as:
\begin{align}
	  h(\mb\beta_i) :=  \begin{cases}
		\mb\beta_i^2 - \frac{\nu_1\lambda}2\abs{\mb\beta_i} & \quad \abs{\mb\beta_i} \geq \nu_1\lambda \\
		\frac{1}{2}\mb\beta_i^2  &\quad \abs{\mb\beta_i} \leq \nu_1\lambda\notag
	\end{cases}, 
\end{align}
then \Cref{lem:max_soft_thresh} tells us since  $\norm{\mb\beta_{\mb\tau\setminus(0)}}\infty \leq \mb\beta_{(1)}$: 
\begin{align}
	\sum_{i\in\mb\tau\setminus(0)} h(\mb\beta_i) &\leq \norm{\mb\beta_{\mb\tau\setminus(0)}}2^2\paren{1-\frac{\nu_1\lambda\mb\beta_{(1)}}{2\mb\beta^2_{(1)}} }\leq  \paren{1+\frac{c_\mu\clog^2}{\abs{\mb\tau}} -\mb\beta_{(0)}^2 }\paren{1-\frac{\nu_1\lambda}{2\mb\beta_{(1)}}} \notag \\
	&\leq  \paren{1-\frac{\nu_1\lambda}{2\mb\beta_{(1)}}}\paren{1-\mb\beta_{(0)}^2} +\frac{c_\mu\clog^2}{\abs{\mb\tau}}, \notag
\end{align}
then condition on the following event using  \Cref{cor:chibeta_ct},
\begin{align}  
	\event_{\ol{\chi}} := \set{\mb\beta_i\mb\chi[\mb\beta]_i \leq \begin{cases} 
 			n\theta \cdot h(\mb\beta_i) +  \frac{c_\mu\theta}{p^{3/2}}\abs{\mb\beta_i},&\quad  \forall i\in\mb\tau\setminus(0)\\
 			n\theta \cdot 4\mb\beta_i^2\theta\abs{\mb\tau} + \frac{c_\mu\theta}{p^{3/2}}\abs{\mb\beta_i}, &\quad\forall i\in\mb\tau^c
 \end{cases}
 }, \notag   
\end{align}
which provides the upper bound of $\mb\beta^*\mb\chi[\mb\beta]$ by applying $5p >\log^{8/3}(p\log^2p) >(\clog^2)^{4/3} $ from lower bound of $\theta$ from \Cref{asm:theta_mu}, $\norm{\mb\beta_{\mb\tau^c}}2\leq \frac{c_\mu\clog }{\sqrt{\theta}\abs{\mb\tau}}$ from \Cref{lem:alpha_beta_tau_lb} , $\abs{\mb\tau} \leq \sqrt p$ from lemma assumption and let $c_\mu<\frac{1}{100}$:  
\begin{align}  
	\mb\beta^*\mb\chi[\mb\beta] &\leq  \mb\chi[\mb\beta]_{(0)}\mb\beta_{(0)} + \sum_{i\in \mb\tau\setminus(0)}\mb\beta_i\mb\chi[\mb\beta]_i + \innerprod{\mb\beta_{\mb\tau^c}}{\mb\chi[\mb\beta]_{\mb\tau^c}}\notag\\
	& \leq \mb\chi[\mb\beta]_{(0)}\mb\beta_{(0)} + n\paren{\theta\sum_{i\in\mb\tau\setminus(0)}h(\mb\beta_i) + 4\theta^2\abs{\mb\tau}\norm{\mb\beta_{\mb\tau^c}}2^2  + \frac{c_\mu\theta}{p^{3/2}}\paren{\sqrt{\abs{\mb\tau}}\norm{\mb\beta_{\mb\tau}}2 + \sqrt p\norm{\mb\beta_{\mb\tau^c}}2} }\notag \\
	&\leq \mb\chi[\mb\beta]_{(0)}\mb\beta_{(0)} +  n\paren{\theta\cdot \eta(1-\mb\beta_{(0)}^2) +\theta\cdot\frac{c_\mu\clog^2}{\abs{\mb\tau}} +   \frac{4\theta^2\abs{\mb\tau}c_\mu^2\clog^2}{\theta\abs{\mb\tau}^2 } + c_\mu\theta \paren{\frac{1+c_\mu}{p^{3/4}\abs{\mb\tau}}  + \frac{c_\mu\clog}{p\sqrt{\theta}\abs{\mb\tau} } } } \notag \\
	&\leq \mb\chi[\mb\beta]_{(0)}\mb\beta_{(0)} +  n\theta \paren{\eta(1-\mb\beta_{(0)}^2) + \frac{6c_\mu\clog^2}{\abs{\mb\tau}}},    \label{eqn:strong_grad_beta_chibeta_ub1}
\end{align}
 where $\eta = 1 - \frac{\nu_1\lambda}{2\mb\beta_{(1)}}$.

\vsni 3. (\ul{Align the gradient with $\ip^*\shift{\mb a_0}{(0)}$}) Base on the definition $\mb\beta$, since $\mb\beta_{(0)} = \innerprod{\mb a}{\ip^*\shift{\mb a_0}{(0)}}$, we can expect that the negative gradient is likely aligned with direction toward one of the candidate solution $\pm \ip^*\shift{\mb a_0}{(0)}$. Wlog assume that both $\mb\beta_{(0)},\mb\beta_{(1)}$ are positive, then expand the gradient and use incoherent property for $\mb a_0$ \Cref{fact:M_entries} we have: 
\begin{align}
	 \innerprod{\ip^*\shift{\mb a_0}{(0)}}{-\mr{grad}_{\varphi_{\ell_1}}[\mb a]} &= \innerprod{\ip^*\shift{\mb a_0}{(0)}}{\ip^*\convmtx{\mb a_0}\paren{\mb\chi[\mb\beta] - \mb\beta^*\mb\chi[\mb\beta]\mb\alpha}} \notag \\
	& \geq \paren{\mb\chi[\mb\beta]_{(0)} -\mb\beta^*\mb\chi[\mb\beta]\mb\alpha_{(0)}} - \mu\norm{\mb\chi[\mb\beta]_{\setminus(0)} - \mb\beta^*\mb\chi[\mb\beta]\mb\alpha_{\setminus(0)} }1, \label{eqn:strong_grad_lb1}
\end{align} 
where $\setminus (0)$ is an abbreviation of the complement set $[\pm 2p_0]\setminus(0)$. The latter part of \eqref{eqn:strong_grad_lb1} has an upper bound using bounds of  $\mb\beta^*\mb\chi[\mb\beta] <  \frac{3n\theta}{2}$, $\norm{\mb\chi[\mb\beta]_{\mb\tau^c}}2 < \frac{n\theta\gamma_2}{20}$ from \eqref{eqn:retract_betachibeta_lb}, and $\norm{\mb\chi[\mb\beta]_{\mb\tau\setminus(0)}}2 \leq n\theta \norm{\mb\beta_{\mb\tau\setminus(0)}}2$ in event $\event_{\ol{\chi}}$, we obtain: 
\begin{align}
	& \phantom{=} \mu\norm{\mb\chi[\mb\beta]_{\setminus(0)} - \mb\beta^*\mb\chi[\mb\beta]\mb\alpha_{\setminus(0)} }1 \notag\\
	&\leq  \mu\paren{\sqrt{\abs{\mb\tau}}\norm{\mb\chi[\mb\beta]_{\mb\tau\setminus(0)}}2 + \mb\beta^*\mb\chi[\mb\beta]\sqrt{\abs{\mb\tau}}\norm{\mb\alpha_{\mb\tau\setminus(0)}}2 + \sqrt{p}\norm{\mb\chi[\mb\beta]_{\mb\tau^c}}2 + \mb\beta^*\mb\chi[\mb\beta]\sqrt{p}\norm{\mb\alpha_{\mb\tau^c}}2 }  \notag \\
	&\leq  n\theta\cdot\brac{\mu\sqrt{\abs{\mb\tau}}\paren{\norm{\mb\beta_{\mb\tau}}2 - \abs{\mb\beta_{(0)}} }+  \mu\sqrt{\abs{\mb\tau}}\paren{\norm{\mb\alpha_{\mb\tau}}2 - \abs{\mb\alpha_{(0)}} }+ \frac{1}{20}\mu\sqrt p\gamma_2 +  \frac32 \mu\sqrt{p}\gamma_2} \notag \\ 
	&\leq n\theta\cdot\frac{c_\mu\clog^2} {4\abs{\mb\tau}}\brac{2\paren{1+c_\mu} - \abs{\mb\beta_{(0)}} - \abs{\mb\alpha_{(0)}} +  \paren{\frac{1}{20} + \frac32}c_\mu }   \notag \\
	&\leq n\theta\cdot\frac{c_\mu\clog^2}{\abs{\mb\tau}}\paren{0.5 + c_\mu -0.5\mb\beta_{(0)}}. \label{eqn:strong_grad_res_ub}
\end{align}

\vsni On the other hand,  the former term of \eqref{eqn:strong_grad_lb1} possesses a lower bound using \eqref{eqn:strong_grad_alpha0_ub}-\eqref{eqn:strong_grad_beta_chibeta_ub1}, $\mb \chi[\mb\beta]_{(0)} > n\theta\paren{\mb\beta_{(0)} - \frac{\nu_1}2\lambda -\frac{c_\mu}{p}}  \geq  n\theta\paren{\mb\beta_{(0)} - 0.51\nu_1\lambda}$ and $\mb\alpha_{(0)}\leq 1$:
\begin{align} \label{eqn:strong_grad_main_lb}
	& \mb\chi[\mb\beta]_{(0)} - \mb\beta^*\mb\chi[\mb\beta]\mb\alpha_{(0)} \notag \\
	&\geq \paren{1-\mb\alpha_{(0)}\mb\beta_{(0)}}\mb\chi[\mb\beta]_{(0)}-n\theta\cdot\brac{ \eta\paren{1-\mb\beta^2_{(0)}} + \frac{6c_\mu\clog^2}{\abs{\mb\tau}} }\mb\alpha_{(0)} \notag  \\
	&\geq n\theta\underbrace{\paren{1-\paren{\mb\beta_{(0)} + \frac{c_\mu\clog^2}{\abs{\mb\tau}} }\mb\beta_{(0)} }\paren{\mb\beta_{(0)}- 0.51\nu_1\lambda} }_{(a)}-n\theta\underbrace{\brac{\eta\paren{1-\mb\beta^2_{(0)}}\paren{\mb\beta_{(0)} + \frac{c_\mu\clog^2}{\abs{\mb\tau}}}  +  \frac{6c_\mu\clog^2}{\abs{\mb\tau}}\mb\alpha_{(0)}   }  }_{(b)} \notag \\
	&\geq   n\theta\brac{\underbrace{\paren{1-\mb\beta_{(0)}^2}\paren{\mb\beta_{(0)} - 0.51\nu_1\lambda} - \frac{c_\mu\clog^2 \mb\beta_{(0)}^2}{\abs{\mb\tau}}}_{(a)}  \underbrace{-\paren{1-\mb\beta_{(0)}^2}\eta\mb\beta_{(0)} -\eta\frac{c_\mu\clog\paren{1-\mb\beta_{(0)}^2}}{\abs{\mb\tau}} -\frac{6 c_\mu\clog^2}{\abs{\mb\tau}}}_{(b)} } \notag \\
	&\geq   n\theta\brac{\paren{1-\mb\beta^2_{(0)}}\paren{\paren{1-\eta}\mb\beta_{(0)} - 0.51\nu_1\lambda}  - \frac{c_\mu\clog^2}{\abs{\mb\tau}}\paren{ (1-\eta) \mb\beta_{(0)}^2 + 7 }},   
\end{align} 
combine \eqref{eqn:strong_grad_lb1} with \eqref{eqn:strong_grad_res_ub}-\eqref{eqn:strong_grad_main_lb} and $\eta > 0$, we have 
\begin{align}
	\eqref{eqn:strong_grad_lb1}
	&\geq n\theta\brac{\paren{1-\mb\beta^2_{(0)}}\paren{\paren{1-\eta}\mb\beta_{(0)} - 0.51\nu_1\lambda}  - \frac{c_\mu\clog^2}{\abs{\mb\tau}}\paren{ (1-\eta) \mb\beta_{(0)}^2 + 7 } } - n\theta\cdot  \frac{c_\mu\clog^2}{\abs{\mb\tau}}\paren{0.5+c_\mu - 0.5\mb\beta_{(0)}} \notag \\ 
	&\geq n\theta\brac{\underbrace{\paren{1-\mb\beta^2_{(0)}}\paren{\frac{\nu_1\lambda}{2\mb\beta_{(1)} }\mb\beta_{(0)} - 0.51\nu_1\lambda}}_{f(\beta)} - \frac{8c_\mu\clog^2}{\abs{\mb\tau}}}.  \label{eqn:strong_grad_main_lb2}  
\end{align}  

\vsni 4. (\ul{Lower bound of $f(\beta)$}) Given a fixed $\mb\beta_{(1)}$, the cubic function $f(\mb\beta_{(0)})$ has zeros set $\mb\beta_{(0)}\in\set{\pm 1,1.02\mb\beta_{(1)}}$  and has negative leading coefficient. Combine with the condition of $\mb\beta_{(0)}\in\set{\beta_{\mr{lb}}, \beta_{\mr{ub}}}$ from \eqref{eqn:strong_grad_beta0_ub}-\eqref{eqn:strong_grad_beta0_lb}, we can observe that  
\begin{align}
	\mb\beta_{(0)} \in  \brac{\beta_{\mr{lb}}, \beta_{\mr{ub}}} = \brac{\frac{5}{4}\mb\beta_{(1)},  1-0.49\mb\beta_{(1)}^2 } \subseteq \brac{ 1.02\mb\beta_{(1)}, 1},   \notag
\end{align}
therefore the cubic term is always positive and minimizer is either one of the boundary point. When $\mb\beta_{(0)} = \beta_{\mr{lb}}$,  use $\paren{1+\frac{25}{16}}\mb\beta_{(1)}^2 < 1.01$, and use $\nu_1\lambda< \frac{\sqrt{\clog}}{2\sqrt{\abs{\mb\tau}}} \leq \frac{1}{2\sqrt 2}$, since $\abs{\mb\tau}\geq 2$,  we have:
\begin{align}
	f(\beta_{\mr{lb}}) &\geq  \paren{1- \beta_{\mr{lb}}^2}\paren{\frac{\nu_1\lambda}{2\mb\beta_{(1)}}\beta_{\mr{lb}} -0.51\nu_1\lambda} \geq\paren{1-0.616}\cdot\paren{\frac58-0.51}\nu_1\lambda \geq \frac{1}{16\sqrt 2} \nu_1\lambda \geq \frac{\clog^2}{32}\lambda^2.    \label{eqn:strong_grad_lb_at_betaub_1}
\end{align}
On the other hand when $\mb\beta_{(0)} = \beta_{\mr{ub}}$:
\begin{align}
	f(\beta_{\mr{ub}}) &\geq \paren{1-\beta_{\mr{ub}}^2}\paren{\frac{\nu_1\lambda}{2\mb\beta_{(1)}}\beta_{\mr{ub}}- 0.51\nu_1\lambda } \geq 0.49\mb\beta_{(1)}^2\cdot \paren{\frac{\nu_1\lambda}{2\mb\beta_{(1)}}\paren{1-0.49\mb\beta_{(1)}^2 } - 0.51\nu_1\lambda }, \notag
\end{align}
which is a cubic function of $\mb\beta_{(1)}$ with negative leading coefficient, whose zeros set is $\set{-0.73,0,2.81}$. Thus it minimizes at the boundary points of $\mb\beta_{(1)} \in\brac{ \frac{\lambda}{4\log\theta^{-1}},1}\subset\brac{0,2.81}$, thus assign $\mb\beta_{(1)} = \frac{\lambda}{4\log\theta^{-1}}$, we have:    
\begin{align}
	f(\beta_{\mr{ub}}) &\geq  0.49\paren{\frac{\lambda}{4\log\theta^{-1}}}^2\cdot \paren{\frac{1}{2}\paren{1-0.49\paren{\frac{\lambda}{4\log\theta^{-1}}}^2} - 0.51\nu_1\lambda} \geq \frac{1}{6}\paren{\frac{\lambda}{4\log\theta^{-1}}}^2\geq \frac{\clog^2}{96}\lambda^2. \label{eqn:strong_grad_lb_at_betaub}  
\end{align} 
Finally combine \eqref{eqn:strong_grad_main_lb2} with the lower bound of cubic function \eqref{eqn:strong_grad_lb_at_betaub_1}-\eqref{eqn:strong_grad_lb_at_betaub} together with condition $c_\mu < \frac{c_\lambda^2}{800}$ and $\nu_1 = \frac{\sqrt{\clog}}{2}$, obtain
\begin{align}
	 \innerprod{\ip^*\shift{\mb a_0}{(0)}}{-\mr{grad}_{\varphi_{\ell_1}}[\mb a]}  &\geq n\theta \cdot \paren{\min\set{f(\beta_{\mr{ub}}), f(\beta_{\mr{lb}}) } - \frac{8c_\mu\clog^2}{\abs{\mb\tau}}} \notag \\
	 &\geq  n\theta\paren{\frac{\clog^2 c_\lambda^2}{96\abs{\mb\tau}} - \frac{8\clog^2 c_\lambda^2}{800\abs{\mb\tau}} } \geq 6\cross10^{-3} n\theta\clog^2 c_\lambda^2. 
\end{align}    
The proof for the case where $\mb\beta_{(0)}$ negative can be derived in the same manner. 
\end{proof}

 \vsni As a consequence, we have that 
\begin{corollary}[Large gradient for $\varphi_{\rho}$]\label{cor:strong_grad} Suppose that $\mb x_0\simiid  \mr{BG}(\theta)$ in $\R^n$, and $k, c_\mu$ such that $(\mb a_0,\theta,k)$ satisfies the sparsity-coherence condition $\mr{SCC}(c_\mu)$. Define $\lambda = c_\lambda/\sqrt{k}$ in $\varphi_{\rho}$ with $c_\lambda\in\brac{\frac15,\frac14}$, then there exists some numerical constants $C,c,c',c'',\ol c > 0$ such that if $\rho$ is $\delta$-smoothed $\ell^1$ function where $\delta \leq c''\lambda\theta^8/p^2\log^2n $ with $n>C p^5\theta^{-2}\log p$ and $c_\mu\leq \ol c$, then with probability at least $1-c'/n$, for every $\mb a\in\cup_{\abs{\mb\tau}\leq k}\goodregion$ satisfying $\tfrac45\abs{\mb\beta_{(0)}}>\abs{\mb\beta_{(1)}} > \frac{1}{4\log\theta^{-1}\lambda} $, 
\begin{align}
	\innerprod{\mb\sigma_{(0)} \ip^*\shift{\mb a_0}{(0)}}{-\mr{grad}[\varphi_{\rho}](\mb a)} \geq cn\theta\paren{\log^{-2}\theta^{-1}}\lambda^2 
\end{align}
where $\mb\sigma_i = \sign(\mb\beta_i)$.  
\end{corollary}
\begin{proof}
	Choose $\ip^*\shift{\mb a_0}{(0)}$ as in \Cref{lem:strong_grad}, and apply \eqref{eqn:smooth_approx_grad} from \Cref{lem:approx_rho_ell1} with the constant multiplier of $\delta$ satisfies $c''^4<c/4$, then utilize $\theta\abs{\mb\tau}\log^2\theta^{-1} < c_\mu$ from \Cref{asm:theta_mu} we have
	\begin{align}
		\innerprod{\mb\sigma_{(0)} \ip^*\shift{\mb a_0}{(0)}}{-\mr{grad}[\varphi_{\rho}](\mb a)} &\geq  cn\theta(\log^{-2}\theta^{-1})\lambda - c''n\theta^2\geq cn\theta(\log^{-2}\theta^{-1})
		\lambda/2 
	\end{align}
\end{proof}

% ========================== %
\vspace{-0.2in}

\subsection{Convex near solutions}
For any $\mb a\in\Sp^{p-1}$ near subspace and the second largest correlation $\mb\beta_{(1)}$ smaller then $\frac{1}{4\log\theta^{-1}}\lambda$, then $\varphi_{\rho}$ will be strongly convex at $\mb a$. We show this in \Cref{lem:strong_convex}, and the $ \varphi_\rho$ version in \Cref{cor:strong_convex} when $\rho$ is properly defined as in \Cref{sec:smooth_approx}. 

  \begin{lemma}[Strong convexity of $\varphi_{\ell^1}$ near shift]\label{lem:strong_convex}  Suppose that  $\mb x_0\simiid  \mr{BG}(\theta)$ in $\R^n$, and $k, c_\mu$ such that $(\mb a_0,\theta,k)$ satisfies the sparsity-coherence condition $\mr{SCC}(c_\mu)$. Define $\lambda = c_\lambda/\sqrt{k}$ in $\varphi_{\ell^1}$  with $c_\lambda\in\brac{\frac14,\frac15}$, then there exists some numerical constants $C,c,c'\ol c>0$ such that if $n>C p^5\theta^{-2}\log p$ and $c_\mu\leq \ol c$, then with probability at least $1-c'/n$, for every $\mb a\in\cup_{\abs{\mb\tau}\leq k}\goodregion$ satisfying $\abs{\mb\beta_{(1)}} < \frac{1}{4\log\theta^{-1}}\lambda$: for all $\mb v \in\Sp^{p-1}\cap \mb v^\perp$, 
\begin{align}\label{eqn:strong_convex_ell1}
	\mb v^*\wt{\mr{Hess}}[\varphi_{\ell^1}](\mb a)\mb v > cn\theta;
\end{align}
furthermore, there exists $\bar{\mb a}$ as an local minimizer such that
\begin{align}\label{eqn:near_shift_ell1}
	\min_\ell\norm{\bar{\mb a} - \shift{\mb a_0}{\ell}}2 \leq \tfrac12\max\set{\mu, p^{-1} }. 
\end{align}
\end{lemma}
\begin{proof} 1. (\ul{Expectation of $\mb\chi$ near shifts}) We will write $\mb x$ as $\mb x_0$ through out this proof.  When $\mb a$ is near one of the shift, the $\mb\chi$ operator shrinks all other smaller entries of correlation vector $\mb\beta_{\setminus(0)}$ in an even larger shrinking ratio. Firstly we can show $\abs{\innerprod{\mb\beta_{\setminus{(0)}}}{\mb x_{\setminus(0)}}}$ is no larger then $\lambda/2$ with probability at least $1-4\theta$, since
\begin{align}\label{eqn:convex_beta_x0_tail}
\prob{\abs{\innerprod{\mb\beta_{\setminus(0)}}{\mb x_{\setminus(0)}} } >\frac{\lambda}2} \leq    \prob{\abs{\innerprod{\mb\beta_{\mb\tau\setminus(0)}}{\mb x_{\mb\tau\setminus(0)}} } >\frac{2\lambda}5} + \prob{\abs{\innerprod{\mb\beta_{\mb\tau^c}}{\mb x_{\mb\tau^c}} } >\frac{\lambda}{10}} \leq 4\theta
\end{align}
via \Cref{cor:tail_beta_x0_tauc} and \Cref{cor:tail_beta_x0_tau}. Now recall from \Cref{lem:chibeta_ublb} and the derivation of \eqref{eqn:exp_chi_is_erf}-\eqref{eqn:chi_ct_exp_ub1_1}, we know for every $i\neq (0)$,
\begin{align}\label{eqn:convex_expect_chi}
	\mb\sigma_i\E\mb\chi[\mb\beta]_i &= n\theta\abs{\mb\beta_i}\E_{\mb s_i}\brac{ 1-\mr{erf}_{\mb\beta_i}\paren{\lambda,\mb s_i} } \notag \\
	&\leq n\theta\abs{\mb\beta_{i}}\E_{g,\mb x_{\setminus i}} \brac{g^2\1_{\set{\abs{\mb\beta_{i}g + \mb\beta_{(0)}\mb x_{(0)} + \mb\beta_{\setminus\set{(0),i}}^*\mb x_{\setminus\set{(0),i}} } > \lambda }} } \notag \\ 
	&\leq n\theta\abs{\mb\beta_i}\paren{\E g^2\1_{\set{\abs{\mb\beta_i g}>\lambda/2}} + \prob{\mb x_{(0)} \neq 0} + \prob{\abs{\innerprod{\mb\beta_{\setminus\set{(0),i}}}{\mb x_{\setminus\set{(0),i}}} } >\lambda/2 }} \notag \\
	&\leq n\theta\abs{\mb\beta_{i}}\paren{\paren{\E g^2}^{1/2}\prob{\abs{\mb\beta_{(1)} g} > \lambda / 2}^{1/2} + \theta + 4\theta }\notag \\ 
	&\leq n\theta\abs{\mb\beta_i}\paren{\exp\paren{-\log^2\theta^{-1}} + 5\theta} \\ \notag
	&\leq 6n\theta^2\abs{\mb\beta_i}   
\end{align}
where the third inequality is derived using union bound; the the fourth inequality is the result of \eqref{eqn:convex_beta_x0_tail}, and the fifth inequality is derived from Gaussian tail bound \cref{lem:gaussian_tail_bound}.  

\vsni 2. (\ul{Local strong convexity})
 Let $\mb\gamma = \convmtx{\mb a_0}^*\ip\mb v$, for any $\norm{\mb v}2 = 1$ we have $\norm{\mb\gamma}2^2 \leq 1+\mu p$. Furthermore: 
\begin{align}
	\abs{\mb\gamma_{(0)}} &=  \abs{\innerprod{\ip^*\shift{\mb a_0}{(0)}}{\mb v}} = \abs{\innerprod{\mb P_{\mb a^\perp}\ip^*\shift{\mb a_0}{(0)} }{\mb v}} = \abs{\innerprod{\ip^*\shift{\mb a_0}{(0)} - \mb\beta_{(0)}\mb a }{\mb v}} \notag\\
	&\leq  \norm{\ip^*\shift{\mb a_0}{(0)}-\mb\beta_{(0)}\mb a }2 \leq  \sqrt{1 - \mb\beta_{(0)}^2}.
\end{align}
Consider any such $\mb v$, the pseudo Hessian can be lower bounded as
\begin{align}\label{eqn:convex_hess_lb1}
	\mb v^*\wt{\nabla}^2\varphi_{\ell^1}(\mb a)\mb v &= -\mb\gamma^* \checkmtx{\mb x}\mb P_{I(\mb a) }\checkmtx{\mb x}\mb \gamma \notag \\
	&\geq - \mb\gamma_{(0)}^2\norm{\mb P_{I(\mb a)}\checkmtx{\mb x}\mb e_{(0)} }2^2 - \sum_{i\neq(0)}\norm{\mb P_{I(\mb a)}\checkmtx{\mb x}\mb  e_i}2^2\mb\gamma_i^2 - 2\sum_{i\neq j}\abs{\mb e_i^*\checkmtx{\mb x}\mb P_{I(\mb a)}\checkmtx{\mb x}\mb e_j}\abs{\mb\gamma_i}\abs{\mb\gamma_j}  \notag \\
	&\geq -\paren{1-\mb\beta_{(0)}^2}\norm{\mb x}2^2 - \max_{i\neq(0)}\norm{\mb P_{I(\mb a)}\shift{\mb x}{-i}}2^2\norm{\mb\gamma}2^2 - 2\max_{i\neq j}\abs{\mb e_i^*\checkmtx{\mb x}\mb P_{I(\mb a)}\checkmtx{\mb x}\mb e_j}\norm{\mb\gamma}1^2,
\end{align}  
where the second term is bounded by using  its expectation derived in \Cref{lem:expect_support}, and utilize $\prob{\abs{\mb s_i} > \lambda/2 } < 4\theta$ from \eqref{eqn:convex_beta_x0_tail},  $\E\mb\chi$ from \eqref{eqn:convex_expect_chi} and regional condition $\abs{\mb\beta_{(1)}}\leq \frac{\lambda}{4\log\theta^{-1}}$ to acquire
\begin{align}  
	\E \norm{\mb P_{I(\mb a)}\shift{\mb x}{-i}}2^2 &= n\theta\brac{1-\E_{\mb s_i}\mr{erf}_{\mb\beta_i}\paren{\lambda,\mb s_i} + \E_{\mb s_i}f_{\mb\beta_i}\paren{\lambda,\mb s_i}  } \notag \\
	&\leq  \frac{\abs{\E\mb\chi[\mb\beta]_i}}{\abs{\mb\beta_i}} + n\theta\cdot\paren{ \max_{\abs{\mb s_i}\leq \frac{\lambda}{2}}f_{\mb\beta_i}(\lambda,\mb s_i) + \prob{\abs{\mb s_i} >\frac{\lambda}{2}} } \notag \\ 
	&\leq 6n\theta^2 + \frac{2n\theta}{\sqrt{2\pi}}\max_{\abs{\mb s_i}\leq \frac\lambda2}\paren{ \frac{\lambda + \abs{\mb s_i}}{\abs{\mb\beta_i}} \cdot\exp\brac{-\frac{(\lambda-\abs{\mb s_i})^2}{2\mb\beta_i^2}}} + 4n\theta^2 \notag \\
	&\leq 10n\theta^2 + n\theta\cdot\log\theta^{-1}\exp\paren{-2\log^2\theta^{-1}} \notag \\
	&\leq 11n\theta^2,
\end{align}
and define the events $\event_{\norm{\mb x}2}$, $\event_{\mr{cross}}$ and $\event_{\mr{pcurv}}$ as follows: 
\begin{align}
	\begin{cases}  
		\event_{\mr{pcurv}} := \set{\forall\,\mb a\in\cup_{\abs{\mb\tau}\leq k}\goodregion,\,\norm{\mb P_{I(\mb a)}\shift{\mb x}{-i}}2^2 \leq 11n\theta^2   + \frac{c_\mu n\theta}{p} }\\
		\event_{\mr{cross}} := \set{\forall\,\mb a\in\cup_{\abs{\mb\tau}\leq k}\goodregion,\abs{\mb\beta_{(1)}}\leq \frac{\lambda}{4\log\theta^{-1}},\,\max_{i\neq j\in[\pm p]} \abs{\mb e_i^*\checkmtx{\mb x}\mb P_{I(\mb a)}\checkmtx{\mb x}\mb e_j} \leq 8n\theta^3 } \\  
		\event_{\norm{\mb x}2} := \set{\norm{\mb x}2^2 \leq n\theta + 3\sqrt{n\theta}\log n } 
	\end{cases}.
\end{align}
For the Hessian term, on the event $\event_{\mr{pcurv}}  \cap\event_{\mr{cross}}\cap\event_{\norm{\mb x}2}$, and use all $\mu p^2\theta^2$, $\mu p\theta\abs{\mb\tau}$ and  $\theta\sqrt p$ are all less then $\frac{c_\mu}{4\log^2\theta^{-1}}$,  from  \Cref{lem:alpha_beta_tau_lb}, and from lemma assumption with sufficiently large $C$ we have $n > \theta^{-1}36\log^2 n $, thus $\mb v^*\wt{\nabla}^2\varphi_{\ell^1}(\mb a)\mb v$ can be lower bounded from \eqref{eqn:convex_hess_lb1} as
\begin{align}
	\mb v^*\wt{\nabla}^2\varphi_{\ell^1}(\mb a)\mb v &\geq -\paren{1-\mb\beta_{(0)}^2}\paren{n\theta + 3\sqrt{n\theta}\log n} - (1+\mu p)\paren{11n\theta^2 + \frac{c_\mu n\theta}{p}}- 8p\paren{1+\mu p}\cdot 8n\theta^3 \notag \\
	&\geq -\frac{1}{2}n\theta\cdot(1-\mb\beta_{(0)}^2) -n\theta\cdot\paren{\frac{11c_\mu}{4} + c_\mu^2 + \frac{64c_\mu}{4} + \frac{64c_\mu}{4}}\notag \\
	&\geq -\frac12n\theta\cdot \paren{1-\mb\beta_{(0)}^2  + 20c_\mu }.
\end{align}
The bounds of $\mb\beta^*\mb\chi[\mb\beta]$ can be derive on the event whose expectation is drawn from \Cref{lem:chibeta_ublb} and \eqref{eqn:convex_expect_chi} as
\begin{align} 
	\event_{\chi} := \set{\begin{cases}  \mb\sigma_i\mb\chi[\mb\beta]_i 
		 \geq    n\theta \soft{\abs{\mb\beta_i}}{\nu_2\lambda} - \frac{c_\mu n\theta}{p},&\quad  \forall\,i\in[\pm p] \notag \\
		\mb\sigma_i\mb\chi[\mb\beta]_i  \leq 6n\theta^2\abs{\mb\beta_i} + \frac{c_\mu n\theta}{p^{3/2}},&\quad \forall\,i\neq(0)
	\end{cases}  },
 \end{align}
 then use $\norm{\mb\beta}1 \leq 1+\frac{\lambda p}{4\log\theta^{-1}} \leq \frac{\lambda p }{2}$, implies:  
\begin{align}
	\mb\beta^*\mb\chi[\mb\beta] &\geq n\theta\abs{\mb\beta_{(0)}}\paren{\abs{\mb\beta_{(0)}} -\nu_2\lambda} - c_\mu\norm{\mb\beta}1\tfrac{n\theta}{p} \notag \\
	&\geq n\theta\paren{\mb\beta_{(0)}^2  -\sqrt{\tfrac2\pi}\lambda - \tfrac{c_\mu  }{2}\lambda } &\notag \\
	&\geq n\theta\paren{\mb\beta_{(0)}^2 - \lambda}. \label{eqn:sc_betachibeta}
\end{align}
Finally via the regional condition $\abs{\mb\beta_{(1)}}\leq \frac{\lambda}{4\log\theta^{-1}}$, the absolute value of leading correlation 
\begin{align}\label{eqn:sc_beta0_lb}
	 \mb\beta^2_{(0)} \geq \norm{\mb\beta_{\mb\tau}}2^2  - \abs{\mb\tau}\mb\beta_{(1)}^2 \geq 1-2c_\mu-0.1^2 > 0.9,
\end{align}
then we collect all above results and obtain:
\begin{align}\label{eqn:convex_smallest_eig}
	\mb v^*\wt{\mr{Hess}}[\varphi_{\ell_1}](\mb a)\mb v &= \mb v^*\wt{\nabla}^2\varphi_{\ell^1}(\mb a)\mb v  - \mb\beta^*\mb\chi[\mb\beta] \geq \paren{1.5\mb\beta_{(0)}^2-0.5 - \lambda - 20c_\mu  }n\theta \geq 0.3n\theta, 
\end{align}  
with probability at least 
\begin{align} 
	1-\underbrace{\prob{\event_{\mr{cross}}^c}}_{\text{\Cref{lem:hess_cross_convex}} } -\underbrace{\prob{\event_{\mr{pcurv}}^c}}_{\text{\Cref{cor:ct_support}}} - \underbrace{\prob{\event_{\norm{\mb x}2}^c}}_{\text{\Cref{lem:x0_bound}}} - \underbrace{\prob{\event_{\chi}^c}}_{\text{\Cref{cor:chibeta_ct}}}\geq 1-c'/n.
\end{align}

\vsni 3. (\ul{Identify local minima}) Wlog let $\mb a_*$ be a local minimum where its gradient is zero that is close to $\mb a_0$. The strong convexity \eqref{eqn:convex_smallest_eig}, provides the upper bound on  $\norm{\mb a_*-\mb a_0}2^2$ via
\begin{align}\label{eqn:grad_bound_diffa}
	& \varphi_{\ell^1}(\mb a_*) \geq \varphi_{\ell^1}(\mb a_0) + \innerprod{\mb a_* - \mb a_0}{\grad[\varphi_{\ell^1}](\mb a_0)} + \tfrac{0.3}{2}n\theta\norm{\mb a_*-\mb a_0}2^2 \notag  \\
	\implies & \norm{\grad[\varphi_{\ell^1}](\mb a_0)}2 \geq 0.15n\theta\norm{\mb a_*-\mb a_0}2
\end{align}
Thus we only require to bound the gradient at $\mb a_0$, whose coefficients $\mb\alpha = \mb e_0$ and correlation $\mb\beta$ has properties $\mb\beta_{0} = 1$ and $\norm{\mb\beta_{\setminus 0}}\infty \leq \mu$ hence $\norm{\mb \beta_{\setminus 0}}\leq \sqrt{2p}\mu$. Expand the gradient term and condition on $\event_\chi$, since $\mu p^2\theta^2\leq \tfrac{c_\mu}4$ and $\theta < \tfrac{c_\mu}{4\sqrt p}$, we can upper bound the gradient at $\mb a_0$ as   
\begin{align}\label{eqn:convex_near_sol}
	\norm{\mr{grad}[\varphi_{\ell^1}](\mb a_0)}2 &= \norm{\ip^*\convmtx{\mb a_0}\paren{\mb\chi\brac{\mb\beta} - \mb\beta^*\mb\chi[\mb\beta]\mb e_0 }}2 \leq \norm{\ip^*\convmtx{\mb a_0}}2\norm{\mb\chi[\mb\beta]_{\setminus 0}}2  \notag \\
	&\leq \sqrt{1+\mu p}\paren{6n\theta^2\norm{\mb\beta_{\setminus 0}}2 + n\theta\cdot \tfrac{c_\mu}{p^{3/2}}\cdot\sqrt{2p} } \notag \\
	&\leq n\theta\sqrt{1+\mu p}\paren{6\mu\sqrt{2p}\cdot \theta  +\tfrac{2c_\mu}{p}} \notag \\
	&\leq n\theta \paren{3c_\mu\mu + 6\mu\cdot\sqrt{2\mu}\cdot p\theta + \tfrac{2c_\mu}{ p} + \tfrac{2c_\mu\sqrt\mu}{\sqrt p}} \notag \\
	&\leq   7\sqrt{c_\mu}n\theta\cdot \max\Brac{\mu,\tfrac{1}{ p}}.
\end{align}
Thus we conclude that with sufficiently small $c_\mu$:
\begin{align}
	\norm{\mb a_*-\mb a_0}2 \leq 50\sqrt{c_\mu}\max\Brac{\mu, p^{-1}} \leq \tfrac{1}{2}\max\set{\mu,p^{-1}}.  
\end{align}
and we complete the proof by generalize this result from minima near $\mb a_0$ to any of its shifts $\shift{\mb a_0}{i}$.
 
\end{proof}

\vsni Similarly, for objective $\varphi_\rho$ we have
\begin{corollary}[Strong convexity of $\varphi_\rho$ of near shift]\label{cor:strong_convex} Suppose that $\mb x_0\simiid  \mr{BG}(\theta)$ in $\R^n$, and $k, c_\mu$ such that $(\mb a_0,\theta,k)$ satisfies the sparsity-coherence condition $\mr{SCC}(c_\mu)$. Define $\lambda = c_\lambda/\sqrt{k}$ in $\varphi_{\rho}$  with $c_\lambda\in\brac{\frac15, \frac14}$, then there exists some numerical constant $C,c,c',c'',\ol c>0$ such that if $\rho$ is $\delta$-smoothed $\ell^1$ function where $\delta \leq c'\lambda\theta^8/p^2\log^2n $ and  $n>C p^5\theta^{-2}\log p$ and $c_\mu\leq \ol c$, then  with probability at least $1-c''/n$, for every $\mb a\in\cup_{\abs{\mb\tau} \leq k}\goodregion$ satisfying $\abs{\mb\beta_{(1)}} < \nu_1\lambda$: for all $\mb v \in\Sp^{p-1}\cap\mb a^\perp$, 
\begin{align}\label{eqn:strong_convex_rho}
	\mb v^*\wt{\mr{Hess}}[\varphi_{\rho}](\mb a)\mb v > cn\theta;
\end{align}
furthermore, there exists $\bar{\mb a}$ as an local minimizer such that
\begin{align}\label{eqn:near_shift_rho}
	\min_{\ell}\norm{\bar{\mb a} - \shift{\mb a_0}{\ell}}2 \leq \tfrac12\max\set{\mu, p^{-1}}
\end{align}  
\end{corollary}
\begin{proof}
The strong convexity \eqref{eqn:strong_convex_rho} is derived by combining \eqref{eqn:strong_convex_ell1} and \eqref{eqn:smooth_approx_hess} by letting constant multiplier of $\delta$ satisfies $c'^{1/4} < 10^{-3}c$. On the other hand the local minimizer near solution \eqref{eqn:near_shift_rho} is derived via combining \eqref{eqn:grad_bound_diffa}, \eqref{eqn:smooth_approx_alpha} and utilize both $\theta\sqrt p<c_\mu$ and $\mu p^2\theta^2 < c_\mu $ such that: 
\begin{align}
	\norm{\mr{grad}[\varphi_{\rho}](\mb a)}2 &\leq \norm{\injector^*\convmtx{\mb a_0}}2\norm{\mb\chi[\mb\beta] - \checkmtx{\mb x_0}\mc S_{\lambda}^{\delta}\brac{\checkmtx{\mb y}\ip\mb a}}2  + \norm{\injector^*\convmtx{\mb a_0}}2\norm{\mb\chi[\mb\beta]_{\setminus0}}2\notag \\
	&\leq \sqrt{1+\mu p}\cdot n\theta^3 + 7\sqrt{c_\mu}n\theta\cdot\max\set{\mu, p^{-1}}\notag \\
	&\leq 8n\theta\sqrt{c_\mu}\cdot\max\set{\mu, p^{-1}}
\end{align}   
\end{proof}

%========================================%

\subsection{Retraction toward subspace} \label{sec:proof_retract}
As in \Cref{fig:side_view}, the function value grows in direction away from subspace $\mc S_{\mb\tau}$, we will illustrate this phenomenon by proving the negative gradient direction $-\mb g$ will point toward the subspace $\mc S_{\mb\tau}$. To show this, we prove for every coefficients of $\mb a$ as $\mb\alpha$, there exists coefficients of $\mb g$ as $\mb\zeta$ satisfies
\begin{align}
	\innerprod{\mb\alpha_{\mb\tau^c}(\mb g)} {\mb\alpha_{\mb\tau^c}(\mb a)}>c\norm{\mb\alpha_{\mb\tau^c}}2\norm{\mb\zeta_{\mb\tau^c}}2 
\end{align}  
whenever $d_\alpha(\mb a,\mc S_{\mb\tau})\in\brac{\tfrac\gamma2,\gamma}$. Apparently, the gradient will decrease $d_\alpha(\mb a,\mc S_{\mb\tau})$, hence being addressed as \emph{retractive toward subspace $\mc S_{\mb\tau}$}. This retractive phenomenon is true for gradient of both $\varphi_{\ell^1}$ and $\varphi_\rho$. 

\begin{lemma}[Retraction of $\varphi_{\ell^1}$  toward subspace]\label{lem:retraction_alpha_tauc}  Suppose that $\mb x_0\simiid  \mr{BG}(\theta)$ in $\R^n$, and $k, c_\mu$ such that $(\mb a_0,\theta,k)$ satisfies the sparsity-coherence condition $\mr{SCC}(c_\mu)$.  Define $\lambda = c_\lambda/\sqrt k$ in $\varphi_{\ell^1}$ with $c_\lambda\in\left(0,\tfrac13\right]$, then there exists some numerical constants $C,c,\ol c>0$  such that if $n>C p^5\theta^{-2}\log p$ and $c_\mu\leq \ol c$, then with probability at least $1-c'/n$, for every $\mb a\in\cup_{\abs{\mb\tau}\leq k}\goodregion$ such that if  
\begin{equation}
 d_{\alpha}(\mb a, \mc S_{\mb\tau} ) \geq \gamma(c_\mu)/2
\end{equation}
then for every $\mb\alpha$ satisfying  $\mb a =  \ip^*\convmtx{\mb a_0}\mb\alpha$, there exists some $\mb\zeta$ satisfying $\grad[\varphi_{\ell^1}](\mb a) = \ip^*\convmtx{\mb a_0}\mb\zeta$ that    
\begin{align}\label{eqn:retract_dist2}
	 \innerprod{\mb\zeta_{\mb\tau^c}}{\mb\alpha_{\mb\tau^c}} \geq \tfrac{1}{4n\theta}\norm{\mb\zeta_{\mb\tau^c}}2^2.
\end{align} 
\end{lemma}

\begin{proof} Write $\gamma = \gamma(c_\mu)$ Recall the gradient can be derived as
\begin{align}\label{eqn:retract_grad}
	\mr{grad}[\varphi_{\ell^1}](\mb a) = \mb -\mb P_{\mb a^\perp}\ip^*\convmtx{\mb a_0}\mb\chi[\mb\beta] = \paren{\mb a\mb a^* - \mb I}\ip^*\convmtx{\mb a_0}\mb\chi[\mb\beta] = \ip^*\convmtx{\mb a_0}\paren{\mb\beta^*\mb\chi[\mb\beta]\mb\alpha - \mb\chi[\mb\beta]}, 
\end{align} 
for every $\mb\alpha$ satisfies $\mb a = \ip^*\convmtx{\mb a_0}\mb\alpha$. Now via  \Cref{cor:chibeta_ct}, condition on the event: 
\begin{align}\label{eqn:event_chi_retract}
	\event_{\chi} := \set{ \sigma_i \mb\chi[\mb\beta]_i \leq \begin{cases} n\theta\cdot \abs{\mb\beta_i} +  \tfrac{c_\mu n\theta}{p}, &\quad \forall\,i\in\mb\tau \\ n\theta\cdot \abs{\mb\beta_i} 4\theta\abs{\mb\tau} + \tfrac{c_\mu n\theta}{p}, &\quad \forall\,i\in\mb\tau^c \end{cases},\quad \sigma_i\mb\chi[\mb\beta]_i \geq n\theta\cdot\soft{\abs{\mb\beta_i}}{\sqrt{2/\pi}\lambda}},   
\end{align}
and on this event, utilize \Cref{lem:alpha_beta_tau_lb},  bounds of $\mb\beta^*\mb\chi[\mb\beta]$ and $\norm{\mb\chi[\mb\beta]_{\mb\tau^c}}2$ can be derived with $ c_\mu<\frac{1}{100} $ as:
\begin{align}  
		&\mb\beta^*\mb\chi[\mb\beta]  \leq n\theta\paren{\norm{\mb\beta_{\mb\tau}}2^2 + 4\theta\abs{\mb\tau}\norm{\mb\beta_{\mb\tau^c}}2^2 + c_\mu   } \geq  n\theta\paren{1+c_\mu + 4c_\mu^2 + c_\mu } \leq \tfrac32n\theta  \\
		&\mb\beta^*\mb\chi[\mb\beta]  \geq n\theta\paren{\norm{\mb\beta_{\mb\tau}}2^2 - \sqrt{2/\pi}\lambda\norm{\mb\beta_{\mb\tau}}1  - c_\mu   } \geq  n\theta\paren{1-4c_\mu - \sqrt{2/\pi}c_\lambda - c_\mu } \geq \tfrac12n\theta  \\
		&\norm{\mb\chi[\mb\beta]_{\mb\tau^c}}2 \leq 4n\theta^2\abs{\mb\tau}\norm{\mb\beta_{\mb\tau^c}}2 + \tfrac{c_\mu n\theta}{p}\sqrt p\leq n\theta\paren{4c_\mu\gamma + c_\mu\gamma} \leq  \tfrac{1}{20} n\theta\gamma.  
\label{eqn:retract_betachibeta_lb}
\end{align}  
Let $\mb \alpha(\mb g) = \mb\beta^*\mb\chi[\mb\beta]\mb\alpha - \mb\chi[\mb\beta]$, derive 
\begin{align}
	&\innerprod{\mb \alpha(\mb g)_{\mb\tau^c}}{\mb\alpha_{\mb\tau^c}} - \tfrac{1}{4n\theta}\norm{\mb \alpha(\mb g)_{\mb\tau^c}}2^2 \notag \\
	&\qquad =\; \mb\beta^*\mb\chi[\mb\beta]\norm{\mb\alpha_{\mb\tau^c}}2^2 - \innerprod{\mb\alpha_{\mb\tau^c}}{\mb\chi[\mb\beta]_{\mb\tau^c}} - \tfrac{1}{4n\theta}\norm{\mb\beta^*\mb\chi[\mb\beta]\mb\alpha_{\mb\tau^c} - \mb\chi[\mb\beta]_{\mb\tau^c}}2^2 \notag \\
	&\qquad\geq\; \mb\beta^*\mb\chi[\mb\beta]\norm{\mb\alpha_{\mb\tau^c}}2^2 - \norm{\mb\alpha_{\mb\tau^c}}2\norm{\mb\chi[\mb\beta]_{\mb\tau^c}}2 - \tfrac{1}{2n\theta}\abs{\mb\beta^*\mb\chi[\mb\beta]}^2\norm{\mb \alpha_{\mb\tau^c}}2^2 - \tfrac{1}{2n\theta}\norm{\mb\chi[\mb\beta]_{\mb\tau^c}}2^2\notag \\
&\qquad\geq \;\paren{\mb\beta^*\mb\chi[\mb\beta] -\tfrac{1}{2n\theta}(\mb\beta^*\mb\chi[\mb\beta])^2 }\norm{\mb\alpha_{\mb\tau^c}}2^2 -\tfrac{1}{20}n\theta\gamma\norm{\mb\alpha_{\mb\tau^c}}2 -\tfrac{1}{1000}n\theta\gamma^2, \label{eqn:retract_lb1}   
\end{align}
notice that this is a quadratic function of $\mb\beta^*\mb\chi[\mb\beta]$ with negative leading coefficient and zeros at $\set{0,2n\theta}$, hence \eqref{eqn:retract_lb1} is minimized when $\mb\beta^*\mb\chi[\mb\beta]=\tfrac{1}{2}n\theta$. Plugging in, 
\begin{align}
	\eqref{eqn:retract_lb1} &\;\geq\;\tfrac38n\theta\norm{\mb\alpha_{\mb\tau^c}}2^2 -\tfrac{1}{20}n\theta\gamma\norm{\mb\alpha_{\mb\tau^c}}2 -\tfrac{1}{1000}n\theta\gamma^2 \label{eqn:retract_lb2}    
\end{align}
then again this is a quadratic function of $\norm{\mb\alpha_{\mb\tau^c}}2$ with positive  leading coefficient and zeros at $\set{0,\tfrac{8}{60}\gamma}$, thus  \eqref{eqn:retract_lb2} is minimized at $\norm{\mb\alpha_{\mb\tau^c}}2 = \tfrac\gamma2$. Plugging in again,
\begin{align}
	\eqref{eqn:retract_lb2} &\;\geq\;\tfrac38n\theta\norm{\mb\alpha_{\mb\tau^c}}2^2 -\tfrac{1}{20}n\theta\gamma\norm{\mb\alpha_{\mb\tau^c}}2 -\tfrac{1}{1000}n\theta\gamma^2 \geq  \paren{\tfrac{3}{32}-\tfrac{1}{80}-\tfrac{1}{1000}}n\theta\gamma^2 >  0\label{eqn:retract_lb3}    
\end{align}
which concludes our proof.
\end{proof}

 \vsni As a consequence, we have that 
\begin{corollary}[Retraction of $\varphi_{\rho}$ toward the subspace]
\label{cor:retraction_alpha_tauc} 
 Suppose that $\mb x_0\simiid  \mr{BG}(\theta)$ in $\R^n$, and $k, c_\mu$ such that $(\mb a_0,\theta,k)$ satisfies the sparsity-coherence condition $\mr{SCC}(c_\mu)$. Define $\lambda = c_\lambda/\sqrt{\abs{k}}$ in $\varphi_{\rho}$ with $c_\lambda\in\left(0,\frac13\right]$, then there exists some numerical constants $C,c,c',c'',\ol c > 0 $ such that if $\rho$ is $\delta$-smoothed $\ell^1$ function where $\delta \leq c''\lambda\theta^8/p^2\log^2n $ and $n > Cp^5\theta^{-2}\log p$ and $c_\mu \leq \ol c$, then  with probability at least $1-c'/n$, for every $\mb a\in\cup_{\abs{\mb\tau}\leq k}\goodregion$ such that if  
\begin{equation}
 d_{\alpha}(\mb a, \mc S_{\mb\tau} ) \geq \gamma(c_\mu)/2  
\end{equation}
then for every $\mb\alpha$ satisfying  $\mb a =  \ip^*\convmtx{\mb a_0}\mb\alpha$, there exists some $\mb\zeta$ satisfying $\grad[\varphi_{\rho}](\mb a) = \ip^*\convmtx{\mb a_0}\mb\zeta$ that    
\begin{align}\label{eqn:retract_dist2}
	 \innerprod{\mb\zeta_{\mb\tau^c}}{\mb\alpha_{\mb\tau^c}} \geq \tfrac{1}{6n\theta}\norm{\mb\zeta_{\mb\tau^c}}2^2.
\end{align} 
\end{corollary}
\begin{proof} Write $\gamma = \gamma(c_\mu)$. Define  
\begin{align}
	\mb\chi_{\ell^1}[\mb\beta] = \checkmtx{\mb x_0}\mc S_\lambda\brac{\wc{\mb a}*\mb y}, \qquad \mb\chi_\rho[\mb\beta]  = \checkmtx{\mb x_0}\mc S_{\lambda}^{\delta}\brac{\wc{\mb a}*\mb y},\notag
\end{align}
which, and on event \eqref{eqn:event_chi_retract} and  \Cref{lem:approx_rho_ell1}, we know
\begin{align}\label{eqn:retract_rho_bd1}
	\mb\beta^*\mb\chi_{\ell^1}[\mb\beta] &\leq \tfrac32n\theta,  \\
	\norm{\mb\chi_{\ell^1}[\mb\beta]_{\mb\tau^c}}2& \leq \tfrac{1}{20}n\theta\gamma,  \\ 
	\norm{\mb\chi_{\ell^1}[\mb\beta] -\mb\chi_{\rho}[\mb\beta] }2 &\leq c_1n\theta^4,\label{eqn:retract_rho_bd3}
\end{align} 
for some constant $c_1 > 0$. Now given any $\mb \alpha$ satisfies $\mb a = \ip^*\convmtx{\mb a_0}\mb\alpha$,  the gradient of both objective can be derived as
\begin{align}\label{eqn:retract_grad}
	\mr{grad}[\varphi_{\ell^1}](\mb a) &= \mb -\mb P_{\mb a^\perp}\ip^*\convmtx{\mb a_0}\prox_{\lambda\norm{\cdot}1}[\wc{\mb a}*\mb y] = \paren{\mb a\mb a^* - \mb I}\ip^*\convmtx{\mb a_0}\mb\chi_{\ell^1}[\mb\beta] \notag \\
	&= \ip^*\convmtx{\mb a_0}\paren{\mb\beta^*\mb\chi_{\ell^1}[\mb\beta]\mb\alpha - \mb\chi_{\ell^1}[\mb\beta]},\\
	\mr{grad}[\varphi_\rho](\mb a) &= \mb -\mb P_{\mb a^\perp}\ip^*\convmtx{\mb a_0}\prox_{\lambda\rho}[\wc{\mb a}*\mb y]  = \paren{\mb a\mb a^* - \mb I}\ip^*\convmtx{\mb a_0}\mb\chi_{\rho}[\mb\beta] \notag \\
	&= \ip^*\convmtx{\mb a_0}\paren{\mb\beta^*\mb\chi_{\rho}[\mb\beta]\mb\alpha - \mb\chi_{\rho}[\mb\beta] }.
\end{align}
In the same spirit, define the coefficient of each gradient vector
\begin{align}
	\mb\zeta_{\ell^1} &=  \mb\beta^*\mb\chi_{\ell^1}[\mb\beta]\mb\alpha - \mb\chi_{\ell^1}[\mb\beta], \\
	\mb\zeta_{\rho} &=  \mb\beta^*\mb\chi_{\rho}[\mb\beta]\mb\alpha - \mb\chi_{\rho}[\mb\beta], 
\end{align}
which, by norm inequality from \eqref{eqn:retract_rho_bd1}-\eqref{eqn:retract_rho_bd3} and \Cref{lem:retraction_alpha_tauc} , we can derive
\begin{align}
	\norm{\mb\zeta_{\ell^1} - \mb\zeta_{\rho}}2 &\leq \norm{(\mb I-\mb\alpha\mb\beta^*)\paren{\mb\chi_\rho[\mb\beta] -\mb\chi_{\ell^1}[\mb\beta] }}2 \leq c_1n\theta^4,  \\   
	\norm{(\mb\zeta_{\ell^1})_{\mb\tau^c}}2 &\geq  \abs{\mb\beta^*\mb\chi_{\ell^1}[\mb\beta]}\norm{\mb\alpha_{\mb\tau^c}}2 - \norm{\mb\chi_{\ell^1}[\mb\beta]_{\mb\tau^c}}2 \geq \tfrac15n\theta\gamma,\\   
	 \innerprod{(\mb\zeta_{\ell^1})_{\mb\tau^c}}{\mb\alpha_{\mb\tau^c}} &\geq \tfrac{1}{4n\theta}\norm{(\mb\zeta_{\ell^1})_{\mb\tau^c}}2^2, 
\end{align}
where the first inequality is derived by observing $(\mb I-\mb\alpha\mb\beta^*)$ is a projection operator, as such:
\begin{align}
	\mb\beta^*\mb\alpha &= \mb a^*\ip^*\convmtx{\mb a_0}\mb \alpha = \mb a^*\mb a = 1, \notag  \\	
	(\mb I - \mb\alpha\mb\beta^*)^2 &= \mb I - 2\mb\alpha\mb\beta^* +  \mb\alpha(\mb\beta^*\mb\alpha)\mb\beta^*  = \mb I - \mb\alpha\mb\beta^*\notag.
\end{align}
Now we are ready to derive \eqref{eqn:retract_dist2}: 
\begin{align}
	\innerprod{(\mb\zeta_{\rho})_{\mb\tau^c}}{\mb\alpha_{\mb\tau^c}} &\geq \innerprod{(\mb\zeta_{\ell^1})_{\mb\tau^c}}{\mb\alpha_{\mb\tau^c}} - \norm{\mb\alpha_{\mb\tau^c}}2\norm{\mb\zeta_{\rho}-\mb\zeta_{\ell^1}}2 \notag\\
	&\geq \tfrac{1}{4n\theta}\norm{(\mb\zeta_{\ell^1})_{\mb\tau^c}}2^2 - c_1n\theta^4\gamma \notag \\
	&\geq \tfrac{1}{12 n\theta}\norm{(\mb\zeta_{\ell^1})_{\mb\tau^c}}2^2  \notag \\
	&\qquad +\tfrac{1}{6n\theta}\paren{\norm{(\mb\zeta_{\rho})_{\mb\tau^c}}2^2 - 2\norm{(\mb\zeta_{\ell^1})_{\mb\tau^c}}2\norm{\mb\zeta_{\ell^1} - \mb\zeta_{\rho}}2  - \norm{\mb\zeta_{\ell^1} - \mb\zeta_{\rho} }2^2 } - c_1n\theta^4\gamma \notag \\
	&\geq  \tfrac{1}{6n\theta}\norm{(\mb\zeta_\rho)_{\mb\tau^c}}2^2 +  \tfrac{1}{12 n\theta}\paren{\tfrac15n\theta\gamma}^2 - \tfrac{1}{3n\theta}\paren{\tfrac{1}{5}n\theta\gamma}\paren{c_1n\theta^4} - \tfrac{1}{6n\theta}\paren{c_1n\theta^4}^2 - c_1n\theta^4\gamma \notag \\
	&\geq  \tfrac{1}{6n\theta} \norm{(\mb\zeta_\rho)_{\mb\tau^c}}2^2.
\end{align}
where the last inequality is true since $\theta^3\ll \gamma$.
\end{proof}

%====================================%

\subsection{Proof of \Cref{thm:three_regions}}\label{sec:proof_three_region}
By collecting result from above, we are ready to prove the acclaimed geometric result in \Cref{thm:three_regions}. It guarantees that for every $\mb a$ near $\mc S_{\mb\tau}$, either one of the following in true
\begin{align}
	\lambda_{\mr{min}}\paren{\mr{Hess}[\varphi_{\rho}](\mb a)} &\leq -c_1n\theta\lambda,\\
	  \innerprod{\mb\sigma_{(0)}\ip^*\shift{\mb a_0}{(0)}}{-\mr{grad}[\varphi_{\rho}](\mb a)} &\geq c_2n\theta\paren{\log^{-2}\theta^{-1}}\lambda^2,\\
	  \mr{Hess}[\varphi_{\rho}](\mb a) &\succ  c_3n\theta\cdot\mb P_{\mb a^\perp},
\end{align}
all local minimizer $\bar{\mb a}$ satisfies for some $\mb a_* \in\set{\pm\ip^*\shift{\mb a}{\ell}\,\big|\,\ell\in[\pm p_0]}$, 
		\begin{align}
			\norm{\bar{\mb a} - \mb a_*}2 \leq c_4 \sqrt{c_\mu} \max\set{\mu, p_0^{-1}},
		\end{align}
and whenever $ \tfrac\gamma2\leq d_\alpha\paren{\mb a,\mc S_{\mb\tau}} \leq \gamma$, coefficient of $\mb a$ and its gradient $\mb g$, $\mb\alpha$, written as $\mb\zeta$, satisfies 
\begin{align}
 	\innerprod{\mb\zeta_{\mb\tau^c}}{\mb\alpha_{\mb\tau^c}} \geq \tfrac{c_5}{n\theta}\norm{\mb\zeta_{\mb\tau^c}}2^2.
\end{align}
To connect the geometric results introduced in  \Cref{lem:neg_curve}, \Cref{lem:strong_grad},  \Cref{lem:strong_convex} and  \Cref{lem:retraction_alpha_tauc}, we are only required to prove the required signal condition claimed in \Cref{thm:three_regions} is necessary from \Cref{asm:theta_mu}. In particular, when the subspace dimension $\abs{\mb\tau}\leq 4p_0\theta$. On top of that, we are also required to show the chosen smooth parameter $\delta$ in the pseudo-Huber penalty $\rho(x) = \sqrt{x^2 + \delta^2}$ approximate $\abs{x}$ sufficiently well, hence results of \Cref{cor:neg_curve}, \Cref{cor:strong_grad},  \Cref{cor:strong_convex} and \Cref{cor:retraction_alpha_tauc} also holds.  

\vspace{.1in}

\begin{proof}
	Firstly we will show when  largest solution subspace dimension $k = 4p_0\theta$, the signal condition of \Cref{asm:theta_mu} will be satisfied. Recall that the signal condition of  \Cref{thm:three_regions} requests
\begin{align}
	\frac{2}{p_0\log^2p_0} \leq \theta \leq \frac{c}{\paren{p_0\sqrt \mu + \sqrt{p_0}}\log^2 p_0},
\end{align}
since $p = 3p_0-2$, this implies the lower bounds for sparsity $\theta$ as:
\begin{align}\label{eqn:mian_geo_asm1}
	\theta \geq \frac{1}{2p_0\paren{\frac{1}{2}\log p_0}^2} \geq  \frac{1}{p\log^2\theta^{-1}};
\end{align}
the upper bound of $\theta$ via $\theta \sqrt{p_0}\log^2p_0 \leq c $ : 
\begin{align}\label{eqn:mian_geo_asm2}
  \theta \leq \frac{9c}{\sqrt{p_0}(3\log p_0)^2} \leq \frac{16c}{\sqrt{p}\log^2\theta^{-1}},\qquad \theta \leq \frac{4c^2}{k\log^4 p_0} \leq \frac{36c^2}{k(3\log p_0)^2} \leq \frac{36c^2}{k\log^2\theta^{-1}} ;
\end{align} 
and the upper bound for coherence  $\mu$ as:
\begin{align}\label{eqn:mian_geo_asm3}
	\mu\max\set{k^2,(p\theta)^2}\log^2\theta^{-1} \leq \mu \max\set{16(p_0\theta)^2 ,9(p_0\theta)^2}\log^2\theta^{-1} \leq 16\paren{\sqrt{\mu}  p_0 \theta}^2 \log^2p_0 \leq 16c.
\end{align}
Therefore \Cref{asm:theta_mu} holds if $\max\set{16c,36c^2} \leq c_\mu / 4$ via \eqref{eqn:mian_geo_asm1}-\eqref{eqn:mian_geo_asm3}.

 Furthermore, we know from lemma assumption all interested $\mb a$ are near subspace $\mc S_{\mb\tau}$ by 
 \begin{align}
 	 d_\alpha(\mb a.\mc S_{\mb\tau}) \leq \frac{c}{\sqrt{p_0}\log^2 \theta^{-1}}\cdot\min\set{\frac{1}{\sqrt\theta},\frac{1}{\sqrt\mu}.\frac{1}{\mu\paren{p_0\theta}^{3/2}}} \leq \frac{c}{\log^2\theta^{-1}}\min\set{\frac{2}{\sqrt k},\frac{1}{\sqrt{p_0\mu}},\frac{4}{\mu p_0\sqrt\theta k}}\leq \gamma
 \end{align}
where $\gamma$ is defined in \Cref{def:gamma} of widened subspace $\goodregion$. 

Lastly, the pseudo-Huber function $\rho(x) = \sqrt{x^2 +\delta^2}$ is an $\ell^1$ smoothed sparse surrogate defined in \Cref{def:smooth_ell1},  by observing that it is convex, smooth, even, whose second order derivative (according to \Cref{tbl:rho_class}) $\nabla^2\rho(x) = \frac{\delta^2}{\paren{x^2 + \delta^2}^{3/2}}$ is monotone decreasing in $\abs x$. More importantly
\begin{align}
	\sup_{x\in\R} \abs{\rho(x)-\abs x} = \abs{\rho(0)- \abs{0}} = \delta.
\end{align}

Hence, by choosing $ \delta  \leq  \frac{c'^4\theta^8}{p^2\log^2n} \lambda $, for some sufficiently small constant $c'$ and letting $\lambda = 0.2\sqrt{k} = 0.1/\sqrt{p_0\theta}$ in $\varphi_\rho$. We obtain the geometrical results in \Cref{cor:neg_curve} when $\abs{\mb\beta_{(1)}}\geq \frac45\abs{\mb\beta_{(0)}}$, \Cref{cor:strong_grad} when $\frac45\abs{\mb\beta_{(0)}} \geq \abs{\mb\beta_{(1)}} \geq \frac{\lambda}{4\log^2\theta^{-1}}$ and \Cref{cor:strong_convex} when $\frac{\lambda}{4\log^2\theta^{-1}}\geq \abs{\mb\beta_{(1)}}$, and the retraction result in \Cref{cor:retraction_alpha_tauc}.
\end{proof}

% !TEX root = ../../BD_DQ.tex
\newcommand{\iJ}{\injector_J}
\newcommand{\iJk}{\injector_{J^{(k)}}}

\section{Analysis of algorithm --- minimization within widened subspace}
In this section, we prove convergence of the first part of our algorithm---minimization of $\varphi_\rho$ near $\mc S_{\mb \tau}$. We begin by proving the initialization method guarantees that $\mb a^{(0)}$ is near $\mc S_{\tau}$, in the sense that 
\begin{align}
	d_\alpha(\mb a^{(0)},\mc S_{\mb\tau})\leq \gamma,
\end{align}
 where the distance $d_\alpha$ is defined in \eqref{eqn:dist_s_tau}. We then demonstrate that small-stepping curvilinear search converges to a desired local minimum of $\varphi_\rho$ at rate $O(1/k)$, where $k$ is the iteration number. To do this, it is important to utilize(i) the \emph{retractive} property to show that the iterates stay near $\mc S_{\mb\tau}$ and (ii) the geometric properties of $\varphi_\rho$ near $\mc S_{\mb \tau}$. 
 
\subsection{Initialization near subspace} \label{sec:proof_init}
The following lemma shows that the initialization $\mb a^{(0)} = \mb P_{\Sp^{p-1}}\brac{\nabla{\varphi_{\ell^1}}(\mb a^{(-1)})} $, where
\begin{align}
	\mb a^{(-1)} = \mb P_{\Sp^{p-1}}\brac{\textstyle\sum_{\ell\in\mb\tau}\mb x_{0\ell}\injector^*_{p_0}\shift{\mb a_0}{\ell}},
\end{align}
and is very close to the subspace $\mc S_{\mb\tau}$:

\begin{lemma}[Initialization from a piece of data]
\label{lem:init_a}  
Let $\ol{\mb x}\in\R^{2p_0-1}$ indexed by $[\pm p_0]$, with $\ol{\mb x}_i \simiid \mr{BG}(\theta)$. Define $\ol{\mb y} = \ol{\mb x}*\mb a_0$, and $\mb a^{(0)}$ as  
\begin{align}\label{eqn:a_init_def}
\mb a^{(0)} = -\mb P_{\bb S^{p-1}} \nabla \varphi_{\ell^1}\left( \mb P_{\bb S^{p-1}} \brac{\mb 0^{p_0-1};[\ol{\mb y}_0;\cdots;\ol{\mb y}_{p_0-1}];\mb 0^{p_0-1} } \right),
\end{align}
with $\lambda = 0.2/\sqrt{p\theta}$ in $\varphi_1$. Set $\mb\tau = \supp(\ol{\mb x})$. Suppose that $(\mb a_0,\theta,k)$ satisfies the sparsity-coherence condition $\mr{SCC}(c_\mu)$ and $\mb a_0$ satisfies $\max_{i\neq j}\abs{\innerprod{\ip_{p_0}^*\shift{\mb a_0}{i}}{\ip_{p_0}^*\shift{\mb a_0}{j}}}\leq \mu$. Then there exists some constant $c,\ol c > 0$ such that if $p_0\theta > 1000c$ and $c_\mu \leq \ol c$, then with probability at least $1-1/c$, we have
\begin{align}
	 d_{\alpha}\left(\mb a^{(0)},\mc S_{\mb\tau} \right) \leq \frac{c_\mu}{4\log^2\theta^{-1}}\min\set{\frac{1}{\sqrt{\abs{\mb\tau}}},\frac{1}{\sqrt{\mu p}},\frac{1}{\mu p\sqrt\theta \abs{\mb\tau}}} . \label{eqn:init_a_gamma_bound}
\end{align}
\end{lemma}

\begin{proof} 1. (\ul{Distance to $\mc S_{\mb\tau} $ from $\mb a^{(0)}$}) Let $\eta = \norm{\injector_{p_0}^*(\mb a_0*\mb x)}2 = \norm{\injector_{p_0}^*\convmtx{\mb a_0}\mb x}2 $ and $\gamma = \gamma(c_\mu)$, as in \eqref{eqn:init_a_gamma_bound}. Expand the expression of $\mb a^{(0)}$ from \eqref{eqn:a_init_def} we have
	\begin{align}
		\mb a^{(0)} &= \mb P_{\Sp^{p-1}} \ip^*\checkmtx{\mb y}\soft{\checkmtx{\mb y}\injector_{p_0}\mb P_{\Sp^{p_0-1}}\injector_{p_0}^*(\mb a_0*\mb x)  }{\lambda} = \mb P_{\Sp^{p-1}}\injector^*\convmtx{\mb a_0}\mb\chi\brac{\tfrac{1}{\eta}\convmtx{\mb a_0}^*\injector_{p_0}\injector_{p_0}^*\convmtx{\mb a_0}\mb x}\label{eqn:a_init_1}
	\end{align}
To relate $\mb a^{(0)}$ to its coefficient, introduce the truncated autocorrelation matrix  $ \wt{\mb M} = \convmtx{\mb a_0}^*\injector_{p_0}\injector_{p_0}^*\convmtx{\mb a_0}$, define $\wt{\mb\alpha},\wt{\mb\beta}$ as
\begin{align}\label{eqn:alpha_beta_tilde}
	\wt{\mb\beta} = \tfrac1\eta\wt{\mb M}\mb x,\quad 
	\wt{\mb\alpha} = \mb\chi\brac{\tfrac1\eta\wt{\mb M}\mb x} = \mb\chi[\wt{\mb\beta}]
\end{align}   
and note that $\wt{\mb M}$ is bounded entrywise as  
\begin{align}\label{eqn:init_wt_M_entries}
	\abs{\wt{\mb M}_{ij}} \leq \begin{cases}
		1 &\quad i=j \in [-p_0+1,p_0-1] \\
		\mu &\quad i\neq j \in [-p_0+1,p_0-1],\;\abs{i-j} < p_0\\
		0 &\quad  \text{otherwise}
	\end{cases}.
\end{align}
From \eqref{eqn:a_init_1}, we can write $\mb a^{(0)} = \mb P_{\Sp^{p-1}}\ip^*\convmtx{\mb a_0}\wt{\mb\alpha}$, meaning that the normalized version of $\wt{\mb\alpha}$ is a valid coefficient vector for $\mb a^{(0)}$.  Let $\mb\tau^c = [\pm 2p_0]\setminus\mb\tau$. The distance  $d_{\mb \alpha}$ to subspace $\mc S_{\mb\tau}$ \eqref{eqn:dist_s_tau} is upper bounded as
\begin{align}
	d_{\alpha}(\mb a^{(0)},\mc S_{\mb\tau}) &\leq  \frac{\norm{\wt{\mb\alpha}_{\mb\tau^c}}2}{\norm{\ip^*\convmtx{\mb a_0}\wt{\mb\alpha}}2}\leq \frac{\norm{\wt{\mb\alpha}_{\mb\tau^c}}2}{\norm{\ip^*\convmtx{\mb a_0}\wt{\mb\alpha}_{\mb\tau}}2 - \norm{\ip^*\convmtx{\mb a_0}\wt{\mb\alpha}_{\mb\tau^c}}2}\notag \\
	&\leq \frac{\norm{\wt{\mb\alpha}_{\mb\tau^c}}2}{\sqrt{1-\mu\abs{\mb\tau}}\norm{\wt{\mb\alpha}_{\mb\tau}}2 - \sqrt{1+\mu p}\norm{\wt{\mb\alpha}_{\mb\tau^c}}2}\notag
\end{align}
where the last inequality is derived with \Cref{fact:M_entries}. Therefore, it is sufficient to show 
\begin{align}\label{eqn:a_init_dist_2}
	\paren{1+\gamma\sqrt{1+\mu p}}\norm{\wt{\mb\alpha}_{\mb\tau^c}}2 \leq \gamma\sqrt{1-\mu\abs{\mb\tau}}\norm{\wt{\mb\alpha}_{\mb\tau}}2
\end{align}
to complete the proof that $d_\alpha(\mb a^{(0)},\mc S_{\mb\tau}) \leq \gamma$. 

\vsni 2. (\ul{Bound $\eta$}) Condition on the following two events
\begin{align}\label{eqn:init_event_tau_2norm}
	\event_\tau :=\set{\abs{\mb\tau} < 4p_0\theta},\quad 
	\event_{\norm{\mb x}2} := \set{\sqrt{p_0\theta} \leq \norm{\mb x}2\leq \sqrt{3p_0\theta}}
\end{align}
and utilize $\mu$ bound from \Cref{lem:alpha_beta_tau_lb} such that $\mu\abs{\mb\tau}<0.1$. An upper bound on $\eta$ can be obtained using properties of $\wt{\mb M}$ of \eqref{eqn:init_wt_M_entries}:
\begin{align}\label{eqn:eta_norm_ub}
	\eta = \norm{\injector_{p_0}^*\convmtx{\mb a_0}\mb x}2 \leq \norm{\ip^*\convmtx{\mb a_0}\mb x}2 \leq \sqrt{1+\mu\abs{\mb\tau}}\norm{\mb x}2 \leq 2\sqrt{p_0\theta} 
\end{align} 
To lower bound $\eta$, use $\eta^2 = \mb g^*\mb P_{\mb\tau}\wt{\mb M}\mb P_{\mb\tau}\mb g$ where $\mb g$ is the standard Gaussian vector. Observe the submatrix of $\wt{\mb M}$ is diagonal dominant:
\begin{align}\label{eqn:M_tilde_diagonal}
\begin{dcases}
	\wt{\mb M}_{ii} = \norm{\injector_{p_0}^*\shift{\mb a_0}{i}}2^2 \in [0,1]\\
	\trace\paren{\wt{\mb M}} = \sum_{i\in[\pm p_0]} \norm{\injector_{p_0}^*\shift{\mb a_0}{i}}2^2 = \norm{\mb a_0}2^2 + \sum_{i=1}^{p_0-1}\paren{\norm{\injector_{p_0}^*\shift{\mb a_0}{i}}2^2 + \norm{\injector_{p_0}^*\shift{\mb a_0}{i-p_0} }2^2  } = p_0
\end{dcases}.
\end{align}
Write $\mb x = \mb g\circ\mb w$ where $\mb w$ and $\mb g$ are Bernoulli and Gaussian vector respectively with $\supp(\mb w) = \mb\tau$, then the trace of $\mb P_{\mb\tau}\wt{\mb M}\mb P_{\mb\tau}$ can be written as sum of independent r.v.s as:
\begin{align}
	\trace\paren{\mb P_{\mb\tau}\wt{\mb M}\mb P_{\mb\tau}} = \sum_{i\in[\pm p_0]} w_i\norm{\injector_{p_0}^*\shift{\mb a_0}{i}}2^2, \notag
\end{align}
Bernstein inequality \Cref{lem:mc_bernstein_scalar} and \eqref{eqn:M_tilde_diagonal} gives
\begin{align}\label{eqn:trace_Mt_lb}
	\prob{\trace\paren{\mb P_{\mb\tau}\wt{\mb M}\mb P_{\mb\tau}} < \frac{3p_0\theta}4} \leq \prob{\trace\paren{\mb P_{\mb\tau}\wt{\mb M}\mb P_{\mb\tau}}-p_0\theta \leq -\frac{p_0\theta}{4} } \leq 2\exp\paren{\frac{-(p_0\theta/4)^2}{2p_0\theta + p_0\theta/2}} \leq 2\exp\paren{\frac{-p_0\theta}{40}},
\end{align}
thus condition on $\mb\omega$ satisfies $\trace\paren{\mb P_{\mb\tau}\wt{\mb M}\mb P_{\mb\tau}} \geq  3p_0\theta/4$ and $\event_{\mb\tau}$, expectation $\eta^2$ has lower bound
\begin{align}
	\E_{\mb g\mid \mb w}\eta^2 = \E_{\mb g\mid\mb w} \brac{\mb {g}^*\mb P_{\mb\tau}\wt{\mb M}\mb P_{\mb\tau}\mb g } = \trace\paren{\mb P_{\mb\tau} \wt{\mb M}\mb P_{\mb\tau}} \geq  \frac{3p_0\theta}{4} \notag
\end{align}
then apply Bernstein inequality again by first writing svd of $\mb P_{\mb\tau}\wt{\mb M}\mb P_{\mb\tau} = \mb U\mb \Sigma\mb U^*$ with $\mb\Sigma$ being rank $\abs{\mb\tau}<4p_0\theta$ and square orthobasis $\mb U$. Let $\mb g' = \mb  U^*\mb g$, then $\mb g'$ is standard i.i.d. Gaussian vector, provides alternative expression $\eta^2 < \sum_{i=1}^{4p_0\theta}{g_i'}^2\sigma_i$ where $\sigma_i \leq  1+\mu\abs{\mb\tau} \leq 1.1$. We obtain probability of $\eta^2$ to be small as
\begin{align}\label{eqn:eta_norm_lb}
	\bb P_{\mb g\mid \mb w}\brac{\eta^2 < \frac{p_0\theta}{2}} & \leq \bb P_{\mb g\mid \mb w}\brac{\eta^2 - \E_{\mb g\mid \mb w}\eta^2 < -\frac{p_0\theta}{4}}\leq 2\exp\paren{\frac{-(p_0\theta/4)^2}{2(1+\mu\abs{\mb\tau})(12p_0\theta + p_0\theta/2) }} \leq 2\exp\paren{\frac{-p_0\theta}{440}} 
\end{align}  
by applying moment bounds $(\sigma^2,R) = \paren{12p_0\theta(1+\mu \abs{\mb\tau}),2(1+\mu\abs{\mb\tau})}$. We thereby define event  
\begin{align}\label{eqn:init_event_eta}
	\event_\eta = \set{\sqrt{p_0\theta/2} \leq \eta \leq 2\sqrt{p_0\theta}},
\end{align}
which holds w.h.p.\ based on \eqref{eqn:init_event_tau_2norm}, \eqref{eqn:trace_Mt_lb} and  \eqref{eqn:eta_norm_lb}.

\vsni 3. (\ul{Bound $\wt{\mb\alpha}$}) Condition on $\event_\eta\cap \event_{\norm{\mb x}2}\cap\event_\tau$. Use definition $\wt{\mb\beta} = \frac{1}{\eta}\wt{\mb M}\mb x$ from \eqref{eqn:alpha_beta_tilde}, and properties of $\wt{\mb M}$ from \eqref{eqn:init_wt_M_entries} we can obtain:  
\begin{align} 
\begin{dcases}
	\|\wt{\mb\beta}_{\mb\tau^c}\|_2 \leq \tfrac{1}{\eta} \norm{\injector^*_{\mb\tau^c}\wt{\mb M}\injector_{\mb\tau}}2\norm{\mb x}2 \leq \tfrac{\mu\sqrt{p_0\abs{\mb\tau}}}{\sqrt{p_0\theta/2}}\cdot\sqrt{3p_0\theta}\leq 3\mu\sqrt{p_0\abs{\mb\tau}} \\
	\|\wt{\mb\beta}_{\mb\tau}\|_2 \geq \tfrac{1}{\eta} \norm{\injector^*_{\mb\tau}\wt{\mb M}\injector_{\mb\tau}}2\norm{\mb x}2 \geq \tfrac{\sqrt{1-\mu\abs{\mb\tau}}}{2\sqrt{p_0\theta}}\cdot\sqrt{p_0\theta} \geq  0.45
\end{dcases}. 
\end{align}     
Use definition  $\norm{\wt{\mb\alpha}}2 = \|\mb\chi[\wt{\mb\beta}]\|_2$, condition on event
\begin{align}
	\event_{\chi} := \set{\begin{cases}  \sigma_i\mb\chi[\mb\beta]_i 
		 \geq    n\theta \soft{\abs{\mb\beta_i}}{\nu_2\lambda} - \tfrac{c_\mu^2 n\theta}{p},&\quad  \forall\,i\in\mb\tau \notag \\
		\sigma_i\mb\chi[\mb\beta]_i  \leq 4n\theta^2\abs{\mb\tau}\abs{\mb\beta_i} + \tfrac{c_\mu n\theta}{p},&\quad \forall\,i\in\mb\tau^c
	\end{cases} },
 \end{align}
also from  \Cref{asm:theta_mu} we have $\mu \paren{p\theta}^{1/2}\abs{\mb\tau}^{3/2} < \frac{c_\mu}{4\log^2\theta^{-1}}$ and from lemma assumption $\lambda = \frac1{5\sqrt{p\theta}}$, provides bounds of $\norm{\wt{\mb\alpha}}2$ via triangle inequality as: 
\begin{align}
\begin{dcases}
	\norm{\wt{\mb\alpha}_{\mb\tau^c}}2\leq   4n\theta^2\abs{\mb\tau}\cdot \|\wt{\mb\beta}_{\mb\tau^c}\|_2 + \tfrac{c_\mu n\theta}{p}\cdot\sqrt{2p_0} \leq 3c_\mu n\theta \paren{\tfrac{\sqrt\theta}{\log^2\theta^{-1}}+\tfrac{c_\mu}p} \\ 
	\norm{\wt{\mb\alpha}_{\mb\tau}}2 \geq  n\theta\paren{\|\wt{\mb\beta}_{\mb\tau}\|_2 - \nu_2\lambda\sqrt{\abs{\mb\tau}} - \tfrac{c_\mu}{p}\sqrt{\abs{\mb\tau}} }  \geq n\theta\paren{0.45-\sqrt{\tfrac{2}{\pi}}\cdot\tfrac15  - c_\mu} \geq 0.2n\theta   
\end{dcases},
\end{align}
since both $\theta\abs{\mb\tau}$, $\mu p\theta\abs{\mb\tau} < c_\mu$, we have  
\begin{align}
	\begin{dcases}
	\sqrt{1+\mu p}\norm{\wt{\mb\alpha}_{\mb\tau^c}}2 \leq 3c_\mu n\theta\sqrt{1+\mu p}\paren{\sqrt\theta + p^{-1}} \leq 6c_\mu n\theta \\
	\norm{\wt{\mb\alpha}_{\mb\tau^c}}2  \leq  \tfrac{6c_\mu^{3/2} n\theta}{\log^2\theta^{-1}}\min\set{\tfrac{1}{\sqrt{\abs{\mb\tau}}},\tfrac{1}{\sqrt{\mu p}}, \tfrac{1}{\mu p\sqrt{ \theta}\abs{\mb\tau}}} \leq  24\sqrt{c_\mu} n\theta\gamma
	\end{dcases},\notag
\end{align}
which satisfies \eqref{eqn:a_init_dist_2}, since $\mu\abs{\mb\tau} < c_\mu < \frac{1}{1000}$,
\begin{align}
	(1+\gamma\sqrt{1+\mu p})\norm{\wt{\mb\alpha}_{\mb\tau^c} }2 \leq \paren{24\sqrt{c_\mu} + 6c_\mu }n\theta \gamma \leq 0.1n\theta\gamma\leq  \gamma\sqrt{1-\mu\abs{\mb\tau}}\norm{\wt{\mb\alpha}_{\mb\tau}}2.
\end{align}
\vsni Finally, given $p_0\theta > 1000c$, this result holds with probability at least 
\begin{align}
	1-\underbrace{\prob{\event^c_{\tau}}}_{\text{\Cref{lem:x0_supp}}} - \underbrace{\prob{\event^c_{\norm{\mb x}2}}}_{\text{\Cref{lem:x0_bound}}}-\underbrace{\prob{\event^c_{\eta}}}_{\text{\eqref{eqn:init_event_eta}} } - \underbrace{\prob{\event^c_{\chi}}}_{\text{\Cref{cor:chibeta_ct}}} \geq 1-\frac{2}{p_0\theta} - \frac1 n - 4\exp\paren{\frac{-p_0\theta}{440}} \geq   1-\frac{1}{c}
\end{align}  
\end{proof}

\subsection{Minimization near subspace (Proof of \Cref{thm:minimization}) }\label{sec:proof_of_minimization}

Before we start the proof of theorem, writing $\mb g = \grad[\varphi_\rho](\mb a)$ and $\mb H = \mr{Hess}[\varphi_\rho](\mb a)$, we will first restate the results of \Cref{thm:three_regions} in simplified terms. The theorem shows that for any $\mb a\in\Sp^{p-1}$ whose distance to subspace $d_\alpha(\mb a,\mc S_{\mb\tau}) \leq \gamma$,  then at least one of the the following statement hold: 
\begin{align}
	& \norm{\mb g}2 \geq  \eta_g \label{eqn:minimize_grad} \\
	& \lambda_{\min}\paren{\mb H} \leq -\eta_{v} \label{eqn:minimize_curve} \\
	& \mb H \succ \eta_c\cdot\mb P_{\mb a^\perp }. \label{eqn:minimize_convex}
\end{align}
Furthermore, $\varphi_\rho$ is retractive near $\mc S_{\mb\tau}$: wherever $d_\alpha(\mb a,\mc S_{\mb\tau}) \geq  \frac\gamma2$, writing $\mb\alpha(\mb a)$, $\mb\alpha(\mb g)$ to be the coefficient of $\mb a$, $\mb g$, we have
\begin{align}\label{eqn:minimize_retract}
	  \innerprod{\mb\alpha(\mb a)_{\mb\tau^c}}{\mb\alpha(\mb g)_{\mb\tau^c}} \geq \eta_r\norm{\mb\alpha(\mb g)_{\mb\tau^c}}2. 
\end{align}
Also, the the gradient, Hessian and the third order derivative are all bounded as follows:
\begin{remark}\label{rmk:bounded_grad_hess}With high probability, for every $\mb a$ whose $d_\alpha(\mb a,\mc S_{\mb\tau}) < \gamma$, its $\max\set{\norm{\mb g}2,\norm{\mb H}2,\norm{\nabla\mb H}2} \leq \ol\eta = \poly(n,p)$.
\end{remark} 
\noindent We state \Cref{rmk:bounded_grad_hess} without explicit proof since its derivation is similar to the proof in \Cref{thm:three_regions}.

We prove that if the negative curvature direction $-\mb v$ is chosen to be the least eigenvector with $\mb v^*\mb H\mb v < -\eta_v$ and $\mb v^*\mb g$ (if cannot, let $\mb v = \mb 0$), then the iterates 
\begin{align}
	\mb a^{(k+1)} = \mb P_{\Sp^{p-1}}\brac{\mb a^{(k)} - t\mb g^{(k)} - t^2\mb v^{(k)}} 
\end{align}  
converges toward the minimizer $\bar{\mb a}$ in $\ell^2$-norm with rate $O(1/k)$. Notice that here all $\eta_g,\,\eta_v,\,\eta_c,\,\eta_r,\,\bar\eta$ are all greater then $0$ and are rational functions of the dimension parameters $n,p$.

Finally, we should note that $\mb a_0$ being $\mu$-truncated shift coherent implies that $\mb a_0$ is at at most $2\mu$-shift coherent. Hence we utilize the usual incoherence condition in the proof.
  
\vsni 

\begin{proof} 
Notice that when $\mb a$ is in the region near some signed shift $\bar{\mb a}$ of $\mb a_0$, the function $\varphi_\rho$ is strongly convex, and the iterates coincide with the Riemannian gradient method, which converges at a linear rate. Indeed, if for all $k$ larger than some $\bar{k}$, $\mb a^{(k)}$ is in this region, then $\norm{\mb a^{(k)} - \bar{\mb a} }{2} \leq (1 - t \eta_c)^{-(k-\bar{k})} \| \mb a^{(\bar{k})} - \bar{\mb a} \|_2$ \cite{absil2009optimization}(Theorem 4.5.6) where the step size $t = \Omega(1/n\theta)$ hence $t\eta_c = \Omega(1)$. We will argue that the iterates $\mb a^{(k)}$ remain close to the subspace $\mc S_{\mb \tau}$ and that after $\bar{k} = \mr{poly}(n,p)$ iterations they indeed remain in the strongly convex region around some $\bar{\mb a}$. 

\vsni 1. (\ul{Existence of Armijo steplength}). First, we show there exists a nontrivial step size $t$ at every iteration, in the sense that for all $\mb a\in\Sp^{p-1}$, there exists $T>0$ such that for all $t\in(0,T)$, the Armijo step condition \eqref{eqn:conv_curvi_armijo} is satisfied.
Note that since $\varphi_\rho$ is a smooth function, $\mb a\to \varphi_\rho\circ\mb P_{\Sp^{p-1}} (\mb a)$ admits a version of Taylor's theorem (see also \cite{absil2009optimization}(Section 7.1.3)): for any $\mb \xi\perp\mb a$, writing $\mb a^+ = \mb P_{\Sp^{p-1}}\brac{\mb a + \mb \xi} $, 
\begin{align}
 	\abs{\varphi_\rho(\mb a^+) -\paren{ \varphi_\rho(\mb a) + \innerprod{\mr{grad}[\varphi_\rho](\mb a)}{\mb \xi} + \tfrac{1}{2}\mb \xi^*\mr{Hess}[\varphi_\rho](\mb a)\mb \xi} } \leq \bar{\eta}\norm{\mb \xi}2^3,
\end{align} using $\norm{\nabla{\mb H}}2 \leq \bar\eta$. Now, let $\mb \xi = - t\mb g - t^2 \mb v$ as in the iterates \eqref{eqn:curvi_steps}. Suppose the Armijo step condition \eqref{eqn:conv_curvi_armijo} does not hold, so 
\begin{align}\label{eqn:curvi_armijo_contrary}
	\varphi_\rho(\mb a^+) > \varphi_\rho(\mb a) -\tfrac12\paren{t\norm{\mb g}2^2 + \tfrac12 t^4 \eta_v\norm{\mb v}2^2}.
\end{align}
Since $\mb g^*\mb v \geq 0$ and $\mb v^*\mb H\mb v \leq  -\eta_v\norm{\mb v}2^2$ or $\mb v = \mb 0$, using $\norm{\mb a + \mb b}2^3 \leq 4\norm{\mb a}2^3 + 4\norm{\mb b}2^3$ (H\"{o}lder's inequality) and $\norm{\mb H}2 < \bar\eta$, we can derive 
\begin{align}
	&\innerprod{\mb g}{-t\mb g - t^2\mb v} + \tfrac{1}{2}(t\mb g + t^2\mb v)^*\mb H\paren{t\mb g + t^2\mb v} + c\norm{t\mb g + t^2\mb v}2^3 >  -\tfrac12\paren{t\norm{\mb g}2^2 + \tfrac12 t^4\eta_v\norm{\mb v}2^2}\notag \\
	\implies & -\tfrac12 t\norm{\mb g}2^2 + \tfrac12t^2\mb g^*\mb H\mb g + t^3\mb v^*\mb H\mb g - \tfrac14 t^4\eta_v\norm{\mb v}2^2 + 4\bar\eta t^3\norm{\mb g}2^3 + 4 \bar\eta t^6\norm{\mb v}2^3 > 0  \notag \\
	\implies  & -\tfrac12 t\norm{\mb g}2^2 + t^2 \paren{\tfrac12 \bar{\eta}\norm{\mb g}2^2 + t\bar\eta\norm{\mb v}2\norm{\mb g}2 + 4 \bar\eta t\norm{\mb g}2^3} - \tfrac14t^4\eta_v\norm{\mb v}2^2 + 4\bar\eta t^6\norm{\mb v}2^3 > 0.  \label{eqn:curvi_armijo_contradiction}  
\end{align}
If
\begin{align}
	t < T = \min\set{\frac{\norm{\mb g}2}{\bar\eta\norm{\mb g}2 + 2\bar\eta t\norm{\mb v}2 + 8\bar\eta t\norm{\mb g}2^2},\,\sqrt{\frac{\eta_v}{16\bar\eta\norm{\mb v}2}}},
\end{align}
then $\text{\eqref{eqn:curvi_armijo_contradiction}} < 0$ contradicting \eqref{eqn:curvi_armijo_contrary}. Using our bounds on $\| \mb g \|_2$, $\bar{\eta}$, $\eta_v$ and $\|\mb v \|$, it follows that $T$ is lower bounded by a polynomial  $\poly\paren{n^{-1},p^{-1}}$.

\vsni 2.(\ul{Bounds on $d_\alpha(\mb g,\mc S_{\mb\tau})$, $d_\alpha(\mb v, \mc S_{\mb\tau})$}) We will show there are numerical constants $c_g$, $c_v$ such that 
\begin{align}\label{eqn:curvi_dalpha_norm_ub}
	d_\alpha(\mb g,\mc S_{\mb\tau}) \leq c_g n\theta\gamma \qquad\text{and}\qquad   d_\alpha(\mb v,\mc S_{\mb\tau}) \leq c_v n\theta p. 
\end{align}
Define  
\begin{align}
	\mb\chi_{\ell^1}[\mb\beta] = \checkmtx{\mb x_0}\prox_{\lambda\ell^1}\brac{\wc{\mb a}*\mb y},\qquad \mb\chi_\rho[\mb\beta]  = \checkmtx{\mb x_0}\prox_{\lambda\rho}\brac{\wc{\mb a}*\mb y},\notag
\end{align}
then the gradient can be written as \eqref{eqn:retract_grad}
\begin{align}
	\mr{grad}[\varphi_{\ell^1}](\mb a) &=  \ip^*\convmtx{\mb a_0}\paren{\mb\beta^*\mb\chi_{\ell^1}[\mb\beta]\mb\alpha - \mb\chi_{\ell^1}[\mb\beta]},\\
	\mr{grad}[\varphi_\rho](\mb a) &=  \ip^*\convmtx{\mb a_0}\paren{\mb\beta^*\mb\chi_{\rho}[\mb\beta]\mb\alpha - \mb\chi_{\rho}[\mb\beta] }. \label{eqn:curvi_alpha_tauc_g} 
\end{align}
Use the following inequalities: 
\begin{align}
	\tfrac12n\theta \leq \abs{\mb\beta^*\mb\chi_{\ell^1}[\mb\beta]}  &\leq \tfrac32 n\theta, \notag \\
	\norm{\mb\chi_{\ell^1}[\mb\beta]_{\mb\tau^c}}2 &\leq   \tfrac{1}{20} n\theta\gamma, \notag \\
	\norm{\mb I - \mb\alpha\mb\beta^*}2  &\leq  4\sqrt p, \notag \\
	\norm{\mb\chi_{\ell^1}[\mb\beta] - \mb\chi_{\rho}[\mb\beta]}2 &\leq n\theta^4,   \notag
\end{align}
where the first and second bounds of $\mb\chi_{\ell^1}[\mb\beta]$ based on event   \eqref{eqn:event_chi_retract}; the third  by observing $\norm{\mb\alpha}2 \leq 2$ and $\norm{\mb \beta}2\leq 2+c_\mu\sqrt p$; the last from  \eqref{eqn:smooth_approx_alpha} of  \Cref{lem:approx_rho_ell1} when $\delta$ is sufficiently small. Hence, by definition of $d_\alpha(\,\cdot,\mc S_{\mb\tau})$  \eqref{eqn:dist_s_tau} and knowing $\mb a$ is close to subspace $\norm{\mb\alpha_{\mb\tau^c}}2 \leq \gamma$,  via triangle inequality, we get
\begin{align}
	d_\alpha(\mb g,\mc S_{\mb\tau}) &\;\leq\; d_\alpha(\mr{grad}[\varphi_{\ell^1}](\mb a),\mc S_{\mb\tau}) + d_\alpha(\mr{grad}[\varphi_\rho](\mb a) - \mr{grad}[\varphi_{\ell^1}](\mb a),\mc S_{\mb\tau})\notag \\
	&\;\leq\; \norm{\mb\beta^*\mb\chi_{\ell^1}[\mb\beta]\mb\alpha_{\mb\tau^c} - \mb\chi_{\ell^1}[\mb\beta]_{\mb\tau^c}}2 + \norm{\paren{\mb I - \mb\alpha\mb\beta^*}\paren{\mb\chi_\rho[\mb\beta] - \mb\chi_{\ell^1}[\mb\beta]}}2.\notag\\
	&\;\leq\; \tfrac32n\theta\gamma \;+\;\tfrac{1}{20}n\theta\gamma \;+\; 4\sqrt p n\theta^4 \notag\\
	&\;\leq\; 3n\theta\gamma.  
\end{align}
To bound the $d_\alpha$ norm of least eigenvector $\mb v$, note that $\mb \beta^*\mb\chi_{\rho}[\mb\beta]>0 $, we can conclude
\begin{align}
	\mb v^*\nabla^2{\varphi_\rho}(\mb a)\mb v \;\leq\;  \mb v^*\mb P_{\mb a^\perp}\nabla^2{\varphi_\rho}(\mb a)\mb P_{\mb a^\perp}\mb v  + \mb \beta^*\mb\chi_{\rho}[\mb\beta] \;= \;\mb v^*\mb H\mb v \;<\; -\eta_v, \notag
\end{align}
expand $\nabla^2\varphi_\rho(\mb a)$ as in \eqref{eqn:smooth_rho_hess}, and since $\mb v$ is the eigenvector of smallest eigenvalue $\lambda_{\mr{min}} < -\eta_v$,   
\begin{align} 
	\mb P_{\mb a^\perp}\nabla^2\varphi_\rho(\mb a)\mb P_{\mb a^\perp}\mb v \;=\;\paren{\mb I - \mb a\mb a^*}\ip^*\convmtx{\mb a_0}\checkmtx{\mb x_0}\nabla\mr{prox}_{\lambda\rho}\brac{\wc{\mb a}*\mb y}\checkmtx{\mb x_0}\convmtx{\mb a_0}^*\ip\mb v \;=\; \lambda_{\mr{min}}\mb v, 
\end{align}
hence there exists $\mb\alpha(\mb v)$ satisfies $\mb v = \ip^*\convmtx{\mb a_0}\mb\alpha(\mb v)$ and 
\begin{align} \label{eqn:curvi_alpha_tauc_v}
	\mb\alpha(\mb v) \;=\; \lambda_{\mr{min}}^{-1}\brac{\checkmtx{\mb x_0}\nabla\mr{prox}_{\lambda\rho}\brac{\wc{\mb a}*\mb y}\checkmtx{\mb x_0}\convmtx{\mb a_0}^*\ip\mb v - \paren{\mb\beta^*\checkmtx{\mb x_0}\nabla\mr{prox}_{\lambda\rho}\brac{\wc{\mb a}*\mb y}\checkmtx{\mb x_0}\convmtx{\mb a_0}^*\ip\mb v}\mb\alpha}. \notag
\end{align}
Now since $\nabla\mr{prox}_{\lambda\rho}\brac{\wc{\mb a}*\mb y}$ is a diagonal matrix with entries in $[0,1]$, 
\begin{align}
	d_\alpha(\mb v,\mc S_{\mb\tau}) \;\leq\; \norm{\mb\alpha(\mb v)}2 \;\leq\;  \abs{\lambda_{\mr{min}}}^{-1}\norm{\ip\convmtx{\mb a_0}}2\norm{\mb x_0}1^2\norm{\mb v}2\paren{1+\norm{\mb\alpha}2\norm{\mb\beta}2} \;<\;  c_v n\theta p, 
\end{align}
where we use upper bound of $\norm{\mb x_0}1 < cn\theta$ from \Cref{lem:x0_bound} and $\abs{\lambda_{\mr{min}}} > \eta_v >cn\theta \lambda$ from \Cref{cor:neg_curve}. 

\vsni 3. (\ul{Iterates stay within widened subspace}). 
Suppose \eqref{eqn:minimize_retract} holds. We will show that whenever  
\begin{align}
	t\leq T' =  \frac{1}{10n\theta},
\end{align}
then setting $\mb a^+ = \mb P_{\Sp^{p-1}}\brac{\mb a - t\mb g - t^2\mb v}$, we have
\begin{align}\label{eqn:curvi_retract_1}
 	\abs{d_\alpha\paren{\mb a^+,\mc S_{\mb\tau}} - d_\alpha\paren{\mb a,\mc S_{\mb\tau}} }\leq\tfrac\gamma2,
\end{align}
and whenever $d_\alpha(\mb a,\mc S_{\mb\tau})\in\brac{\tfrac\gamma2,\gamma}$  
\begin{align}\label{eqn:curvi_retract_2}
		d_\alpha^2\paren{\mb a^+,\mc S_{\mb\tau}} \leq d_\alpha^2\paren{\mb a,\mc S_{\mb\tau}} - t\cdot c'n\theta \gamma^2.
\end{align}
If both \eqref{eqn:curvi_retract_1} and \eqref{eqn:curvi_retract_2} hold, then all iterates $\mb a^{(k)}$ will stay near the subspace: $d_\alpha(\mb a^{(k)}, \mc S_{\mb\tau}) < \gamma$.

To derive \eqref{eqn:curvi_retract_1}, since both $\mb g\perp\mb a$ and $\mb v\perp\mb a$ we have $\norm{\mb a-t\mb g-t^2\mb v}2^2 = \norm{\mb a}2^2 + \norm{t\mb g + t^2\mb v}2^2 > 1$, and since $d_\alpha(\cdot,\mc S_{\mb\tau})$ is a seminorm \Cref{lem:d_alpha_norm}: 
\begin{align} 
	d_\alpha\paren{\mb a^+,\mc S_{\mb\tau}} &= d_\alpha(\mb P_{\Sp^{p-1}}\brac{\mb a - t\mb g - t^2\mb v},\mc S_{\mb\tau}) \leq d_\alpha\paren{\mb a - t\mb g - t^2\mb v,\mc S_{\mb\tau}} \notag \\
	&\leq d_\alpha(\mb a,\mc S_{\mb\tau}) + t d_\alpha(\mb g,\mc S_{\mb\tau}) + t^2d_\alpha(\mb v,\mc S_{\mb\tau})
\end{align}
suggests \eqref{eqn:curvi_retract_1} holds via  \eqref{eqn:curvi_dalpha_norm_ub} and let $n > Cp^5\theta^{-2}$, we have    
\begin{align}
	 t d_\alpha(\mb g,\mc S_{\mb\tau}) + t^2d_\alpha(\mb v,\mc S_{\mb\tau}) \leq \tfrac{c_g n\theta\gamma}{10n\theta} + \tfrac{c_v n\theta p}{(10n\theta)^2} < \tfrac{\gamma}{2}
\end{align}
with sufficiently large $C$. 

To derive \eqref{eqn:curvi_retract_2}, let $\mb\alpha(\mb a)$ to be a coefficient vector satisfying $d_\alpha(\mb a,\mc S_{\mb\tau}) = \norm{\mb\alpha(\mb a)_{\mb\tau^c}}2$, and based on \eqref{eqn:curvi_alpha_tauc_g} and \eqref{eqn:curvi_alpha_tauc_v}, define  
\begin{align}
	\mb\alpha(\mb g) &= \mb\beta^*\mb\chi_{\rho}[\mb\beta]\mb\alpha(\mb a) - \mb\chi_{\rho}[\mb\beta] \\
	\mb\alpha(\mb v) &= \lambda_{\mr{min}}^{-1}\checkmtx{\mb x_0}\nabla\mr{prox}_{\lambda\rho}\brac{\wc{\mb a}*\mb y}\checkmtx{\mb x_0}\convmtx{\mb a_0}^*\ip\mb v. 
\end{align}
By the retraction property and norm bounds,
\begin{align}
	\innerprod{\mb\alpha(\mb a)_{\mb\tau^c}}{\mb\alpha(\mb g)_{\mb\tau^c}} &\geq  \tfrac{1}{6n\theta}\norm{\mb\alpha(\mb g)_{\mb\tau^c}}2^2\\
	\norm{\mb\alpha(\mb a)_{\mb\tau^c}}2 &\leq  \gamma \\
	\norm{\mb\alpha(\mb v)}2 &\leq c_v n \theta p.  
\end{align}
Since $\norm{\mb\alpha_{\mb\tau^c}}2 > \tfrac\gamma2$,  
\begin{align}
	\norm{\mb a(\mb g )_{\mb\tau^c}}2 &\geq 	 \norm{\mb\beta^*\mb\chi_{\ell^1}[\mb\beta]\mb\alpha_{\mb\tau^c} \;-\; \mb\chi_{\ell^1}[\mb\beta]_{\mb\tau^c}}2 - \norm{\paren{\mb I - \mb\alpha\mb\beta^*}\paren{\mb\chi_\rho[\mb\beta] - \mb\chi_{\ell^1}[\mb\beta]}}2\notag\\
	&\geq \abs{\mb\beta^*\mb\chi_{\ell^1}[\mb\beta]}\norm{\mb\alpha_{\mb\tau^c}}2 \;-\; \norm{\mb\chi_{\ell^1}[\mb\beta]_{\mb\tau^c}}2 \;-\; \norm{\paren{\mb I - \mb\alpha\mb\beta^*}}2\norm{\paren{\mb\chi_\rho[\mb\beta] - \mb\chi_{\ell^1}[\mb\beta]}}2\notag\\
	&\geq \tfrac12n\theta\times\tfrac{\gamma}2 \;-\;\tfrac{1}{20}n\theta\gamma \;+\; 2n\theta^4 \notag\\
	&\geq \tfrac{1}{10}n\theta\gamma.    
\end{align}
Finally, we can bound $d_\alpha(\mb a^+,\mc S_{\mb\tau})$ as 
\begin{align}
 	d^2_\alpha(\mb a^+,\mc S_{\mb\tau}) &\leq d^2_\alpha(\mb a-t\mb g-t^2\mb v,\mc S_{\mb\tau}) \notag \\
 	&\leq  \norm{\big[ \mb\alpha(\mb a) -t\mb\alpha(\mb g) - t^2\mb\alpha(\mb v) \big]_{\mb\tau^c} }2^2\notag \\
 	&= \norm{\mb\alpha(\mb a)_{\mb\tau^c}}2^2 \;-\; 2t\innerprod{\mb\alpha(\mb a)_{\mb\tau^c}}{\big[\mb\alpha(\mb g) + t\mb\alpha(\mb v)\big]_{\mb\tau^c}} \;+\; t^2\norm{\big[\mb\alpha(\mb g) + t\mb\alpha(\mb v)\big]_{\mb\tau^c}}2^2\notag \\
 	&\leq \norm{\mb\alpha(\mb a)_{\mb\tau^c}}2^2 \;-\; 2t\innerprod{\mb\alpha(\mb a)_{\mb\tau^c}}{\mb\alpha(\mb g)_{\mb\tau^c}} \;+\; 2t^2\norm{\mb \alpha(\mb a)_{\mb\tau^c}}2\norm{\mb \alpha(\mb v)}2 \;+\;  2t^2\norm{\mb\alpha(\mb g)_{\mb\tau^c}}2^2 \;+\; 2t^4\norm{\mb\alpha(\mb v)}2^2\notag \\
 	&\leq d^2(\mb a,\mc S_{\mb\tau}) - 2t\brac{\paren{\tfrac{1}{3n\theta}-t}\norm{\mb\alpha(\mb g)_{\mb\tau^c}}2^2 -  tn\theta p\gamma - t^3(c_vn\theta p)^2 }\notag \\
 	&\leq d^2(\mb a,\mc S_{\mb\tau}) -t\cdot c'n\theta\gamma^2 
\end{align}
where the last inequality holds when $t < \frac{0.1}{n\theta}$ with sufficiently large $n$.

\vsni 4. (\ul{Polynomial time convergence}) The iterates $\mb a^{(k)}$ remain within a $\gamma$ neighborhood of $\mc S_{\mb \tau}$ for all $k$. At any iteration $k$, $\mb a^{(k)}$ is in at least one of three regions: strong gradient, negative curvature, or strong convexity. In the gradient and curvature regions, we obtain a decrease in the function value which is at least some (nonzero) rational function of $n$ and $p$. On the strongly convex region, the function value does not increase. The suboptimality at initialization is bounded by a polynomial in $n$ and $p$,$\poly(n,p)$, and hence the total number of steps in the gradient and curvature regions is bounded by a polynomial in $n,p$. After the iterates reach the strongly convex region, the number of additional steps required to achieve $\| \mb a^{(k)} - \bar{\mb a} \|_2 \le \eps$ is bounded by $\mr{poly}(n,p) \log \eps^{-1}$. In particular, the number of iterations required to achieve $\| \mb a^{(k)} - \bar{\mb a} \|_2 \le \mu + 1/p$ is bounded by a polynomial in $(n,p)$, as claimed.
\end{proof}

% !TEX root = ../../BD_DQ.tex
\section{Analysis of algorithm --- local refinement} \label{sec:altmin}
In this section, we describe and analyze an algorithm which refines an estimate $\mb a^{(0)} \approx \mb a_0$ of the kernel to exactly recover $(\mb a_0,\mb x_0)$. Set 
\begin{align}
	\lambda^{(0)} \gets 5\kappa_I\mut \qquad \text{and}  \qquad  I^{(0)} \gets \supp(\soft{\convmtx{\mb a^{(0)}}^*\mb y}{\lambda}),
\end{align}
where as each iteration of the algorithm consists of the following key steps:
\begin{itemize}
\item {\bf Sparse Estimation using Reweighted Lasso:} Set 
\begin{align}
\mb x^{(k+1)} & \;\gets\; \argmin_{\mb x}  \tfrac{1}{2}\|\mb a^{(k)} * \mb x - \mb y\|_2^2 +\sum_{i\,\not\in\, I^{(k)}}\lambda^{(k)} \abs{\mb x_i};
\end{align}   
\item {\bf Kernel Estimation using Least Squares:} Set 
\begin{align}
\mb a^{\left(k+1\right)} \;\gets\; \mb P_{\Sp^{p-1}}\big[\argmin_{\mb a} \tfrac{1}{2} \|\mb a \ast \mb x^{(k+1)} - \mb y\|_2^2\big];
\end{align} 
\item {\bf Continuation and reweighting by decreasing sparsity regularizer:} Set 
\begin{align} \lambda^{(k+1)} \gets\; \tfrac{1}{2} \lambda^{(k)} \qquad \text{and}\qquad  	I^{(k+1)} &\;\gets\; \supp(\mb x^{(k+1)}).
\end{align}
\end{itemize}
Our analysis will show that $\mb a^{(k)}$ converges to $\mb a_0$ at a linear rate. In the remainder of this section, we describe the assumptions of our analysis. In subsequent sections, we prove key lemmas analyzing each of the three main steps of the algorithm.

 \paragraph{Modified coherence and rate assumptions}
 
 Below, we will write 
 \begin{equation}
 \mut = \max\set{\mu, p^{-1} }. 
 \end{equation}
 Our refinement algorithm will demand an initialization satisfying 
 \begin{equation}
 \|\mb a^{(0)} - \mb a_0\|_2 \le \mut. 
 \end{equation}

 \paragraph{Support density of \texorpdfstring{$\mb x_0$}{the True Sparse Signal}}

 Our goal is to show that the proposed annealing algorithm exactly solves the SaS deconvolution problem, i.e., exactly recovers $(\mb a_0,\mb x_0)$ up to a signed shift. We will denote the support sets of true sparse vector $\mb x_0$ and recovered $\mb x^{(k)}$ in the intermediate $k$-th steps as 
 \begin{align}
 	 I = \supp(\mb x_0) ,\qquad \qquad I^{(k)} = \supp(\mb x^{(k)}).  
 \end{align} 
It should be clear that exact recovery is unlikely if $\mb x_0$ contains many consecutive nonzero entries: in this situation, even {\em non-blind} deconvolution fails. We introduce the notation $\kappa_I$ as an upper bound for number of nonzero entries of $\mb x_0$ in a length-$p$ window: 
\begin{equation}
\kappa_I = 6 \max\set{ \theta p, \log n},
\end{equation} 
then in the Bernoulli-Gaussian model, with high probability,
 \begin{equation}\label{}
\max_{\ell} \left| I \cap \paren{[p]+\ell} \right| \leq\kappa_I.
 \end{equation}
 Here, indexing and addition should be interpreted modulo $n$.
 The $\log n$ term reflects the fact that as $n$ becomes enormous (exponential in $p$) eventually it becomes likely that some length-$p$ window of $\mb x_0$ is densely occupied. In our main theorem statement, we preclude this possibility by putting an upper bound on $n$ (w.r.t $\mut$). We find it useful to also track the maximum $\ell^2$ norm of $\mb x_0$ over any length-$p$ window:  
\begin{equation}
\norm{\mb x_0}{\square} := \max_{\ell} \norm{ \mb P_{\paren{[p]+\ell}} \mb x_0 }2.
\end{equation}
Below, we will sometimes work with the $\square$-induced operator norm:
\begin{equation}
\norm{\mb M}{\square \to \square} = \sup_{\|\mb x\|_\square \le 1} \norm{\mb M\mb x}{\square}
\end{equation}
For now, we note that in the Bernoulli-Gaussian model, $\norm{\mb x_0}{\square}$ is typically not large
\begin{equation}
\norm{\mb x_0}{\square} \;\le\; \sqrt{\kappa_I}.
\end{equation}

\subsection{Reweighted Lasso finds the large entries of \texorpdfstring{$\mb x_0$}{the True Sparse Signal}}
 
 The following lemma asserts that when $\mb a$ is close to $\mb a_0$, the reweighted Lasso finds all of the large entries of $\mb x_0$. Our reweighted Lasso is modified version from \cite{candes2008enhancing},  we only penalize $\mb x$ on entries outside of its known support subset. We write $T$ to be the subset of true support $I$, and define the sparsity surrogate as
 \begin{align}
 	\sum_{i\,\in T^c}\abs{\mb x_i}
 \end{align}
The reweighted Lasso recovers more accurate $\mb x$ on set $T$ compares to the vanilla Lasso problem, it turns out to be very helpful in our analysis which proves convergence of the proposed alternating minimization.  
 
 \begin{lemma}[Accuracy of reweighted Lasso estimate] \label{lem:lasso-big-entries} Suppose that $\mb y=\mb a_0*\mb x_0$ with $\mb a_0$ is $
 \mut$-shift coherent and  $\norm{\mb x_0}\square \leq \sqrt{\kappa_I}$ with $\kappa_I\geq 1$. If $\mut\kappa_I^2 \leq c_\mu$, then for every  $T\subseteq I$ and $\mb a$ satisfying $\|\mb a - \mb a_0 \|_2 \le \mut $, the solution $\mb x^+$ to the optimization problem
 \begin{equation}
 \min_{\mb x} \Big\{\, \tfrac{1}{2}\|\mb a * \mb x - \mb y\|_2^2 +\lambda\sum_{i\,\in T^c} \abs{\mb x_i} \,\Big\},
 \end{equation}
with
\begin{equation}
\lambda \;>\; 5\kappa_I\|\mb a - \mb a_0\|_2,
\end{equation}
is unique with the form
\begin{align}\label{eqn:refine_lasso_sol}
	\mb x^+ \;=\; \injector_J\paren{\convmtx{\mb aJ}^*\convmtx{\mb a J}}^{-1}\injector_J^*\paren{\convmtx{\mb a}^*\mb y - \lambda\mb P_{J\setminus T}\mb\sigma}
\end{align} 
where $\mb\sigma = \sign(\mb x^+)$. Its support set $J$ satisfies 
\begin{align}
 	(\,T\cup I_{\geq 3\lambda}\,) \;\subseteq\; J \;\subseteq\; I
\end{align} 
and its entrywise error is bounded as
 \begin{equation}
 \norm{ \mb x^+ - \mb x_0 }{\infty} \;\le\; 3\lambda.
 \end{equation}
 Above, $c_\mu > 0$ is a positive numerical constant.
\end{lemma}

\noindent We prove \Cref{lem:lasso-big-entries} below. The proof relies heavily on the fact that when $\mb a_0$ is shift-incoherent and $\mb a \approx \mb a_0$, $\mb a$ is also shift-incoherent, an observation which is formalized in a sequence of calculations in  \Cref{sec:appx-supp-lemma-refine}. \\

\begin{proof} 1. (\ul{Restricted support Lasso problem}). We first consider the restricted problem 

\begin{equation} \label{eqn:lasso-restricted}
\min_{\mb w\in\R^{\abs I }} \Big\{ \tfrac{1}{2} \norm{ \mb a \ast \injector_I \mb w - \mb y }{2}^2 + \lambda \sum_{i\in T^c}\abs{(\injector_I\mb w)_i} \Big\}.
\end{equation}
Under our assumptions, provided $c < \tfrac{1}{9}$, \Cref{lem:ca-spectral} implies that 
\begin{equation} 
\injector_I^* \mb C_{\mb a}^* \mb C_{\mb a} \injector_I = [\convmtx{\mb a}^*\convmtx{\mb a}]_{I,I} \succ \mb 0,
\end{equation}  
and the restricted problem is strongly convex and its solution is unique. The KKT conditions imply that a vector $\mb w_\star$ is the unique optimal solution to this problem if and only if 
\begin{equation} \label{eqn:kkt-restricted-i}
\injector_I^* \mb C_{\mb a}^* \mb C_{\mb a} \injector_I \mb w_\star \in \injector_I^* \mb C_{\mb a}^* \mb y - \lambda\, \partial\norm{\,\mb P_{T^c}\brac{\cdot}\, }1( \mb w_\star ).
\end{equation} 

Writing $J = \supp( \injector_I \mb w_\star ) \subseteq I$, $\,{\mb C_{\mb a}}_J = \mb C_{\mb a} \injector_J$, $\,\mb w_J = \injector_J^* \injector_I \mb w_\star$ the corresponding sub-vector containing the nonzero entries of $\mb w_\star$ and $\mb \sigma_{J\setminus T} = \injector_J^*\mb P_{T^c}\brac{\mr{sign}(\injector_I\mb w_*)}$, the condition \eqref{eqn:kkt-restricted-i} is satisfied if and only if 
\begin{align}
&{\mb C_{\mb a}}_J^*{\mb C_{\mb a}}_J \mb w_J = {\mb C_{\mb a}}_J^* \mb y - \lambda \mb \sigma_{J\setminus T}, \label{eqn:kkt-restricted-ii} \\
&\| {\mb C_{\mb a}}_{I \setminus J}^* \left({\mb C_{\mb a}}_J \mb w_J - \mb y \right)\|_\infty \le \lambda. \label{eqn:kkt-restricted-iii} 
\end{align}
We will argue that under our assumptions, $J$ necessarily contains all of the large entries of $\mb x_0$:
\begin{equation}
I_{> 3 \lambda} = \set{ \ell \in I \mid |\mb x_{0\ell}| > 3\lambda } \subseteq J. 
\end{equation} 
We show this by contradiction -- namely, if some large entry of $\mb x_0$ is not in $J$, then the dual condition \eqref{eqn:kkt-restricted-iii} is violated, contradicting the optimality of $\mb w_\star$. To this end, note that by  \Cref{cor:ca-sub-spectral}, ${\mb C_{\mb a}}_J^*{\mb C_{\mb a}}_J$ has full rank. From \eqref{eqn:kkt-restricted-ii}, 
\begin{equation}
\mb w_J = \left[ {\mb C_{\mb a}}_J^* {\mb C_{\mb a}}_J\right]^{-1} \left[ {\mb C_{\mb a}}_J^* \mb y - \lambda \mb \sigma_{J\setminus T} \right].
\end{equation} 
Write ${\mb x_0}_J = \injector_J^* \mb x_0$  and $(\mb x_0)_{I \setminus J} = \mb P_{I \setminus J} \mb x_0$. We can further notice that 
\begin{eqnarray}
\lefteqn{ \convmtx{\mb aJ} \mb w_J - \mb y \; = \; \left( {\mb C_{\mb a}}_J  \left[ {\mb C_{\mb a}}_J^* {\mb C_{\mb a}}_J\right]^{-1}  {\mb C_{\mb a}}_J^* - \mb I \right) \mb y - \lambda {\mb C_{\mb a}}_J \left[ {\mb C_{\mb a}}_J^* {\mb C_{\mb a}}_J\right]^{-1} \mb \sigma_{J\setminus T} } \nonumber \\
&=&  \left( {\mb C_{\mb a}}_J  \left[ {\mb C_{\mb a}}_J^* {\mb C_{\mb a}}_J\right]^{-1}  {\mb C_{\mb a}}_J^* - \mb I \right) {\mb C_{\mb a_0}}_J {\mb x_0}_J + \left( {\mb C_{\mb a}}_J  \left[ {\mb C_{\mb a}}_J^* {\mb C_{\mb a}}_J\right]^{-1}  {\mb C_{\mb a}}_J^* - \mb I \right) \mb C_{\mb a_0 I\setminus J} (\mb x_0)_{I \setminus J} \nonumber \\ && \quad - \; \lambda {\mb C_{\mb a}}_J \left[ {\mb C_{\mb a}}_{J}^* {\mb C_{\mb a}}_J\right]^{-1} \mb \sigma_{J\setminus T} \nonumber \\
&=&  \left( {\mb C_{\mb a}}_J  \left[ {\mb C_{\mb a}}_J^* {\mb C_{\mb a}}_J\right]^{-1}  {\mb C_{\mb a}}_J^* - \mb I \right) {\mb C_{\mb a_0 - \mb a}}_J {\mb x_0}_J + \left( {\mb C_{\mb a}}_J  \left[ {\mb C_{\mb a}}_J^* {\mb C_{\mb a}}_J\right]^{-1}  {\mb C_{\mb a}}_J^* - \mb I \right) \mb C_{\mb a_0 I\setminus J} (\mb x_0)_{I \setminus J} \nonumber \\ && \quad - \; \lambda {\mb C_{\mb a}}_J \left[ {\mb C_{\mb a}}_J^* {\mb C_{\mb a}}_J\right]^{-1} \mb \sigma_{J\setminus T}, \label{eqn:residual-expansion}
\end{eqnarray}
where in the final line we have used that 
\begin{equation}
 \left( {\mb C_{\mb a}}_J  \left[ {\mb C_{\mb a}}_J^* {\mb C_{\mb a}}_J\right]^{-1}  {\mb C_{\mb a}}_J^* - \mb I \right) {\mb C_{\mb a}}_J
 = \mb 0. 
 \end{equation} 

Suppose that $J$ is a strict subset of $I$ (otherwise, if $J=I$, we are done). Take any $i \in I\setminus J$ such that $|\mb x_{0i}| = \norm{(\mb x_0)_{I \setminus J} }{\infty}$, and let $\xi = \mr{sign}(\mb x_{0i})$. Using \eqref{eqn:residual-expansion}, \Cref{cor:ca-sub-spectral} and \Cref{lem:refine_conv_a0-a}, we have 
\begin{eqnarray}
\lefteqn{ -\xi \shift{\mb a}{i}^* \left( \convmtx{\mb aJ} \mb w_J - \mb y \right) \;=\; \xi \shift{\mb a}{i}^* \left( \mb I - { \mb C_{\mb a}}_J \left[ {\mb C_{\mb a}}_J^* { \mb C_{\mb a}}_J \right]^{-1} {\mb C_{\mb a}}_J^*  \right) s_i[\mb a_0] \mb x_{0i} } \nonumber \\
&& \quad + \; \xi \shift{\mb a}{i}^* \left( \mb I -  { \mb C_{\mb a}}_J \left[ {\mb C_{\mb a}}_J^* { \mb C_{\mb a}}_J \right]^{-1} {\mb C_{\mb a}}_J^* \right) \mb C_{\mb a_0} (\mb x_0)_{I \setminus ( J \cup \set{i} )} \nonumber \\
&& \quad + \; \xi \shift{\mb a}{i}^*  \left( \mb I - { \mb C_{\mb a}}_J \left[ {\mb C_{\mb a}}_J^* { \mb C_{\mb a}}_J \right]^{-1} {\mb C_{\mb a}}_J^*  \right) {\mb C_{\mb a_0 - \mb a}}_J {\mb x_0}_J \nonumber \\
&& \quad + \; \xi \lambda \shift{\mb a}{i}^* {\mb C_{\mb a}}_J \left[ {\mb C_{\mb a}}_J^* {\mb C_{\mb a}}_J \right]^{-1} \mb \sigma_{J\setminus T}  \\
&\ge&  \paren{\innerprod{ s_i[\mb a] }{ s_i[\mb a_0] } \;-\; \norm{\shift{\mb a}{i}^* {\mb C_{\mb a}}_J }{1} \norm{ \left[ {\mb C_{\mb a}}_J^* {\mb C_{\mb a}}_J \right]^{-1} }{\infty \to \infty} \norm{ {\mb C_{\mb a}}_J^* s_i[\mb a_0] }{\infty} }\norm{(\mb x_0)_{I \setminus J}}{\infty} \nonumber \\
&& \quad - \; \paren{\norm{\shift{\mb a}{i}^* {\mb C_{\mb a_0}}_{I \setminus\set{i}} }{1} \; + \; \norm{\shift{\mb a}{i}^* {\mb C_{\mb a}}_J }{1} \norm{ \left[ {\mb C_{\mb a}}_J^* {\mb C_{\mb a}}_J \right]^{-1} }{\infty \to \infty} \norm{ {\mb C_{\mb a}}_J^* {\mb C_{\mb a_0}}_{I \setminus J} }{\infty \to \infty} }\norm{ (\mb x_0)_{I \setminus J} }{\infty} \nonumber \\
&& \quad - \; \paren{\norm{ \shift{\mb a}{i}^* {\mb  C_{\mb a_0 - \mb a}}_J }{2} \; + \; \norm{ \shift{\mb a}{i}^* {\mb C_{\mb a}}_J }{2}   \norm{\left[ {\mb C_{\mb a}}_J^* {\mb C_{\mb a}}_J \right]^{-1}}{\square \to \square} \norm{ {\mb C_{\mb a}}_J^* {\mb C_{\mb a_0 - \mb a}}_J }{\square \to \square}}\sqrt 2\norm{ \mb x_0}{\square} \nonumber \\
&& \quad - \; \lambda \norm{ \shift{\mb a}{i}^* {\mb C_{\mb a}}_J }{1} \norm{\left[ {\mb C_{\mb a}}_J^* {\mb C_{\mb a}}_J \right]^{-1}}{\infty \to \infty} \norm{\mb \sigma_{J\setminus T}}{\infty} \\
&\ge& \Big(\left( 1 - \|\mb a - \mb a_0 \|_2 \right) \;-\; C_1 \kappa_I \mut \times 1 \times  \mut \Big)\norm{ (\mb x_0)_{I \setminus J} }{\infty} \nonumber \\
&& \quad - \; C_2\Big( \kappa_I \mut  \;+\;  \kappa_I \mut \times 1 \times  \kappa_I \mut  \Big)\norm{ (\mb x_0)_{I \setminus J} }{\infty} \nonumber \\  
&& \quad -\; \Big(2\sqrt{\kappa_I}\| \mb a - \mb a_0 \|_2  \;+\; C_3\sqrt{\kappa_I}\mut  \times 1 \times \kappa_I \|\mb a - \mb a_0 \|_2 \Big)\norm{\mb x_0}{\square} \nonumber \\
&& \quad -\; \lambda \,C_4  \kappa_I \mut \\  
&\ge& \left( 1 \,-\, C_1' \kappa_I \mut - C_2 \left(\kappa_I \mut \right)^2 \right) \| (\mb x_0)_{I\setminus J} \|_\infty \nonumber  \\
&& \quad - \;  2 \kappa_I \| \mb a - \mb a_0 \|_2  - \paren{ C_3 \kappa_I^{3/2}\mut } \kappa_I \norm{\mb a - \mb a_0}{2} - \paren{C_4 \kappa_I \mut} \lambda \qquad \; \\
&\ge& \tfrac{1}{2} \norm{ (\mb x_0)_{I \setminus J} }{\infty} - \lambda / 2,
\end{eqnarray}
where the last line holds provided $\mut\kappa_I^2 \leq c_\mu $ to be a sufficiently small numerical constants. If $\| (\mb x_0)_{I\setminus J} \|_\infty > 3 \lambda$, this is strictly larger than $\lambda$, implying that 
$\left| \mb a_i^* \left( \convmtx{\mb aJ} \mb w_J - \mb y \right) \right| > \lambda$, and contradicting the KKT conditions for the restricted problem. Hence, under our assumptions
\begin{equation}
\norm{( \mb x_0 )_{I \setminus J}}{\infty} \le 3 \lambda.
\end{equation}

\vsni 2. (\ul{Solution of Full Lasso problem}) We next argue that the solution of the restricted support Lasso problem, $\mb w_J$, when  extended to $\R^n$ as $\mb x^+ = \injector_J \mb w_J$, is the unique optimal solution to the {\em full} Lasso problem
\begin{equation} \label{eqn:lasso-full}
\min_{\mb x}\, \varphi_{\mr{lasso}}(\mb x) \;\equiv\;   \tfrac{1}{2} \norm{ \mb a * \mb x - \mb y }{2}^2 + \lambda \sum_{i\in T^c}\abs{\mb x_i}.
\end{equation}
To prove that $\mb x^+$ is the unique optimal solution, it suffices to show that for every $i \in I^c$, 
\begin{equation} \label{eqn:expanded-dual-cond}
| \,\shift{\mb a}{i}^* ( \mb a \ast \mb x^+ - \mb y ) \, | < \lambda.
\end{equation}
Indeed, suppose that this inequality is in force. Write $\eps = \lambda - \max_{i \in I^c} \left| \shift{\mb a}{i}^* ( \mb a \ast \mb x^+ - \mb y ) \right|$, and notice that from the KKT conditions for the restricted problem, 
\begin{equation}
\mb 0 \,\in\, \mb P_I \partial_{\mb x} \varphi_{\mr{lasso}}(\mb x)
\end{equation}
Combining with \eqref{eqn:expanded-dual-cond}, we have that for every vector $\mb \zeta$ with $\mr{supp}(\mb \zeta) \subseteq I^c$ and $\|\mb \zeta \|_\infty \le 1$, then $\eps \mb \zeta \in \partial \varphi_{\mr{lasso}}( \mb x^+)$. Let $\mb x'$ be any vector with $\mb x'_{I^c} \ne \mb 0$ and set $\mb \zeta = \mc P_{I^c} \mr{sign}( \mb x' )$, then from the subgradient inequality,
\begin{align}
\varphi_{\mr{lasso}}( \mb x' ) &\ge \varphi_{\mr{lasso}}(\mb x^+ ) + \innerprod{ \eps \mb \zeta }{ \mb x' - \mb x^+ } \nonumber \\
 &\ge \varphi_{\mr{lasso}}(\mb x^+ ) + \eps \norm{\mb x'_{I^c}}{1},
\end{align}
which is strictly larger than $\varphi_{\mr{lasso}}(\mb x^+)$. Hence, when \eqref{eqn:expanded-dual-cond} holds, any optimal solution $\bar{\mb x}$ to the full Lasso problem must satisfy $\mr{supp}(\bar{\mb x}) \subseteq I$. By strong convexity of the restricted problem, the solution to \eqref{eqn:lasso-full} is unique and equal to $\mb x^+$.

We finish by showing \eqref{eqn:expanded-dual-cond}. Using the same expansion as above, we obtain 
\begin{eqnarray}
 \lefteqn{|\shift{\mb a}{i}^* ( \convmtx{\mb aJ}\mb w_J - \mb y ) |  \;\;\le\;\; \left| \shift{\mb a}{i}^* \left( \mb I -  { \mb C_{\mb a}}_J \left[ {\mb C_{\mb a}}_J^* { \mb C_{\mb a}}_J \right]^{-1} {\mb C_{\mb a}}_J^* \right) \mb C_{\mb a_0 I\setminus J} (\mb x_0)_{I \setminus J } \right|}  \nonumber \\
&& \quad\quad + \;\;  \left| \shift{\mb a}{i}^*  \left( \mb I - { \mb C_{\mb a}}_J \left[ {\mb C_{\mb a}}_J^* { \mb C_{\mb a}}_J \right]^{-1} {\mb C_{\mb a}}_J^*  \right) {\mb C_{\mb a_0 - \mb a}}_J {\mb x_0}_J \right| \nonumber \\
&& \quad\quad  + \;\; \lambda \left|\shift{\mb a}{i}^* {\mb C_{\mb a}}_J \left[ {\mb C_{\mb a}}_J^* {\mb C_{\mb a}}_J \right]^{-1} \mb \sigma_{J\setminus T} \right|  \\
&\le & \paren{\norm{ \shift{\mb a}{i}^* {\mb C_{\mb a_0}}_{I \setminus J} }{1}  \;+\; \norm{\shift{\mb a}{i}^* {\mb C_{\mb a}}_J}{1} \norm{\left[ {\mb C_{\mb a}}_J^* {\mb C_{\mb a}}_J \right]^{-1}}{\infty \to \infty} \norm{ {\mb C_{\mb a}}_J^* {\mb C_{\mb a_0}}_{I \setminus J} }{\infty\to\infty} }\norm{(\mb x_0)_{I \setminus J}}{\infty}  \nonumber \\
&& \quad\quad +\; \paren{ \norm{ \shift{\mb a}{i}^* {\mb C_{\mb a_0 - \mb a}}_J }{2} \; + \; \norm{ \shift{\mb a}{i}^* {\mb C_{\mb a}}_J }{2} \norm{\left[ {\mb C_{\mb a}}_J^* {\mb C_{\mb a}}_J \right]^{-1}}{\square \to \square} \norm{ {\mb C_{\mb a}}_J^* {\mb C_{\mb a_0 - \mb a}}_J }{\square \to \square} }\sqrt 2\norm{\mb x_0}{\square}  \nonumber \\
&& \quad\quad + \;\; \lambda \norm{ \shift{\mb a}{i}^* {\mb C_{\mb a}}_J}{1} \norm{ \left[ {\mb C_{\mb a}}_J^* {\mb C_{\mb a}}_J \right]^{-1} }{\infty \to \infty} \norm{\mb \sigma_{J\setminus T}  }{\infty} \\
&\le& C_1\paren{\mut \kappa_I  \;+\;  \mut \kappa_I \times 1 \times  \mut \kappa_I }\times 2\lambda  \nonumber \\
&& \quad\quad + \;\;   \paren{2\sqrt{ \kappa_I}\| \mb a - \mb a_0 \|_2  \;+ \; C_2 \sqrt{\kappa_I} \mut \times 1 \times  \kappa_I \|\mb a - \mb a_0 \|_2 }\times \sqrt{\kappa_I} \nonumber \\
&& \quad\quad + \;\; \lambda C_3 \times  \mut \kappa_I   \\
&\leq & \big( \paren{C_1+C_3}\mut\kappa_I \;+\; C_1(\mut\kappa_I)^2  \big)\lambda \;\;+\;\; \big(2 \;+\; C_2\mut\kappa_I \big)\kappa_I\norm{\mb a-\mb a_0}2  \\
&<&  \lambda,
 \end{eqnarray}  
where the last line holds as long as  $c_\mu$ is a sufficiently small numerical constant. This establishes that $\mb x^+$ is the unique optimal solution to the full Lasso problem. 

\vsni 3. (\ul{Entrywise difference to $\mb x_0$}) Finally we will be controlling $\norm{\mb x^+_J - (\mb x_0)_J}{\infty}$. Indeed, from \label{cor:ca-sub-spectral} \Cref{lem:refine_conv_a0-a}, 
\begin{eqnarray}
\norm{\mb x^+_J - (\mb x_0)_J}{\infty} &=& \norm{ \left[{\mb C_{\mb a}}_J^* {\mb C_{\mb a}}_J\right]^{-1} {\mb C_{\mb a}}_J^* \mb C_{\mb a_0} \mb x_0 - \lambda \left[{\mb C_{\mb a}}_J^* {\mb C_{\mb a}}_J\right]^{-1} \mb \sigma_{J\setminus T} - (\mb x_0)_J }{\infty} \nonumber \\
&\le& \norm{ [{\mb C_{\mb a}}_J^* {\mb C_{\mb a}}_J]^{-1} {\mb C_{\mb a}}_J^* {\mb C_{\mb a_0 - \mb a }}_J (\mb x_0)_J }{\infty} + \lambda \norm{  [{\mb C_{\mb a}}_J^* {\mb C_{\mb a}}_J]^{-1} \mb \sigma_{J\setminus T} }{\infty}  \nonumber \\ && \qquad + \;\; \norm{ [{\mb C_{\mb a}}_J^* {\mb C_{\mb a}}_J]^{-1} {\mb C_{\mb a}}_J^* {\mb C_{\mb a }}_{I\setminus J} (\mb x_0)_{I\setminus J} }{\infty} \nonumber \\
&\le& 2 \norm{{\mb C_{\mb a}}_J^* {\mb C_{\mb a_0 - \mb a }}_J}{\square\to\infty}\norm{(\mb x_0)_J}\square \;+\; 2 \lambda \;+\; 2 \norm{{\mb C_{\mb a}}_J^* {\mb C_{\mb a }}_{I\setminus J} }{\infty\to\infty}\norm{(\mb x_0)_{I\setminus J}}\infty  \nonumber \\
&\le& 2 \sqrt{2\kappa_I}\norm{\mb a-\mb a_0}2\norm{\mb x_0}{\square} \;+\; 2 \lambda  \;+\;  2\times 3\mut \times 2\kappa_{I\setminus J}\times 3\lambda  \nonumber  \\
&\le& 3 \kappa_I \| \mb a - \mb a_0 \|_2  \;+\; 2 \lambda \;+\; 36 \lambda \mut \kappa_I   \nonumber \\
&\le& 3 \lambda,
\end{eqnarray}
establishing the claim.
\end{proof}

\subsection{Least squares solution \texorpdfstring{$\mb a^{(k)}$}{kernel iterates} contracts} 
\paragraph{Approximation of least squares solution.} In this section, given $\mb x$ to be the solution to the reweighted Lasso from $\mb a$, we will show the solution of the least squares problem 
\begin{align}
	\mb a^+\;\gets\; \argmin_{\mb a'\in\R^p}\tfrac12\norm{\mb a'*\mb x\,-\,\mb y}2^2
\end{align}
is closer to $\mb a_0$ compared to $\mb a$. 
Observe that in \Cref{lem:lasso-big-entries}, the solution of \eqref{eqn:refine_lasso_sol}
\begin{align}
	\mb x \;=\; \injector_J\paren{\convmtx{\mb aJ}^*\convmtx{\mb a J}}^{-1}\injector_J^*\paren{\convmtx{\mb a}^*\convmtx{\mb a_0}\mb x_0 - \lambda\mb P_{J\setminus T}\mb\sigma},
\end{align}
by assuming $\convmtx{\mb aJ}^*\convmtx{\mb aJ} \approx \mb I$, $\,\mb a\approx\mb a_0\,$ and $J\setminus T\approx\emptyset$, is a good approximation to the true sparse map $\mb x_0$
\begin{align}
	\mb x\;\approx\; \mb I\paren{\mb x_0-\mb 0} \;=\; \mb x_0\,;
\end{align} 
furthermore, its difference to the true sparse map $\norm{\mb x_0-\mb x}2$ is proportional to $\norm{\mb a_0-\mb a}2$ as
\begin{align}
 	\mb x-\mb x_0\;\approx\; \mb P_I\paren{\convmtx{\mb a}^*\convmtx{\mb a_0}\mb x_0-\convmtx{\mb a}^*\convmtx{\mb a}\mb x_0} \;\approx\; \mb P_I\brac{\,\convmtx{\mb a_0}^*\convmtx{\mb x_0}\ip(\mb a_0-\mb a)\,}.
\end{align}
To this end, since we know the solution of least square problem $\mb a^+$ is simply
\begin{align}
	\mb a^+ \;=\;\paren{\ip^*\convmtx{\mb x}^*\convmtx{\mb x}\ip}^{-1}\paren{\ip^*\convmtx{\mb x}^*\convmtx{\mb x_0}\ip\mb a_0 }, 
\end{align}
this implies the difference between the new $\mb a^+$ and $\mb a_0$, has the relationship with $\mb a-\mb a_0$ roughly
\begin{align} 
	\mb a^+-\mb a_0 &\;=\; \paren{\ip^*\convmtx{\mb x}^*\convmtx{\mb x}\ip}^{-1}\paren{\ip^*\convmtx{\mb x}^*\convmtx{\mb x_0}\ip\mb a_0 \,-\, \ip^*\convmtx{\mb x}^*\convmtx{\mb x}\ip\mb a_0} \;\approx\; (n\theta)^{-1}\,\injector^*\convmtx{\mb x_0}^*\convmtx{\mb a_0}(\mb x_0-\mb x)\notag \\
	&\;\approx\; (n\theta)^{-1}\, \ip^*\convmtx{\mb x_0}^*\convmtx{\mb a_0}\mb P_I\convmtx{\mb a_0}^*\convmtx{\mb x_0}\ip(\mb a-\mb a_0).
\end{align}
To make this point precise, we introduce the following lemma:

\begin{lemma}[Approximation of least square estimate]\label{lem:refine_approx_least_square} Given $\mb a_0\in\R^{p_0}$ to be $\mut$-shift coherent and $\mb x_0\sim \mr{BG}(\theta)\in\R^n$. There exists some constants $C,C',c,c',c_\mu$ such that if $\lambda < c'\mut\kappa_I$,  $\mut\kappa_I^2 \leq c_\mu $ and  $n > Cp^2\log p$, then with probability at least $1-c/n$, for every $\mb a$ satisfying $\norm{\mb a-\mb a_0}2\leq\mut$ and $\mb x$ of the form 
\begin{align}
	\mb x \;=\; \injector_J\paren{\convmtx{\mb aJ}^*\convmtx{\mb a J}}^{-1}\injector_J^*\paren{\convmtx{\mb a}^*\mb y - \lambda\mb P_{J\setminus T}\mb\sigma}
\end{align} 
where the set $J,T$ satisfies $I_{ > 6\lambda}\subseteq T \subseteq J\subseteq I$, we have  
\begin{align}
	\frac{1}{n\theta }\norm{\,\ip^*\convmtx{\mb x}^*\convmtx{\mb x-\mb x_0}\ip \mb a_0 \;\;-\;\;\ip^*\convmtx{\mb x_0}^*\convmtx{\mb a_0}\mb P_I\convmtx{\mb a_0}^*\convmtx{\mb x_0}\ip(\mb a_0-\mb a)\, }2 \;\;\leq\;\;  C' \lambda \paren{\lambdat + \mut\kappa_I} + \frac{1}{32}\norm{\mb a-\mb a_0}2 
\end{align}
with $\lambdat = \lambda + \frac{\log n}{\sqrt{n\theta^2} }$. 
\end{lemma}
\begin{proof} We will begin with listing the conditions we use for both $\mb x $ and $\mb x_0$. First, we know from  \Cref{lem:lasso-big-entries} and our assumptions on the set $T$, then $\mb x$  approximates $\mb x_0$ in the sense that
\begin{align}
	\norm{\mb x-\mb x_0}\infty &\;\leq\; 3\lambda \\
	\norm{(\mb x_0)_{I\setminus J}}\infty &\;\leq\; 3\lambda \label{eqn:refine_ls_approx1_cond1} \\
	\norm{(\mb x_0)_{I\setminus T}}\infty &\;\leq\; 6\lambda. 
\end{align}
Write $\mb x_0 = \mb g\circ \mb\omega$ with $\mb g$ iid standard normal, $\mb \omega$ iid Bernoulli and $\mb g$ and $\mb \omega$ independent. From \eqref{eqn:refine_ls_approx1_cond1}  we know $\abs{I\setminus J} = \abs{\set{\,i\,|\,\abs{\mb g_i}\leq 3\lambda,\, \mb\omega_i\neq 0\,}}$. Since $\prob{\mb\omega_i \neq 0} = \theta$ and $\prob{\abs{\mb g_i}\leq 3\lambda}  \leq 3\lambda $, \Cref{lem:x0_supp} implies that with probability at least $1-2/n$:
\begin{align}	
	\abs{I\setminus J} &\;\leq\; 3\lambda n\theta + 6\sqrt{\lambda n\theta}\log n \;\leq\; 3\lambdat n\theta  \label{eqn:refine_set_i_j} \\
	\abs{I\setminus T} &\;\leq\; 6\lambda n\theta + 12\sqrt{\lambda n\theta}\log n\;\leq\; 6\lambdat  n\theta, 
\end{align}
and 
\begin{align}
	\abs{\paren{I\setminus J}\cap\shift{I}{\ell}} \leq 3\lambda n\theta^2 + 6\sqrt{\lambda n\theta^2}\log n \leq 3\lambdat n\theta^2;\label{eqn:refine_set_i_j_and_sI} 
\end{align}
together with  base on properties of Bernoulli-Gaussian vector $\mb x_0$ from \Cref{sec:basic_bg_bounds} and we conclude with probability at least $1-c/n$, all the following events hold: 
\begin{align}
\tfrac12n\theta \;\leq\; \abs{I} &\;\leq\; 2n\theta,\\
\max_{\ell\neq 0}\abs{I\cap\shift{I}{\ell}} &\;\leq\; 2n\theta^2 \\
 \max_{\ell\neq 0}\abs{(I\setminus J)\cap\shift{I}{\ell}} &\;\leq\; 6\lambdat n\theta^2,\\
\norm{\mb x_0}\square^2 & \;\leq\; \kappa_I, \\
\norm{\wc{\mb a}_0*\mb x_0}\square^2 &\;\leq\; \kappa_I, \\
\norm{\mb x_0}2^2 &\;\leq \; 2n\theta,  \\
\norm{\mb x_0}1 &\;\leq \; 2n\theta,  \\
\max_{\ell\neq 0} \|\mb P_{I\cap\shift{I}{\ell}}\mb x_0\|_2^2 &\;\leq\; 2n\theta^2, \\
 \max_{\ell\neq 0}\norm{\mb P_{I\cap\shift{I\setminus J}{\ell}}\mb x_0}1 &\;\leq\; 12\lambdat n\theta^2, \\
\norm{\convmtx{\mb x_0}\ip}2^2 &\;\leq\; 3n\theta,
\end{align} 
provided by $n \geq C\theta^{-2}\log p$ for sufficiently large constant $C$.

\vsni 1. (\ul{Approximate $\convmtx{\mb x}$ with $\convmtx{\mb x_0}$}) Since
\begin{align} 
	\ip^*\convmtx{\mb x}^*\convmtx{\mb x-\mb x_0}\ip \mb a_0 & \;=\; \ip^*\convmtx{\mb x_0}^*\convmtx{\mb x-\mb x_0}\ip \mb a_0 \;+\; \ip^*\convmtx{\mb x-\mb x_0}^*\convmtx{\mb x-\mb x_0}\ip \mb a_0 \label{eqn:refine_reduction_main1} 
\end{align}
where 
\begin{align}
	\norm{\ip^*\convmtx{\mb x-\mb x_0}^*\convmtx{\mb x-\mb x_0}\ip \mb a_0}2 &\;\leq\;  \norm{\mb a_0}2\norm{\mb x-\mb x_0}2^2 \;+\;  \norm{\convmtx{\mb a_0}\ip}2\sqrt{2p}\max_{\ell\neq 0} \abs{\innerprod{\shift{\mb x-\mb x_0}{\ell}}{\mb x-\mb x_0}}  \notag \\
	&\;\leq\;\norm{\mb x-\mb x_0}\infty^2\times \abs{I} \;+\;\sqrt{2\mut p^2}\paren{\norm{\mb x-\mb x_0}\infty^2\times \max_{\ell\neq 0}\abs{I\cap\shift{I}{\ell}}} \notag \\
	&\;\leq\;  C_1\paren{\lambda^2n\theta \;+\;  \sqrt{2\mut p^2}\paren{\lambda^2 n\theta^2} } \notag \\
	&\;\leq\;   2C_1\lambda^2 n\theta, \label{eqn:refine_reduction_main2} 
\end{align}
we have that
\begin{align} 
	\|\,\ip^*\convmtx{\mb x}^*\convmtx{\mb x-\mb x_0}\ip \mb a_0 & \;-\; \ip^*\convmtx{\mb x_0}^*\convmtx{\mb x-\mb x_0}\ip \mb a_0\,\|_2 \;\leq\;2C_1\lambda^2n\theta. \label{eqn:refine_reduction_main3}
\end{align}

\vsni 2. (\ul{Extract the $\mb a_0-\mb a$ term}) Observe that
\begin{align}
	&\ip^*\convmtx{\mb x_0}^*\convmtx{\mb x-\mb x_0}\ip \mb a_0 & \notag \\
	&=\;\ip^*\convmtx{\mb x_0}^*\convmtx{\mb a_0}(\mb x-\mb x_0) \notag \\
	&=\;  \ip^*\convmtx{\mb x_0}^*\convmtx{\mb a_0}\paren{\injector_J\paren{\convmtx{\mb aJ}^*\convmtx{\mb a J}}^{-1}\injector_J^*\paren{\convmtx{\mb a}^*\convmtx{\mb a_0}\mb x_0 - \lambda\mb P_{J\setminus T}\mb\sigma}-\injector_J\paren{\convmtx{\mb aJ}^*\convmtx{\mb a J}}^{-1}\paren{\convmtx{\mb aJ}^*\convmtx{\mb a J}}(\mb x_0)_J - \mb P_{I\setminus J}\mb x_0} \notag \\
	&=\; \ip^*\convmtx{\mb x_0}^*\convmtx{\mb a_0J}(\convmtx{\mb aJ}^*\convmtx{\mb aJ})^{-1}\convmtx{\mb aJ}^*\paren{\convmtx{\mb a_0-\mb a}\mb x_0 }\notag \\
	&\qquad +\; \ip^*\convmtx{\mb x_0}^*\convmtx{\mb a_0J}(\convmtx{\mb aJ}^*\convmtx{\mb aJ})^{-1}\convmtx{\mb aJ}^*\paren{\convmtx{\mb a}\mb x_0 - \convmtx{\mb aJ}(\mb x_0)_J} \notag \\
	&\qquad  -\; \ip^*\convmtx{\mb x_0}^*\convmtx{\mb a_0}\mb P_{I\setminus J}\mb x_0 \notag \\
	&\qquad -\; \lambda\, \ip^*\convmtx{\mb x_0}^*\convmtx{\mb a_0J}(\convmtx{\mb aJ}^*\convmtx{\mb aJ})^{-1}\injector_J^*\mb P_{J\setminus T}\mb\sigma,\label{eqn:refine_reduction_main4}    
\end{align}
where, the second term in \eqref{eqn:refine_reduction_main4} is bounded as
\begin{align}
	&\norm{\ip^*\convmtx{\mb x_0}^*\convmtx{\mb a_0J}(\convmtx{\mb aJ}^*\convmtx{\mb aJ})^{-1}\convmtx{\mb aJ}^*\paren{\convmtx{\mb a}\mb x_0 - \convmtx{\mb aJ}(\mb x_0)_J}}2 \notag \\
	&\qquad\leq\; \norm{\convmtx{\mb x_0}\ip}2\;\times\;\norm{\convmtx{\mb a_0J}}2\norm{(\convmtx{\mb aJ}^*\convmtx{\mb aJ})^{-1}}2\;\times\;\norm{\convmtx{\mb aJ}^*\convmtx{\mb aI\setminus J}}2\;\times \;\norm{(\mb x_0)_{I\setminus J}}2 \notag \\
	&\qquad\leq\;C_2\paren{\sqrt{n\theta} \;\times \;3\;\times\;  \mut\kappa_I\;\times\;\lambda\sqrt{\lambdat n\theta}}\notag \\
	&\qquad\leq\; 3C_2\mut\kappa_I\lambda n\theta; \label{eqn:refine_reduction_main4-1}     
\end{align}
the third term in \eqref{eqn:refine_reduction_main4} is bounded as
\begin{align}
	\norm{\ip^*\convmtx{\mb x_0}^*\convmtx{\mb a_0}\mb P_{I\setminus J}\mb x_0}2 & \;=\;  \norm{\ip^*\convmtx{\mb a_0}\paren{\mb P_{[\pm p]\setminus 0}+\mb e_0\mb e_0^*}\convmtx{\mb x_0}^*\mb P_{I\setminus J}\mb x_0}2\notag \\
	& \;\leq\; \norm{\mb a_0}2\norm{(\mb x_0)_{I\setminus J}}2^2 \;\;+\;\; \norm{\convmtx{
	\mb a_0}\ip}2 \times \sqrt{2 p }\times \max_{\ell\neq 0}\norm{\mb P_{I\cap\shift{I\setminus J}{\ell}}\mb x_0}1\times \norm{(\mb x_0)_{I\setminus J}}\infty \notag \\
	&\;\leq\; C_3\paren{\lambda^2\times \lambdat n\theta \;+\; \sqrt{\mut p^2}\times \lambdat n\theta^2 \times \lambda } \notag \\
	&\;\leq\; 2C_3\lambdat\lambda n\theta; \label{eqn:refine_reduction_main4-2}    
\end{align} 
and finally, write $\mb\Delta = (\convmtx{\mb aJ}^*\convmtx{\mb aJ})^{-1}-\mb I$, then the forth term in \eqref{eqn:refine_reduction_main4} is bounded as
\begin{align}
	&\lambda\norm{\ip^*\convmtx{\mb x_0}^*\convmtx{\mb a_0}\injector_J(\convmtx{\mb aJ}^*\convmtx{\mb aJ})^{-1}\injector_J^*\mb P_{J\setminus T}\mb\sigma }2 \notag \\ 
	&\qquad =\; \lambda\norm{\ip^*\convmtx{\mb a_0}\paren{\mb P_{[\pm p]\setminus 0} + \mb e_0\mb e_0^*}\convmtx{\mb x_0}^*\injector_J\paren{ \mb I +\mb\Delta }\injector_J^*\mb P_{J\setminus T}\mb\sigma  }2 \notag \\
	&\qquad \leq\; \lambda \norm{\convmtx{\mb a_0}^*\ip}2\sqrt{2p}\max_{\ell\neq 0}\norm{\mb P_{I\cap \shift{I\setminus T}{\ell} } \mb x_0 }1 \;+\; \lambda\norm{\mb a_0}2\norm{\mb P_{I\setminus T}\mb x_0}1 \notag \\
	&\qquad\quad  \;+\;\lambda \norm{\convmtx{\mb a_0}^*\injector}2\sqrt{2p}\norm{\mb P_{I\cap\shift{I}{\ell}}\mb x_0}1\norm{\mb\Delta}{\infty\to\infty}\;+\; \lambda\norm{\mb a_0}2\norm{\mb x_0}2\norm{\mb\Delta}2\sqrt{\abs{J\setminus T}}  \notag \\
	&\qquad \leq \;C_4\lambda \paren{\sqrt{\mut p^2}\times \lambdat n\theta^2 \;\;+\;\; \lambda \lambdat  n\theta \;\;+\;\; \sqrt{\mut p^2}\times n\theta^2 \times \mut\kappa_I \;\;+\;\;\sqrt{n\theta}\times\mut\kappa_I\sqrt{\lambdat n\theta} }  \notag \\
	&\qquad \leq 2C_4\paren{\lambdat + \mut\kappa_I }\lambda n\theta.\label{eqn:refine_reduction_main4-3}  
\end{align}
Therefore, combining \eqref{eqn:refine_reduction_main4-1}-\eqref{eqn:refine_reduction_main4-3} we obtain
\begin{align}
	\norm{\ip^*\convmtx{\mb x_0}^*\convmtx{\mb x-\mb x_0}\ip \mb a_0 \;-\; \ip^*\convmtx{\mb x_0}^*\convmtx{\mb a_0J}(\convmtx{\mb aJ}^*\convmtx{\mb aJ})^{-1}\convmtx{\mb aJ}^*\convmtx{\mb a_0-\mb a}\mb x_0 }2 \;\;\leq \;\; C_5\paren{\lambdat + \mut\kappa_I}\lambda n\theta.\label{eqn:refine_reduction_main5} 
\end{align}

\vsni 3. (\ul{Extract the set $J$}) Lastly, we will further simplify the term with $\mb a-\mb a_0$ in \eqref{eqn:refine_reduction_main5} by extracting the set $J$:
\begin{align}
	&\ip^*\convmtx{\mb x_0}^*\convmtx{\mb a_0J}(\convmtx{\mb aJ}^*\convmtx{\mb aJ})^{-1}\convmtx{\mb aJ}^*\convmtx{\mb a_0-\mb a}\mb x_0  \notag \\
	&\qquad =\;  \ip^*\convmtx{\mb x_0}^*\convmtx{\mb a_0J}\paren{\mb I +\mb\Delta}\convmtx{\mb a_0 + (\mb a-\mb a_0) J}^*\convmtx{\mb x_0}\ip \paren{\mb a_0-\mb a} \notag \\
	&\qquad   =\;  \ip^*\convmtx{\mb x_0}^*\convmtx{\mb a_0}\mb P_I\convmtx{\mb a_0}^*\convmtx{\mb x_0}\ip(\mb a_0-\mb a)\notag \\
	&\qquad\qquad +\;\ip^*\convmtx{\mb x_0}^*\convmtx{\mb a_0J}\mb\Delta \convmtx{\mb a_0J}^*\convmtx{\mb x_0}\ip(\mb a_0-\mb a) \;+\; \ip^*\convmtx{\mb x_0}^*\convmtx{\mb a_0J}\paren{\convmtx{\mb aJ}^*\convmtx{\mb aJ}}^{-1}\convmtx{\mb a-\mb a_0J}^*\convmtx{\mb x_0}\ip(\mb a_0-\mb a )\notag \\
	&\qquad\qquad -\;\ip^*\convmtx{\mb x_0}^*\convmtx{\mb a_0}\mb P_{I\setminus J}\convmtx{\mb a_0}^*\convmtx{\mb x_0}\ip(\mb a_0-\mb a),\label{eqn:refine_reduction_main6}
\end{align}   
where, the latter terms in \eqref{eqn:refine_reduction_main6} are bounded as 
\begin{align}
	&\norm{\ip^*\convmtx{\mb x_0}^*\convmtx{\mb a_0J}\mb\Delta \convmtx{\mb a_0J}^*\convmtx{\mb x_0}\ip}2 \;\leq\; \norm{\convmtx{\mb x_0}\ip}2^2\norm{\convmtx{\mb a_0J}}2^2\norm{\mb\Delta}2 \;\leq\; C_6\mut\kappa_I n\theta \notag \\
	&\norm{\ip^*\convmtx{\mb x_0}^*\convmtx{\mb a_0J}\paren{\convmtx{\mb aJ}^*\convmtx{\mb aJ}}^{-1}\convmtx{\mb a-\mb a_0J}^*\convmtx{\mb x_0}\ip}2 \;\le\; \norm{\convmtx{\mb x_0}\ip}2^2\norm{\convmtx{\mb a_0J}}2\norm{(\convmtx{\mb aJ}^*\convmtx{\mb aJ})^{-1}}2\norm{\convmtx{\mb a_0-\mb a}\injector_J}2 \;\leq\; C_7 \mut\sqrt{\kappa_I}n\theta\notag \\
	&\norm{\mb P_{I\setminus J}\convmtx{\mb a_0}^*\convmtx{\mb x_0}\ip }2^2 \;\leq \;  \abs{I\setminus J}\norm{\wc{\mb a}_0*\mb x_0}\square^2  \;\leq\; C_8\lambdat n\theta\times \kappa_I  \;\leq\; C_8\paren{\lambda\kappa_I + \tfrac{\kappa_I \log n}{\sqrt{n\theta^2}}}n\theta , \label{eqn:refine_reduction_main7}
\end{align}
whence we conclude, that since $c_\mu\kappa_I^2\leq c_\mu$ and $\lambda\kappa_I\leq 5c_\mu$, as long as $c_\mu < \frac{1}{100}\paren{\frac{1}{C_6}+\frac{1}{C_7} + \frac{1}{5C_8}}$ and $n >10^6 C_8^2\theta^{-2}\kappa_I^2\log^2 n$,  we gain: 
\begin{align}
	&\norm{\ip^*\convmtx{\mb x_0}^*\convmtx{\mb a_0J}(\convmtx{\mb aJ}^*\convmtx{\mb aJ})^{-1}\convmtx{\mb aJ}^*\convmtx{\mb a_0-\mb a}\mb x_0 \;-\;  \ip^*\convmtx{\mb x_0}^*\convmtx{\mb a_0}\mb P_I\convmtx{\mb a_0}^*\convmtx{\mb x_0}\ip(\mb a_0-\mb a)}2\notag \\
	&\qquad \leq\; \paren{\tfrac{3}{100} + \tfrac{1}{1000} }n\theta\norm{\mb a_0-\mb a}2 \notag \\
	&\qquad \leq\; \tfrac{1}{32}n\theta\norm{\mb a_0-\mb a}2. \label{eqn:refine_reduction_main8}
\end{align} 
The claimed result therefore is followed by combining \eqref{eqn:refine_reduction_main3}, \eqref{eqn:refine_reduction_main5} and \eqref{eqn:refine_reduction_main8}.
\end{proof}

\vspace{.1in}

\paragraph{Contraction of least square estimate of $\mb a$ toward $\mb a_0$.}   The next thing is to show the operator
\begin{align}
	(n\theta)^{-1}\paren{\ip^*\convmtx{\mb x_0}^*\convmtx{\mb a_0}\mb P_I\convmtx{\mb a_0}^*\convmtx{\mb x_0}\ip}
\end{align}
contracts $\mb a$ toward $\mb a_0$. We first will show that
\begin{align}
	(n\theta)^{-1}\paren{\ip^*\convmtx{\mb x_0}^*\convmtx{\mb a_0}\mb P_I\convmtx{\mb a_0}^*\convmtx{\mb x_0}\ip} \;\approx\;\mb a_0\mb a_0^*
\end{align}
by seeing $\ip^*\convmtx{\mb x_0}^*\mb P_I\convmtx{\mb x_0}\ip \,\approx\, (n\theta)\,\mb e_0\mb e_0^*$ via sparsity of $\mb x_0$. Finally since the local perturbation on sphere is close to a quadratic function in $\ell^2$-norm of difference, we have 
\begin{align}
	\abs{\innerprod{\mb a_0}{\mb a-\mb a_0}} \;\leq\; \tfrac12\norm{\mb a-\mb a_0}2^2.
\end{align}
Again, we introduce the following lemma to solidify our claim:

\begin{lemma}[Contraction of $\mb a$ to $\mb a_0$]\label{lem:refine_contract_a} Given $\mb a_0\in\R^{p_0}$ to be $\mut$-shift coherent and $\mb x_0\sim \mr{BG}(\theta)\in\R^n$. There exists some constants $C,C',c,c',c_\mu$ such that if $\lambda < c'\mut\kappa_I$,  $\mut\kappa_I^2 \leq c_\mu $ and $n > C \theta^{-2}p^2\log p$, then with probability at least $1-c/n$, for every $\norm{\mb a-\mb a_0}2\leq\mut$, 
\begin{align}
	\norm{\,\ip^*\convmtx{\mb x_0}^*\convmtx{\mb a_0}\mb P_I\convmtx{\mb a_0}^*\convmtx{\mb x_0}\ip(\mb a_0-\mb a)\, }2\;\leq\; \frac{1}{32}\norm{\mb a-\mb a_0}2 n\theta. 
\end{align}
\end{lemma}
\begin{proof}  Since $\E\innerprod{\mb P_I\shift{\mb x_0}{i}}{\shift{\mb x_0}{j}} = 0$ for all $i\neq j$ and set $I$, we calculate
\begin{align}
	\E\brac{\injector_{[\pm p]}^*\convmtx{\mb x_0}^*\mb P_I\convmtx{\mb x_0}\injector_{[\pm p]} } &\;=\; \sum_{i\in[\pm p]} \E\brac{\mb e_i^*\convmtx{\mb x_0}^*\mb P_I\convmtx{\mb x_0}\mb e_i}\mb e_i\mb e_i^* \;=\; \E\norm{\mb x_0}2^2\mb e_0\mb e_0^*\;+\;\sum_{i\in[\pm p]\setminus 0}\E\norm{\mb P_I\shift{\mb x_0}{i}}2^2\mb e_i\mb e_i^*\notag \\
	&\;=\; n\theta\mb e_0\mb e_0^* \,+\, n\theta^2\mb P_{[\pm p]\setminus 0} \;=\; n\theta^2\mb I\,+\,n\theta(1-\theta)\,\mb e_0\mb e_0^*.
\end{align}
whence
\begin{align}
	\E\brac{\ip^*\convmtx{\mb x_0}^*\convmtx{\mb a_0}\mb P_I\convmtx{\mb a_0}^*\convmtx{\mb x_0}\ip} &\;=\; \injector^*\convmtx{\mb a_0}^*\E\brac{\convmtx{\mb x_0}^*\mb P_I\convmtx{\mb x_0}}\convmtx{\mb a_0}\ip \;=\; n\theta^2 \ip^*\convmtx{\mb a_0}^*\convmtx{\mb a_0}\ip \,+\, n\theta(1-\theta )\mb a_0\mb a_0^*,
\end{align} 
implying the expectation  is a contraction mapping for $\mb a_0-\mb a$ when $c_\mu < \tfrac{1}{200}$:
\begin{align}
	\norm{\,\E\brac{\,\ip^*\convmtx{\mb x_0}^*\convmtx{\mb a_0}\mb P_I\convmtx{\mb a_0}^*\convmtx{\mb x_0}\ip}(\mb a_0-\mb a)\, }2 &\;\leq \; n\theta^2\norm{\ip^*\convmtx{\mb a_0}^*\convmtx{\mb a_0}\ip}2\norm{\mb a_0-\mb a}2 \,+\, n\theta \norm{\mb a_0}2\abs{\innerprod{\mb a_0}{\mb a_0-\mb a}} \notag \\
	&\;\leq\; n\theta^2 \times 2\mut p\times \norm{\mb a_0-\mb a}2 + \tfrac{1}{2}n\theta\norm{\mb a_0-\mb a}2^2\notag \\
	&\;\leq\; \paren{2c_\mu  + \tfrac{1}{2}c_\mu}\norm{\mb a_0-\mb a}2n\theta \notag \\
	&\;\leq\; \tfrac{1}{64}\norm{\mb a_0-\mb a}2 n\theta.\label{eqn:refine_contract_exp}
\end{align}
For each entry of $\convmtx{\mb x_0}^*\mb P_I\convmtx{\mb x_0}$, again from \Cref{sec:basic_bg_bounds} we know with probability at least $1-c/n$:
\begin{align}
	\abs{\mb e_i^*\convmtx{\mb x_0}^*\mb P_I\convmtx{\mb x_0}\mb e_j - \E\brac{\mb e_i^*\convmtx{\mb x_0}^*\mb P_I\convmtx{\mb x_0}\mb e_j} } &\;\leq \; \caseof{ C'\sqrt{n\theta\log n} &\; i=j=0 \\ C'\sqrt{n\theta^2\log n} &\; \text{otherwise}}\notag.
\end{align}
Thus via Gershgorin disc theorem, when $n > 10^3C'^2\theta^{-2}p^2\log n$:
\begin{align}
	\lambda_{\mr{max}}\paren{\injector^*_{[\pm p]}\convmtx{\mb x_0}^*\mb P_I\convmtx{\mb x_0}\injector_{[\pm p]} - \E\brac{\injector^*_{[\pm p]}\convmtx{\mb x_0}^*\mb P_I\convmtx{\mb x_0}\injector_{[\pm p]}} } \,\leq\, C'p\sqrt{n\theta^2\log n}\,\leq\, \tfrac{1}{64}n\theta^2. \label{eqn:refine_contract_fs}
\end{align}
Finally we combine \eqref{eqn:refine_contract_exp}, \eqref{eqn:refine_contract_fs} and get 
\begin{align}
	\norm{\,\,\ip^*\convmtx{\mb x_0}^*\convmtx{\mb a_0}\mb P_I\convmtx{\mb a_0}^*\convmtx{\mb x_0}\ip(\mb a_0-\mb a)\, }2 \;\leq\; \paren{\tfrac{1}{64}n\theta + \tfrac{1}{64}n\theta^2\norm{\convmtx{\mb a_0}\ip_{\pm p}}2^2 }\norm{\mb a_0-\mb a}2 \;\leq\;   \tfrac{1}{32}\norm{\mb a_0-\mb a}2n\theta.  
\end{align}

\end{proof}

\vsni \Cref{lem:lasso-big-entries}-\ref{lem:refine_contract_a} together implies the single iterate of alternating minimization contracts $\mb a$  toward $\mb a_0$. We show it with the following lemma:

\begin{lemma}[Contraction of least square estimate]\label{lem:refine_contract} Given $\mb a_0\in\R^{p_0}$ to be $\mut$-shift coherent and $\mb x_0\sim \mr{BG}(\theta)\in\R^n$. There exists some constants $C,C',c,c_\mu$ such that if  $\mut\kappa_I^2 \leq c_\mu $ and $n > C \theta^{-2}p^2\log n$, then with probability at least $1-c/n$, for every $\lambda$ and $\mb a$ satisfying
\begin{equation} 
5\mut\kappa_I \;\geq\; \lambda \;\geq\; 5\kappa_I\norm{\mb a-\mb a_0}2,
\end{equation}
and suppose $\mb x^+$ has the form of \eqref{eqn:refine_lasso_sol}, 
then the solution $\mb a^+$ to
\begin{equation}\label{eqn:refine_least_square}
\min_{\mb a' \in \R^p} \set{ \norm{\mb a' \ast \mb x^+ - \mb y}{2}^2 }
\end{equation} 
is unique and satisfies 
\begin{equation} \label{eqn:aplus-norm-dev}
\norm{ \mb P_{\Sp^{p-1}}\brac{\mb a^+}  - \mb a_0 }{2} \;\;\le\;\; \frac12\norm{\mb a-\mb a_0}2.
\end{equation}
\end{lemma}
\begin{proof} Write $\mb x$ as  $\mb x^+$, then
\begin{align}
\lambda_p\left( \injector^* \mb C_{\mb x}^* \mb C_{\mb x} \injector \right) &\;=\; \sigma_{\min}^2 \left( \mb C_{\mb x_0} \injector + \mb C_{\mb x - \mb x_0} \injector \right)  \nonumber \\
&\;\ge\; \Bigl[ \sigma_{\min}( \mb C_{\mb x_0} \injector ) -  \| \mb C_{\mb x - \mb x_0} \injector \|  \Bigr]_+^2 \nonumber  \\
&\;\ge\; \Bigl[ \sigma_{\min}( \mb C_{\mb x_0} \injector ) - 2\sqrt{\kappa_I}\norm{\mb x-\mb x_0}2\Bigr]_+^2 \nonumber  \\
&\;\ge\; \Bigl[ \tfrac23 \sqrt{\theta n} -  8\lambda \sqrt{\kappa_I} \sqrt{\theta n} \Bigr]_+^2 \nonumber  \\
&\;\ge\; \tfrac{1}{2} \theta n,\label{eqn:refine_ls_main1}
\end{align}
where the fourth inequality is derived from using the upper bound of sparse convolution matrix from  \Cref{rmk:convolution_x0}, and the last line holds by knowing  $\lambda < 5c_\mu\kappa_I^{-1}$. From \eqref{eqn:refine_ls_main1} we know the least square problem of \eqref{eqn:refine_least_square} has unique solution $\mb a^+$, written as 
\begin{align}  
	\mb a^+ \;=\; \paren{\injector^*\convmtx{\mb x}^*\convmtx{\mb x}\injector}^{-1}\injector\convmtx{\mb x}^*\mb y,  
\end{align} 
whence
\begin{align}\label{eqn:refine_ls_main2}
	\mb a^+ -\mb a_0 \;=\; \paren{\injector^*\convmtx{\mb x}^*\convmtx{\mb x}\injector}^{-1}\paren{\ip^*\convmtx{\mb x}^*\convmtx{\mb x_0}\ip}\mb a_0 - \mb a_0 \;=\; \paren{\injector^*\convmtx{\mb x}^*\convmtx{\mb x}\injector}^{-1}\paren{\ip^*\convmtx{\mb x}^*\convmtx{\mb x_0-\mb x} \ip}\mb a_0 . 
\end{align}
Combine \Cref{lem:refine_approx_least_square} and \Cref{lem:refine_contract_a}, we know
\begin{align}\label{eqn:refine_ls_main3}
	\norm{\,\injector^*\convmtx{\mb x}^*\convmtx{\mb x_0-\mb x}\ip\,}2 \;\leq\; \paren{ C_1\lambda \paren{\lambdat + \mut\kappa_I} + \tfrac{1}{16}\norm{\mb a-\mb a_0}2 }n\theta
\end{align}
for some constant $C_1$. Combine \eqref{eqn:refine_ls_main1}, \eqref{eqn:refine_ls_main2}, \eqref{eqn:refine_ls_main3} and since $\lambda < \mut\kappa_I$, by letting $c_\mu < \frac{1}{4C_1} $, we gain 
\begin{align}
	\norm{\mb a^+-\mb a_0}2 &\;\leq\; \frac{\norm{\injector^*\convmtx{\mb x}^*\convmtx{\mb x_0-\mb x}\injector }2 }{\lambda_p(\injector^*\convmtx{\mb x}^*\convmtx{\mb x}\injector)} \;\leq\; 2C_1\lambda\paren{\lambdat+\mut\kappa_I} + \frac18\norm{\mb a-\mb a_0}2 \;\leq\;\frac14.
\end{align}
For the final bound,
\begin{align}
\norm{\frac{\mb a^+}{\| \mb a^+ \|_2} - \mb a_0 }{2} &\;\le\; \frac{\norm{\mb a^+-\mb a_0}2 + \abs{\norm{\mb a^+}2-1}}{\norm{\mb a^+}2}  \;\le\;  \frac{2\norm{\mb a^+-\mb a_0}2}{1-\norm{\mb a^+-\mb a_0}2} \; \leq \;   \frac{8}{3}\norm{\mb a^+-\mb a_0}2, \notag \\
&\;\leq\;C_2\lambda \paren{\lambdat+\mut\kappa_I}\;+\;\frac13\norm{\mb a-\mb a_0}2,\label{eqn:contract_least_square_main}
\end{align} 
and since $\lambda > \kappa_I\norm{\mb a-\mb a_0}2$, finally we gain 
 \begin{align}
 	\eqref{eqn:contract_least_square_main}&\;\leq\; C_2\paren{\lambda\kappa_I+\frac{p\kappa_I\log n}{n\theta}+\mut\kappa_I^2}\norm{\mb a-\mb a_0}2 \;+\; \frac13\norm{\mb a-\mb a_0}2 \notag \\
 	&\;\leq\; \frac{1}{2}\norm{\mb a-\mb a_0}2 
 \end{align}
 as long as $n > 20C_2\theta^{-1}p\kappa_I\log n$ and $c_\mu < \frac{1}{20C_2}$. 

\end{proof}

\subsection{Linear convergence of alternating minimization (Proof of \Cref{thm:altmin} )}\label{sec:proof_altmin}
In the first two sections we have shown the iterate contract $\mb a$ toward $\mb a_0$,  under our signal assumption. We tie up these result by  showing the following theorem which proves that the iterates produced by alternating minimization converge linearly to $\mb a_0$:

\begin{proof} 
We will prove our claim by induction on $k$. Clearly, when $k=0$, we have $5\,\kappa_I\norm{\mb a^{(0)}-\mb a_0}2 \leq \lambda^{(0)} = 5\mut\kappa_I$ and $I^{(0)} = \set{i : \abs{\shift{\mb a^{(0)}}{i}^*\ip^*\convmtx{\mb a_0}\mb x_0} > \lambda^{(0)} }$. Then for all $\abs{\mb x_j}> 6\lambda^{(0)}$, we have
\begin{align}
	\abs{ s_j\big[\mb a^{(0)}\big]^*\convmtx{\mb a_0}\mb x_0} &\;\geq \; \paren{1-\big|\langle{\mb a^{(0)}}{\mb a_0}\rangle\big|}\abs{\mb x_j} \;-\; \norm{\mb P_{[\pm p]\setminus\set j}\convmtx{\mb a_0}^*\ip s_j\big[\mb a^{(0)}\big]}2 \times \sqrt 2\norm{\mb x_0}\square \notag \\
	&\;\geq\;(1-2\mut)\,6\lambda^{(0)} \;-\; 2\mut\sqrt{\kappa_I}\times\sqrt{2\kappa_I} \notag \\
	&\;\geq\; 5\lambda^{(0)} \;-\; 4\lambda^{(0)}\notag \\
	&\;=\; \lambda^{(0)}.  
\end{align}
hence $I_{>6\lambda^{(0)}}\subseteq I^{(0)}$, therefore the condition of  \Cref{lem:refine_contract} is satisfied,  implies \eqref{eqn:refine_main_bound} holds for $k=0$.  

Suppose it is true for $1,2,\dots, k-1$, such that 
\begin{equation}
\kappa_I\big\|\mb a^{(k)}-\mb a_0\big\|_2 \;\leq\;\tfrac12\lambda^{(k-1)} \;=\;\lambda^{(k)},\qquad\text{and}\qquad  I_{>3\lambda^{(k-1)}}\subseteq I^{(k)}
\end{equation}
and since $I_{>6\lambda^{(k)}} = I_{>3\lambda^{(k-1)}}\subseteq I^{(k)}$, we can again apply \Cref{lem:refine_contract}, resulting 
\begin{equation}
	\kappa_I\big\|\mb a^{(k+1)}-\mb a\big\|_2 \;\leq\; \tfrac{1}{2} \kappa_I\big\|\mb a^{(k)}-\mb a_0\big\|_2 \;\leq\;\tfrac{1}{2}\lambda^{(k)}
\end{equation}
as claimed. 
\end{proof}

\subsection{Supporting lemmas for refinement} \label{sec:appx-supp-lemma-refine}   
The following lemma controls the shift coherence of $\mb a$: 

 \begin{lemma}[Coherence of $\mb a$ near $\mb a_0$ ] Suppose that $\mb a_0$ is $\mut$-shift coherent, and $\norm{\mb a - \mb a_0 }{2} \le \mut$. Then 
 \begin{align}
 \norm{ \mr{off}\left[ \mb C_{\mb a}^* \mb C_{\mb a_0 } \right] }{ \infty } &\le 2 \mut \\
  \norm{ \mr{off}\left[ \mb C_{\mb a}^* \mb C_{\mb a } \right] }{ \infty } &\le 3 \mut
 \end{align} 
 \end{lemma}
 \begin{proof}
Notice that for any $\ell \ne 0$, $| \innerprod{ \mb a }{ s_\ell[\mb a_0 ] } | \le | \innerprod{ \mb a_0 }{ s_{\ell}[\mb a_0] } | + | \innerprod{ \mb a - \mb a_0 }{s_{\ell}[\mb a_0]} | \le \mut + \| \mb a_0 - \mb a \|_2 \le 2 \mut$. 
Similarly, $| \innerprod{ \mb a}{ s_\ell[\mb a ] } | \le | \innerprod{ \mb a - \mb a_0 }{ s_\ell[\mb a_0 ] } | + | \innerprod{ \mb a }{ s_\ell[\mb a_0 ] } |  \le \|\mb a- \mb a_0\|_2 + 2 \mut \le 3 \mut,$ as claimed.
 \end{proof}

\vsni From this we obtain the following spectral control on $\mb C_{\mb a}^* \mb C_{\mb a}$, to simply the notations, we will write
\begin{align}
	\convmtx{\mb aI}^*\convmtx{\mb aI} = \injector_I^*\convmtx{\mb a}^*\convmtx{\mb a}\injector_I = [\convmtx{\mb a}^*\convmtx{\mb a}]_{I,I}
\end{align}
in the latter part of this section.

 \begin{lemma}[Off-diagonals of {$[\convmtx{\mb a}^*\convmtx{\mb a}]_{I,I}$} ] \label{lem:ca-spectral} 
 Suppose that $\mb a_0$ is $\mut$-shift coherent and $\norm{\mb a - \mb a_0}{2} \le \mut$.  Then 
 \begin{equation}
 \norm{ \left[ \mb C_{\mb a}^* \mb C_{\mb a} - \mb I \right]_{I,I} }{2 } \le 9 \kappa_I \mut.
 \end{equation}
 \end{lemma}
 
 \noindent We prove this lemma by noting that $\mb C_{\mb a}^* \mb C_{\mb a} = \mb C_{\mb r_{\mb a, \mb a}}$ is the convolution matrix associated with the autocorrelation $\mb r_{\mb a, \mb a}$ of $\mb a$. Since $\mr{supp}( \mb r_{\mb a, \mb a} ) \subseteq \set{-p+1,\dots,p-1}$ is confined to a (cyclic) stripe of width $2p-1$, we can tightly control the norm of this matrix by dividing it into three block-diagonal submatrices with blocks of size $p\times p$. Formally:
 
 \vspace{.1in}
 
 \begin{proof}
Divide $I$ into $r = \lceil n/p \rceil$ subsets $I_0, \dots, I_{r-1}$ such that for all $\ell = 0,\ldots,r-1$:
\begin{align}
I_\ell = I \cap\set{p \ell,\,p \ell+1,\,\ldots,\,p\ell+(p-1)} = I\cap([p]+p\ell).\notag
\end{align}
Notice that for each $\ell$:
\begin{equation}
\mr{supp}\left( [\mb C_{\mb a}^* \mb C_{\mb a} ]_{I_\ell,I} \right) \subseteq I_\ell \times \Bigl( I_{\ell-1} \uplus I_{\ell} \uplus I_{\ell+1} \Bigr),\notag
\end{equation}
where $\ell+1$ and $\ell-1$ are interpreted cyclically modulo $r$. 

For an arbitrary $\mb v \in \R^{|I|}$, we calculate
 \begin{align}
 \norm{ \left[ \mb C_{\mb a}^* \mb C_{\mb a} - \mb I \right]_{I,I} \mb v }{2}^2 &= \sum_{\ell = 0}^{ r-1 } \norm{ \left[ \mb C_{\mb a}^* \mb C_{\mb a} - \mb I \right]_{I_{\ell}, I} \mb v }{2}^2  \\
&= \sum_{\ell = 0}^{ r-1 } \norm{ \left[ \mb C_{\mb a}^* \mb C_{\mb a} - \mb I \right]_{I_{\ell}, I_{\ell-1} \uplus I_{\ell} \uplus I_{\ell+1} } \mb v_{I_{\ell-1} \uplus I_{\ell} \uplus I_{\ell+1}} }{2}^2  \\
&\le \sum_{\ell = 0}^{r-1} \norm{ \left[ \mb C_{\mb a}^* \mb C_{\mb a} - \mb I \right]_{I_{\ell}, I_{\ell-1} \uplus I_{\ell} \uplus I_{\ell+1} } }{F}^2 \norm{ \mb v_{I_{\ell-1} \uplus I_{\ell} \uplus I_{\ell+1}} }{2}^2 \\
&\le 3 \kappa_I^2\times \paren{3\mut}^2\times \sum_{\ell = 0}^{r-1} \norm{ \mb v_{I_{\ell-1} \uplus I_{\ell} \uplus I_{\ell+1}} }{2}^2 \\
&\le 3 \kappa_I^2 \times 9 \mut^2 \times 3 \norm{\mb v}{2}^2,
\end{align}
giving the claimed result. 
 \end{proof}  
 
 \vsni As a consequence, we have that 
 \begin{corollary}[Inverse of {$[\convmtx{\mb a}^*\convmtx{\mb a}]_{J,J}$}] \label{cor:ca-sub-spectral} Suppose that $\mb a_0$ is $\mu$-shift coherent, that $\norm{\mb a - \mb a_0}{2} \le \mut$ and that $\kappa_I \mut < \tfrac{1}{18}$. Then for every $J \subseteq I$ and any norm $\norm{\cdot}\diamondsuit \in \set{\,\norm{\cdot}{\square\to\square},\,\norm{\cdot}{\infty\to\infty},\,\norm{\cdot}2\,}$, we have 
 \begin{align}
   \norm{ \left[ \mb C_{\mb a}^* \mb C_{\mb a} - \mb I \right]_{J,J} }{\diamondsuit} &\;\le\; 9 \kappa_I \mut\\
     \norm{ \left[ \mb C_{\mb a}^* \mb C_{\mb a} \right]_{J,J}^{-1} - \mb I }{\diamondsuit} &\;\le\; 18 \kappa_I \mut\\
    \norm{ \left[ \mb C_{\mb a}^* \mb C_{\mb a} \right]_{J,J}^{-1} }{\diamondsuit} &\;\le\;2. \label{eqn:ca-j-inv-spectral}
 \end{align}  
 \end{corollary}
 \begin{proof} First we prove
 \begin{align}
 	  \norm{ \left[ \mb C_{\mb a}^* \mb C_{\mb a} - \mb I \right]_{J,J} }{2 } \le 9 \kappa_I \mut,\quad \norm{ \left[ \mb C_{\mb a}^* \mb C_{\mb a} - \mb I \right]_{J,J} }{\infty \to \infty } \le 6 \kappa_I \mut, \quad \norm{ \left[ \mb C_{\mb a}^* \mb C_{\mb a} - \mb I \right]_{J,J} }{\square \to \square } \le 6\kappa_I \mut 
\end{align}
Where the first claim follows from  \Cref{lem:ca-spectral}. The second follows by noting that the $\ell^\infty$ operator norm is the maximum row $\ell^1$ norm, and that each row has at most $2 \kappa_I$ entries, of size at most $3 \mut$. The last follows by noting that  
\begin{align}
\norm{ \left[ \mb C_{\mb a}^* \mb C_{\mb a} - \mb I \right]_{J,J} }{\square \to \square} &\le  \max_{\ell,\ell'} \norm{ \left[ \mb C_{\mb a}^* \mb C_{\mb a} - \mb I \right]_{J \cap([p]+\ell),\,J \cap([2p]+\ell')} }F \nonumber \\ &\le 6\kappa_I \mut.  
\end{align}
Then we prove 
\begin{align}
	\norm{ \left[ \mb C_{\mb a}^* \mb C_{\mb a} \right]_{J,J}^{-1} - \mb I }{2} \le 18 \kappa_I \mut,\quad \norm{ \left[ \mb C_{\mb a}^* \mb C_{\mb a} \right]_{J,J}^{-1} - \mb I }{\infty \to \infty } \le 12 \kappa_I \mut,\quad \norm{ \left[ \mb C_{\mb a}^* \mb C_{\mb a}  \right]_{J,J}^{-1} - \mb I }{\square \to \square } \le 12\kappa_I \mut, 
\end{align}
which are followed from the fact that if $\| \cdot \|_\diamondsuit$ is a matrix norm and $\norm{\mb \Delta}{\diamondsuit} < 1$, then $$\norm{ (\mb I + \mb \Delta)^{-1} - \mb I }{\diamondsuit} \le \frac{ \norm{\mb \Delta }{\diamondsuit} }{ 1 - \norm{\mb \Delta }{\diamondsuit} }.$$
Finally, \eqref{eqn:ca-j-inv-spectral} follows from the triangle inequality. 
\end{proof}

\vsni Also, we need to bound the convolution of $\mb a_0-\mb a$ with $\norm{\mb a_0-\mb a}2$ requiring for bounds of the lasso solution:
\begin{lemma}[Convolution of $\mb a_0-\mb a$]\label{lem:refine_conv_a0-a}  Suppose that $\mb a_0$ is $\mu$-shift coherent and $\norm{\mb a - \mb a_0}{2} \le \mut$, then for every $J\subseteq I$,   
\begin{align}
	\norm{[\convmtx{\mb a}^*\convmtx{\mb a_0-\mb a}]_{J,J}}{\square\to\infty} &\;\leq\;  \sqrt{2\kappa_I}\norm{\mb a-\mb a_0}2 \\
	\norm{[\convmtx{\mb a}^*\convmtx{\mb a_0-\mb a}]_{J,J}}{\square\to\square} &\;\leq\;  \sqrt2\kappa_I\norm{\mb a-\mb a_0}2
\end{align}
\end{lemma}
\begin{proof} For the first inequality, we have
\begin{align}
	\norm{[\convmtx{\mb a}^*\convmtx{\mb a_0-\mb a}]_{J,J}\mb v}{\square\to\infty} &= \max_{j\in J,\,\norm{\mb v}\square = 1}\abs{\innerprod{\shift{\mb a}{j}}{(\mb a_0-\mb a)*\mb v}} \notag \\
	&\leq \max_{j\in[n],\,\norm{\mb v}\square = 1}\norm{\mb P_{[p]+j}\brac{(\mb a_0-\mb a)*\mb v}}2 \notag \\
	&\leq  \norm{\mb a-\mb a_0}2\times \max_{j\in[n],\,\norm{\mb v}\square = 1} \norm{\mb P_{[\pm p]+j}\mb v}1 \notag \\
	&\leq \sqrt{2\kappa_I}\norm{\mb a_0-\mb a}2 
\end{align}
The second inequality is derived by 
\begin{align}
	\norm{[\convmtx{\mb a}^*\convmtx{\mb a_0-\mb a}]_{J,J}}{\square\to\square} &\leq \max_{\ell,\ell'} \norm{[\convmtx{\mb a}^*\convmtx{\mb a_0-\mb a}]_{J\cap([p]+\ell),J\cap([2p]+\ell')}}F  \notag \\
	&\leq \sqrt{2\kappa_I^2\textstyle\max_{i,j}\abs{\innerprod{\shift{\mb a}{i}}{\shift{\mb a_0-\mb a}{j}}}^2} \notag \\
	&\leq  \sqrt 2\kappa_I\norm{\mb a-\mb a_0}2,    
\end{align}
finishing the proof.
\end{proof}

\vsni Again, using a variant of the argument for \Cref{lem:ca-spectral}, we have the following: 
 
 \begin{lemma}[Off-diagonal of submatrix of $\convmtx{\mb a}^*\convmtx{\mb a_0}$]\label{lem:ca-spectral-off} Suppose that $\mb a_0$ is $\mu$-shift coherent and $\norm{\mb a - \mb a_0}{2} \le \mut$. For any $J \subset I$, if 
 \begin{align}
 \kappa_J &\;=\; \max_{\ell} \left| J \cap \set{ \ell, \ell + 1, \dots, \ell + p - 1} \right| \\
 \kappa_{I\setminus J} &\;=\; \max_{\ell} \left| (I \setminus J) \cap \set{ \ell, \ell + 1, \dots, \ell + p - 1} \right|
 \end{align}
 Then 
 \begin{equation} 
 \norm{ \left[ \mb C_{\mb a}^* \mb C_{\mb a_0} \right]_{J,I \setminus J} }{2} \;\le\; 6 \sqrt{\kappa_J \kappa_{I\setminus J} } \mut. 
 \end{equation} 
 \end{lemma} 
 \begin{proof}
Take $r = \lceil n/p \rceil$ and for $\ell=0,\ldots,r-1$, write
\begin{align}
J_\ell = J \cap([p]+p\ell), \qquad L_\ell = (I\setminus J) \cap ([p]+p\ell),  \nonumber 
\end{align}
Take $\mb v \in \R^{|I \setminus J|}$ arbitrary and notice that 
\begin{align}
\norm{ \left[ \mb C_{\mb a}^* \mb C_{\mb a_0} \right]_{J,I\setminus J} \mb v}{2}^2 &= \sum_{\ell = 0}^{r-1} \norm{ \left[ \mb C_{\mb a}^* \mb C_{\mb a_0} \right]_{J_\ell,I\setminus J} \mb v}{2}^2 \nonumber \\
&= \sum_{\ell = 0}^{r-1} \norm{ \left[ \mb C_{\mb a}^* \mb C_{\mb a_0} \right]_{J_\ell,L_{\ell-1} \cup L_\ell \cup L_{\ell+1}} \mb v_{L_{\ell-1} \cup L_\ell \cup L_{\ell+1}}}{2}^2 \nonumber \\
&\le 4 \mut^2 \times \kappa_J \times 3 \kappa_{I \setminus J} \times \sum_{\ell = 0}^{r-1} \norm{ \mb v_{L_{\ell-1} \cup L_\ell \cup L_{\ell+1}}}{2}^2 \nonumber \\
&\le 4 \mut^2 \times \kappa_J \times 3 \kappa_{I \setminus J} \times 3 \| \mb v \|_2^2,
\end{align}
giving the result. 
 \end{proof}

\begin{lemma}[Perturbation of vector over sphere] If both $\mb a,\mb a_0$ are unit vectors in inner product space, then 
\begin{align}
	\abs{\innerprod{\mb a}{\mb a-\mb a_0}} \;\leq\;\tfrac{1}{2}\norm{\mb a-\mb a_0}2^2.
\end{align}
\end{lemma}
\begin{proof} Via simple norm inequalities:
\begin{align}
	\tfrac12\norm{\mb a-\mb a_0}2^2 \;=\; 1-\innerprod{\mb a}{\mb a_0} \;=\; 1-\innerprod{\mb a}{\mb a_0-\mb a+\mb a} \; =\; \innerprod{\mb a}{\mb a-\mb a_0} \;>\; 0 
\end{align}
\end{proof}

\begin{lemma}[Convolution of short and sparse] Suppose $\mb\delta\in\R^{p}$, and $\mb v\in\R^n$ where $\supp(\mb v) = I$ satisfies
\begin{align}
	\max_{\ell\in [n]}\abs{\,I\cap\paren{[p]+\ell}\,} \;\leq\;\kappa
\end{align}
then 
\begin{align}
	\norm{\mb \delta * \mb v}2\;\leq\;\sqrt{2\kappa}\norm{\mb\delta}2\norm{\mb v}2 
\end{align}
\end{lemma} 
\begin{proof} Since every $p$-contiguous segment of $I$ has at most $\kappa$ elements, by splitting $I = I_1 \uplus I_2 \uplus ,\ldots,\uplus I_{\kappa}\uplus R $ such that each sets $I_i$  are $p$-separated:
\begin{align}
I_1 &\;=\; \set{ i_1, i_{\kappa+1}, i_{2\kappa+1}, \dots } \cap \set{ 0, \dots, n-p-1}, \nonumber \\
I_2 &\;=\; \set{ i_2, i_{\kappa+2}, i_{2\kappa+2}, \dots } \cap \set{ 0, \dots, n-p-1}, \nonumber \\
&\;\;\vdots \nonumber \\
I_{\kappa} &\;=\; \set{ i_{\kappa}, i_{2\kappa}, i_{3\kappa}, \dots } \cap \set{ 0, \dots, n - p -1}, \\
R &\;=\; I \cap \set{ n-p, \dots, n-1 }.
\end{align}
Then the p-separating property gives $\norm{\mb\delta*\mb P_{I_i}\mb v}2 = \norm{\mb\delta}2\norm{\mb P_{I_i}\mb v}2$. Hence:
\begin{align}
	\norm{\mb \delta*\mb P_I\mb v}2 &\;=\; \norm{\sum_{i\in\kappa}\mb \delta*\mb P_{I_i}\mb v + \mb\delta *\mb P_R\mb v }2 \;\leq\; \sum_{i\in\kappa}\norm{\mb \delta*\mb P_{I_i}\mb v}2 + \norm{\mb\delta*\mb P_R\mb v}\notag \\
	& \;=\; \norm{\mb \delta}2\sum_{i\in\kappa}\norm{\mb v_{I_i}}2 \;+\; \norm{\mb\delta}2\norm{\mb P_R\mb v}1   \notag \\
	&\;\leq\; \sqrt{\kappa}\norm{\mb v_{I_1,\uplus,\ldots,\uplus I_{\kappa}}}2\norm{\mb \delta}2 \;+\; \sqrt{\kappa}\norm{\mb v_R}2\norm{\mb\delta}2 \notag \\
	&\;\leq\; \sqrt{2\kappa}\norm{\mb v}2\norm{\mb\delta}2,
\end{align}
where the last two inequalities were coming from Cauchy-Schwartz.
\end{proof}
% !TEX root = ../../BD_DQ.tex

\section{Finite sample approximation}
In this section we collect several major components of proof about large sample deviation. In particular, the concentration for shift space gradient $\mb\chi(\mb\beta)_i$, shift space Hessian diagonals $ \norm{\mb P_{I(\mb a)}\shift{\mb x_0}{-i}}2$, and the set of gradients discontinuity entries $\abs{J_B(\mb a)}$.

\subsection{Proof of \Cref{cor:chibeta_ct}}\label{sec:proof_chibeta_ct} 
 
\begin{proof} 1. (\ul{$\eps$-net})  Write $\mb x$ as $\mb x_0$ and $\norm{\mb\beta}2 = \eta$ through out this proof, firstly from \Cref{asm:theta_mu} for every $\mb a\in\cup_{\abs{\mb\tau}\leq k}\goodregion$, we know $\eta \leq 1+c_\mu+\frac{c_\mu}{\sqrt\theta k\log\theta^{-1}}\leq \sqrt p$.  Define $\eps = \frac{c_2}{2n^{3/2}p^{3/2}} $ and consider the $\eps$-net $\epsnet$ for sphere of radius $\eta$. From \Cref{lem:epsnet} we know for any $c_2 < 1$:
\begin{align}
	\abs{\epsnet} \leq \paren{\frac{3\eta}{\eps} }^{2p} \leq \paren{\frac{3n^{3/2}p^2}{c_2}}^{2p} \leq  \paren{\frac{3n p^2}{c_2}}^{3p}   
\end{align}
for each $i\in[n]$ define such net as $\mc N_{\eps,i}$, and define an event such that all center of subsets in $\mc N_{\eps,i}$ are being well-behaved: 
\begin{align}\label{eqn:chi_ct_epsnet}
	\eventnet := \set{\forall\,i\in[n],\quad \mb \sigma_in^{-1}\mb\chi[\mb\beta_\eps]_i- \mb \sigma_in^{-1}\ol{\E \mb\chi[\mb\beta_\eps]}_i < \frac{c_1\theta}{p^{3/2}}  \quad \forall\,\mb \beta_\eps\in\mc N_{\eps,i},}
\end{align}   

\vsni 2. (\ul{Lipschitz constant}) The Lipschitz constant $L$ of $\mb\chi[\cdot]_i$ w.r.t $\mb\beta$ is bounded in terms of $\mb x$ regardless of entry $i$: 
\begin{align}
	\abs{\mb\chi[\mb\beta]_i - \mb\chi[\mb\beta']_i}&\leq  \abs{\mb  e_i^*\checkmtx{\mb x}\soft{\checkmtx{\mb x}\mb\beta}{\lambda} - \mb e_i^*\checkmtx{\mb x}\soft{\checkmtx{\mb x}\mb\beta'}{\lambda}}  \leq \norm{\mb x}2\norm{\soft{\checkmtx{\mb x}\mb\beta}{\lambda} - \soft{\checkmtx{\mb x}\mb\beta'}{\lambda}   }2 \notag \\
	&\leq \norm{\mb x}2\sqrt{\sum_{j\in[n]}\abs{\soft{\checkmtx{\mb x}\mb\beta}{\lambda}_j - \soft{\checkmtx{\mb x}\mb\beta'}{\lambda}_j}^2} \leq \norm{\mb x}2\norm{\checkmtx{\mb x}\mb\beta - \checkmtx{\mb x}\mb\beta'}2 \notag \\
	&\leq \norm{\mb x}2\cdot\norm{\mb x}1\cdot\norm{\mb \beta-\mb\beta'}2 =: L\norm{\mb\beta-\mb\beta'}2
\end{align}
Define the  event that $\mb \chi[\mb\beta]_i$ that has small Lipschitz constant as 
\begin{align}
	\eventlip := \set{L < 2n^{3/2}\theta}
\end{align}
on the event $\eventlip$, for every points in $\goodregion$ and $i\in[n]$, there exists some $\mb \beta_\eps\in\mc N_{\eps,i}$  such that
\begin{align} \label{eqn:chi_ct_lip}
	\abs{\paren{\mb \sigma_i n^{-1}\mb\chi[\mb\beta]_i  - \mb \sigma_i n^{-1}\ol{\E\mb\chi[\mb\beta]}_i} - \paren{\mb \sigma_in^{-1}\mb\chi[\mb\beta_\eps]_i - \mb \sigma_in^{-1}\ol{\E\mb\chi[\mb\beta_\eps]}_i } } &\leq 2L\eps \leq \frac{c_2\theta}{p^{3/2}}
\end{align}  
On event $\eventlip\cap\eventnet$, \eqref{eqn:chi_ct_epsnet}, \eqref{eqn:chi_ct_lip} implies $\mb\chi[\mb\beta]$ is well concentrated entrywise and  anywhere in $\cup_{\abs{\mb\tau}\leq k}\goodregion$:
\begin{align}
	\abs{\mb \sigma_i n^{-1}\mb\chi[\mb\beta]_i - \mb \sigma_i n^{-1}\ol{\E\mb\chi[\mb\beta]}_i} \leq \frac{(c_1 + c_2)\theta}{p^{3/2}}, \quad \forall\,\mb a\in\cup_{k\leq k}\goodregion,\;\forall\,i\in[n] 
\end{align}  
as desired, where, using \Cref{lem:x0_bound}, 
\begin{align}
	\prob{\eventlip^c} \leq \prob{\norm{\mb x}2^2 > 2n\theta} \leq 1/n;
\end{align}
and using union bound,
\begin{align}
	\prob{\eventnet^c} &\leq  \prob{\max_{\substack{\mb a_\eps\in \mc N_{\eps,i}\\ i\in[n]}}\mb \sigma_i n^{-1}\mb\chi[\mb\beta_\eps]_i - \mb \sigma_i n^{-1}\ol{\E\mb\chi[\mb\beta_\eps]}_i>\frac{c_1\theta}{p^{3/2}} } \notag \\
	&\leq n\abs{\mc N_\eps}\prob{\mb \sigma_0n^{-1}\mb\chi[\mb\beta_\eps]_0 - \mb \sigma_0 n^{-1}\E\mb\chi[\mb\beta_\eps]_0 > \frac{c_1\theta}{p^{3/2}}}.\label{eqn:chi_ct_epsnet_be_ub1} 
\end{align}

\vsni 3. (\ul{ Bound $\prob{\eventnet^c}$}) Wlog write $n = t\cdot(2p)$ for some integer $t$ and $2p \geq 4p_0-3$ and replace $\mb x_0$ with $\mb x$. Observe that $\mb Z_j(\mb\beta)$ from \eqref{eqn:chibeta_Zj} is independent of $\mb Z_{j+2p}(\mb\beta)$ for all $j\in[n]$ while all $\mb Z_j$ are identical distributed. We write $\mb\chi[\mb\beta]_0$ as sum of iid r.v.s. as
\begin{align}
	\mb\chi[\mb\beta]_0 = \sum_{j\in[n]} \mb Z_j(\mb\beta) = \sum_{k\in[2p]} \paren{\sum_{t=0}^{n/2p - 1}\mb Z_{k+2tp}(\mb\beta) } \notag
\end{align}
wlog let $\mb\sigma_0 =1$ and split the independent r.v.s, write $\E \mb Z_0 = \E \mb Z$,  bound the tail probability of $\mb\chi[\mb\beta]_0$ as 
\begin{align} 
	\prob{ n^{-1}\mb\chi[\mb\beta]_0 > n^{-1}\ol{\E \mb\chi(\mb\beta)}_0 + \frac{c_1\theta}{p^{3/2}} } \leq 2p\cdot \prob{\sum_{t=0}^{n/2p-1}\mb Z_{2tp}(\mb\beta) > \frac{n}{2p}\E \mb Z(\mb\beta) + \frac{c_1n\theta}{2p^{5/2}}  }  \label{eqn:chi_ct_epsnet_be_ub2} 
\end{align}
The moments of $\mb Z_0$ can be bounded by using $\abs{\mb Z_0(\mb\beta)} \leq\abs{\mb x_0}\abs{\mb \beta_0 \mb x_0+\mb s_0} \leq \mb \beta_0 \mb x_0^2 + \abs{\mb x_0}\abs{\mb s_0}$ where $\mb s_0 = \sum_{\ell\neq 0}\mb x_\ell\mb \beta_\ell $, write $\mb x = \mb \omega \circ \mb g \simiid \mr{BG}(\theta)$. For the 2-norm we know
\begin{align}
		\E\abs{\mb s_0}^2  &=  \E\abs{\sum_\ell \mb x_\ell\mb \beta_\ell}^2 \leq  \theta\norm{\mb\beta}2^2 \leq \theta\paren{1+c_\mu + \frac{c_\mu}{\theta k^2}}\leq \frac12\label{eqn:ct_chi_exp_s02}  
\end{align}
As for the $q$-norm, use the moment generating function bound, such that for all $t\geq 0$:
\begin{align}
		\E\abs{\mb s_0}^q &\leq  q!t^{-q}\E \exp\brac{t\abs{\mb s_0}} \leq q!t^{-q}\prod_{\ell}\E_{\mb\omega_\ell,\mb g_\ell}\exp\brac{t\mb \omega_\ell \abs{\mb g_\ell} \abs{\mb\beta_\ell}} \leq 2q! t^{-q}\prod_\ell\E_{\mb \omega_\ell} \exp\brac{\mb\omega_\ell t^2\mb\beta^2_\ell/2} \notag \\
		&\leq 2q!t^{-q}\prod_\ell\paren{1 -\theta + \theta\exp\brac{ t^2\mb\beta_\ell^2/2}}\label{eqn:chi_ct_Es0q_1}
\end{align}
notice that the entrywise twice derivative of \eqref{eqn:chi_ct_Es0q_1} w.r.t. $\mb\beta^2_\ell$'s are always positive, this function is convex for all $\beta_\ell^2$. Constrain on the polytope $\sum_\ell \mb\beta_\ell^2 \leq \norm{\mb\beta}2^2$, the maximizer of \eqref{eqn:chi_ct_Es0q_1} w.r.t. $\mb\beta^2_\ell$'s occurs and a vertex point where $\mb\beta_0^2 = \norm{\mb\beta}2^2$. Thus
\begin{align}
	\eqref{eqn:chi_ct_Es0q_1} \leq 2q! t^{-q}\paren{1-\theta + \theta\exp\brac{t^2\norm{\mb\beta}2^2/2}}\prod_{\ell\neq 0}(1-\theta + \theta e^0) \leq  2q!t^{-q}(1 + \theta\exp[\norm{\mb\beta}2^2t^2/2]). \notag
\end{align}
Choose  $t = \sqrt q /\norm{\mb\beta}2$, use $q!! > \paren{q!/2}\cdot \paren{e/q}^{q/2}$, we have
\begin{align}
	 	\E\abs{\mb s_0}^q  &\leq 2q!q^{-q/2}\norm{\mb\beta}2^q\paren{1+\theta\exp\brac{q/2}} \leq 8\norm{\mb\beta}2^q \max\set{e^{-q/2},\theta}q!!. \label{eqn:ct_chi_exp_s0q}
\end{align} 
Apply Jensen's inequality $\paren{\sum_{i=1}^{N}\mb z_i}^q \leq N^{q-1}\sum_{i=1}^{N}\mb z_i^q$, use Gaussian moment \Cref{lem:gaussian_moment} , \eqref{eqn:ct_chi_exp_s02} and \eqref{eqn:ct_chi_exp_s0q}, obtain  for $q\geq 3$,
\begin{align}
	\E Z(\mb\beta)^2 &\leq  \E \paren{\mb\beta_0 \mb x_0^2 + \abs{\mb x_0}\abs{\mb s_0}}^2 \leq 2\E\brac{\mb \beta_0^2\mb x_0^4 + \mb x_0^2 \mb s_0^2}\leq 6\theta +2\theta^2\norm{\mb\beta}2^2 \leq 7\theta,  \notag \\
	\E \mb Z(\mb\beta)^q & \leq \E\paren{\mb \beta_0 \mb x_0^2 + \abs{\mb x_0}\abs{\mb s_0}}^q \leq 2^{q-1}\paren{\E \mb x_0^{2q} + \E\abs{\mb x_0}^q\E \abs{\mb s_0}^q} \notag \\
	&\leq \theta 2^{q-1}(2q-1)!! +  \theta 2^{q-1}(q-1)!! \paren{8\norm{\mb\beta}2^q\max\set{e^{-q/2},\theta}q!!} \notag \\
	&\leq \theta 4^qq! + \theta  2^q \norm{\mb\beta}2^q q!. \notag   
\end{align}
Thus, recall that $\norm{\mb\beta}2 = \eta$, use $(\sigma^2,R) = (8\theta\eta^2, 4\eta)$,  from \eqref{eqn:chi_ct_epsnet_be_ub1}-\eqref{eqn:chi_ct_epsnet_be_ub2}, apply Bernstein inequality  \Cref{lem:mc_bernstein_scalar} with $n \geq   C p^5\theta^{-2}\log p$, and  $c_1,c_2\in[0,1]$  we have
\begin{align}  
	\prob{\eventnet^c} &\leq 2np\abs{\mc N_\eps}\cdot \prob{\sum_{t=0}^{n/2p-1}\mb Z_{2tp}(\mb\beta) > \frac{n}{2p}\E \mb Z(\mb\beta) + \frac{c_1n\theta}{2p^{5/2}}} \leq 2np\paren{\frac{3np^2}{c_2}}^{3p}  \exp \paren{ \frac{-\paren{c_1n\theta/2p^{5/2}}^2}{16n\theta\eta^2/2p + 8\eta  c_1n\theta/2p^{5/2}} }  \notag  \\ 
	&\leq \exp \left( 4p\log\paren{\frac{3np^2}{c_2}} - \frac{\paren{c_1n\theta/2p^{5/2}}^2 }{16n\theta\eta^2/p} \right)\leq \exp \left( 4p\log\paren{\frac{3np^2}{c_2}}  - \frac{c_1^2n\theta^2}{64p^4}\right)\notag \\
	&\leq \exp\paren{\frac{-c_1^2n\theta^2}{100p^4}}\leq  \frac1n       
\end{align}
when $\frac{C}{\log C} > \frac{10^5}{c_1^2c_2}$. The proof of lower bound and negative $\mb \beta_0$ is derived in the same manner.  
\end{proof}

%% ================================== %%

\subsection{Proof of \Cref{cor:ct_support}}\label{sec:proof_ct_supp}

\begin{proof} Write $\mb x$ as $\mb x_0$ though our this proof. Write  $\mb \beta_i\mb x_j + \mb s_j = \sum_{\ell\in[\pm p]}\mb \beta_\ell \mb x_{\ell - i + j} = \innerprod{\mb\beta}{\mb x_{[\pm p]-i+j }}$, and  the support w.r.t. some $\mb a$ as $I(\mb\beta)$. Define the random variable $\mb Z_{ij}(\mb\beta)$ as  
\begin{align}
	\norm{\mb P_{I(\mb\beta)}\shift{\mb x}{-i}}2^2 = \sum_{j\in[n]} \mb x_j^2\1_{\set{\abs{\innerprod{\mb\beta}{\mb x_{[\pm p]-i+j}}} > \lambda}} =: \sum_{j\in[n]} \mb Z_{ij}(\mb\beta)
\end{align}
and define $\set{\ol{\mb Z}_{ij}(\mb\beta)}_{j\in[n]}$ that are independent r.v.s. and as a upper bounding function of $\mb Z_{ij}(\mb\beta)$ as    
\begin{align}
	\ol{\mb Z}_{ij}(\mb\beta) := \begin{cases}
		\mb x_j^2, &\quad \abs{\innerprod{\mb\beta}{\mb x_{[\pm p]-i+j}}} > \lambda \\
		0, &\quad \abs{\innerprod{\mb\beta}{\mb x_{[\pm p]-i+j}} } < \lambda/2 \\
		\frac{\mb x_j^2}{\lambda/2}\paren{\abs{\innerprod{\mb\beta}{\mb x_{[\pm p]-i+j}}}-\lambda/2}, &\quad  \text{otherwise}
	\end{cases},
\end{align}  
Similar to proof of \Cref{cor:chibeta_ct}. Let $\norm{\mb\beta}2 \leq \eta \leq \sqrt p$. Define $\eps = \frac{c_2'\lambda}{24np\sqrt{p\theta\log n\log\theta^{-1}}} $ for some $c'_2 > 0$   and consider the $\eps$-net $\epsnet$ for sphere of radius $\eta$. From \Cref{lem:epsnet} we know  
\begin{align}
	\abs{\epsnet} \leq \paren{\frac{3\eta}\eps }^{2p} \leq \paren{\frac{72}{c_2'c_\lambda}np^2\sqrt{\theta\abs{\mb\tau}\log n\log\theta^{-1}}}^{2p} \leq \paren{\frac{72}{c_2'c_\lambda}np^2\log n}^{2p}, 
\end{align}
for each $i\in[n]$ define such net as $\mc N_{\eps,i}$,  and define an event such that all center of subsets in $\mc N_{\eps,i}$ are being well-behaved: 
\begin{align}\label{eqn:ct_suppI_epsnet}
	\eventnet := \set{\forall\,i\in[n],\quad \abs{n^{-1}\sum_{j\in[n]}\ol{\mb Z}_{ij}(\mb\beta_\eps) -\E \ol{\mb Z}_{i}(\mb\beta_\eps)}\leq \frac{c_1'\theta}{p}  \quad \forall\,\mb \beta_\eps\in\mc N_{\eps,i}},
\end{align}  
Also, $\sum_{j}\ol{\mb Z}_{ij}(\mb\beta)$  is a Lipchitz function over $\mb\beta$ for every $i\in[n]$ as 
\begin{align}
	\abs{\sum_{j\in[n]} \ol{\mb Z}_{ij}(\mb\beta) - \sum_{j\in[n]} \ol{\mb Z}_{ij}(\mb\beta')} &\leq \sum_{j\in[n]}\frac{\mb x_j^2}{\lambda/2}\abs{\innerprod{\mb\beta-\mb\beta'}{\mb x_{[\pm p]-i+j}}}  \leq \sum_{j\in[n]}\frac{\mb x_j^2\norm{\mb x_{[\pm p]-i+j}}2 }{\lambda/2}\norm{\mb\beta - \mb\beta'}2, \notag \\
	&\leq  \frac{1}{\lambda/2}\norm{\mb x}2^2\cdot\max_{j\in[n]}\norm{\mb x_{[\pm p]+j}}2\cdot\norm{\mb\beta-\mb\beta'}2 := L\norm{\mb\beta - \mb\beta'}2,
\end{align}

\vsni and define event $\event_{\mr{Lip}}$ such that the Lipchitz constant is bounded as  
\begin{align}
	\event_{\mr{Lip}} := \set{ L \leq 12n\theta\sqrt{p\theta\log n\log\theta^{-1}}\lambda^{-1} },  
\end{align}  
then on event $\event_{\mr{Lip}}$, for any points $\mb\beta$ in $\goodregion$ and $i\in[n]$, there exists some $\mb\beta_\eps$ in $\mc N_{\eps,i}$ with $\norm{\mb\beta-\mb\beta_\eps}2\leq \eps$, and thus
\begin{align}
	 \abs{\paren{n^{-1}\sum_{j\in[n]}\ol{\mb Z}_{ij}(\mb\beta) - \E \ol{\mb Z}_{i}(\mb\beta)} - \paren{n^{-1}\sum_{j\in[n]}\ol{\mb Z}_{ij}(\mb\beta_\eps) - \E \ol{\mb Z}_{i}(\mb\beta_\eps)}} \leq 2L\eps \leq \frac{c_2' \theta}{p}. \label{eqn:ct_suppI_eventlip}
\end{align}
On event $\event_{\mr{Lip}}\cap\event_{\mr{Net}}$, from \eqref{eqn:ct_suppI_epsnet}, \eqref{eqn:ct_suppI_eventlip}, we can conclude that for all $\mb\beta\in\goodregion$ and $i\in[n]$ that: 
\begin{align}
	 n^{-1}\norm{\mb P_{I(\mb\beta)}\shift{\mb x_0}{-i}}2^2 - n^{-1}\E \norm{\mb P_{I(\mb\beta)}\shift{\mb x_0}{-i}}2^2  &\leq  n^{-1}\sum_{j\in[n]}\ol{\mb Z}_{ij}(\mb\beta) - \E \ol{\mb Z}_{i}(\mb\beta) \leq \frac{(c_1'+c_2')\theta}{p}
\end{align}  
as desired, where the error probability of $\eventlip^c$ is bounded using \Cref{lem:x0_bound} and \Cref{lem:x0_subvec_bound}, which give
\begin{align}
	\prob{\event_{\mr{Lip}}^c} \leq  \prob{\norm{\mb x}2^2 > 2n\theta } + \prob{\max_{j\in[n]}\norm{\mb x_{[\pm p]+j}}2  > 3\sqrt{p\theta\log n\log\theta^{-1}} }  \leq  3/n,
\end{align}  
when $n > 10^3 \theta^{-1}$. As for $\eventnet^c$ use union bound and split the r.v.s since $\mb Z_j,\mb Z_{j+2p}$ are independent for all $j$:
\begin{align}
	\prob{\eventnet^c} &\leq 2np\cdot\abs{\epsnet}\cdot\prob{\abs{\sum_{k}^{n/2p}\ol{\mb Z}_{i,2kj}(\mb\beta) - \frac{n}{2p}\E\ol{\mb Z}_i(\mb\beta)} \geq \frac{c_1'n\theta}{2p^2} }.  \notag  
\end{align}
Now we calculate the variance and $L^q$-norm of $\sum_k\ol{\mb Z}_{i,2kj}$ for $q\geq 3$:
\begin{align}
	\begin{cases}
		\E \ol{\mb Z}_{i,j}^2 \leq  \E \mb x_j^4 \leq  3\theta \\
		\E \ol{\mb Z}_{i,j}^q \leq  \E \mb x_j^{2q} \leq \theta (2q-1)!! \leq  \frac12 \cdot(3\theta)\cdot 2^{q-2} q!  
	\end{cases}
\end{align}
and apply Bernstein inequality with $(\sigma^2,R) = (3\theta,2)$, then use $n \geq  C p^4\theta^{-1}\log p$ and $c_1',c_2'<1$ to obtain
\begin{align}
	2np\abs{\epsnet}\prob{\abs{\sum_{k}^{n/2p}\ol{\mb Z}_{i,2kj}(\mb\beta ) - \frac{n}{2p^2} \E\ol{\mb Z}_i} \geq \frac{c_1'n\theta}{2p^2} } &\leq \exp\brac{\log(2np) + 2p\log\paren{\frac{72}{c_2'c_\lambda}np^2\log n}   - \frac{(c_1'n\theta/2p^2)^2}{6n\theta/2p + 4c_1'n\theta/2p^2 } }\notag \\
	&\leq \exp\brac{ 3p\log\paren{\frac{72}{c_2'c_\lambda}np^2\log n}  - \frac{c_1'^2n\theta}{24p^3}} \notag \\ 
	&\leq \exp[-c_1'^2n\theta/(50p^3)] \leq 1/n, 
\end{align}
where the last two inequalities holds when $\frac{C}{\log C} \geq \frac{10^5}{c_1'^2c_2'c_\lambda}$. 
The other side of inequality of \eqref{eqn:ct_support} can be derived by defining $\ul{\mb Z}_{ij}$ as
\begin{align}
	\ul{\mb Z}_{ij}(\mb\beta) := \begin{cases}
		\mb x_j^2, &\quad \abs{\innerprod{\mb\beta}{\mb x_{[\pm p]-i+j}}} > 3\lambda/2 \\
		0, &\quad \abs{\innerprod{\mb\beta}{\mb x_{[\pm p]-i+j}} } < \lambda \\
		\frac{\mb x_j^2}{\lambda/2}\paren{\abs{\innerprod{\mb\beta}{\mb x_{[\pm p]-i+j}}}-\lambda}, &\quad  \text{otherwise} 
	\end{cases},
\end{align}
and define $\eventnet$, $\eventlip$ similarly, such that on intersection of these events,
\begin{align}
	 n^{-1}\norm{\mb P_{I(\mb\beta)}\shift{\mb x}{-i}}2^2 - n^{-1}\E \norm{\mb P_{I(\mb\beta)}\shift{\mb x}{-i}}2^2  &\geq  n^{-1}\sum_{j\in[n]}\ul{\mb Z}_{ij}(\mb\beta) - \E \ul{\mb Z}_{i}(\mb\beta) \geq \frac{(c_1'+c_2')\theta}{p}  
\end{align}  
 as desired.
\end{proof}

%% ================================== %%

\subsection{Proof of \Cref{lem:num_entries_onlambda} }\label{sec:proof_entries_onlambda}
   
\begin{proof} 1. (\ul{Expectation upper bound}) We will write $\mb x$ as $\mb x_0$. Similar to proof of \Cref{cor:chibeta_ct} let $\norm{\mb\beta}2 \leq \eta \leq \sqrt p$. For each $i\in[n]$, define the random variable
\begin{align}
	{\mb X}_i(\mb \beta) = \1_{\set{\abs{\innerprod{\shift{\mb x}{i}}{\mb \beta} - \lambda} \leq B}} +  \1_{\set{\abs{\innerprod{\shift{\mb x}{i}}{\mb \beta} + \lambda} \leq B}},
\end{align} 
then number of indices for vector $\mb x*\wc{\mb\beta}$ that are within $B$ of $\pm\lambda$ is a random variable $ \sum_{i\in[n]}\mb X_i(\mb \beta)$. For each of the $\mb X_i(\mb \beta)$'s consider an upper bound  $\ol{\mb X}_i(\mb \beta)$ defined as 
\begin{align}
	\ol{\mb X}_i(\mb \beta) = \begin{cases}	\frac{1}{M}\paren{\abs{\innerprod{\shift{\mb x}i}{\mb \beta}} - (\lambda -B - M )} & \abs{\innerprod{\shift{\mb x}i}{\mb \beta}} \in [\lambda - B - M, \lambda - B] \\ 
	1 & \abs{\innerprod{\shift{\mb x}i}{\mb \beta}} \in [\lambda - B,\lambda+ B] \\
	\frac{1}{M}\paren{  (\lambda + B + M) - \abs{\innerprod{\shift{\mb x}i}{\mb \beta}}  } & \abs{\innerprod{\shift{\mb x}i}{\mb \beta}} \in [\lambda + B, \lambda + B + M ] \\
	0 & \text{else}	
	\end{cases}
\end{align}
where $B < M =  c\lambda\theta^2/\paren{p\log n} \leq \lambda/4$ for some constant $0<c<1$. 

Notice that $\mb x \simiid \mr{BG}(\theta)$ is equal in distribution to $\mb P_{I(\mb a)} \mb g$, where $\mb g \simiid \mc N(0,1)$, and $I(\mb a) \subseteq [n]$ is an independent Bernoulli subset. Conditioned on $I(\mb a)$, $\innerprod{ \mb x }{\mb \beta } = \innerprod{ \mb g }{\mb P_{I(\mb a)} \mb \beta } \sim \mc N(0,\norm{\mb P_{I(\mb a)} \mb \beta }{2}^2 )$. For all realizations of $I(\mb a)$, the variance $\norm{\mb P_{I(\mb a)} \mb \beta }{2}^2$ is bounded by $\norm{\mb P_{I(\mb a)} \mb \beta }{2}^2 \le \norm{\mb \beta}{2}^2 \le p$. Using these observations, and letting $ f_{\sigma}(t) = \paren{\sqrt{2\pi} \sigma}^{-1}\exp\paren{ - t^2/2\sigma^2}$ denote the pdf of an $\mc N(0,\sigma^2)$ random variable, the expectation of $\sum_i \ol{\mb X}_i(\mb \beta)$ can be upper bounded as 
\begin{align}
	\sum_{i\in[n]}  \expect{\ol{\mb X}_i(\mb \beta)} &\leq (2n)\cdot\prob{\innerprod{\mb x}{\mb \beta} \in \brac{\lambda - B - M, \lambda + B + M}}  \label{eqn:B_bound_expect_bound} \nonumber \\
	& \leq (2n) \cdot  2 ( B + M ) \sup_{\sigma^2 \in (0, p]} \max_{t \in \brac{\lambda - B - M, \lambda + B + M} } f_{\sigma}( t ) \nonumber \\
	& \le 4 n ( B + M ) \sup_{\sigma^2 \in (0, p]} f_{\sigma}\left( \lambda - B - M \right) \nonumber \\
	& \le 4 n ( B + M ) \sup_{\sigma^2 \in (0,p]} f_{\sigma}\left( \lambda / 2 \right).
\end{align}
Notice that
\begin{align}
	\frac{d}{d\sigma}f_{\sigma}\left( \frac{\lambda }{2} \right) = \frac{d}{d\sigma}\frac{1}{\sqrt{2\pi} \sigma}\exp\paren{-\frac{\lambda^2}{8\sigma^2}} = \frac{\lambda^2 - 4\sigma^2}{4\sqrt{2\pi}\sigma^4}\exp\left({-\frac{\lambda^2}{8\sigma^2}} \right), \notag
\end{align}
and hence $f_{\sigma}( \lambda / 2 )$ is maximized at either $\sigma^2 = 0$, $\sigma^2 = p $ or $\sigma^2 = \lambda^2 / 4$. Comparing values at these points, we obtain that 
\begin{align}
	\sup_{\sigma^2 \in (0,p]}f_{\sigma}(\lambda /2) & \quad \leq \quad f_{\lambda / 2}(\lambda/2) 
	 \quad \leq \quad \frac{1}{\sqrt{2\pi}(\lambda/2 )} \exp\paren{-\frac{1}{2}} \quad \leq \quad \frac{1}{2\lambda}, \label{eqn:ct_y_pdf}
\end{align} 
whence, by letting $B \leq c \lambda\theta^2/\paren{p\log n}$, the upper bound of expectation become:
\begin{align}\label{eqn:ct_y_expX}
	\sum_{i\in[n]} \expect{\ol{\mb X}_i(\mb \beta)}\leq \frac{4n}{2\lambda}(B+M) \leq  \frac{4cn\theta^2}{p\log n}  =: n\E\ol{\mb X(\mb\beta)}.
\end{align}

\vsni 2. (\ul{$\eps$-net}) Define $\eps = \frac{c^2\lambda\theta^{3.5}}{3p^{2.5}\log^{2.5}n\log^{0.5}\theta^{-1}}$. Write $\lambda = c_\lambda/\sqrt{\abs{\mb\tau}}$  and consider the $\eps$-net $\epsnet$ for sphere of radius $\eta\leq \sqrt p$. From \Cref{lem:epsnet} we know  
\begin{align}
	\abs{\epsnet} \leq \paren{\frac{3\eta}{\eps} }^{2p} \leq  \paren{\frac{81\abs{\mb\tau}p^6\log^5n\log\theta^{-1}}{c^4 c_\lambda^2 \theta^{7}}}^{p} \leq \paren{\frac{2p\log n}{c\cdot c_\lambda}}^{13p}         
\end{align} 
and define an event such that all center of subsets in $\mc N_{\eps}$ are being well-behaved: 
\begin{align}\label{eqn:ct_y_epsnet}
	\eventnet := \set{ \sum_{i\in[n]} \ol{\mb X}_i(\mb\beta_\eps)-  n\E \ol{\mb X}(\mb\beta_\eps) < \frac{18cn\theta^2}{p\log n}   \quad \forall\,\mb \beta_\eps\in\mc N_{\eps},}
\end{align}   

\vsni 3. (\ul{Lipschitz constant}) Furthermore, the function $\sum_i^n\ol{\mb X}_i(\mb \beta)$ is Lipchitz over $\mb \beta$ such that  
\begin{align}
	\abs{\sum_{i\in[n]} \ol{\mb X}_i(\mb\beta) - \sum_{i\in[n]}\ol{\mb X}_i(\mb\beta')} \leq \sum_{i\in[n]}^n\frac{1}{M}\abs{\innerprod{\shift{\mb x}{i}}{\mb\beta-\mb\beta'} } \leq \frac{n}{M}\max_{i\in[n]}\norm{\mb P_{[\pm p] + i}\mb x}2\norm{\mb\beta-\mb\beta'}2 =: L\norm{\mb\beta-\mb\beta'}2   \notag
\end{align}  
define the set $\epsnet$ where Lipschitz constant is well bounded:   
\begin{align}
	\eventlip := \set{L \leq \frac{3n \sqrt{p\theta\log n\log\theta^{-1}}}{M} },\notag
\end{align}
then on event $\event_{\mr{Lip}}$, for every $\mb\beta$ in $\goodregion$, there exists some $\mb\beta_\eps$ in $\mc N_{\eps,i}$ with $\norm{\mb\beta-\mb\beta_\eps}2\leq \eps$, thus
\begin{align}
	 \abs{\paren{\sum_{i\in[n]}\ol{\mb X}_i(\mb\beta) - n\E \ol{\mb X}(\mb\beta)} - \paren{\sum_{i\in[n]}\ol{\mb X}_i(\mb\beta_\eps) - n\E \ol{\mb X}(\mb\beta_\eps)}} \leq 2L\eps \leq \frac{2c n \theta^2}{p\log n}. \label{eqn:ct_y_eventlip}
\end{align} 
On event $\event_{\mr{Lip}}\cap\event_{\mr{Net}}$, from \eqref{eqn:ct_y_expX}, \eqref{eqn:ct_y_epsnet} and \eqref{eqn:ct_y_eventlip}, we can conclude that for every $\mb\beta\in\goodregion$ and $i\in[n]$, 
\begin{align}
	\sum_{i\in[n]} \ol{\mb X}_i(\mb\beta) \leq \frac{24cn\theta^2}{p\log n} 
\end{align}  
as desired, where the error probability of $\eventlip^c$ is bounded using  \Cref{lem:x0_subvec_bound}, which gives
\begin{align}
	\prob{\event_{\mr{Lip}}^c} \leq \prob{\max_{j\in[n]}\norm{\mb x_{[\pm p]+j}}2  > 3\sqrt{p\theta\log n\log\theta^{-1} }}  \leq  2/n,
\end{align}

\vsni 4. (\ul{Bound $\prob{\eventnet^c}$}) Wlog let us assume that $2p$ divides $n$. By applying union bound and observing that $\ol{\mb X}_i(\mb \beta)$ is independent of $\ol{\mb X}_{i+2p}(\mb \beta)$ for any $i\in[n]$, we split $\sum_i\ol{\mb X}_i(\mb \beta)$ into $n/2p$ independent sums of r.v.s, we have
\begin{align} 
	\prob{\eventnet^c} 
	   &\leq  2p\abs{\epsnet}\cdot \prob{ \sum_{j=0}^{n/2p-1}\paren{\ol{\mb X}_{2pj}(\mb \beta) -  \expect{\ol{\mb X}(\mb \beta)}} > \frac{9cn\theta^2}{p^2\log n}},\notag 
\end{align}
where each summand has bounded variance and $L^q$-norm derived similarly as its expectation such that
\begin{align}
	\E \ol{\mb X}_i(\mb\beta)^q \leq  2\cdot \prob{\innerprod{\shift{\mb x}i}{\mb \beta} \in \left[\lambda-B-M,\lambda + B + M\right]} \leq  2\cdot\frac{1}{2\lambda}\cdot 2(B+M) \leq \frac{4c\theta^2}{p \log n} , \notag
\end{align}
and apply Bernstein inequality \Cref{lem:mc_bernstein_scalar} with $(\sigma^2,R) = (4c\theta^2/\paren{p\log n}, 1)$,  obtains
\begin{align}
	\prob{ \sum_{j=0}^{n/2p-1}\paren{\ol{\mb X}_{2pj}(\mb \beta) -  \expect{\ol{\mb X}(\mb \beta)}} > \frac{9cn\theta^2}{p^2\log n}} \leq \exp\brac{\frac{-(9cn\theta^2/p^2\log n)^2}{2cn\theta^2/p^2\log n + 2(9cn\theta^2/p^2\log n)}} \leq \exp\brac{\frac{-4cn\theta^2}{p^2\log n}}, \notag
\end{align}
thus when $n = Cp^5\theta^{-2}\log p$: 
\begin{align}
	\prob{\eventnet^c} \leq \exp\brac{\log(2p) + 13p\log\paren{\frac{2p\log n}{c\cdot c_\lambda }} -\frac{4 cn\theta^2}{p^2\log n} } \leq  1/n
\end{align} 
as long as $\frac{C}{\log C} > 10^5/\paren{c^2\cdot c_\lambda}$.

\end{proof}

% !TEX root = ../../BD_DQ.tex
\section{Tools}

\begin{lemma}[Tail bound for Gaussian r.v.]\label{lem:gaussian_tail_bound} If $X\sim \mc N(0,\sigma^2)$, then its tail bound for $t>0$ 
	can be
	\begin{align}
		\prob{X>t} \leq \frac{\sigma}{t\sqrt{2\pi}} \exp\paren{-\frac{t^2}{2\sigma^2}}
	\end{align}
\end{lemma} 
 
\begin{lemma}[Moments of the Gaussian random variables] \label{lem:gaussian_moment}
If $X \sim \mc N\left(0, \sigma^2\right)$, then ifor all integer $p \geq 1$,
\begin{align}
\expect{\abs{X}^p} \leq \sigma^p \paren{p -1}!!. 
\end{align}
\end{lemma}

\begin{lemma}[Gaussian concentration inequality]\label{lem:gaussian_concentration} Let $\mb x = (\mb x_1,\ldots,\mb x_n)$ be a vector of $n$ independent standard normal variables. Let $f:\R^n\to\R$ be an $L$-Lipschitz function. Then for all $t>0$,
\begin{align}
	\prob{\abs{f(\mb x) - \E f(\mb x)} \geq t} \leq 2\exp\left(-\frac{t^2}{2L^2} \right). 
\end{align}
\end{lemma}

\begin{lemma}[Moment control Bernstein inequality for scalar r.v.s]\emph{(\cite{foucart2013mathematical}, Theorem 7.30)}\label{lem:mc_bernstein_scalar}
Let $\mb x_1, \dots, \mb x_n$ be independent real-valued random variables. Suppose that there exist some positive number $R$ and $\sigma^2$ such that $\frac1n\sum_{i=1}^n\expect{\mb X_i^2}\leq \sigma^2$ and
\begin{align*}
\tfrac1n\textstyle\sum_{i=1}^n\expect{\abs{\mb x_k}^p} \leq \frac{1}{2}\sigma^2 R^{p-2}p!, \; \; \text{for all integers}\; p \ge 3.
\end{align*}
Let $S \doteq \sum_{i=1}^n \mb x_i$, then for all $t > 0$, it holds  that 
\begin{align}
\prob{\abs{S - \expect{S}} \ge t} \leq 2\exp\left(-\frac{t^2}{2n\sigma^2 + 2Rt}\right).   
\end{align}
\end{lemma}

\begin{lemma}[$\eps$-net on sphere]\label{lem:epsnet}\emph{\cite{vershynin2010introduction}}  Let $(X, d)$ be a metric space and let $\eps > 0$.
A subset $\epsnet$ of $X$ is called an $\eps$-net of $X$ if for every point $x\in X$ there exists some point $y \in \mc N_\eps$ so that $d(x, y) \leq\eps$. There exists an $\eps$-net $\epsnet$ for the sphere $\bb S^{n-1}$ of size 
	$ \abs{\epsnet } \leq \paren{3/\eps}^n$.
\end{lemma}

\begin{lemma}[Hanson-Wright]\emph{\cite{rudelson2013hanson}} \label{lem:Hanson-wright} Let $\mb x_1,\ldots,\mb x_n$ be independent, subgaussian random variables with subgaussian norm $\sup_{p\geq 1 }p^{-1/2}\paren{\E\abs{x_i^p}}^{1/p} \leq \sigma $. Let $\mb A\in\R^{n\times n}$, then for every $t>0$,
\begin{align}
	\prob{\abs{\mb x^*\mb A\mb x - \E\mb x^*\mb A\mb x} \geq t} \leq 2\exp\paren{-c\min\paren{\frac{t^2}{64\,\sigma^4\norm{\mb A}F^2},\frac{t}{8\sqrt 2\,\sigma^2 \norm{\mb A}2}}}.
\end{align} 
	
\end{lemma}

\begin{lemma}[Maximum of separable convex function] \label{lem:max_soft_thresh} Let $f:\R_+\to \R_+$ be a convex function of the form $f(x) = x - s(x)$ with $s:\R_+\to\R_+$ satisfying
\begin{align}
	\frac{s(x)}{x} \leq  \frac{s(y)}{y},\;\; \text{for all $x\geq y > 0$}. \notag
\end{align}
Then for $n\in\N$ and $0<N \le nL$,
\begin{align}
	\max_{ 0\leq \mb x \leq L,\; \norm{\mb x}1 \leq N }\sum_{i=1}^n f(\mb x_i) \leq N\paren{1 - \frac{s(L)}{L}}
\end{align}
\end{lemma}
\begin{proof} Since the feasible set is a convex polytope; the convex function $\sum_{i=1}^n f(\mb x_i)$ is maximized at a vertex, and that its vertices consist of $0$ and permutations of the vector $\big[\underbrace{L,\ldots,L}_{\floor{N/L}},r,0,\dots,0\big]$, where $
 	r = N - \floor{ N/L } L \leq L$. Then the function value at the maximizing vector $\mb x_*$ can be derived as:
 \begin{align}
 	\sum_{i=1}^n f(\mb x_{*i}) &=  \floor{\tfrac{N}{L}}f(L) + f(r) = \tfrac{N-r}{L}\paren{L-s(L)} + \paren{r-s(r)}\notag\\
 	&= N\paren{1-\tfrac{s(L)}{L}} +r\paren{\tfrac{s(L)}{L} - \tfrac{s(r)}{r} } \leq N\paren{1-\tfrac{s(L)}{L}}\notag
 \end{align}
\end{proof}


\newcommand{\etalchar}[1]{$^{#1}$}
\begin{thebibliography}{YWHM10}

\bibitem[AD88]{ayers1988iterative}
GR~Ayers and J~Christopher Dainty.
\newblock Iterative blind deconvolution method and its applications.
\newblock {\em Optics letters}, 13(7):547--549, 1988.

\bibitem[AMS09]{absil2009optimization}
P-A Absil, Robert Mahony, and Rodolphe Sepulchre.
\newblock {\em Optimization algorithms on matrix manifolds}.
\newblock Princeton University Press, 2009.

\bibitem[ARR14]{ahmed2014blind}
Ali Ahmed, Benjamin Recht, and Justin Romberg.
\newblock Blind deconvolution using convex programming.
\newblock {\em IEEE Transactions on Information Theory}, 60(3):1711--1732,
  2014.

\bibitem[BC11]{BC11}
Heinz~H. Bauschke and Patrick~L. Combettes.
\newblock {\em Convex Analysis and Monotone Operator Theory in Hilbert Spaces}.
\newblock Springer Publishing Company, Incorporated, 1st edition, 2011.

\bibitem[BDH{\etalchar{+}}13]{briers2013laser}
David Briers, Donald~D Duncan, Evan~R Hirst, Sean~J Kirkpatrick, Marcus
  Larsson, Wiendelt Steenbergen, Tomas Stromberg, and Oliver~B Thompson.
\newblock Laser speckle contrast imaging: theoretical and practical
  limitations.
\newblock {\em Journal of biomedical optics}, 18(6):066018, 2013.

\bibitem[BK02]{baker2002limits}
Simon Baker and Takeo Kanade.
\newblock Limits on super-resolution and how to break them.
\newblock {\em IEEE Transactions on Pattern Analysis and Machine Intelligence},
  24(9):1167--1183, 2002.

\bibitem[BPSW95]{bones1995deconvolution}
PJ~Bones, CR~Parker, BL~Satherley, and RW~Watson.
\newblock Deconvolution and phase retrieval with use of zero sheets.
\newblock {\em JOSA A}, 12(9):1842--1857, 1995.

\bibitem[BS95]{bell1995information}
Anthony~J Bell and Terrence~J Sejnowski.
\newblock An information-maximization approach to blind separation and blind
  deconvolution.
\newblock {\em Neural computation}, 7(6):1129--1159, 1995.

\bibitem[BT09]{beck2009fast}
Amir Beck and Marc Teboulle.
\newblock A fast iterative shrinkage-thresholding algorithm for linear inverse
  problems.
\newblock {\em SIAM journal on imaging sciences}, 2(1):183--202, 2009.

\bibitem[BVG13]{benichoux2013fundamental}
Alexis Benichoux, Emmanuel Vincent, and R{\'e}mi Gribonval.
\newblock A fundamental pitfall in blind deconvolution with sparse and
  shift-invariant priors.
\newblock In {\em ICASSP-38th International Conference on Acoustics, Speech,
  and Signal Processing-2013}, 2013.

\bibitem[Can76]{cannon1976blind}
Michael Cannon.
\newblock Blind deconvolution of spatially invariant image blurs with phase.
\newblock {\em IEEE Transactions on Acoustics, Speech, and Signal Processing},
  24(1):58--63, 1976.

\bibitem[Car01]{carasso2001direct}
Alfred~S Carasso.
\newblock Direct blind deconvolution.
\newblock {\em SIAM Journal on Applied Mathematics}, 61(6):1980--2007, 2001.

\bibitem[CE16]{campisi2016blind}
Patrizio Campisi and Karen Egiazarian.
\newblock {\em Blind image deconvolution: theory and applications}.
\newblock CRC press, 2016.

\bibitem[Chi16]{Chi2016-TIP}
Yuejie Chi.
\newblock Guaranteed blind sparse spikes deconvolution via lifting and convex
  optimization.
\newblock {\em IEEE Journal of Selected Topics in Signal Processing},
  10(4):782--794, June 2016.

\bibitem[CLC{\etalchar{+}}17]{Cheung17-Nature}
Sky Cheung, Yenson Lau, Zhengyu Chen, Ju~Sun, Yuqian Zhang, John Wright, and
  Abhay Pasupathy.
\newblock Beyond the fourier transform: A nonconvex optimization approach to
  microscopy analysis.
\newblock {\em Submitted}, 2017.

\bibitem[CM15]{choudhary2015fundamental}
Sunav Choudhary and Urbashi Mitra.
\newblock Fundamental limits of blind deconvolution part ii: Sparsity-ambiguity
  trade-offs.
\newblock {\em arXiv preprint arXiv:1503.03184}, 2015.

\bibitem[CW98]{chan1998total}
Tony~F Chan and Chiu-Kwong Wong.
\newblock Total variation blind deconvolution.
\newblock {\em IEEE transactions on Image Processing}, 7(3):370--375, 1998.

\bibitem[CWB08]{candes2008enhancing}
Emmanuel~J Candes, Michael~B Wakin, and Stephen~P Boyd.
\newblock Enhancing sparsity by reweighted ℓ 1 minimization.
\newblock {\em Journal of Fourier analysis and applications}, 14(5-6):877--905,
  2008.

\bibitem[DZSW11]{dong2011image}
Weisheng Dong, Lei Zhang, Guangming Shi, and Xiaolin Wu.
\newblock Image deblurring and super-resolution by adaptive sparse domain
  selection and adaptive regularization.
\newblock {\em IEEE Transactions on Image Processing}, 20(7):1838--1857, 2011.

\bibitem[EHJ{\etalchar{+}}04]{efron2004least}
Bradley Efron, Trevor Hastie, Iain Johnstone, Robert Tibshirani, et~al.
\newblock Least angle regression.
\newblock {\em The Annals of statistics}, 32(2):407--499, 2004.

\bibitem[ETS11]{Ekanadham2011-NIPS}
Chaitanya Ekanadham, Daniel Tranchina, and Eero~P. Simoncelli.
\newblock A blind sparse deconvolution method for neural spike identification.
\newblock In {\em Advances in Neural Information Processing Systems 24}, pages
  1440--1448. 2011.

\bibitem[FR13]{foucart2013mathematical}
Simon Foucart and Holger Rauhut.
\newblock {\em A Mathematical Introduction to Compressive Sensing}.
\newblock Springer, 2013.

\bibitem[FSH{\etalchar{+}}06]{fergus2006removing}
Rob Fergus, Barun Singh, Aaron Hertzmann, Sam~T Roweis, and William~T Freeman.
\newblock Removing camera shake from a single photograph.
\newblock In {\em ACM transactions on graphics (TOG)}, volume~25, pages
  787--794. ACM, 2006.

\bibitem[GHJY15]{ge2015escaping}
Rong Ge, Furong Huang, Chi Jin, and Yang Yuan.
\newblock Escaping from saddle points—online stochastic gradient for tensor
  decomposition.
\newblock In {\em Conference on Learning Theory}, pages 797--842, 2015.

\bibitem[GMWZ17]{goldfarb2017using}
Donald Goldfarb, Cun Mu, John Wright, and Chaoxu Zhou.
\newblock Using negative curvature in solving nonlinear programs.
\newblock {\em Computational Optimization and Applications}, 68(3):479--502,
  2017.

\bibitem[Gol80]{goldfarb1980curvilinear}
Donald Goldfarb.
\newblock Curvilinear path steplength algorithms for minimization which use
  directions of negative curvature.
\newblock {\em Mathematical programming}, 18(1):31--40, 1980.

\bibitem[HHSS09]{harmeling2009online}
Stefan Harmeling, Michael Hirsch, Suvrit Sra, and Berhard Scholkopf.
\newblock Online blind deconvolution for astronomical imaging.
\newblock In {\em 2009 IEEE International Conference on Computational
  Photography (ICCP 2009)}, pages 1--7. IEEE, 2009.

\bibitem[JSE{\etalchar{+}}98]{johnson1998blind}
Richard Johnson, Philip Schniter, Thomas~J Endres, James~D Behm, Donald~R
  Brown, and Ra{\'u}l~A Casas.
\newblock Blind equalization using the constant modulus criterion: A review.
\newblock {\em Proceedings of the IEEE}, 86(10):1927--1950, 1998.

\bibitem[JSK08]{joshi2008psf}
Neel Joshi, Richard Szeliski, and David~J Kriegman.
\newblock Psf estimation using sharp edge prediction.
\newblock In {\em Computer Vision and Pattern Recognition, 2008. CVPR 2008.
  IEEE Conference on}, pages 1--8. IEEE, 2008.

\bibitem[KH96]{kundur1996blind}
Deepa Kundur and Dimitrios Hatzinakos.
\newblock Blind image deconvolution.
\newblock {\em IEEE signal processing magazine}, 13(3):43--64, 1996.

\bibitem[KK17]{kech2017optimal}
Michael Kech and Felix Krahmer.
\newblock Optimal injectivity conditions for bilinear inverse problems with
  applications to identifiability of deconvolution problems.
\newblock {\em SIAM Journal on Applied Algebra and Geometry}, 1(1):20--37,
  2017.

\bibitem[KT98]{kaaresen1998multichannel}
Kjetil~F Kaaresen and Tofinn Taxt.
\newblock Multichannel blind deconvolution of seismic signals.
\newblock {\em Geophysics}, 63(6):2093--2107, 1998.

\bibitem[KTF11]{krishnan2011blind}
Dilip Krishnan, Terence Tay, and Rob Fergus.
\newblock Blind deconvolution using a normalized sparsity measure.
\newblock In {\em Computer Vision and Pattern Recognition (CVPR), 2011 IEEE
  Conference on}, pages 233--240. IEEE, 2011.

\bibitem[Lan92]{lane1992blind}
Richard~G Lane.
\newblock Blind deconvolution of speckle images.
\newblock {\em JOSA A}, 9(9):1508--1514, 1992.

\bibitem[LB87]{lane1987automatic}
RG~Lane and RHT Bates.
\newblock Automatic multidimensional deconvolution.
\newblock {\em JOSA A}, 4(1):180--188, 1987.

\bibitem[LB18]{Li18-multiBD}
Yanjun Li and Yoram Bresler.
\newblock Global geometry of multichannel sparse blind deconvolution on the
  sphere.
\newblock {\em arXiv preprint arXiv:1404.4104}, 2018.

\bibitem[Lew98]{lewicki1998review}
Michael~S Lewicki.
\newblock A review of methods for spike sorting: the detection and
  classification of neural action potentials.
\newblock {\em Network: Computation in Neural Systems}, 9(4):R53--R78, 1998.

\bibitem[LFDF07]{levin2007deconvolution}
Anat Levin, Rob Fergus, Fredo Durand, and William~T Freeman.
\newblock Deconvolution using natural image priors.
\newblock {\em Massachusetts Institute of Technology, Computer Science and
  Artificial Intelligence Laboratory}, 3, 2007.

\bibitem[LLB16]{li2016identifiability}
Yanjun Li, Kiryung Lee, and Yoram Bresler.
\newblock Identifiability in blind deconvolution with subspace or sparsity
  constraints.
\newblock {\em IEEE Transactions on Information Theory}, 62(7):4266--4275,
  2016.

\bibitem[LLB17]{Li17-IT}
Yanjun Li, Kiryung Lee, and Yoram Bresler.
\newblock Identifiability and stability in blind deconvolution under minimal
  assumptions.
\newblock {\em IEEE Transaction of Information Theory}, 2017.

\bibitem[LS15]{Ling2015-IP}
Shuyang Ling and Thomas Strohmer.
\newblock Self-calibration and biconvex compressive sensing.
\newblock {\em Inverse Problems}, 31(11):115002, 2015.

\bibitem[LS17]{Ling2017-IT}
Shuyang Ling and Thomas Strohmer.
\newblock Blind deconvolution meets blind demixing: Algorithms and performance
  bounds.
\newblock {\em IEEE Transactions on Information Theory}, 63(7):4497--4520,
  2017.

\bibitem[LWDF11]{levin2011understanding}
Anat Levin, Yair Weiss, Fredo Durand, and William~T Freeman.
\newblock Understanding blind deconvolution algorithms.
\newblock {\em IEEE transactions on pattern analysis and machine intelligence},
  33(12):2354--2367, 2011.

\bibitem[MC99]{markham1999parametric}
Joanne Markham and Jos{\'e}-Angel Conchello.
\newblock Parametric blind deconvolution: a robust method for the simultaneous
  estimation of image and blur.
\newblock {\em JOSA A}, 16(10):2377--2391, 1999.

\bibitem[MK88]{miyoshi1988inverse}
Masato Miyoshi and Yutaka Kaneda.
\newblock Inverse filtering of room acoustics.
\newblock {\em IEEE Transactions on acoustics, speech, and signal processing},
  36(2):145--152, 1988.

\bibitem[NG10]{naylor2010speech}
Patrick~A Naylor and Nikolay~D Gaubitch.
\newblock {\em Speech dereverberation}.
\newblock Springer Science \& Business Media, 2010.

\bibitem[OPT00]{osborne2000new}
Michael~R Osborne, Brett Presnell, and Berwin~A Turlach.
\newblock A new approach to variable selection in least squares problems.
\newblock {\em IMA journal of numerical analysis}, 20(3):389--403, 2000.

\bibitem[PF14]{perrone2014total}
Daniele Perrone and Paolo Favaro.
\newblock Total variation blind deconvolution: The devil is in the details.
\newblock In {\em Proceedings of the IEEE Conference on Computer Vision and
  Pattern Recognition}, pages 2909--2916, 2014.

\bibitem[PSG{\etalchar{+}}16]{pnevmatikakis2016simultaneous}
Eftychios~A Pnevmatikakis, Daniel Soudry, Yuanjun Gao, Timothy~A Machado, Josh
  Merel, David Pfau, Thomas Reardon, Yu~Mu, Clay Lacefield, Weijian Yang,
  et~al.
\newblock Simultaneous denoising, deconvolution, and demixing of calcium
  imaging data.
\newblock {\em Neuron}, 89(2):285--299, 2016.

\bibitem[RV{\etalchar{+}}13]{rudelson2013hanson}
Mark Rudelson, Roman Vershynin, et~al.
\newblock Hanson-wright inequality and sub-gaussian concentration.
\newblock {\em Electronic Communications in Probability}, 18, 2013.

\bibitem[Sah07]{saha2007diffraction}
Swapan~K Saha.
\newblock {\em Diffraction-limited imaging with large and moderate telescopes}.
\newblock World Scientific, 2007.

\bibitem[Sat75]{sato1975method}
Yoichi Sato.
\newblock A method of self-recovering equalization for multilevel
  amplitude-modulation systems.
\newblock {\em IEEE Transactions on communications}, 23(6):679--682, 1975.

\bibitem[SCI75]{stockham1975blind}
Thomas~G Stockham, Thomas~M Cannon, and Robert~B Ingebretsen.
\newblock Blind deconvolution through digital signal processing.
\newblock {\em Proceedings of the IEEE}, 63(4):678--692, 1975.

\bibitem[SGG{\etalchar{+}}09]{shtengel2009interferometric}
Gleb Shtengel, James~A Galbraith, Catherine~G Galbraith, Jennifer
  Lippincott-Schwartz, Jennifer~M Gillette, Suliana Manley, Rachid Sougrat,
  Clare~M Waterman, Pakorn Kanchanawong, Michael~W Davidson, et~al.
\newblock Interferometric fluorescent super-resolution microscopy resolves 3d
  cellular ultrastructure.
\newblock {\em Proceedings of the National Academy of Sciences},
  106(9):3125--3130, 2009.

\bibitem[SJA08]{shan2008high}
Qi~Shan, Jiaya Jia, and Aseem Agarwala.
\newblock High-quality motion deblurring from a single image.
\newblock In {\em Acm transactions on graphics (tog)}, volume~27, page~73. ACM,
  2008.

\bibitem[SQW17]{sun2017complete}
Ju~Sun, Qing Qu, and John Wright.
\newblock Complete dictionary recovery over the sphere ii: Recovery by
  riemannian trust-region method.
\newblock {\em IEEE Transactions on Information Theory}, 63(2):885--914, 2017.

\bibitem[SW90]{shalvi1990new}
Ofir Shalvi and Ehud Weinstein.
\newblock New criteria for blind deconvolution of nonminimum phase systems
  (channels).
\newblock {\em IEEE Transactions on information theory}, 36(2):312--321, 1990.

\bibitem[Ver10]{vershynin2010introduction}
Roman Vershynin.
\newblock Introduction to the non-asymptotic analysis of random matrices.
\newblock {\em arXiv preprint arXiv:1011.3027}, 2010.

\bibitem[WC16]{wang2016blind}
Liming Wang and Yuejie Chi.
\newblock Blind deconvolution from multiple sparse inputs.
\newblock {\em IEEE Signal Processing Letters}, 23(10):1384--1388, 2016.

\bibitem[WJPH17]{walk2017blind}
Philipp Walk, Peter Jung, G{\"o}tz~E Pfander, and Babak Hassibi.
\newblock Blind deconvolution with additional autocorrelations via convex
  programs.
\newblock {\em arXiv preprint arXiv:1701.04890}, 2017.

\bibitem[WY13]{wen2013feasible}
Zaiwen Wen and Wotao Yin.
\newblock A feasible method for optimization with orthogonality constraints.
\newblock {\em Mathematical Programming}, 142(1-2):397--434, 2013.

\bibitem[WZ14]{wipf2014revisiting}
David Wipf and Haichao Zhang.
\newblock Revisiting bayesian blind deconvolution.
\newblock {\em The Journal of Machine Learning Research}, 15(1):3595--3634,
  2014.

\bibitem[XJ10]{xu2010two}
Li~Xu and Jiaya Jia.
\newblock Two-phase kernel estimation for robust motion deblurring.
\newblock In {\em European conference on computer vision}, pages 157--170.
  Springer, 2010.

\bibitem[YK96]{you1996anisotropic}
Yu-Li You and Mostafa Kaveh.
\newblock Anisotropic blind image restoration.
\newblock In {\em Image Processing, 1996. Proceedings., International
  Conference on}, volume~2, pages 461--464. IEEE, 1996.

\bibitem[YWHM10]{yang2010image}
Jianchao Yang, John Wright, Thomas~S Huang, and Yi~Ma.
\newblock Image super-resolution via sparse representation.
\newblock {\em IEEE transactions on image processing}, 19(11):2861--2873, 2010.

\bibitem[ZKW18]{zhang2018structured}
Yuqian Zhang, Han-Wen Kuo, and John Wright.
\newblock Structured local optima in sparse blind deconvolution.
\newblock {\em arXiv preprint arXiv:1806.00338}, 2018.

\bibitem[ZLK{\etalchar{+}}17]{zhang2017global}
Yuqian Zhang, Yenson Lau, Han-wen Kuo, Sky Cheung, Abhay Pasupathy, and John
  Wright.
\newblock On the global geometry of sphere-constrained sparse blind
  deconvolution.
\newblock In {\em Proceedings of the IEEE Conference on Computer Vision and
  Pattern Recognition}, pages 4894--4902, 2017.

\end{thebibliography}
\end{document}